\documentclass[pdftex,twocolumn,epjc3]{svjour3}
\usepackage{t1enc}
\usepackage[latin2,utf8]{inputenc}

\usepackage{graphicx}
\graphicspath{ {./figs/} }

\RequirePackage{graphicx}
\RequirePackage{mathptmx}      
\RequirePackage{flushend}
\RequirePackage[numbers,sort&compress]{natbib}
\RequirePackage[colorlinks,citecolor=blue,urlcolor=blue,linkcolor=blue]{hyperref}

\RequirePackage{amsfonts}
\RequirePackage{amsmath,amssymb}
\RequirePackage{bm}

\usepackage[utf8]{inputenc}

\usepackage[normalem]{ulem}

\newcommand{\notimplies}{%
  \mathrel{{\ooalign{\hidewidth$\not\phantom{=}$\hidewidth\cr$\implies$}}}}

\journalname{Eur. Phys. J. C}

\begin{document}


\title{Evidence of Odderon-exchange from scaling properties\\ of elastic scattering at TeV energies}

\titlerunning{Evidence of Odderon-exchange}        

\author{
        T. Cs\"{o}rg\H{o}\, \thanksref{e1,addr1,addr2,addr3}
        \and
        T. Nov\'ak\thanksref{e2,addr2}
        \and
        R. Pasechnik\thanksref{e3,addr4}
       \and
        A. Ster\, \thanksref{e4,addr1}
        \and
        I. Szanyi\, \thanksref{e5,addr1,addr6}
}

\thankstext{e1}{e-mail: tcsorgo@cern.ch}
\thankstext{e2}{e-mail: novak.tamas@uni-mate.hu \\ Before February 1, 2021: \\ Szent Istv\'an University, K\'aroly R\'obert Campus}
\thankstext{e3}{e-mail: roman.pasechnik@thep.lu.se}
\thankstext{e4}{e-mail: ster.andras@wigner.hu}
\thankstext{e5}{e-mail: istvan.szanyi@cern.ch}

\institute{Wigner FK, H-1525 Budapest 114, POB 49, Hungary\label{addr1}
          \and
          MATE Institute of Technology, K\'aroly R\'obert Campus, H-3200 Gy\"ongy\"os, M\'atrai \'ut 36, Hungary\label{addr2}
          \and
          CERN, CH-1211 Geneva 23, Switzerland\label{addr3}
	      \and
          Department of Astronomy and Theoretical Physics, Lund University, SE-223 62 Lund, Sweden\label{addr4}
           \and
          E\"otv\"os University, H - 1117 Budapest, P\'azm\'any P. s. 1/A, Hungary\label{addr6}
}


\maketitle

\begin{abstract}
\vspace{0.25cm}
We study the scaling properties of the differential cross section of elastic proton-proton ($pp$) and proton-antiproton ($p\bar p$) collisions at high energies. We introduce a new scaling function, that scales -- within the experimental errors -- all the ISR data on elastic $pp$ scattering from $\sqrt{s} = 23.5$ to $62.5$ GeV to the same universal curve. We explore the scaling properties of the differential cross-sections of the elastic $pp$ and $p\bar p$ collisions in a limited TeV energy range. Rescaling the TOTEM $pp$ data from $\sqrt{s} = 7$ TeV to $2.76$ and $1.96$ TeV, and comparing it to D0 $p\bar p$ data at $1.96$ TeV, our results provide an evidence for a $t$-channel Odderon exchange at TeV energies, with a significance of at least 6.26$\sigma$. We complete this work with a model-dependent evaluation of the domain of validity of the new scaling and its violations. We find that the $H(x)$ scaling is valid, model dependently, within  $200$ GeV $ \leq \sqrt{s} \leq$ $ 8$ TeV, with a $-t$ range gradually narrowing with decreasing colliding energies.
\vspace{0.25cm}
\end{abstract}

\maketitle

\section{Introduction}
\label{s:intro}

One of the most important and critical tests of quantum chromodynamics (QCD) in the infrared regime 
is provided by the ongoing studies of elastic differential hadron-hadron scattering cross section at 
various energies and momentum transfers. The characteristics of the elastic amplitude, its both real and imaginary 
parts, carry a plenty of information about the inner proton structure, the proton profile in the impact parameter space and its energy dependence, as well as about the properties of QCD exchange interaction at low momentum transfers. 

The first and most precise measurement of the total, elastic and differential cross sections of elastic $pp$ collisions, together with the $\rho$-parameter, has recently been performed by the TOTEM Collaboration at the Large Hadron Collider (LHC) at CERN at the highest energy frontier of $\sqrt{s} = 13$ TeV (for the corresponding recent TOTEM publications, see Refs.~\cite{Antchev:2017dia,Antchev:2017yns,Antchev:2018edk,Antchev:2018rec}). A correct theoretical interpretation of the LHC data, together with the lower-energy Tevatron and ISR data, is a subject of intense debates and ongoing research development in the literature, see e.g.~Refs.~\cite{Samokhin:2017kde,Khoze:2018kna}. Among the important recent advances,  data by the TOTEM Collaboration~\cite{Antchev:2018rec}
 for the first time have indicated the presence of an odd-under-crossing (or C-odd) contribution to the elastic scattering amplitude known as the Odderon~\cite{Lukaszuk:1973nt}. In particular, a comparison of the differential cross-section of elastic proton-proton ($pp$) scattering obtained by the TOTEM Collaboration at $\sqrt{s} = 2.76$ TeV with D0 results on elastic proton-antiproton ($p\bar p$) scattering at 1.96 TeV~\cite{Abazov:2012qb} indicates important qualitative differences that can be attributed to the Odderon 
effect~\cite{Csorgo:2018uyp,Antchev:2018rec}. In the  more rigorous language of QCD, an Odderon exchange is usually associated with a quarkless odd-gluon (e.g. three-gluon, to the lowest order) bound state such as a vector glueball, and a vast literature is devoted to theoretical understanding of its implications. 
An increase of the total cross section, $\sigma_{\rm tot}(s)$, associated with 
a decrease of the real-to-imaginary ratio, $\rho(s)$, with energy, first identified 
at $\sqrt{s} = 13$ TeV \cite{Antchev:2017dia,Antchev:2017yns}, also indicated a 
possible Odderon effect. 

The TOTEM measurements have recently triggered intense theoretical studies 
in the literature. In particular, the Phillips-Barger parameterisation of the elastic amplitude 
has been found to describe the recent $pp$ data in Refs.~\cite{Ster:2015esa, Goncalves:2018nsp}. Several other Regge parameterisations have also been found to describe 
the LHC data reasonably well (see e.g.~Refs.~\cite{Khoze:2017swe,Khoze:2018kna,Selyugin:2018uob}), 
while the Pomeron dominance has been explored in a generic Regge theory set-up in Refs.~\cite{Broilo:2018els,Broilo:2018qqs}. In Ref.~\cite{Samokhin:2017kde}, 
a new feature of the second diffractive cone in the differential cross-section of elastic scattering at large $t$ and $s$ has been identified arguing about the existence of two stationary points in $d\sigma/dt$ at the LHC energies and relating those to the two-scale structure of protons at these energies. Remarkably, this rules out the dominance of perturbative exchanges of a few non-inter\-acting gluons pointing towards a core-like proton substructure found also in the framework of the so-called L\'evy imaging technique in Refs.~\cite{Csorgo:2018uyp,Csorgo:2019egs}. For a thorough discussion of general properties of the $s$-dependence of $\rho(s)$ in the light of the TOTEM data and its connections to the growing energy dependence of the elastic-to-total cross–sections ratio, see Ref.~\cite{Troshin:2018ihb}. 
A number of studies based upon a QCD-based analysis of the Odderon signatures considering the non-linear QCD evolution have also been triggered recently (see e.g. Refs. \cite{Gotsman:2018buo,Gotsman:2020mkd,Hagiwara:2020mqb,Contreras:2020lrh}).

Important statements about the maximal nature of the Odderon effect were made in Refs.~\cite{Martynov:2018nyb,Martynov:2018sga,Shabelski:2018jfq,Khoze:2018kna}  but apparently these studies still lack a  rigorous statistical significance analysis.  Although the $s$-dependence of both $\sigma_{\rm tot}(s)$ and $ \rho(s)$ is consistent with an Odderon effect, this indication is not a unique Odderon signal as the same effect can also be attributed to the secondary Reggeon effects~\cite{Gotsman:2018buo}, reinforcing the elusiveness of the Odderon.
As it was argued in Ref.~\cite{Pancheri} any conclusions about the magnitude of the Odderon effects based upon the $\rho(s)$ measurement 
alone have to be made with special care  due to a zero in the real part of the elastic amplitude at very small $t$, as the latter can affect the Coulomb-Nuclear Interference (CNI) region at high energies.

In earlier studies of Refs.~\cite{Csorgo:2018ruk,Csorgo:2019rsr}, the Odderon signatures have been identified and qualitatively described in a model-independent way using the power of the L\'evy imaging technique \cite{Csorgo:2018uyp}. One of such signatures concern the presence of a dip-and-bump structure in the differential cross section of elastic $pp$ collisions and the lack of such a structure in elastic $p\bar p$ collisions. The latter effectively emerges in the $t$-dependence of the elastic slope $B(t)$, that crosses zero for elastic $pp$ collisions and remains non-negative for all values of $t$ in elastic $p\bar p$ collisions. Besides, Ref.~\cite{Csorgo:2018uyp} noted that the position of the node of the nuclear phase $\phi(t)$, as reconstructed with the help of the L\'evy expansion method, is characteristically and qualitatively different for elastic $pp$ from $p\bar p$ collisions, thus, indicating the Odderon exchange.
In addition, the presence of a smaller substructure of the proton has been revealed in the data that is imprinted in the behaviour of the $t$-dependent elastic slope $B(t)$, apparent at large values of $t$. In particular, in Refs.~\cite{Csorgo:2018uyp,Csorgo:2018ruk,Csorgo:2019rsr,Csorgo:2019egs} two substructures of  two distinct sizes have been identified in the low (a few tens of GeV) and high (a few TeV) energy domains, respectively. Besides, a new statistically significant feature in the $b$-dependent shadow (or inelasticity) profile has been found at the maximal available energy $\sqrt{s} = 13$ TeV and represents a long-debated hollowness, or ``black-ring'' effect that emerges instead of the conventionally anticipated ``black-disk'' regime~\cite{Csorgo:2018ruk,Csorgo:2019egs}.

In this paper, in order to further unveil the important characteristics of elastic hadron-hadron scattering we study the scaling properties of the existing data sets available from the ISR and Tevatron colliders as well as those provided by the TOTEM Collaboration in a TeV energy 
range~\cite{Antchev:2013gaa,Antchev:2017dia,Antchev:2017yns,Antchev:2018edk,Antchev:2018rec}.
We investigate a generic scaling behavior of elastic differential proton-(anti)proton scattering cross section, with the goal of transforming out the trivial colliding energy dependent variation of the key observables like that of the total and elastic cross-sections $\sigma_{\rm tot}(s)$ and $\sigma_{\rm el}(s)$, the elastic slope $B(s)$ and the real-to-imaginary ratio $\rho(s)$. We search successfully for a universal scaling function and the associated data-collapsing behaviour that is valid not only in the low-$|t|$ domain, but also in the dip-and-bump region. We discuss the physics implications of such a scaling behaviour and explore its consequences for understanding of the Odderon effect as well as the high-energy behaviour of the proton structure.

The paper is organised as follows. In section~\ref{s:formalism}, we recapitulate the formalism that is utilized for evaluation of the observables of elastic proton-(anti)proton scattering in the TeV energy range. In section~\ref{s:Odderon-search}, we connect this formalism to a more general strategy of the experimental Odderon search, namely, to the search for a crossing-odd component in the differential cross-section of elastic proton-(anti)proton scattering. In section~\ref{s:Scalings}, we study some of the scaling functions of elastic scattering already existing in the literature as well as propose a new scaling function denoted as $H(x)$ that is readily measurable in $pp$ and $p \bar p$ collisions. In particular, in  subsection~\ref{ss:Hxcone} we introduce a new scaling function for the diffractive cone or low values of the square of the four-momentum  ($-t$) region. We generalize this scaling function for larger values of $-t$ in subsection ~\ref{ss:Hx-dip-bump} and present a first test of the $H(x)$ scaling in the ISR energy range of 23.5 -- 62.5 GeV in the same subsection. Subsequently, in section~\ref{s:results-TeV} we extend these studies to the TeV (Tevatron and LHC) energy range, where the possible residual effects of Reggeon exchange are expected to be below the scale of the experimental errors~\cite{Broniowski:2018xbg}. In section~\ref{s:quantification}, we present a method of how to quantify the significance of our findings, giving the formulas that are used to evaluate $\chi^2$, confidence level (CL), and significance in terms of the standard deviation, $\sigma$. In section~\ref{s:extrapolations}, we discuss how to employ the newly found scaling behavior of the differential cross-section to extrapolate the differential cross-sections of elastic $pp$ scattering within the domain of the validity of the new $H(x)$ scaling. Let us note, that this method of comparing differential cross-sections is a possible strategy for the Odderon search. However, as we detail later, the overall normalization uncertainties are large and reduce the statistical significance of these kind of results: practically it is a better strategy to compare scaling functions, evaluated from the differential cross-sections in such a way, that the overall normalization constants (including their large errors) cancel. In section~\ref{s:results}, we present further, more detailed results of our studies with the help of $H(x)$ and compare such a scaling function for $pp$ differential cross-sections at the LHC energies with the $p\bar p$ scaling function at the Tevatron energy. In section~\ref{s:Odderon-significance} we evaluate the significance of the Odderon-effect, and find that it is at least a 6.26$\sigma$-significant effect, taking into account also the improvements detailed in \ref{app:A}.
Subsequently, we present several cross-checks in section~\ref{s:cross-checks} and discuss the main results and its implications in section~\ref{s:discussion}. Finally, we summarize and conclude our work in section~\ref{s:summary}. 

This manuscript is completed with several Appendices that highlight various aspects of this analysis. \ref{app:A} details the robustness and symmetry properties of the $\chi^2$ definition and provides the final Odderon significance of at least $6.26\sigma$ from a model-independent comparison of the $H(x)$ scaling functions of already published  data. In~\ref{app:B} we discuss the model-independent properties of the Pomeron and Odderon exchanges at the TeV energy scale, under the condition that this energy is sufficiently large: as the effects from the exchange of known hadronic resonances decreases as an inverse power of $s$, at large enough energies Pomeron and Odderon exchanges can be identified with the crossing-even and the crossing-odd contributions to the elastic scattering, respectively. We demonstrate here that $S$-matrix unitarity  constrains the possible form of the impact-parameter dependence of the Pomeron and Odderon amplitudes. In \ref{app:C}, we discuss  model-dependent properties of the Pomeron and Odderon exchanges at the TeV energy scale and
derive, how the $H(x)$ scaling emerges within a specific model, defined in Ref.~\cite{Nemes:2015iia}. This model is one of the possible models in the class considered in~\ref{app:B}.
We evaluate the experimentally observable consequences of the $H(x)$ scaling in~\ref{app:D}, where  we  estimate the domain of validity of the $H(x)$ scaling also in a model-dependent manner, based on Ref.~\cite{Nemes:2015iia}.
Finally, in \ref{app:E} we cross-check the stability and robustness of the Odderon signal for the variation of the
$x$-range, the domain or support in $x$ where the signal is determined.  We also identify here a minimal set of only 8 out of 17 D0 datapoints, 
close to the diffractive interference region, that alone carry an at least 5 $\sigma$ Odderon signal, when compared to the TOTEM datapoints in the same region.

\section{Formalism}\label{s:formalism}

For the sake of completeness and clarity, let us start first with recapitulating the connection between the scattering amplitude and the key observables of elastic scattering, following the conventions of Refs.~\cite{Bialas:2006qf,Nemes:2012cp,CsorgO:2013kua,Nemes:2015iia}.

The Mandelstam variables $s$ and $t$ are defined as usual $s = (p_1 + p_2)^2$, $t = (p_1 - p_3)^2$ for an elastic scattering of particles $a$ and $b$ with incoming four-momenta $p_1$ and $p_2$, and outgoing four-momenta $p_3$ and $p_4$, respectively.

The elastic cross-section is given as an integral of the differential cross-section of elastic scattering:
\begin{equation}
     \sigma_{\rm el}(s) = \int_{0}^\infty d|t| \frac{d\sigma(s,t)}{dt} 
     \label{e:sigmael}
\end{equation}

The elastic differential cross section is
\begin{equation}
\frac{d\sigma(s,t)}{dt}   =   \frac{1}{4\pi}|T_{\rm el}(s,\Delta)|^2 \,, \qquad \Delta=\sqrt{|t|}\, .
\label{e:dsigmadt-Tel}
\end{equation}

The $t$-dependent slope parameter $B(s,t)$ is defined as
\begin{equation}
    B(s,t) = \frac{d}{dt} \ln \frac{d\sigma(s,t)}{dt} 
    \label{e:Bst}
\end{equation}
and in the experimentally accessible low-$t$ region this function is frequently assumed or found 
within errors to be a constant. In this case, a $t$-independent slope parameter $B(s)$ is introduced as
\begin{equation}
    B(s) \equiv B_0(s) \, = \, \lim_{t\rightarrow 0} B(s,t) , \label{e:Bs}
\end{equation}
where the $t\rightarrow 0$ limit is taken within the experimentally probed region. Actually, experimentally the optical $t=0$ point can only be approached by extrapolations from the measurements 
in various $-t > 0$ kinematically accessible regions that depend on the optics and various  settings of the particle accelerators and colliding beams.

According to the optical theorem, the total cross section is also found by a similar extrapolation. Its value is given by
\begin{equation}
\sigma_{\rm tot}(s) \equiv 2\,{\rm Im}\, T_{\rm el}(\Delta=0,s) \,,
    \label{e:sigmatot}
\end{equation}
while the inelastic cross-section is defined by
\begin{equation}
\sigma_{\rm in}(s) =  \sigma_{\rm tot}(s)-\sigma_{\rm el}(s) . \label{eq:inelastic_cross_section}
\end{equation}
The ratio of the real to imaginary parts of the elastic amplitude is found as
\begin{equation}
\rho(s,t)\equiv \frac{{\rm Re}\, T_{\rm el}(s,\Delta)}{{\rm Im}\, T_{\rm el}(s,\Delta)} \label{e:rhost}
\end{equation}
and its measured value at $t=0$ reads
\begin{equation}
    \rho(s) \equiv \rho_0(s) \, = \, \lim_{t\rightarrow 0} \rho(s,t) \label{e:rhos} \,.
\end{equation}
Here, the $t\rightarrow 0$ limit is taken typically as an extrapolation in dedicated differential cross section measurements at very low $-t$, where the parameter $\rho_0$ can be measured using various CNI methods. The differential cross section at the optical $(t = 0)$ point is thus represented as
\begin{equation}
\frac{d\sigma(s)}{dt}\Big|_{t\to 0}=\frac{1+\rho_0^2(s)}{16\pi}\, \sigma_{\rm tot}^2(s) \, .
\label{e:optical-point}
\end{equation}

In the impact-parameter $b$-space, we have the following relations:
\begin{eqnarray}\nonumber
	t_{\rm el}(s,b) & = & \int \frac{d^2\Delta}{(2\pi)^2}\, e^{-i{\bm \Delta}{\bm b}}\,T_{\rm el}(s,\Delta) \, = \\
	\null & = & \frac{1}{2\pi} \int J_0(\Delta\,b)\,T_{\rm el}(s, \Delta)\,\Delta\, d\Delta \,, \label{e:tel-b} \\ 
	\Delta & \equiv & |{\bm \Delta}|\,, \quad b\equiv|{\bm b}|\,. \label{e:Delta}
\end{eqnarray}
This Fourier-transformed elastic amplitude $t_{el}(s,b)$ can be represented in the eikonal form
\begin{eqnarray}
	t_{\rm el}(s,b) & = & i\left[ 1 - e^{-\Omega(s,b)} \right] \,,
	\label{e:tel-eikonal}
\end{eqnarray}
where $\Omega(s,b)$ is the so-called opacity function (known also as the eikonal function), which is complex in general. The shadow profile function is then defined as
\begin{eqnarray}
	P(s,b) & = & 1-\left|e^{-\Omega(s,b)}\right|^2 \,.
    \label{e:shadow}
\end{eqnarray}

For clarity, let us note that other conventions are also used in the literature and for example the shadow profile $P(b,s)$ is also referred to as the inelasticity profile function since it corresponds to the probability distribution of inelastic proton-proton collisions in the impact parameter $b$ with $0\le P(b,s) \le 1$. When the real part of the scattering amplitude is neglected, $P(b,s)$ is frequently denoted as $G_{\rm inel}(s,b)$, see for example Refs.~\cite{Petrov:2018rbi,Dremin:2013qua,Dremin:2014spa,Dremin:2018urc,Dremin:2019tgm}.

\section{Looking for Odderon effects in the differential cross-section of elastic scattering}\label{s:Odderon-search}

As noted in Refs.~\cite{Jenkovszky:2011hu,Ster:2015esa}, the only direct way to see the Odderon is by comparing the particle and antiparticle scattering at sufficiently high energies provided that the high-energy $pp$ or $p\bar p$ elastic scattering amplitude is a sum or a difference of even and odd C-parity contributions, respectively,
\begin{eqnarray}
T_{\rm el}^{pp}(s,t) & = & T_{\rm el}^{+}(s,t) - T_{\rm el}^{-}(s,t), \label{e:tel-pp}\\
T_{\rm el}^{p\overline{p}}(s,t) & = & T_{\rm el}^{+}(s,t) + T_{\rm el}^{-}(s,t) \label{e:tel-pbarp} , \\
 T_{\rm el}^{+}(s,t) & = & T_{\rm el}^{P}(s,t) + T_{\rm el}^{f}(s,t),\\
 T_{\rm el}^{-}(s,t) & = & T_{\rm el}^{O}(s,t) + T_{\rm el}^{\omega}(s,t) \,.
\end{eqnarray}
Here, the even-under-crossing part consists of the Pomeron $P$ and the Reggeon $f$ trajectories, while the odd-under-crossing part contains the Odderon $O$ and a contribution from the Reggeon $\omega$. 

At sufficiently high collision energies $\sqrt{s}$, the relative contributions from secondary Regge trajectories are suppressed since they decay as negative powers of $\sqrt{s}$. In Ref.~\cite{Ster:2015esa}, the authors argued that the LHC energy scale is already sufficiently large to suppress the Reggeon contributions, and they presented the $(s,t)$-dependent contributions of an Odderon exchange to the differential and total cross-sections at typical LHC energies. More recently, this observation was confirmed in Ref.~\cite{Broniowski:2018xbg}, suggesting that indeed the relative contribution of the Reggeon trajectories is well below the experimental precision in elastic $pp$ scattering in the TeV energy range. The analysis of Ref.~\cite{Ster:2015esa} relies on a model-dependent, phenomenological picture formulated in the framework of the Phillips-Barger model~\cite{Phillips:1974vt} and is focused primarily on fitting the dip region of elastic $pp$ scattering, but without a detailed analysis of the tail and cone regions. In Ref.~\cite{Broniowski:2018xbg}, a phenomenological Reggeon + Pomeron + Odderon exchange model is employed to study, in particular, the possible hollowness effect in the high-energy elastic $pp$ collisions. A similar study of the Philips-Barger model was performed in Ref.~\cite{Goncalves:2018nsp} using the most recent TOTEM data on elastic $pp$ scattering. Similarly, Ref.~\cite{Lebiedowicz:2018eui} has also argued that the currently highest LHC energy of $\sqrt{s} = $ 13 TeV is sufficiently high to observe the Odderon effect.

In this paper, we follow Refs.~\cite{Ster:2015esa,Broniowski:2018xbg,Lebiedowicz:2018eui} 
and assume that the Reggeon contributions to the elastic scattering amplitudes for $\sqrt{s} \geq$ 1.96 TeV and at higher energies are negligibly small. We search for an odd-under-crossing contribution to the scattering amplitude, in a model independent way, and find that such a non-vanishing contribution is present at a TeV scale that is recognised as an Odderon effect. The vanishing nature of the Reggeon contributions offers a direct way of extracting the Odderon as well as the Pomeron contributions, $T_{\rm el}^{O}(s,t)$ and $T_{\rm el}^{P}(s,t)$, respectively, from the elastic $pp$ and $p\bar p$ scattering data at sufficiently high colliding energies as follows
\begin{eqnarray}
 T_{\rm el}^{P}(s,t) & = & \frac{1}{2} \left(T_{\rm el}^{pp}(s,t) + T_{\rm el}^{p\overline{p}}(s,t)\right) \,\, \mbox{ \rm for}\,\, \sqrt{s}\ge 1 \,\, \mbox{\rm TeV} , \label{e:Tel-P} \\
  T_{\rm el}^{O}(s,t) & = & \frac{1}{2} \left(T_{\rm el}^{p\overline{p}}(s,t) - T_{\rm el}^{pp}(s,t)\right) \,\, \mbox{ \rm for}\,\, \sqrt{s}\ge 1 \,\, \mbox{\rm TeV} \,.\label{e:Tel-O}
\end{eqnarray}

These kind of studies rely on the extrapolation  of the fitted model parameters of $pp$ and $p\bar p$ reactions to an exactly the same energy, given that the elastic $pp$ and $p\bar p$ scattering data have not been measured at the same (or close enough) energies in the TeV region so far. Another problem is a lack of precision data at the low- and high-$|t|$, primarily, in $p\bar p$ collisions. Recently, the TOTEM Collaboration noted in Ref.~\cite{Antchev:2018rec} that ``{\it Under the condition that the effects due to the energy difference between TOTEM and D0 can be neglected, the result}" (namely the differential cross-section
measured by TOTEM at $\sqrt{s} = 2.76 $ TeV) "{\it provides evidence for a colourless 3-gluon
bound state exchange in the t-channel of the proton-proton elastic scattering}". In other words, if the effects due to the energy difference between TOTEM and D0 measurements can be neglected, the direct comparison of the differential cross section of elastic $pp$ scattering at $\sqrt{s} = 2.76$ with that of $p\bar p$ scattering at $\sqrt{s}= 1.96$ TeV provides a {\it conditional} evidence for a colourless three-gluon state exchange in the $t$-channel. 

In this paper, we show that the conditional evidence stated by TOTEM can be turned to an unconditional evidence, 
i.e. a discovery of the Odderon, by closing the energy gap as much as possible at present, 
without a direct measurement, based on a re-analysis of already published
TOTEM and D0 data. 

Our main result, an at least 6.26$\sigma$ Odderon effect, is obtained by taking the data at a face value as given in published sources,
without any attempt  to extrapolate them with a help of a model, or using phenomenological, $s$-dependent parameters and extrapolating them
towards their unmeasured values (in unexplored energy domains). Nevertheless, we have tested what happens if  one employs this kind of model as detailed in a different manuscript, see Ref.~\cite{Csorgo:2020wmw}. These model-dependent results lead to a higher than 7.08$\sigma$ combined significance for the Odderon effect, based on the results of ~\ref{app:C}. The experimentally observable signs of the newly found $H(x)$ scaling are detailed in~\ref{app:D}, where we also determine the model-dependent domain of validity of this new scaling and find that this domain of validity is model-dependently, but sufficiently large
so that the Odderon signal remains well above the discovery threshold of a 5$\sigma$ effect, as detailed in~\ref{app:E}. As the 7.08$\sigma$ combined significance is based only on model-dependent results, evaluated and combined at two energies, $\sqrt{s}=1.96$ and $2.76$ TeV (detailed in both ~\ref{app:C} and ~\ref{app:E}), we find that the model-independent approach, summarized in the body of this manuscript and detailed in~\ref{app:A} and \ref{app:B}, provides a more conservative, 6.26 $\sigma$ estimate for the Odderon significance.

Our main result is based on the validity of  a new kind of scaling relation, called as the $H(x)$-scaling.  
We test this scaling on the experimental data and show their data-collapsing behaviour in a limited energy range. 
We demonstrate that such a data-collapsing behaviour can be used to close the small energy gap between the 
highest-energy elastic $p\bar p$ collisions, $\sqrt{s}= 1.96$ TeV 
and the lowest-energy elastic $pp$ collisions at the LHC where the public data are available, 
$\sqrt{s} = 2.76$ TeV. We investigate the stability of this result on the $x$-range or the domain of validity of
the $H(x)$ scaling in~\ref{app:E}. We find that the result is extremely stable for the removal of data points at the beginning or at the end of the 
acceptance of the D0 experiment. Namely, 9 out of the 17 D0 data points can be removed without decreasing the significance of the Odderon signal
below the 5$\sigma$ discovery threshold. 

We look for the even-under-crossing and odd-under-crossing contributions by 
comparing the scaling functions of $pp$ and $p\bar p$ collisions in the TeV energy range. 
In other words, we look for and find a robust Odderon signature in the difference of the scaling functions of the 
elastic differential cross-section between $pp$ and $p\bar p$ collisions. 
We thus discuss the Odderon features that can be extracted in a model-independent manner 
by directly comparing the corresponding data sets to one another.

Let us start with three general remarks as direct consequences of Eqs.~(\ref{e:Tel-P},\ref{e:Tel-O}):
\begin{itemize}
    \item If the Odderon exchange effect is negligibly small (within errors, equal to zero) or if it does not interfere with that of the Pomeron at a given energy, then the differential cross sections of the elastic $pp$ and $p\bar p$ scattering have to be equal:
    \begin{equation}
        T_{\rm el}^O(s,t) = 0  
        \implies \frac{d\sigma^{pp}}{dt} = \frac{d\sigma^{p\bar p}}{dt} \,\,\, \mbox{ \rm for}\,\, \sqrt{s}\ge 1 \,\, \mbox{\rm TeV}.
    \end{equation}
    \item If the  differential cross sections of elastic $pp$ and $p\bar p$ collisions are equal within the experimental errors, this does not imply that the Odderon contribution has to be equal to zero. Indeed, the equality of cross sections does not require the equality of complex amplitudes:
    \begin{equation}
        \frac{d\sigma^{pp}}{dt} = \frac{d\sigma^{p\bar p}}{dt} \,\,\, \mbox{ \rm for}\,\, \sqrt{s}\ge 1 \,\, \mbox{\rm TeV}
        \notimplies 
                T_{\rm el}^O(s,t) = 0  \, .
    \end{equation}
    \item If the $pp$ differential cross sections differ from that of $p\bar p$ scattering at the same value of $s$ in a TeV energy domain, then the Odderon contribution to the scattering amplitude cannot be equal to zero, i.e.
     \begin{equation}
        \frac{d\sigma^{pp}}{dt} \neq  \frac{d\sigma^{p\bar p}}{dt} \,\,\, \mbox{ \rm for}\,\, \sqrt{s}\ge 1 \,\, \mbox{\rm TeV}
        \implies 
                T_{\rm el}^O(s,t) \neq 0  \, .
    \end{equation}
\end{itemize}
Such a difference is thus a clear-cut signal for the Odderon-exchange, if the differential cross sections were measured at exactly the same energies. However, currently such data are lacking in the TeV energy range.
Our research strategy in this paper is to scale out the known $s$-dependencies of the differential cross section by scaling out its 
dependencies on $\sigma_{\rm tot}(s)$, $\sigma_{\rm el}(s)$, $B(s)$ and $\rho(s)$ functions. 
The residual scaling functions will be compared for the $pp$ and $p\bar p$ elastic scattering to see if any difference remains. 


In what follows, we introduce and discuss the newly found scaling function $H(x)$ in section~\ref{s:Scalings} and subsequently evaluate the significance of these observations as detailed in sections~\ref{s:quantification} and \ref{s:Odderon-significance}.

\section{Possible scaling relations at low values of $|t|$}
\label{s:Scalings}

In this section, let us first investigate the scaling properties of the experimental data based on 
a simple Gaussian model elaborating on the discussion presented in Ref.~\cite{Csorgo:2019fbf}.
The motivation for this investigation is that we would like to work out a scaling law that works at least 
in the simplest, exponential diffractive cone approximation, and scales out the trivial $s$-dependencies 
of $\sigma_{\rm tot}(s)$, $\sigma_{\rm el}(s)$, $\rho(s)$, and $B(s)$. Based on the results of such a frequently used exponential approximation, we gain some intuition and experience 
on how to generalize such scaling laws for realistic non-exponential differential cross sections.

Experimentally, the low-$|t|$ part of the measured distribution is usually approximated with an exponential,
\begin{equation}
    \frac{d\sigma}{dt} = A(s) \, \exp\left[ B(s) t\right] \, , \label{e:dsdt-exp}
\end{equation}
where it is explicitly indicated that both the normalization parameter $A  \equiv A(s) $ and the slope parameter $B \equiv B(s)$ are the functions of the center-of-mass energy squared $s$. If the data deviate from such an exponential shape, that can be described if one allows for a $t$-dependence of the slope parameter $B \equiv B(s,t)$ as defined in Eq.~(\ref{e:Bst}). For simplicity, we would like to scale out the energy dependence of the elastic slope $B(s) \equiv B(s,t=0)$ from the differential cross section of elastic scattering, together with the energy dependence of the elastic and total cross sections, $\sigma_{\rm el}(s)$ and $\sigma_{\rm tot}(s)$, as detailed below. For this purpose, let us follow the lines of a similar derivation in Refs.~\cite{Broniowski:2018xbg,Csorgo:2019fbf}.

It is clear that Eq.~(\ref{e:dsdt-exp}) corresponds to an exponential ``diffractive cone'' approximation, that may be valid in the low-$t$ domain only. This equation corresponds to the so called ``Grey Gaussian'' approximation that suggests a relationship between the nuclear slope parameter $B(s)$, the real-to-imagi\-nary ratio $\rho_0(s)$, the total cross section $\sigma_{\rm tot}(s)$, and the elastic cross section
$\sigma_{\rm el}(s)$ as follows~\cite{Block:2006hy,Fagundes:2011hv,Broniowski:2018xbg}:
\begin{eqnarray}
    A(s) & = & B(s) \, \sigma_{\rm el}(s) \, = \, \frac{1+\rho_0^2(s)}{16 \, \pi}\, \sigma_{\rm tot}^2(s), \label{e:Asigma}\\
    B(s) & = & \frac{1+\rho_0^2(s)}{16 \, \pi }\, \frac{\sigma_{\rm tot}^2(s)}{\sigma_{\rm el}(s)} \,.~\label{e:Bsigma}
\end{eqnarray}
Such relations for $A$ and $B$ parameters in terms of the elastic and total cross sections are particularly useful when studying the shadow profile function as detailed below. The above relationships, in a slightly modified form, have been utilized by TOTEM to measure the total cross section at $\sqrt{s} = $ 2.76, 7, 8 and 13 TeV in Refs.~\cite{Nemes:2017gut,Antchev:2013iaa,Antchev:2013paa,Antchev:2017dia}, using the luminosity independent method. In what follows, we do not suppress the $s$-dependence of the observables, i.e. $\sigma_{\rm tot} \equiv \sigma_{\rm tot}(s)$, $\sigma_{\rm el} \equiv \sigma_{\rm el}(s)$.

\subsection{\label{ss:shadow-profile} Scaling properties of the shadow profiles}

In the exponential approximation given by Eqs.~(\ref{e:dsdt-exp},\ref{e:Asigma},\ref{e:Bsigma}), 
the shadow profile function introduced in Eq.~(\ref{e:shadow}) has a remarkable and very 
interesting scaling behaviour, as anticipated in Ref.~\cite{Broniowski:2018xbg}:
\begin{eqnarray}
 P(b,s)  & = &  1 - \Big[ 1 - r(s) \, \exp\Big( - \frac{b^2}{2 B(s)}\Big)\Big]^2 \, - \nonumber \\
  && \,\,\,\qquad  - \, \rho_0^2(s)  r^2(s) \, 
                \exp\Big( - \frac{b^2}{ B(s)}\Big) , \\ \label{e:Pbs}
                    r(s)   & \equiv & 4\, \frac{ \sigma_{\rm el}(s)}{\sigma_{\rm tot}(s)} .  \label{e:rs}
\end{eqnarray}

Thus, the shadow profile at the center, $P_0(s) \equiv P(b=0,s)$ reads as
\begin{equation}
P_0(s) \, = \,  \frac{1}{1+\rho_0^2(s)} \, - \, \left[1+\rho_0^2(s)\right] \, \Big[ r(s) - \frac{1}{1+\rho_0^2(s)}\Big]^2 \,,
\end{equation}
which cannot become maximally absorptive (or black), i.e. $P_0(s) = 1$ is not reached at those colliding energies, where $\rho_0$ is not negligibly small. The maximal absorption corresponds to $P_0(s) \, = \,  \frac{1}{1+\rho_0^2(s)}$, which is rather independent of the detailed $b$-dependent shape of the inelastic collisions \cite{Broniowski:2018xbg}. It is achieved when $r(s)$ of Eq.~(\ref{e:rs}) approaches the value $r(s) = 1/(1+\rho_0^2(s))$. Thus, at such a threshold, we have the following critical value of the ratio
 \begin{equation}
      \left. 
      \frac{\sigma_{\rm el}(s)}{\sigma_{\rm tot}(s)}
      \right\vert_{\mbox{\rm threshold}} = \frac{1}{4 \left[1 + \rho_0^2(s)\right] } \,.
\label{crit-ratio}
 \end{equation}
 
As $\rho_0 \le 0.15$ for the existing measurements and $\rho_0(s)$ seems to decrease with increasing energies at least in the 8 $\le \sqrt{s} \le 13$ TeV region, the critical value of the elastic-to-total cross section ratio (\ref{crit-ratio}) corresponds to, roughly, $\sigma_{\rm el}/\sigma_{\rm tot} \approx 24.5-25.0 $ \%. Evaluating the second derivative of $P(b,s)$ at $b=0$, one may also observe that it changes sign from a negative to a positive one exactly at the same threshold given by Eq.~(\ref{crit-ratio}). Such a change of sign can be interpreted as an onset of the hollowness effect~\cite{Broniowski:2018xbg}. 
The investigation of such a hollowness at $b=0$ is a hotly debated topic in the literature. For early papers on this fundamental feature of $pp$ scattering at the LHC and asymptotic energies, see Refs.~\cite{Troshin:2007fq,Fagundes:2011hv,Dremin:2013qua,Alkin:2014rfa,Troshin:2014rva,Dremin:2014spa,Anisovich:2014wha}, as well as Refs.~\cite{RuizArriola:2016ihz,Troshin:2016frs,Albacete:2016pmp,Broniowski:2017aaf,Broniowski:2017rhz,Troshin:2017ucy,Dremin:2018orv,Campos:2018tsb,Dremin:2018urc,Broniowski:2018xbg,Petrov:2018rbi,Dremin:2019tgm} for more recent theoretical discussions.

As pointed out in Ref.~\cite{Csorgo:2019fbf}, the threshold (\ref{crit-ratio}), within errors, is reached approximately already at $\sqrt{s} = 2.76 $ TeV. The threshold behavior saturates somewhere between 2.76 and 7 TeV and a transition may happen around the threshold energy of $\sqrt{s_{\rm th}} \approx 2.76 - 4 $ TeV. The elastic-to-total cross section ratio becomes significantly larger than the threshold value at $\sqrt{s} = 13 $ TeV. As a result, the shadow profile function of the proton undergoes a qualitative change in the region of $2.76 < \sqrt{s} < 7 $ TeV energies. At high energies, with $\sigma_{\rm el} \ge \sigma_{\rm tot}/4$, the hollowness effect may become a generic property of the impact parameter distribution of inelastic scatterings. However, the expansion at low impact parameters corresponds to the large-$|t|$ region of elastic scattering, where the diffractive cone approximation of Eqs.~(\ref{e:dsdt-exp},\ref{e:Asigma},\ref{e:Bsigma}) technically breaks down, and more refined studies are necessary (see below). For the most recent, significant and model-independent analysis of the hollowness effect at the LHC and its extraction directly from the TOTEM data, see Ref.~\cite{Csorgo:2019egs}.

\subsection{Scaling functions for testing the black-disc limit}

When discussing the scaling properties of the differential cross section of elastic scattering, let us mention that various scaling laws have been proposed to describe certain features and  data-collapsing behaviour of elastic proton-proton scattering already in the 1970s. One of the early proposals was the so called geometric scaling property of the inelastic overlap function~\cite{DiasDeDeus:1987njz,Buras:1973km}. The concept of geometric scaling was based on a negligibly small ratio of the real-to-imaginary parts of the scattering amplitude at $t=0$, $\rho_0 \le 0.01$ and resulted in an $s$-independent ratio of the elastic-to-total cross-sections, $\sigma_{\rm el}/\sigma_{\rm tot} \approx {\rm const}(s)$, while at the LHC energies, $\rho_0$ is not negligibly small and the elastic-to-total cross section ratio is a strongly rising function of $s$. Here, we just note about the geometric scaling as one of the earliest proposals to have a data-collapsing behavior in elastic scattering, but we look in detail for other kind of scaling laws that are more in harmony and consistency with the recent LHC measurements~\cite{Csorgo:2019fbf}.

Let us first detail the following two dimensionless scaling functions proposed in Ref.~\cite{CsorgO:2013kua} and denoted as $F(y)$ and $G(z)$ in what follows. These scaling functions were introduced in order to cross-check if elastic $pp$ collisions at the LHC energies approach the so-called black-disc limit, expected at ultra-high energies, or not. In a strong sense, the black disc limit corresponds to the shadow profile $P(b) = \theta(R_b - b)$ that results in  $\sigma_{\rm el} /\sigma_{\rm tot} = 1/2$, independently of the black disc radius $R_b$. This limit is clearly not yet approached at LHC energies,  but in a weak sense, a black-disc limit is considered to be reached also if the shadow profile function at $b=0$ reaches unity, i.e. $P(b=0) = 1$,
corresponding to black disc scattering at zero impact parameter. This kind of black disc scattering might have been approached 
at $\sqrt{s} = 7$ TeV LHC energy~\cite{Nemes:2015iia}.

The first scaling function of the differential cross-section is defined as follows:
\begin{eqnarray}
F(y) & = & \frac{|t|}{\sigma_{\rm tot}} \frac{d \sigma}{d t} \, , \\
y & = & |t| \sigma_{\rm tot} \,,
\end{eqnarray}
In the diffractive cone approximation, the $s$-dependence in $F(y)$ does not cancel, but it can be approximately written as
\begin{eqnarray}
F(y) & \simeq & \frac{1 + \rho_0^2(s)}{16 \pi}  \exp\left[ - \frac{1 + \rho_0^2(s)}{16 \pi} \, \frac{\sigma_{tot}(s)}{\sigma_{el}(s)} \, y
        \right]  \\
B(s) t & = & - \frac{B(s) }{   \sigma_{\rm tot}(s)} \, y \, .
\end{eqnarray}
This result clearly indicates that in the diffractive cone, generally the $F(y)$ scaling is violated by energy-dependent factors, while in the black-disc limit of elastic scattering, corresponding to $\frac{\sigma_{tot}(s)}{\sigma_{el}(s)} \rightarrow 2$ and $\rho_0(s) \rightarrow 0$, the $F(y)$ scaling becomes valid as detailed and discussed in Ref.~\cite{CsorgO:2013kua}.
Indeed, the aim to introduce the scaling function $F(y)$ was to clarify that even at the highest LHC energies we do not reach the black-disk limit (in the strong sense). As discussed in the previous section, the deviations from the black-disc limit might be due to the effects of the real part and the hollowness, i.e.~reaching a black-ring limit instead of a black-disc one at the top LHC energies.

Since in the $F(y)$ scaling function the position of the diffractive minimum (dip) remains $s$-dependent, yet another scaling function denoted as $G(z)$ was proposed to transform out such $s$-dependence of the dip. This function was introduced also in Ref.~\cite{CsorgO:2013kua} as follows:
\begin{eqnarray}
G(z) & = & \frac{z |t_{\rm dip}(s)|}{\sigma_{\rm tot}(s)}  \left. \frac{d \sigma}{d t}\right\vert_{t = z |t_{\rm dip}(s)|}, \\
z & = & \frac{t}{|t_{\rm dip}(s)|}.
\end{eqnarray}
In principle, all black-disc scatterings, regardless of the value of the total cross section, should show a data-collapsing behaviour to the same $G(z)$ scaling function. As observed in Ref.~\cite{CsorgO:2013kua}, such an asymptotic form of the $G(z)$ scaling function is somewhat better approached at the LHC energies as compared to the lower ISR energies but still not reproduced it exactly. This is one of the key indications the black-disc limit in the elastic $pp$ scattering is not achieved at the LHC, up to $\sqrt{s} = 13$ TeV. This may have several other important implications. For example, this result indicates that in simulations of relativistic heavy-ion collisions at the LHC energies, more realistic profile functions have to be used to describe the impact parameter dependence of the inelastic $pp$ collisions: a simple gray or black-disc approximation for the inelastic interactions neglects the key features of elastic $pp$ collisions at the TeV energy scales.

One advantage of the scaling variables $y$ and $z$ mentioned above is that they are dimensionless. Numerically, $G(z)$ corresponds to the $F(y)$ function if the scaling variable $y$ is rescaled to $z$. As indicated in Fig.~23 of Ref.~\cite{CsorgO:2013kua}, indeed the main difference between $F(y)$ and $G(z)$ is that the diffractive minimum is rescaled in $G(z)$ to the $z=1$ position, so $G(z)$ has less evolution with $s$ as compared to $F(y)$. However, as it is clear from the above discussion, the function
\begin{eqnarray}
G(z) & \simeq & \frac{\sigma_{\rm el}(s)}{\sigma_{\rm tot}(s)} B(s) z |t_{\rm dip}(s)|\left. \frac{d \sigma}{d t}\right\vert_{t = z |t_{\rm dip}(s)|}, \\
B(s) t & = &  B(s) t_{\rm dip}(s) \, z,
\end{eqnarray}
is well-defined only for $pp$ elastic scattering, where a unique dip structure is observed experimentally. 

Even the dip region is not always measurable in $pp$ reactions if the experimental acceptance is limited to the cone region, which is a sufficient condition for the total cross section measurements. If the acceptance was not large enough in $|t|$ to observe the diffractive minimum, or, in the case when the diffractive minimum did not clearly exist, then neither  the $F(y)$ nor the $G(z)$ scaling functions would be usable. So, the major disadvantage of these scaling functions for extracting the Odderon signatures from the data is that in $p\bar p$ collisions no significant diffractive minimum is found by the D0 collaboration at 1.96 TeV \cite{Abazov:2012qb}. Besides, even if $z$ variable were defined, the above expressions indicate, in agreement with Fig.~23 of Ref.~\cite{CsorgO:2013kua}, that the $G(z)$ scaling function has a non-trivial energy-dependent evolution in the cone ($z \ll 1$) region. Due to these reasons, variables $z$ and  $y$ are not appropriate scaling variables for a scale-invariant analysis of the crossing-symmetry violations at high energies.

Having recapitulated the considerations in Ref.~\cite{Csorgo:2019fbf}, with an emphasis on the $s$-dependence of the parameters, let us now consider, how these $s$-dependencies can be scaled out at low values of $|t|$, where the diffraction cone approximation is valid, by evaluating the scaling properties of the experimental data on the differential elastic $pp$ and $p\bar p$ cross sections. For this purpose, let us
look into the scaling properties of the differential cross sections and their implications related to the Odderon discovery in a new way.

\subsection{A new scaling function for the elastic cone}
\label{ss:Hxcone}

In the elastic cone region, all the $pp$ and $p\bar p$ differential cross sections
can be rescaled to a straight line in a linear-logarithmic plot, when the horizontal axis is scaled by the slope parameter to $-t B(s)$ while the vertical axis is simultaneously rescaled by $B(s) \sigma_{\rm el}(s)$, namely,
\begin{equation}
    \frac{1}{B(s) \sigma_{\rm el}(s)} \frac{d\sigma}{d t} = \exp\left[ t B(s)\right] \qquad \mbox{\rm versus}\quad x = - t B(s) \, .
\end{equation}
This representation, in the diffractive cone, scales out the $s$-dependencies of the total and elastic cross section, $\sigma_{\rm tot}(s)$ and $\sigma_{\rm el}(s)$, and also that of the slope parameter, $B(s)$. As a function of the scaling variable $x = - tB$, it will correspond to the plot of $\exp(-x)$ i.e. a straight line with slope $-1$ on a linear-logarithmic plot. It is well-known that the elastic scattering is only approximately exponential in the diffractive cone, but by scaling out this exponential feature one may more clearly see the scaling violations on this simple scaling plot. We will argue that such a scaling out of the trivial energy-dependent terms can be used as a powerful method in the search for the elusive Odderon effects in the comparison of elastic $pp$ and $p\bar p$ data in the TeV energy range.

In what follows, we investigate the scaling properties of the new scaling function,
\begin{eqnarray}
    H(x) & \equiv  & \frac{1}{B(s) \sigma_{\rm el}(s)} \frac{d\sigma}{d t}, \\
    x    & = &  - t B(s) \, .
\end{eqnarray}
This simple function has four further advantages summarized as follows:
\begin{enumerate}
\item{}
First of all, it satisfies a sum-rule or normalization condition rather trivially, $\int dx H(x) = 1$, as follows from the definition of the elastic cross section. 
\item{} 
Secondly, if almost all of the elastically scattered particles belong to the diffractive cone, the differential cross-section at the optical point is also given by $ \left. \frac{d\sigma}{dt}\right\vert_{t=0} \, = \, A(s)\, = \, B(s) \sigma_{\rm el}(s)$, and in these experimentally realized cases we have another (approximate) normalization condition, namely, $H(0) = 1.$
\item{} 
Third, in the diffractive cone, all the energy dependence is scaled out from this function, i.e., $H(x) = \exp(-x)$ that shows up as a straight line on a linear-logarithmic plot 
with a trivial slope $-1$.
\item{} 
Last, but not least, the slope parameter $B(s)$ is readily measurable not only for $pp$ but also for $p\bar p$ collisions, hence the $pp$ and the $p\bar p$ data can be scaled to the same curve without any experimental difficulties.
\end{enumerate}

\begin{figure*}[hbt]
\begin{center}
\begin{minipage}{0.9\textwidth}
 \centerline{\includegraphics[width=0.9\textwidth]{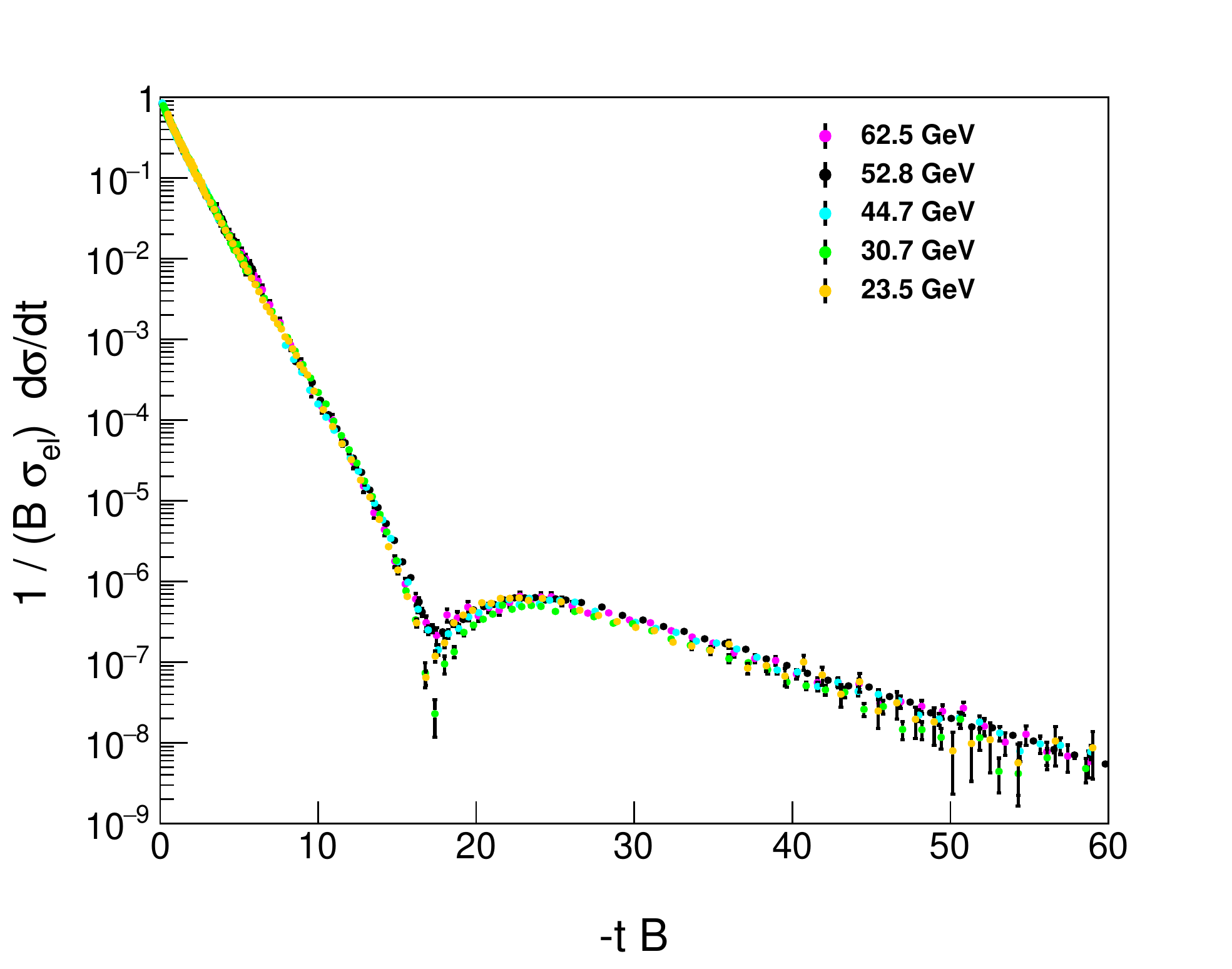}}
\end{minipage}    
\end{center}
\caption{Scaling behaviour of the differential cross section $d\sigma/dt$ of elastic $pp$ collisions in the ISR energy range of $\sqrt{s} = 23.5 $ -- $62.5$ GeV. The measured differential cross section data are taken from Ref.~\cite{Amaldi:1979kd} and references therein. These data are rescaled to $H(x) = \frac{1}{B\sigma_{\rm el}} \frac{d\sigma}{dt}$ as a function of $x = - t B$. This figure indicates a clear, better than expected data-collapsing behaviour.
}
\label{fig:scaling-ISR-x}
\end{figure*}

Let us first test these ideas by using the ISR data in the energy range of $\sqrt{s} = 23.5 - 62.5$ GeV. The results are shown in Fig.~\ref{fig:scaling-ISR-x} which indicates that the ISR data indeed show a data-collapsing behaviour.

At low values of $x$, the scaling function is indeed, approximately, $H(x) \simeq \exp(-x)$, that remains a valid approximation over, at least, five orders of magnitude in the decrease of the differential cross section. However, at the ISR energies, the scaling seems to be valid, within the experimental uncertainties, not only at low values of  $x = - B t$, but extended to the whole four-momentum transfer region, including the dip and bump region $(15 \le x \le 30)$ as well. Even at large-$|t|$ after the bump region, corresponding to $x \ge 30$, the data can approximately be scaled to the same, non-exponential scaling function: $H(x) \neq \exp(-x)$ in the tails of the distribution. Thus, Fig.~\ref{fig:scaling-ISR-x} indeed indicates a non-trivial data-collapsing behaviour to the same, non-trivial scaling function at the ISR energy range of $\sqrt{s} = 23.5 - 62.5$ GeV. 

This observation motivated us to generalize the derivation presented above in this section, to arbitrary positively definite non-exponential scaling functions $H(x)$. Such a generalisation is performed in the next subsection, in order to give a possible explanation of the data-collapsing behaviour in Fig.~\ref{fig:scaling-ISR-x}.

\subsection{Generalized scaling functions for non-exponential 
differential cross-sections}
\label{ss:Hx-dip-bump}

In this section, we search for a novel type of scaling functions of $pp$ elastic data that 
may be valid not only in the diffractive cone, but also in the crucial dip and bump region, as well. 
In Fig.~\ref{fig:scaling-ISR-x}, we have noticed that the data-collapsing behaviour may extend 
well above the small $x = - tB$ region significantly beyond the diffractive maximum, indicating 
a clear deviation of the scaling function $H(x)$ from the exponential shape.

In addition, a recent detailed study of the low-$|t|$ behaviour of the differential elastic $pp$ cross section at $\sqrt{s} = 8$ TeV observed a more than 7$\sigma$-significant deviation from the exponential shape~\cite{Antchev:2015zza,Csorgo:2016qyr}, which also corresponds to a non-exponentiality in the scaling function $H(x)$ even in the low-$|t|$, or small $x$, range.

In this section, we thus further generalize the derivation of the $H(x) = \exp(-x)$ scaling function, in order to allow for arbitrary positively definite functions with $H(x=0) = 1$ normalisation, 
and to develop a physical interpretation of the experimental observations.

Let us start the derivation from the relation of the elastic scattering amplitude 
in the impact parameter space $t_{\rm el}(s,b)$ and the complex opacity function $\Omega(s,b)$ 
based on Eq.~(\ref{e:tel-eikonal}), using the same notation as in Ref.~\cite{Nemes:2015iia}:
\begin{equation}
    t_{\rm el}(s,b) = i \left[1 - \exp(-i \, \mbox{\cal Im}\,
    \Omega(s,b))\sqrt{1 - \tilde\sigma_{\rm in}(s,b)} \right] \,.
    \label{e:telsb-Omega}
\end{equation}
The shadow profile function $P(s,b)$ is equal to the inelastic scattering 
profile $\tilde\sigma_{in}(s,b)$ as follows from Eq.~(\ref{e:shadow}), 
$P(s,b) = \tilde\sigma_{\rm in}(s,b)$.
The imaginary part of the opacity function $\Omega$ is generally not known or less constrained by the data, but it is experimentally known that $\rho_0(s)$ is relatively small at high energies: at all the measured LHC energies and below, $\rho_0 \le 0.15$, hence, $\rho^2 \le 2.3 $ \%.

Here, we thus follow the choice of Ref.~\cite{Nemes:2015iia}, that has demonstrated that the ansatz
\begin{equation}
    \mbox{\cal Im}  \, \Omega(s,b) = - \frac{\rho_0(s) }{2} \tilde\sigma(s,b)
\end{equation}
gives a satisfactory description of the experimental data in the $-t \le 2.5$ GeV$^2$ region, with a small coefficient of proportionality that was denoted in Ref.~\cite{Nemes:2015iia} by $\alpha \propto \rho_0$ parameter. This ansatz assumes that the inelastic collisions at low four-momentum transfers correspond to the cases when the parts of proton suffer elastic scattering but these parts are scattered to different directions, not parallel to one another. This physical interpretation is actually due to $\rho_0 \ll 1$ and $\mbox{\cal Im} \, \Omega(s,b) \ll 1$. We will use this approximation below to demonstrate that the $H(x)$ scaling function can have more complex shapes, that differ from $ H(x) = \exp(-x)$.

Based on the results of the previous section obtained in the diffractive cone in the $\rho_0 \ll 1$ and $\tilde\sigma(s,b) \ll 1$ limit, we have the following scaling property of the opacity function:
\begin{eqnarray}
    \mbox{\cal Re}  \, \exp\left[-\Omega(s,b)\right] & = & 1 -  r(s) E( \tilde{x}), \\
    \mbox{\cal Im}  \, \exp\left[-\Omega(s,b)\right] & = & \rho_0(s) \, r(s) E( \tilde{x}), \\
    \tilde{x }  & = & b / R(s), \label{e:Hx} \\
    R(s) & = & \sqrt{B(s)} \,, \label{e:RB}
\end{eqnarray}
where $r(s)$ is four times the ratio of the elastic to the total cross section, as given in Eq.~(\ref{e:rs}), and $E(\tilde{x})$ describes the distribution of the inelastic collisions as a function of the dimensionless impact parameter $b$ normalised to $\sqrt{B(s)}$, the characteristic length-scale
of the $pp$ collisions at a given value of the center-of-mass energy $\sqrt{s}$.

This ansatz allows for a general shape of the impact parameter $b$-dependent scattering amplitude, that leads to a $H(x)$ scaling.
Under the assumption that the $b$-dependence may occur only through the two-dimensional scaling variable $\tilde{x}$, as described by the scaling function $E(\tilde{x})$,
\begin{equation}
    t_{\rm el}(s,{b})  =  \left( i + \rho_0(s)\right) \, r(s) E(\tilde{x}) \, , \label{e:tel-scaling}
\end{equation}
a general form of the $H(x)$ scaling can be obtained.
Here we assume that $E(\tilde{x})$ is a real function that depends on the modulus of the dimensionless impact parameter $\tilde{x} = b/R(s)$. For normalization, we choose that the Fourier-trans\-form $\tilde E({0}) = 1$, which also corresponds to the condition 
\begin{equation}
         \int \, d^2\tilde{x} \, E(\tilde{x}) = 1 \,, \label{e:E-norm-new}
\end{equation} 
keeping in mind that we have two-dimensional Fourier-trans\-form 
which at zero is equal to the integral over the two different 
directions in the impact-para\-meter space.

Let us investigate first the consequences of the scaling ansatz of Eq.~(\ref{e:tel-scaling}) for the shadow profile function $P(s,b)$. The algebra is really very similar to that of the exponential cone approximation that was implemented above. 
We obtain the following result:
\begin{eqnarray}
    P(s,b) & = & \frac{1}{1+\rho_0^2(s)} - \nonumber \\
    & - & (1 + \rho_0^2(s)) \left[ r(s)E\left(\frac{b}{R(s)}\right) - \frac{1}{1+\rho_0^2(s)}\right]^2 \, .
    \label{e:Psb-any-H}
\end{eqnarray}
Evaluating the above relation at $b=0$ and using the normalization condition $E({0}) = 1$, we obtain again 
that the shadow profile at zero impact parameter value has a maximum that is slightly less than unity: 
$P(s,0) \le 1/(1+\rho_0^2)$. It is interesting to note that the maximum in the profile function 
is reached at the same threshold (\ref{crit-ratio}) as in the case of the exponential cone approximation, 
corresponding to 
\begin{eqnarray}
\left. r(s)\right\vert_{\rm threshold} & = & \frac{1}{1+\rho_0^2(s)} \, , \label{e:rthreshold}\\
\left. \frac{\sigma_{\rm el}}{\sigma_{\rm tot}}\right\vert_{\rm threshold} & = & \frac{1}{4(1 +\rho_0^2(s))} \,. \label{e:sigmaelpertot-threshold}
\end{eqnarray}
Thus a threshold-crossing behaviour seems to happen if the elastic-to-total cross-section ratio
exceeds $0.25$. Remarkably, in the domain of validity of our derivation, this threshold crossing point is independent 
of the detailed shape of the $H(x)$ scaling function for a broad class of models. However, it is also clear from Eq.~(\ref{e:Psb-any-H}) that the shape of $E(\tilde x)$ function plays an important role in determining the
hollowness effect, so a detailed precision shape analysis is necessary to obtain the significance of this effect.

Starting from the definition, Eq.~(\ref{e:dsigmadt-Tel}), the scattering amplitude in the $b$-space (\ref{e:tel-scaling}) yields the following form of the differential cross section in the momentum space:
\begin{equation} \label{dsigdtscaling}
    \frac{d\sigma}{dt} = \frac{1 +\rho_0^2(s)}{4 \pi} r^2(s) R^4(s) |\tilde E(R(s) \Delta)|^2 \,.
\end{equation}
Utilizing Eq.~(\ref{e:RB}), we find that this form of the differential cross section is dependent 
on the four-momentum transfer squared, $t$, indeed only through the variable 
$x \equiv - B(s) t = R^2(s) \Delta^2$, so it is a promising candidate to be a scaling variable.

Now, if we consider the function (\ref{dsigdtscaling}) at the optical point, $t = 0$, we find 
\begin{equation}
    A(s) = \left. \frac{d\sigma}{dt}\right\vert_{t=0} \, = \, \frac{1 +\rho_0^2(s)}{4 \pi} r^2(s) R^4(s) |\tilde E(0)|^2 \,. \label{e:As-optical}
\end{equation}
If the impact parameter dependent elastic amplitude has an 
$s$-dependent internal scale and $s$-dependent strength,  we thus obtain the following generalized scaling relation
for arbitrary elastic scattering amplitudes that 
satisfy Eq.~(\ref{e:tel-scaling}): 
\begin{equation}
    \frac{1}{A(s)} \frac{d\sigma}{dt}  \equiv \, H(x) \, = \, 
    \frac{|\tilde E(\sqrt{x})|^2}{|\tilde E(x=0)|^2} 
    \, . \label{e:Hx-general}
\end{equation}
This scaling is derived for $\rho_0 \ll 1$ and $\tilde \sigma(s,b) \ll 1$, and it indicates that the $H(x)$ with a non-exponential
scaling function is a very interesting theoretical possibility. Further generalizations of this derivation are possible and interesting
but go clearly well beyond the scope of this manuscript, that aims to look for Odderon effects using the experimentally available
information on this $H(x)$ scaling and its possible violations.

In addition to providing an insight to the meaning of the non-exponential behaviour in the interference (dip and bump) region,
the above derivation also clarifies meaning of the normalization of $H(x)$. In particular, 
the normalization of $H(x)$ scaling 
function on the left hand side of Eq.~(\ref{e:Hx-general}) should be made by the value of the differential
cross section at the optical ($t = 0$) point as given by Eq.~(\ref{e:As-optical}). This value for differential 
cross sections with nearly exponential diffractive cone is indeed approximately equal to 
$A(s) = B(s) \sigma_{\rm el}(s)$. In this case, the normalization condition $H(0) = 1$ is maintained, 
while the integral of $H(x)$ becomes unity only for differential cross sections dominated by the exponential 
cone (i.e. when the integral contribution from the non-exponential tails is several orders of magnitude smaller as compared to the integral of the cone region).

For the total cross section, we find from Eq.~(\ref{e:sigmatot})
\begin{equation}
    \sigma_{\rm tot}(s) = 2 r(s) R^2(s) \tilde E(0) = \sqrt{ \frac{16 \, \pi \, A(s)}{1 + \rho^2_0(s)}} \,.
\end{equation}
Note that here we have indicated the normalization just for clarity, but one should keep in mind that in our normalization, $\tilde E(0) = 1$, and correspondingly, $H(x=0)=1$ by definition.

As clarified by Eq.~(\ref{e:Hx-general}), the scaling function $H(x)$ coincides with the modulus squared of the normalized Fourier-trans\-form of the scaling function $E(\tilde{x})$, if the elastic amplitude depends on the impact parameter $b$ only through its scale invariant combination $x = \frac{b}{R(s)}$ and if $\rho(s,t) \equiv \rho_0(s)$. In this case, the $H(x)$ scaling is directly connected to the impact parameter dependence of the elastic amplitude and transforms out the trivial $s$-dependencies coming from $\sigma_{\rm tot}(s)$, $\sigma_{\rm el}(s)$, $B(s)$, and $\rho_0(s)$ functions. This approximation has enabled us to establish possible physical reasons of this new scaling, and to derive non-exponential shapes for the $H(x)$ scaling function and to connect violations of the $H(x)$ scaling to the hollowness effect in the shadow profile function of the proton at ultra-high energies.  
At the time of closing this manuscript, the generalization of the above derivation to a $t$-dependent $\rho(s,t)$ function is still incomplete, and will be the subject of a separate study.
Nevertheless, in our numerical analysis of the $H(x)$ scaling, detailed in the subsequent sections, 
in the comparisons of the scaled differential cross-sections and the deduced Odderon significance we have not imposed any $\rho(s,t) \equiv \rho(s)$ condition. Our analysis is generic and has been done using the published experimental data sets only, without imposing any theoretical assumptions such as a $t$-independent $\rho(s,t)$ etc.

The above derivation also indicates that it is a promising possibility to evaluate the $H(x)$ scaling function directly from the experimental data. It has a clear normalization condition, $H(0) = 1$. Furthermore, in the diffractive cone, for nearly exponential cone distributions, $H(x) \approx \exp(-x)$. We have shown  in this section, that even if one neglects the possible $t$ dependence of $\rho(s,t)$, arbitrary positively definite $H(x)$ scaling functions can be introduced if the elastic amplitude is a product of $s$-dependent functions, and its impact parameter dependence originates only through an $s$-dependent scaling variable which can be conveniently 
defined as $\tilde{x}^2 = \frac{b^2}{B(s)}$. Thus, the violations of the $H(x)$ scaling may happen if not only the slope parameter $B(s)$, the real-to-imaginary ratio $\rho_0(s)$ and the integrated elastic and total cross sections $\sigma_{\rm el}(s)$ and  $\sigma_{\rm tot}(s)$ depend on $s$, but also the $b$-dependence of the elastic scattering amplitude starts to change noticeably. Na\-mely, the $H(x)$ scaling breaks if the scaling relation $t_{\rm el}(b,s) = C(s) E(b/R(s))$ gets violated in the above mentioned case.

Let us also note that the leading-order exponential shape of $H(x) \approx \exp(-x)$ can be derived as a consequence of the analyticity of $T_{\rm el}(s,\Delta)$ at $\Delta = 0$ corresponding to the $t =0$ optical point, as follows. By leading order we mean the result of a first-order Taylor series expansion at $x = 0$, so that $H(x) \approx \exp(-x) \approx 1 - x$, although beyond this approximation the functional behaviour of the $H(x)$ function cannot be determined from analyticity. If $T_{\mathrm el}(s,\Delta)$ is an analytic function  at $\Delta = 0$, then its leading-order behaviour is $T_{\mathrm el}(s,0) + c(s) \Delta$, where $c(s)$ is a complex coefficient that is in general dependent on $s$. Hence, in this approximation the differential cross-section behaves as $d\sigma/dt \simeq A(s) \exp\left(B(s) t\right) \approx A(s) (1 + B(s) t + \dots)$ corresponding to the scaling function $H(x) \approx  \exp(-x)$ in the diffractive cone. Similar considerations, related to (non)-analyticity of modulus squared amplitudes and L\'evy stable source distributions
were introduced to Bose-Einstein correlations in high energy physics in Ref.~\cite{Csorgo:2003uv}.

On the other hand, our recent analysis of the differential elastic cross sections in the LHC energy
range~\cite{Csorgo:2018uyp,Csorgo:2018ruk} suggests that the approximation $H(x) \approx \exp(-x)$ breaks down since the TOTEM experiment observed a significant non-exponential behaviour already in the diffractive cone. In this case, at low values of $|t|$, nearly L\'evy stable source distributions can be introduced,
that lead to an approximate $H(x)\propto \exp(-x^{\alpha})$ behaviour, where $\alpha = \alpha_{\rm Levy}/2 \le 1.$ 
In this case, the leading order behaviour is non-analytic, $H(x) \approx 1 -x^{\alpha}$. We have shown in Refs.~\cite{Csorgo:2018uyp,Csorgo:2018ruk}, at low $|t|$, such a stretched exponential form with $\alpha \simeq 0.9$ describes the elastic scattering data from ISR to LHC energies
reasonably well in a very broad energy range from $23.5$ GeV to $13$ TeV.

The main limitation of the above derivation is that although it leads to a $H(x)$ scaling, the real-to-imaginary ratio $\rho(s,t) \rightarrow \rho_0(s)$ is independent of $t$ in this approximation. 
So let us consider  a generalization, where the real to imaginary ratio is not only $s$ but also $t$ dependent. We will discuss, model independently,
such a scenario in terms of the impact parameter dependent elastic scattering amplitude in ~\ref{app:B}.
Such a $t$ dependence of $\rho(s,t) $ can actually be realized in a number of physical models. In greater details, we consider one particular model,
that has a $H(x,s)$ type of scaling limit and the $s$-dependent scaling violations are related to the $s$-dependence of the opacity parameter
in this model. We discuss the emergence of the $H(x)$ scaling within a physical model, the so-called Real Extended Bialas-Bzdak model of Refs.~\cite{Bialas:2006kw,Bialas:2006qf,Bialas:2007eg,Bzdak:2007qq,Csorgo:2013bwa,CsorgO:2013kua,Nemes:2015iia} in~\ref{app:C}. We evaluate the domain of validity of this ReBB model in $(s,x = -tB)$ in ~\ref{app:D}, in order to determine if this domain is including (or not) a kinematic region, where the $H(x)$ scaling indicates the Odderon signal.

%
\section{Results in the TeV energy range} 
\label{s:results-TeV}
%

\begin{figure*}[!hbt]
\begin{minipage}{0.5\textwidth}
  \centerline{\includegraphics[width=0.98\textwidth]{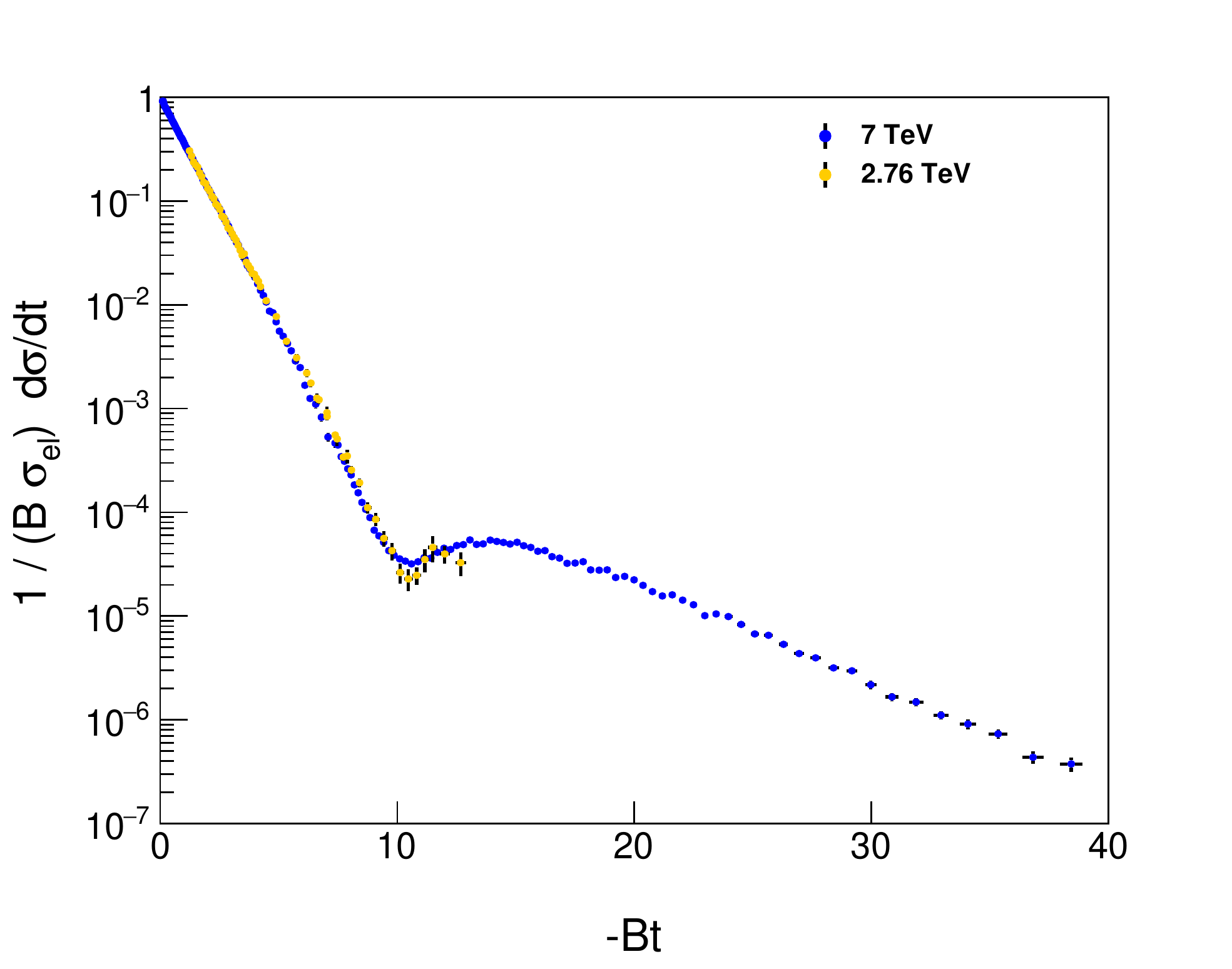}}
\end{minipage}
\begin{minipage}{0.5\textwidth}
  \centerline{\includegraphics[width=0.98\textwidth]{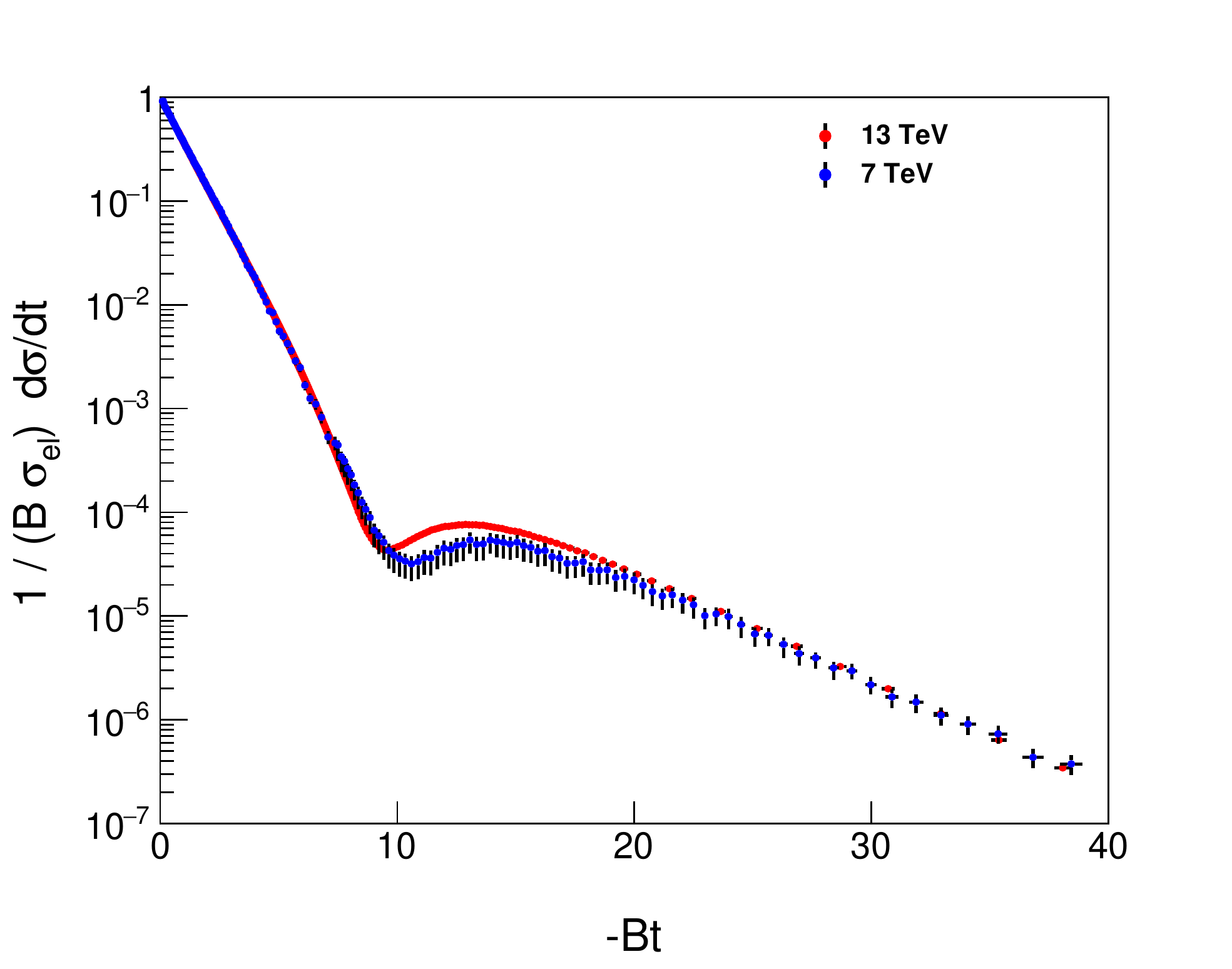}}
\end{minipage}
\caption{Scaling behaviour of the differential cross section $d\sigma/dt$ of elastic $pp$ collisions at LHC energies. Elastic scattering data are measured by the TOTEM Collaboration at $\sqrt{s} = 13 $  TeV~\cite{Antchev:2017dia}, at $\sqrt{s}= 7$ TeV~\cite{Antchev:2013gaa}, and at $\sqrt{s} = 2.76 $ TeV~\cite{Antchev:2018rec}. Left panel shows the $2.76 $ and $7$ TeV data points with statistical errors only, while the right panel shows the $7.0$ and $13.0$ TeV data with statistical and $t$-dependent systematic errors added in quadrature.
The left panel indicates, that the $H(x)$ scaling is within statistical errors valid between $\sqrt{s} = 2.76$ TeV and $7.0$ TeV, so the $H(x)$ scaling works from $7$ TeV downwards. The right panel indicates that the $H(x)$ scaling is violated, 
when the colliding energy is increased from  $\sqrt{s} = 7.0$ to $13$ TeV: 
the right panel indicates scaling violations that go well beyond the combined statistical and systematic errors.
}
\label{fig:scaling-LHC}
\end{figure*}

We established that the $H(x)$ scaling holds within experimental errors at the ISR center-of-mass energies varying from $23.5$ to $62.5$ GeV, i.e. less than by a factor of three. Let us also investigate the same scaling function at the LHC energies, where the TOTEM measurements span, on a logarithmic scale, a similar energy range, from $2.76$ TeV to $13$ TeV, i.e.~slightly more than by a factor of four.
The TOTEM data at 13, 7 and $2.76$ TeV are collected from Refs.~\cite{Antchev:2017dia},
\cite{Antchev:2013gaa}, and Ref.~\cite{Antchev:2018rec}, respectively, and plotted in Fig.~\ref{fig:scaling-LHC}. Note that the possible scaling violating terms are small in the 
$\sqrt{s} = 2.76 - 7$ TeV region: they are within the statistical errors, when increasing $\sqrt{s}$ 
from 2.76 to 7 TeV, i.e. by about a factor of 2.5. Let us also stress that we do not claim the validity of the $H(x)$ scaling up to the 
top LHC energy of $\sqrt{s} = 13$ TeV, as scaling violating terms start to be significant at that energy, in particular, close to the
diffractive dip region. 

Let us look into the scaling behaviour in the energy range  of $\sqrt{s} = 2.76 - 7$ TeV   in more detail.

The left panel of Fig.~\ref{fig:scaling-LHC} indicates that the $H(x)$ scaling valid within statistical errors in the $\sqrt{s} = 2.76 - 7$ TeV energy range. The confidence level of this comparison corresponds to a CL = 99 \% (statistical errors only). 
The right panel of the same Fig.~\ref{fig:scaling-LHC} indicates
that this scaling is violated, beyond systematic errors, if the $\sqrt{s} = 13$ TeV data are also included into this comparison:
the  violation of the $H(x)$ scaling by the 13 TeV data is
focused to the region of the diffractive dip. However, in the $x < 10$ region, the $H(x) $
scaling is approximately valid at each of these LHC energies of $\sqrt{s} = 2.76$, $7$ and $13$ TeV.
Instead of being approximately valid in the whole measurable $x$ region, at the LHC this scaling remains valid at all these three LHC energies 
only through about 3-4 orders of magnitude drop in the differential cross-section at lower values of $x$. 
The so called ``swing'' effect becomes clear at $\sqrt{s} = 13$ TeV: 
the scaling function starts to decrease faster than exponential before the diffractive mimimum, 
and also the diffractive minimum moves to lower values in $x$ as compared to its position at lower LHC energies. 
This swing effect, apparent in Fig.~\ref{fig:scaling-LHC}, can be interpreted in terms of changes 
in the shadow profile of protons at the LHC energies as the energy range increases from $2.76$ through $7$ to $13$ TeV. 
Indeed, such small $s$-dependent scaling violations in the $H(x)$ scaling function 
show the same qualitative picture as what has been observed by the direct reconstruction 
of the $P(s,b)$ shadow profiles in the TeV energy range in several earlier papers, 
see for example Refs.~\cite{Kohara:2017ats,Dremin:2018urc,Dremin:2019tgm} or our Refs.~\cite{Nemes:2015iia,Csorgo:2018uyp,Csorgo:2018ruk}. 

Inspecting the left panel of Fig.~\ref{fig:scaling-LHC}, we find, that the $H(x)$ scaling functions agree within statistical errors,
if the colliding energy is increased from $\sqrt{s} = 2.76$ TeV to 7 TeV. 
The right panel of the same figure shows that these data change significantly if the colliding energy increases further to $\sqrt{s } = 13$ TeV.
This implies that the possible scaling violating terms are small  
as they are within the statistical errors, when increasing $\sqrt{s}$ from 2.76 to 7 TeV, by about a factor of 2.5. 
We have checked that TOTEM preliminary data at $\sqrt{s}$ $=$ 8 TeV also satisfy this $H(x)$ scaling~\cite{Kaspar:2018ISMD,Csorgo:2020rlb}.


However, this $H(x)$ scaling is violated by $s$-dependent terms when increasing $\sqrt{s}$ from 8 to 13 TeV, and such a scaling violation is significantly larger than the quadratically (maximally) added statistical and $t$-dependent systematic errors, as indicated on the right panel of
Fig.~\ref{fig:scaling-LHC}.

This behaviour may happen due to approaching a new domain, where the shadow profile function of $pp$ scattering changes from a nearly Gaussian form to a saturated shape, that in turn may develop hollowness at 13 TeV and higher energies. The experimental indications of such a threshold-crossing behaviour were summarized recently in Ref.~\cite{Csorgo:2019fbf}, and are also described above: a new domain may be indicated by a sudden change of $B(s)$ in between 2.76 and 7 TeV and, similarly, the crossing of the critical $\sigma_{\rm el}(s)/\sigma_{\rm tot}(s) = 1/4$ line in multi-TeV range of energies, somewhere between 2.76 and 7 TeV.  From the theoretical side, we have previously noted such as drastic change in the size of the proton substructure between the ISR and LHC energy domains from a dressed quark-like to a dressed di-quark type of a substructure~\cite{Csorgo:2018uyp,Csorgo:2018ruk} which may be, in principle, connected to such a dramatic change in the scaling behaviour of the elastic cross section. However, in this work we focus on the scaling properties of the experimental data, and do not intend to draw  model-dependent conclusions. Nevertheless, we use the model-dependent results as well in order to cross-check our model-independent conclusions. Some details of the model-independent calculations are summarized in~\ref{app:A} and~\ref{app:B}, while our model-dependent estimates are described in~\ref{app:C},~\ref{app:D} and~\ref{app:E}.

\begin{figure*}[!hbt]
\begin{minipage}{0.48\textwidth}
  \centerline{\includegraphics[width=0.98\textwidth]{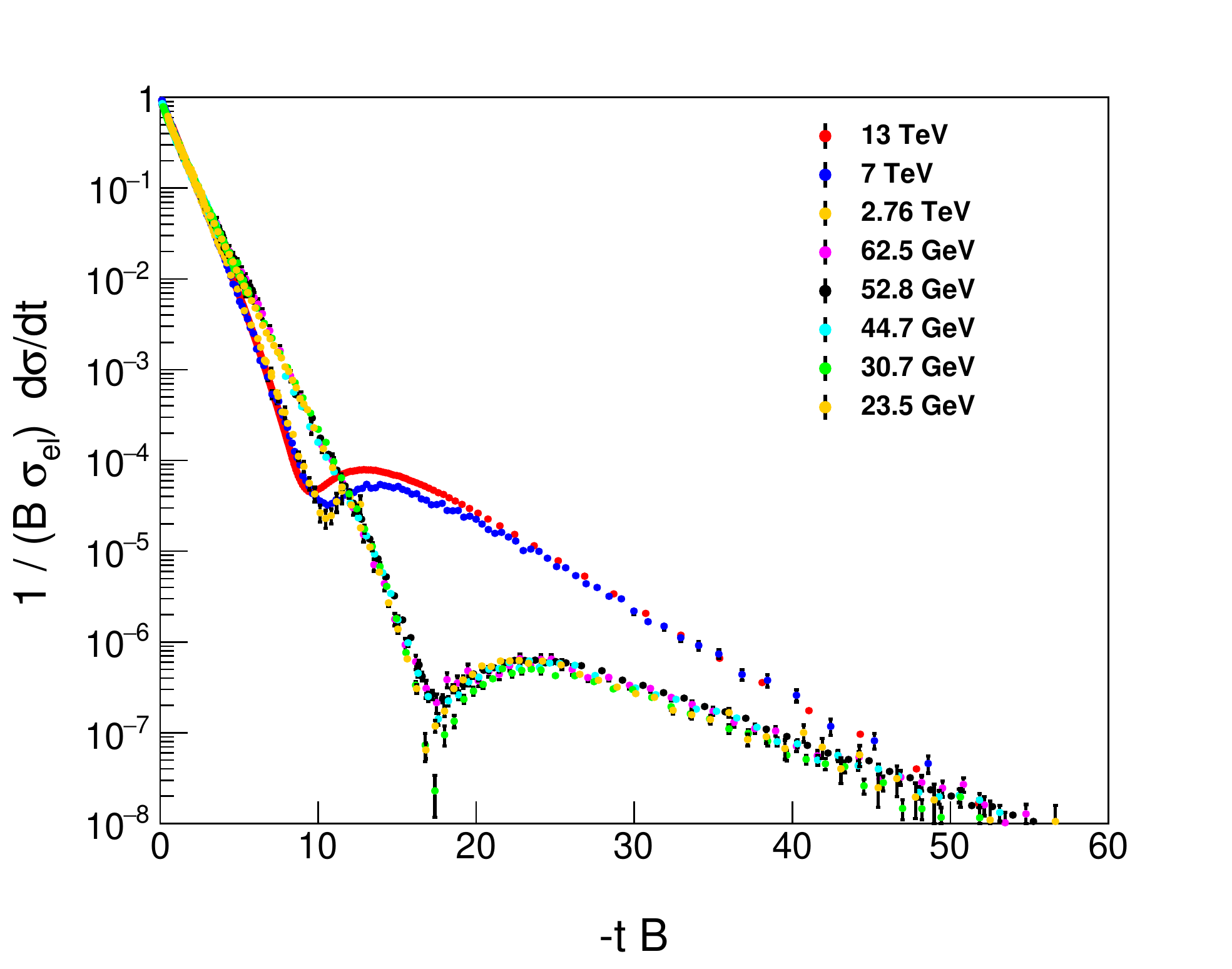}}
\end{minipage}
\begin{minipage}{0.48\textwidth}
  \centerline{\includegraphics[width=0.98\textwidth]{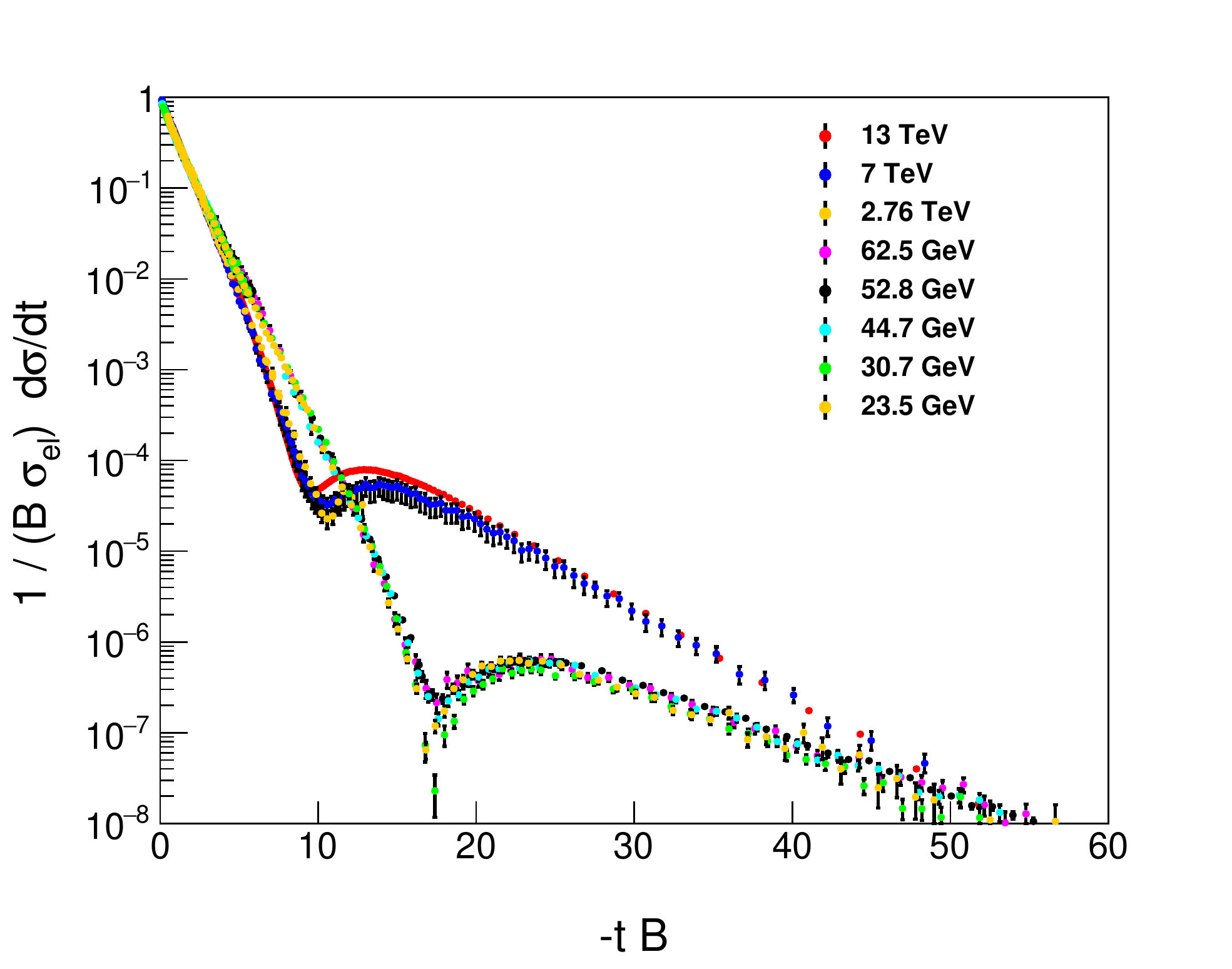}}
\end{minipage}
\caption{Scaling behaviour of the differential cross section $d\sigma/dt$ of elastic $pp$ collisions from ISR to LHC energies. Data points are the same as shown in Figs~\ref{fig:scaling-ISR-x} and \ref{fig:scaling-LHC}. 
{\it (Left panel):} Data points are shown with statistical errors only. 
{\it (Right panel):} Same data set, but now showing both statistical and $t$-dependent systematic errors added in quadrature.  
}
\label{fig:scaling-LHC-vs-ISR}
\end{figure*}

In Fig.~\ref{fig:scaling-LHC-vs-ISR} we directly compare the $H(x)$ scaling functions of the differential cross sections, using the same ISR and LHC data, as in Figs.~\ref{fig:scaling-ISR-x} and \ref{fig:scaling-LHC}, respectively. This range of data now spans nearly a factor of about 500, about a three orders of magnitude increase in the range of available colliding energies, from 23.5 GeV to 13 TeV. As can be seen in the corresponding Fig.~\ref{fig:scaling-LHC-vs-ISR}, the scaling works approximately in the diffractive cone, however, the $H(x)$ scaling function cannot be considered as an approximately constant if such a huge change in the colliding energies is considered.

Comparing Figs.~\ref{fig:scaling-ISR-x}, \ref{fig:scaling-LHC} and \ref{fig:scaling-LHC-vs-ISR}, 
we find that the $s$-dependence of the $H(x)$ scaling functions is rather weak if $s$ changes within 
a factor of two, however, there are very significant changes if the range of energies is
changing by a factor of a few hundred, from the ISR energy range of $\sqrt{s} = 23.5 - 62.5$ GeV to the LHC
energy range of 2.76 -- 7.0 -- 13.0 TeV.

\begin{figure*}[!hbt]
\begin{center}
\begin{minipage}{0.98\textwidth}
\includegraphics[width=0.48\textwidth]{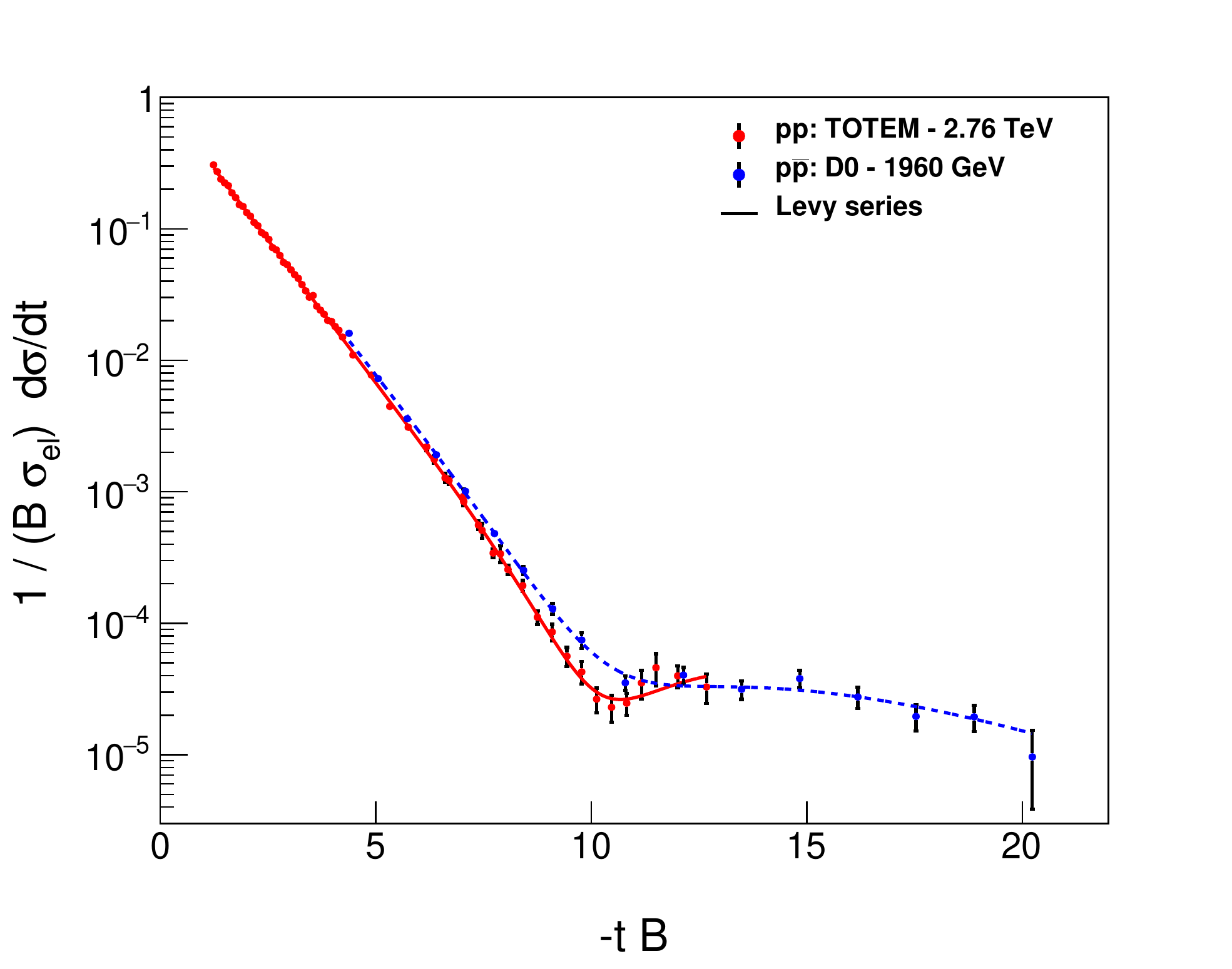}
\includegraphics[width=0.48\textwidth]{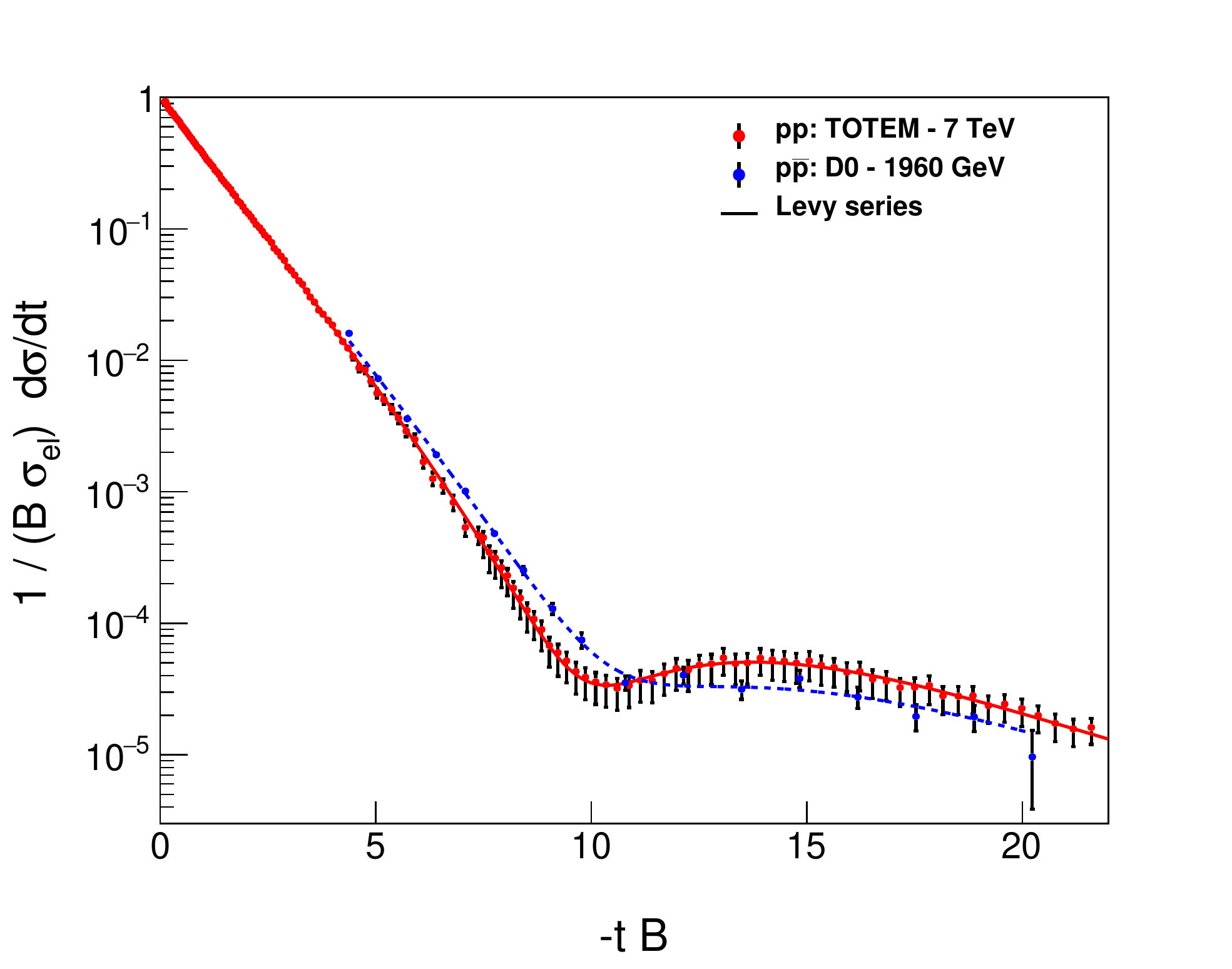}
\end{minipage}
\end{center}
\caption{
{\it Left panel:} Scaling function
$H(x) =  \frac{1}{B \sigma_{\rm el} }\frac{d\sigma}{dt}$ of the differential cross section
of elastic $pp$ collisions at $\sqrt{s} = 2.76$ TeV LHC (red), as compared to that 
of the elastic $p\overline{p}$ collisions at the Tevatron energy of $\sqrt{s} = 1.96$ TeV (blue),
shown as a function of $x = -tB$.
{\it Right panel:}
Same as the left panel, but now using elastic $pp$ data at
$\sqrt{s} = 7$ TeV (red), as compared to  elastic $p\overline{p}$ collisions at $\sqrt{s} = 1.96$ TeV (blue). 
On both panels, statistical errors and $t$-dependent systematic errors are added in quadrature. 
Lines are shown to guide the eye, corresponding to fits with the model-independent L\'evy series from Refs.~\cite{Csorgo:2018uyp,Csorgo:2018ruk}. 
}
\label{fig:scaling-LHC-2.76-vs-D0}
\end{figure*}

In the left panel of Fig.~\ref{fig:scaling-LHC-2.76-vs-D0}, the $H(x)$ function of the $\sqrt{s} = 2.76 $ TeV TOTEM data set of Ref.~\cite{Antchev:2018rec} is compared with that of the $p\bar p$ collisions measured by the D0 collaboration at $\sqrt{s} = 1.96 $ TeV Tevatron energy~\cite{Abazov:2012qb}. The right panel of Fig.~\ref{fig:scaling-LHC-2.76-vs-D0} compares the $H(x)$ scaling functions of elastic $pp$ collision at $\sqrt{s} = 7$ TeV LHC energy~\cite{Antchev:2011zz,Antchev:2013gaa} to that of the elastic $p\overline{p}$ collisions at the Tevatron energy, $\sqrt{s} = 1.96$ TeV. On both panels, the statistical errors and $t$-dependent systematic errors are added in quadrature. Lines are shown to guide the eye corresponding to fits with the model-independent L\'evy series studied in Refs.~\cite{Csorgo:2018uyp,Csorgo:2018ruk}. These plots suggest that the comparison of the $H(x)$ scaling functions or elastic $pp$ to $p\bar p$ collisions in the TeV energy range is a promising method for the Odderon search, and a precise quantification of the difference between the $H(x)$ scaling functions for $pp$ to $p\bar p$ collisions data sets is important. But how big 
is the difference between the $H(x)$ scaling functions of elastic $pp$ collisions at similar energies? 

\begin{figure*}[!hbt]
\begin{minipage}{0.5\textwidth}
  \centerline{\includegraphics[width=0.98\textwidth]{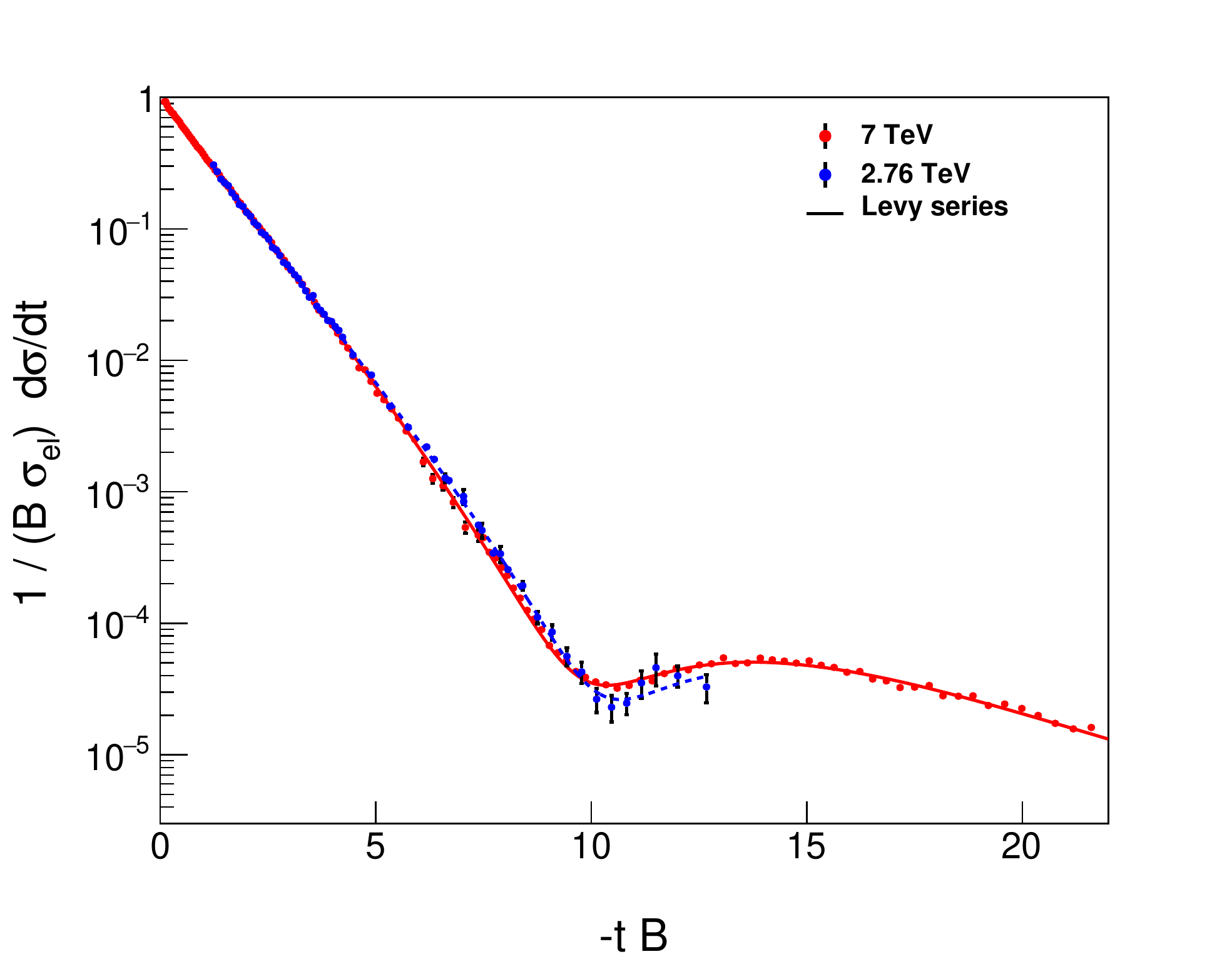}}
\end{minipage}
\begin{minipage}{0.5\textwidth}
  \centerline{\includegraphics[width=0.98\textwidth]{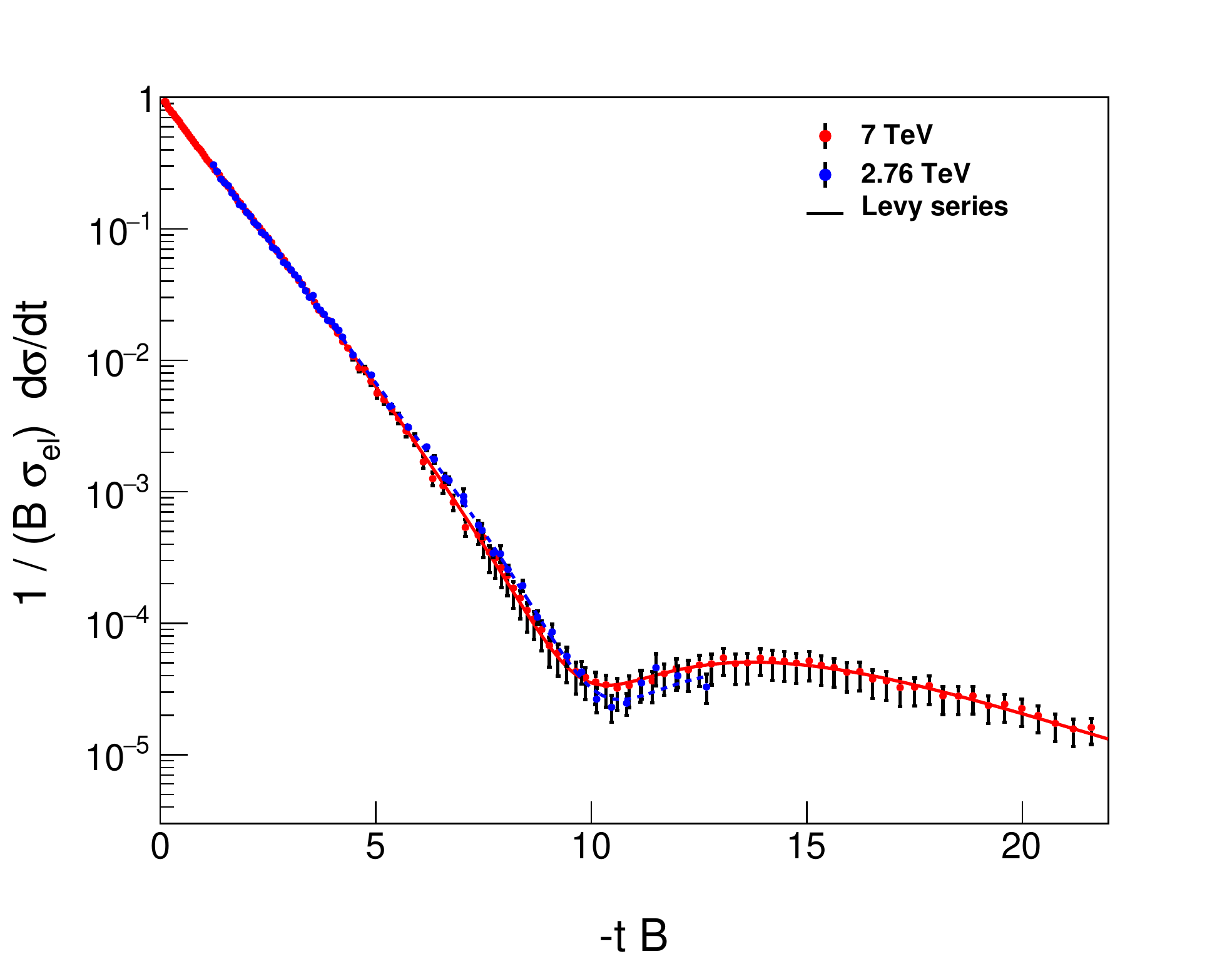}}
\end{minipage}
\caption{Same as Fig.~\ref{fig:scaling-LHC-2.76-vs-D0}, but now the $H(x)$ scaling of the differential cross section $d\sigma/dt$ of elastic $pp$ collisions is compared at the nearby $\sqrt{s} =  2.76$ and $7$ TeV LHC energies. Left panel shows the data with statistical errors only, while on the right panel, statistical errors and $t$-dependent systematic errors are added in quadrature. The two $H(x)$ scaling functions are, within statistical errors, apparently the same.
}
\label{fig:scaling-LHC-7-vs-2.76-Hx}
\end{figure*}

The $H(x)$ scaling of the differential cross section $d\sigma/dt$ of elastic $pp$ collisions is compared at the nearby $\sqrt{s} =  2.76$ and $7$ TeV LHC energies in Fig.~\ref{fig:scaling-LHC-7-vs-2.76-Hx}. These plots are similar to the panels of Fig.~\ref{fig:scaling-LHC-2.76-vs-D0}. The $H(x)$ scaling functions are remarkably similar, in fact, they are the same within the statistical errors of these measurements. Due to their great similarity, it is important to quantify precisely how statistically significant their difference is. 

We stress in particular that the possible scaling violations are small, apparently within the statistical errors, when $pp$ results are compared at LHC energies and $\sqrt{s}$ is increased from 2.76 to 7 TeV, by about a factor of 2.5. This makes it very interesting to compare the differential cross-sections of $pp$ and $p\bar p$ elastic scattering at the nearest measured energies in the TeV range, where crossing-odd components are associated with Odderon effects. Actually, the largest $\sqrt{s}$ of $p\bar p$ elastic scattering data is 1.96 TeV~\cite{Abazov:2012qb} while at the LHC the public data set on the elastic $pp$ scattering is available at $\sqrt{s} = 2.76$ TeV~\cite{Antchev:2018rec}, corresponding to a change in $\sqrt{s}$ by a factor of $2.76/1.96 \approx 1.4$. This is a rather small multiplicative factor on the logarithmic scale, relevant to describe changes both in high energy $pp$ and $p\bar p$ collisions. Given that the $H(x)$ scaling function is nearly constant between 2.76 TeV and 7 TeV within the statistical errors of these
data sets, we will search for a significant difference between the $H(x)$ scaling function of elastic $pp$ collisions at $\sqrt{s} = 2.76 $ and $7$ TeV as well as that of the elastic $p\bar p$ scattering at $\sqrt{s} = 1.96 $ TeV. If such a difference is observed, then there must be a crossing-odd (Odderon) component in the scattering amplitude of elastic $pp$ and $p\bar p$ scatterings.

\begin{figure}[!hbt]
\begin{center}
\begin{minipage}{0.48\textwidth}
  \centerline{\includegraphics[width=0.95\textwidth]{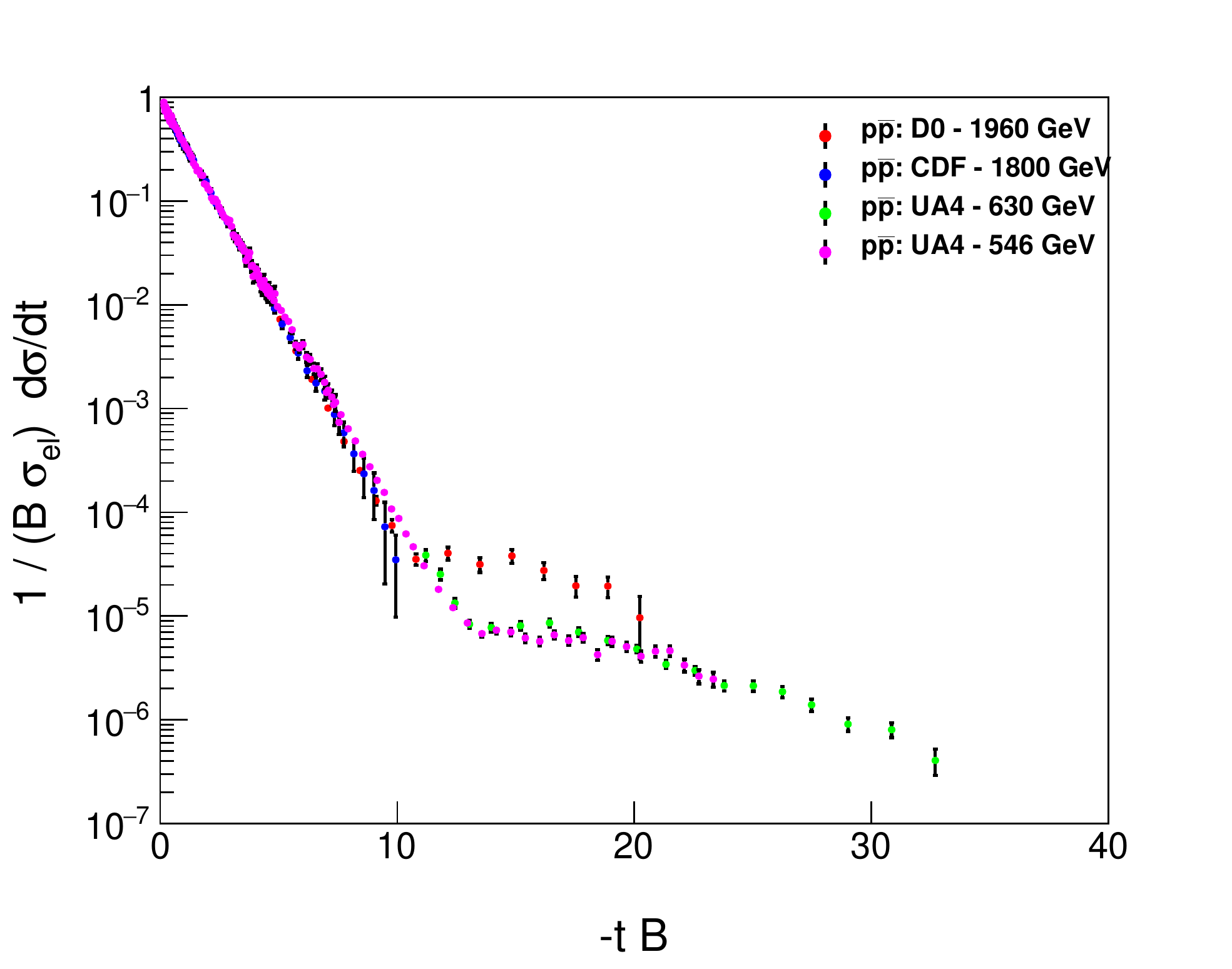}}
\end{minipage}
\end{center}
\caption{
Approximate $H(x) = \frac{1}{B \sigma_{\rm el}} \frac{d \sigma}{dt}$ 
scaling of the differential cross section $d\sigma/dt$ of elastic $p\overline{p}$ collisions at $\sqrt{s} =  0.546$ to 1.96 TeV. The scaling behaviour is valid in the exponential cone region, with the scaling function $H(x) = \exp(-x)$. The scaling domain starts at $x = 0$ and extends up to $x = -tB \simeq 10$. Scaling violations are evident in the $-t B \ge 10$ region, when the colliding energy increases from 546 GeV to 1.96 TeV, nearly by a factor of four. 
}
\label{fig:scaling-antiprotons}
\end{figure}

Let us now consider Fig.~\ref{fig:scaling-antiprotons}. This plot compares the $H(x)$ scaling functions for $p\bar p$ collisions at various energies from $\sqrt{s} = 546$ GeV to 1.96 TeV. Within experimental errors, an exponential cone is seen that extends to $x = - t B \approx 10$ at each measured energies, while for larger values of $x$ the scaling law breaks down in an energy dependent manner. At lower energies, the exponential region extends to larger values of $x \approx 13$, and the tail regions are apparently changing with varying colliding energies. Due to this reason, in this paper we do not scale the differential cross section of elastic $p\bar p$ collisions to different values of $\sqrt{s}$ as this cannot be done model-independently.
This property of elastic $p\bar p$ collisions is in contrast to that of the elastic $p p$ collisions, where we have demonstrated in Figs. \ref{fig:scaling-ISR-x},\ref{fig:scaling-LHC} that in a limited energy range between $\sqrt{s} = 23.5$  and $62.5$ GeV, as well as at the LHC in the energy range between $\sqrt{s} = 2.76$  and $7$ TeV, the $H(x)$ scaling works well. Due to these experimental facts and the apparent violations of the $H(x)$ scaling for $p\bar p$ collisions in the $x = -t B \ge 10$ region, in this paper we do not attempt to evaluate the energy dependence of the differential cross sections for $p\bar p$ collisions. However, based on the observed $H(x)$ scaling in $pp$ collisions, we do find a model-independent possibility to rescale the differential cross sections of elastic $pp$ collisions in limited energy ranges.

After the above qualitative discussion of $H(x)$ scaling for both $pp$ and $p\bar p$ elastic collisions, let us work out the details of the possibility of rescaling the measured differential cross sections to other energies in the domain where $H(x)$ indicates a scaling behaviour within experimental errors.

\begin{figure*}[hbt]
\begin{center}
\begin{minipage}{0.98\textwidth}
    \begin{center}
    \includegraphics[width=0.33\textwidth]{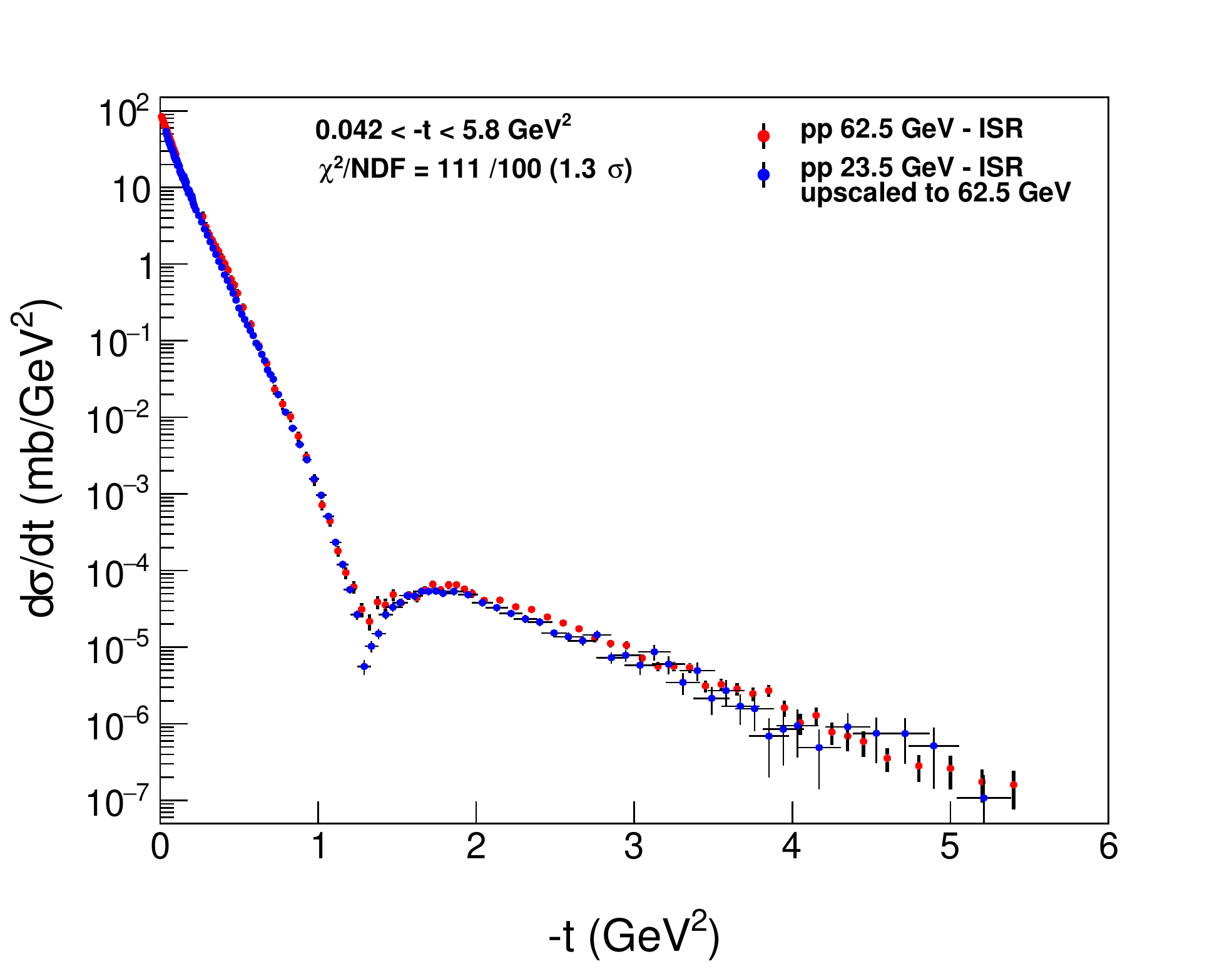}  
    \includegraphics[width=0.33\textwidth]{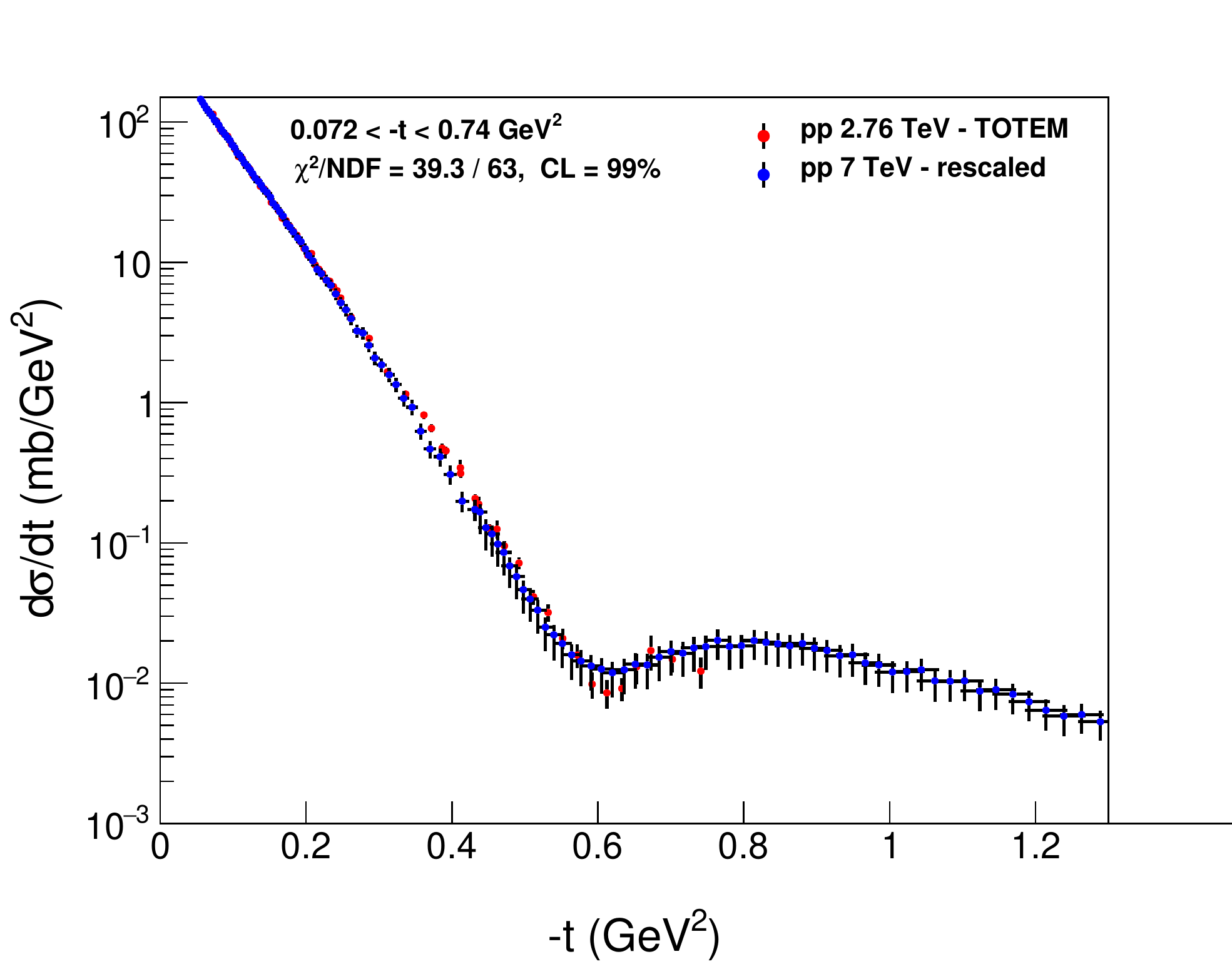}
    \includegraphics[width=0.33\textwidth]{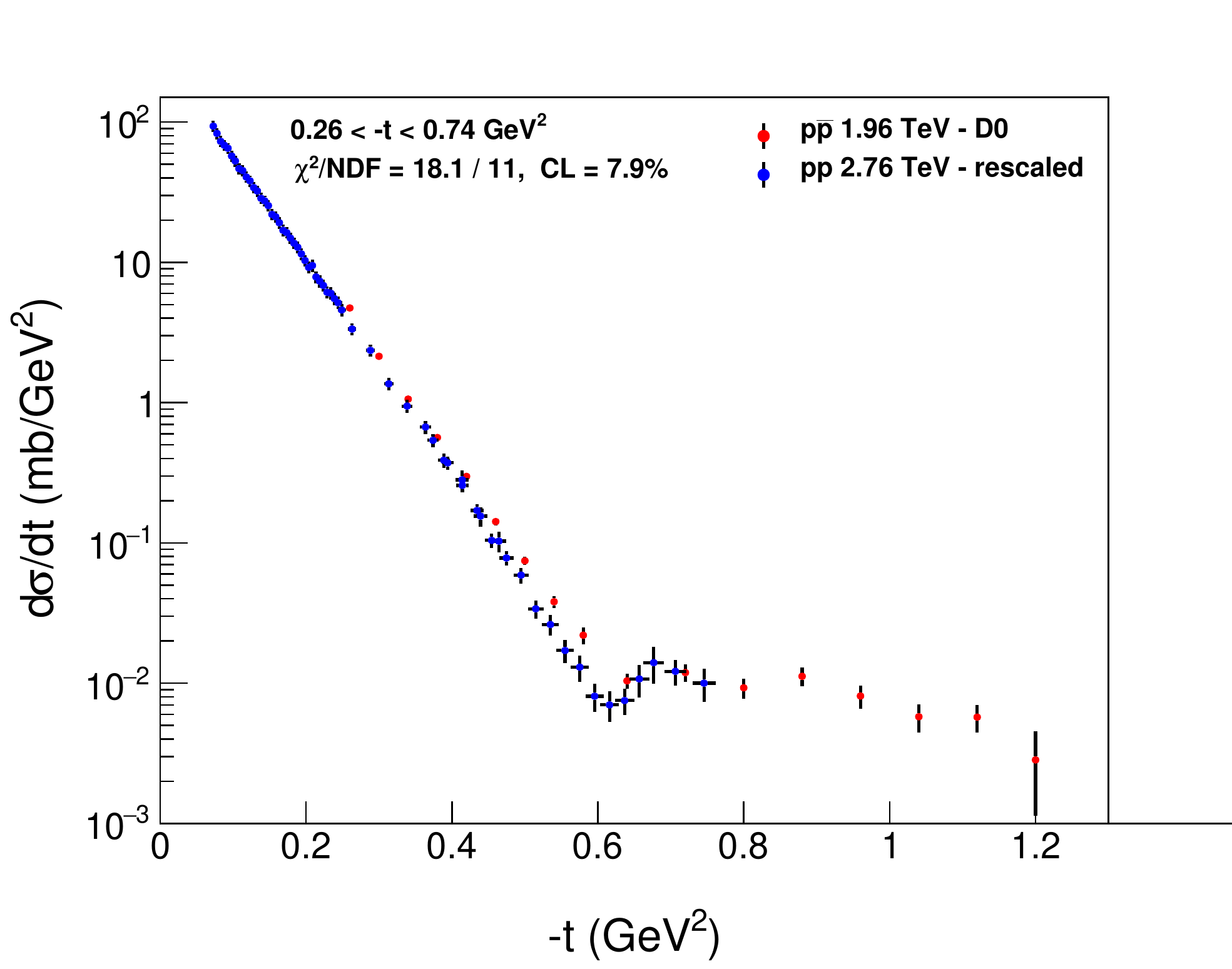}
    \end{center}
\end{minipage}    
\end{center}
\caption{
Rescaling of the differential cross section of elastic $pp$ collisions at the ISR and LHC energies, using  Eq.~(\ref{e:dsdt-rescaling}). This demonstrates that our method can also be used to get 
the differential cross sections at other energies by such a rescaling procedure, provided that 
the nuclear slope and the elastic cross sections are known at the new energy as well as 
at the energy from where we start to rescale the differential cross section.
In all panels, we have evaluated the level of agreement between
the rescaled  and measured data  with the help of Eq.~(\ref{e:chi2-data}).
{\it Left panel:} Rescaling of the differential cross sections from the lowest ISR energy of
$\sqrt{s} = 23.5 $ to the highest ISR energy of $62.5$ GeV.
The level of agreement between the rescaled 23.5 GeV $pp$ data and the measured 62.5 GeV $pp$ data 
corresponds to $\chi^2/{\rm NDF} = 111.0/110$ with a CL = 21.3 \% , that indicates an agreement within 1.3$\sigma$. 
{\it Middle panel:} Rescaling of the differential cross section of elastic $pp$ collisions from the energy 
of $\sqrt{s} = 7$ TeV~\cite{Antchev:2011zz,Antchev:2013gaa} down to $2.76$ TeV~\cite{Antchev:2018rec}.
The level of agreement between the rescaled 7.0 TeV $pp$ data and the measured 2.76 TeV $pp$ data 
corresponds to $\chi^2/{\rm NDF} = 39.3/63$ with a CL = 99.2 \% , that indicates an agreement, within 0.01$\sigma$, corresponding to a nearly vanishing deviation.
{\it Right panel:} Rescaling of the differential cross section of elastic $pp$ collisions from the energy 
of $\sqrt{s} = 2.76$ TeV, measured by TOTEM~\cite{Antchev:2018rec}, down to $1.96$ TeV, where it is compared to the D0 dataset of Ref.~\cite{Abazov:2012qb}. The level of agreement between the rescaled 2.76  TeV $pp$ data and the measured 1.96 TeV $p\overline{p}$ data is quantified by a $\chi^2/{\rm NDF} = 18.1/11$ and a  CL = 7.9 \% , that indicates an agreement within 1.76$\sigma$. 
}
\label{fig:rescaling-of-dsigma-dt-at-ISR-and-LHC}
\end{figure*}

The left panel of Fig.~\ref{fig:rescaling-of-dsigma-dt-at-ISR-and-LHC} indicates the result of rescaling of the differential cross sections of elastic $pp$ scattering from the lowest  $\sqrt{s} = 23.5$ GeV to the highest  $62.5$ GeV ISR energy, using Eq.~(\ref{e:dsdt-rescaling}). We have evaluated the level of agreement of the rescaled 23.5 GeV $pp$ data with the measured 62.5 GeV $pp$ data with the help of Eq.~(\ref{e:chi2-data}). The result indicates that the data measured at $\sqrt{s} = 23.5$ GeV and duly rescaled to 62.5 GeV are, within the errors of the measurements, consistent with the differential cross section of elastic $pp$ collisions as measured at $\sqrt{s} = 62.5$ GeV. This demonstrates that our method can also be used to extrapolate the differential cross sections at other energies by rescaling, provided that the $H(x)$ scaling is not violated in that energy range and that the nuclear slope and the elastic cross sections are known at a new energy as well as at the energy from where such a rescaling starts.

A similar method is applied at the LHC energies in the middle panel of Fig.~\ref{fig:rescaling-of-dsigma-dt-at-ISR-and-LHC}. This plot also indicates a clear agreement between the 2.76 TeV data and the rescaled 7 TeV data, which corresponds to a $\chi^2/{\rm NDF} = 39.3/63$ and a CL of 99.2 \% and a deviation on the 0.01 $\sigma$ level only. This suggests that indeed the rescaling of the differential cross section of elastic scattering can be utilized not only in the few tens of GeV range but also in the few TeV energy range. Most importantly, this plot indicates that there is a scaling regime in elastic $pp$ collisions, that includes
the energies of $\sqrt{s} = $ 2.76 and 7 TeV at LHC, where the $H(x)$ scaling is within errors, not violated. This is in a qualitative contrast to the elastic $p\bar{p}$ collisions at TeV energies, where the validity of the $H(x)$ scaling is limited only to the diffractive cone region with $x \le 10$, while at larger values of $x$, the $H(x)$ scaling is violated.

The right panel of Fig.~\ref{fig:rescaling-of-dsigma-dt-at-ISR-and-LHC} indicates a surprising agreement: 
after rescaling of the differential cross section of elastic $pp$ collisions from 2.76 TeV to 1.96 TeV, we find no significant difference between the rescaled 2.76 TeV $pp$ data with the $p\bar p$ data at the same energy, $\sqrt{s} = 1.96 $ TeV. The agreement between the extrapolated $pp$ and the measured $p\bar p$ differential cross sections correspond to an agreement at a CL of 7.9 \%, i.e. a surprising agreement at the $1.76\sigma$ level. It can be seen on the right panel of Fig.~\ref{fig:rescaling-of-dsigma-dt-at-ISR-and-LHC} that in the swing region, before the dip, the rescaled $pp$ differential cross section seems to differ qualitatively with the $p\bar p$ collisions data. However, according to our $\chi^2$ analysis that also takes into account the horizontal errors of the TOTEM data, we find that this apparent qualitative difference between these two data sets is quantitatively not significiant: it is characterized as an agreement within less than 2$\sigma$.

These plots suggest that the $H(x)$ scaling functions of elastic $pp$ and $p\bar p$ collisions differ at similar energies, 
while the same scaling functions for elastic $pp$ collisions are similar at similar energies, thus the comparison of 
the $H(x)$ scaling functions of elastic $pp$ and $p\bar p$ collisions is a promising candidate for an Odderon search.
Due to this reason, it is important to quantify how significant is this difference, given that the $H(x)$ scaling 
functions scale out the dominant $s$-dependent terms, that arise from the energy-dependent $\sigma_{\rm el}(s)$ 
and $B(s)$ functions. Such a quantification is the subject of the next section.

Before going into more details, we can already comment on a new Odderon effect qualitatively. When comparing the $H(x)$ scaling function of the differential cross section of elastic $pp$ collisions at 2.76 and 7.0 TeV colliding energies, we see no qualitative difference. By extrapolation, we expect that the $H(x)$ scaling function may be approximately energy independent 
in a bit broader interval, that extends down to 1.96 TeV. Such a lack of energy evolution of the $H(x)$ scaling function 
of the $pp$ collisions is in a qualitative contrast with the evolution of the $H(x)$ scaling functions 
of $p\bar p$ collisions at energies of $\sqrt{s} = 0.546 - 1.96$ TeV, where a qualitative and significant energy 
evolution is seen in the $x = -t B > 10 $ kinematic range. Thus, our aim is to quantify the Odderon effect 
in particular in this kinematic range of $x = -t B > 10 $ in order to evaluate the significance of this qualitative difference between elastic $pp$ and $p\bar p$ collisions.

\begin{figure*}[hbt]
\begin{center}
\begin{minipage}{0.8\textwidth}
 \centerline{\includegraphics[width=0.8\textwidth]{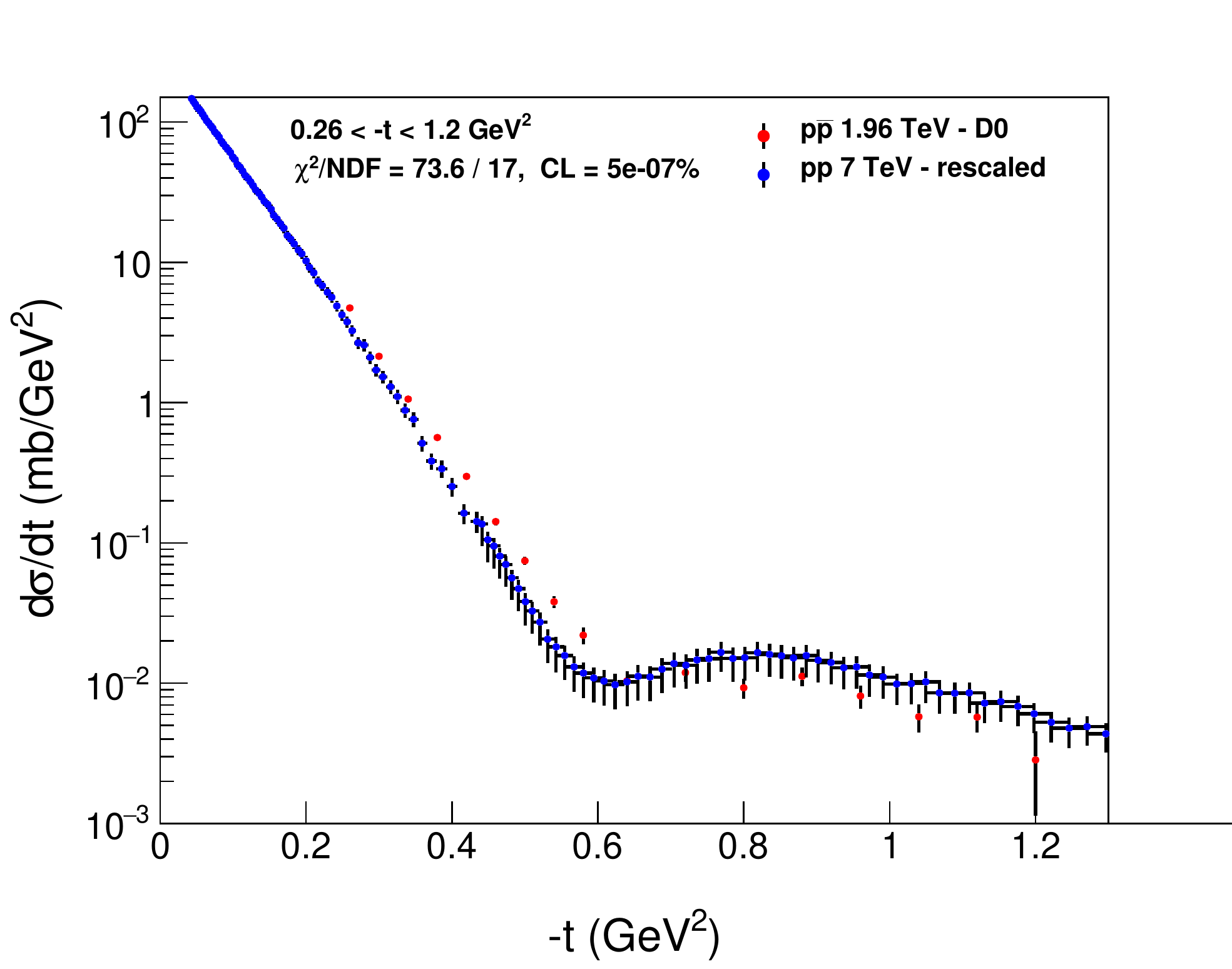}}
\end{minipage}    
\end{center}
\caption{
Rescaling of the differential cross section of elastic $pp$ collisions from the energy of
$\sqrt{s} = 7$ to $1.96$ TeV using Eq.~(\ref{e:dsdt-rescaling}). We have evaluated the confidence level 
of the comparison between the rescaled 7 TeV $pp$ data set and the 1.96 TeV $p\bar p$ data set with the help 
of Eq.~(\ref{e:chi2-data}), that does not take into account the horizontal errors of $x$ coming from the slopes $B$ and the type C point-to-point correlated errors on the vertical scale. Without these important effects, the difference between the datasets
provides a $\chi^2/{\rm NDF} = 73.6/17 $, equivalent to a confidence level of CL $ = $ $5.13 \times 10^{-7}$\%  and a statistically significant, $5.84\sigma$ effect. 
}
\label{fig:rescaling-from-7-to-1.96TeV}
\end{figure*}

\section{Quantification with interpolations}
\label{s:quantification}

In this section, we investigate the question of how to compare the two different scaling functions 
$H(x)  = \frac{1}{B\sigma_{el}}\frac{d\sigma}{dt}$ with $x = - t B$ introduced above measured at two distinct energies. 
We would like to determine if two different measurements correspond to significantly different scaling functions
$H(x)$, or not. In what follows, we introduce and describe a model-independent, simple and robust method, 
that enables us to quantify the difference of datasets or $H(x)$ measurements. The proposed method takes into account 
the fact that the two distinct measurements may have partially overlapping acceptance in $x$ and their binning 
might be different, so the datasets may correspond to two different sets of $x$ values.

Let us first consider two different datasets denoted as $D_i$, with $i = 1, 2$. In the considered case, 
$D_i = \big\{x_i(j), H_i(j), e_i(j)\big\}$, $j = 1, ... n_i$ consists of a set of data points located 
on the horizontal axis at $n_i$ different values of $x_i$, ordered as $x_i(1) < x_i(2) < ... < x_i(n_i)$, 
$H_i(j) \equiv H_i(x_i(j))$ are the measured values of $H(x)$ at $x=x_i(j)$ points, and $e_i(j)\equiv e_i(x_i(j))$ 
is the corresponding error found at $x_i(j)$ point. 

In general, two different measurements have data points at different values of $x$. Let us denote as 
$X_1 = \big\{x_1(1), ... x_1(n_1)\big\}$ the domain of $D_1$, and similarly 
$X_2 = \big\{x_2(1), ... , x_2(n_2)\big\}$ stands for the domain of $D_2$.
Let us choose the dataset $D_1$ which corresponds to $x_1(1) < x_2(1)$.
In other words, $D_1$ is the dataset that starts at a smaller value of the scaling 
variable $x$ as compared to the second dataset $D_2$. If the first dataset ends before 
the second one starts, i.e. when $x_1(n_1) < x_2(1)$, their acceptances would not overlap. 
In this limiting case, the two datasets cannot be compared with our method. 
Fortunately, however, the relevant cases e.g. the D0 data on elastic $p\overline{p}$ collisions 
at $\sqrt{s} = 1.96 $ TeV have an overlapping acceptance in $x$ with the elastic $pp$ 
collisions of TOTEM at $\sqrt{s} = 2.76$, 7 and 13 TeV. So from now on we consider 
the case with $x_1(n_1) > x_2(1)$. 

If the last datapoint in $D_2$ satisfies $x_2(n_2) < x_1(n_1)$, then $D_2$ is within the acceptance of $D_1$. 
In this case, let us introduce $f_2 = n_2$ as the final point with the largest value of $x_f$ from $D_2$.
If $D_2$ has $x_2(n_2) > x_1(n_1)$, then the overlapping acceptance ends at the largest (final) value 
of index $f_2$ such that $x_2(f_2) < x_1(n_1) < x_2(f_2+1)$. This means that the point $f_2$ of $D_2$ 
is below the largest value of $x$ in $D_1$, but the next point in $D_2$ is already above the final, 
largest value of $x(n_1)$ in $D_1$.

The beginning of the overlapping acceptance can be found in a similar manner. 
Due to our choice of $D_1$ as being a dataset that starts at a lower value, $x_1(1) < x_2(1)$, 
let us determine the initial point $i_1$ in $D_1$ that already belongs to the acceptance domain 
of $D_2$. This is imposed by the criterion that $x_1(i_1-1) < x_2(1) < x_1(i_1)$.

We compare the $D_1$ and $D_2$ datasets in the region of their overlapping acceptance, defined above, either in a one-way or in a two-way  projection method. The projection $1 \rightarrow 2$ has the number of degrees of freedom NDF$(1 \rightarrow 2)$ equal to  the number of points  of $D_2$ in the overlapping acceptance. For any of such a point $x_i(2)$, we used linear interpolation of the nearest points from $D_1$ such that $x_j(1)  < x_i(2) \le x_{j+1}(1)$ in order to evaluate the data and the errors of $D_1$ at this particular value of $x = x_i(2)$. This is done employing a default (linear, exponential) scale in the $(x, H(x))$ plane, that is expected to work well in the diffraction cone, where the exponential cone is a straight line. However, for safety and due to the unknown exact structure at the dip and bump region, we have also tested the linear interpolation utilizing the (linear, linear) scales in the $(x, H(x))$ plane.

Similarly, the projection $2 \rightarrow 1$ has the number of degrees of freedom  NDF$(2\rightarrow 1)$ as the number of points of dataset $D_1$ that fell into the overlapping common acceptance. A linear extrapolation was used for each  $x_i(1)$ points in this overlapping acceptance, so that $x_j(2)  < x_i(1) \le x_{j+1}(2)$, using both (linear, exponential) and (linear, linear) scales in the $(x,H(x))$-planes. For the two-way projections, for example $1 \longleftrightarrow 2$, the number of degrees of freedom is the sum of the points of $D_1$ and $D_2$ in the overlapping acceptance, defined as NDF$(1\longleftrightarrow 2)$ = NDF$(1\rightarrow 2)$ + NDF$(2\rightarrow 1)$.
 
Let us describe the two-way projections in more detail as the one-way projections can be 
considered as special cases of this method. A common domain $X_{12} = \big\{ x_{12}(1), ... , x_{12}(n_{12})\big\}$ in the 
region of the overlap of the $X_1$ and $X_2$ domains can be introduced as follows. Take the data points in the interval 
$[i_1\dots n_1]$ from the $D_1$ set and the data points in the interval 
$[1\dots f_2]$ from the $D_2$ set. This selection procedure provides a total of $n_{12} = n_1+f_2-i_1 + 1$ points. 
Let us order this new set of points and denote such a united domain as $X_{12}$. This domain corresponds 
to a common acceptance region which has $n_{12}$ data points on the horizontal axis denoted as 
$\big\{ x_{12}(1), ... , x_{12}(n_{12})\big\}$.

In order to compare the datasets $D_1$ and $D_2$, one needs to build two analog datasets that are both 
extrapolated to the same common domain $X_{12}$ starting from $D_1$ and $D_2$ as if the data in both analog 
datasets were measured at the same values of $x$. So far, either $D_1$ or $D_2$ has some data value on any element of the domain $X_{12}$, but only one of them is determined.

Let us take first those points from $X_{12}$ that belong to $D_1$, and label them with $j$ index. There are $n_1 - i_1 +1 $ such points. For such points, the data and error-bars of the extrapolated data set $D_{12}$ will be taken from $D_1$:
$d_{12}(x_{12}(j)) = d_1(x_1(j))$, $e_{12}(x_{12}(j) = e_1(x_1(j))$. However, for the same points, $D_2$ has no measured value.
But we need to compare the data of $D_1$ and $D_2$ at common values of $x$. So $D_2$ data and errors can be interpolated using linear
or more sophisticated interpolation methods. If the binning is fine enough, linear interpolation between the neighbouring datapoints
can be used.

At this point, let us consider that in the diffractive cone, when an exponential approximation to the differential
cross section can be validated, the shape of the scaling function is known to be $H(x) \approx \exp(-x)$. This function
is linear on a (linear, logarithmic) plot of $(x, H(x))$. In what follows, we will test both a (linear, exponential)
interpolation in the $(x, H(x))$ plots (that is expected to give the best results in the diffractive cone) and
a (linear, linear) interpolation that has the least assumptions and that may work better than the (linear, exponential)
interpolation technique around the diffractive minimum. These two different interpolation methods also allow us to estimate
the systematic error that comes from the interpolation procedure itself. If the data points are measured densely enough in
the $(x, H(x))$ plot, both methods are expected to yield similar results. We present our final results using both techniques
and note that indeed we find similar results with both methods.

Suppose that for the $j$-th point of data set $D_{12}$ and for  some $i$ value of $D_2$, $x_2(i) < x_{12}(j) < x_2(i+1)$.
Then a linear interpolation between the $i$-th and $i+1$-th point of $D_2$ yields the following formula:
\begin{equation}
d_{12}(j) = d_2(i) + (d_2(i+1) - d_2(i)) \frac{x_{12}(j) - x_2(i)}{x_2(i+1)-x_2(i)}.
\end{equation}
Similarly, the errors can also be determined by linear interpolation as
\begin{equation}
e_{12}(j) = e_2(i) + (e_2(i+1) - e_2(i)) \frac{x_{12}(j) - x_2(i)}{x_2(i+1)-x_2(i)} \,.
\end{equation}
This way, one extends $D_2$ to the domain $X_{12}$, corresponding to the overlapping acceptance of two measurements.
If there is a measured value in $D_2$, we use that value and its error bar. If there is no measurement in $D_2$ precisely
at that given value of $x$ that is part of the overlapping acceptance (corresponding to a value $x$ from $D_1$) then
we use the two neighbouring points from $D_2$ and use a (linear) interpolation to estimate the value at this intermediate
point. This method works if the binning of both data sets is sufficiently fine so that non-linear structures are 
well resolved.

This way, for those $j= 1, ... , n_1 - i_1 +1 $  points from $X_{12}$ that belonged to $D_1$, we have defined the data values from $D_1$ by identity and defined the data points from $D_2$ by linear interpolation from the neighbouring bins, so for these points both data sets are defined.

A similar procedure works for the remaining points in $D_{12}$ that originate from $D_2$. The number of such points is $f_2$. Let us index them with $k = 1, ... , f_2$. For these points, data and error-bars of the extrapolated data set $D_{12}$ will be taken from $D_2$: $d_{21}(x_{12}(k)) = d_2(x_2(k))$, while the errors are given as $e_{12}(x_{12}(k)) = e_2(x_2(k))$. However, for the same points, $D_1$ has no measured value. As we need to compare the data of $D_1$ and $D_2$ at common values of $x$, for these points,  $D_1$ data and errors can be extrapolated using the linear
or more sophisticated interpolation methods based on the nearest measured points. If the binning is fine enough, linear interpolation between the neighbouring data-points can be appropriately used. For broader bins, more sophisticated interpolation techniques may also be used that take into account non-linear interpolations based on more than two nearby bins, for example interpolations using Levy series expansion techniques
of Ref.~\cite{Csorgo:2018uyp}. However, in the present manuscript such refinements are not necessary as the (linear, linear) and the (linear, exponential) interpolations in ($x, H(x)$) give similar results.

Consider now that for the $k$-th point of data set $D_{12}$ and for some $l$-th value of $D_2$, $x_1(l) < x_{12}(k) < x_1(l+1)$.
Then linear interpolation between the $l$-th and $l+1$-th point of $D_2$ yields the following formula:
\begin{equation}
    d_{21}(k) = d_1(l) + (d_1(l+1) - d_1(l)) \frac{x_{12}(k) - x_1(l)}{x_1(l+1)-x_1(l)} \,.
    \label{e:interpolation}
\end{equation}
Similarly, the errors can also be determined by linear interpolation as
\begin{equation}
    e_{21}(k) = e_1(l) + (e_1(l+1) - e_1(l)) \frac{x_{12}(k) - x_1(l)}{x_1(l+1)-x_1(l)} \,.
    \label{e:error-E}
\end{equation}

This way, using the linear interpolation techniques between the neighbouring data points, we can now compare the extended $D_1$ and $D_2$ to their common kinematic range: $D_1$ was embedded and extrapolated to data points and errors denoted as $d_{12}(x_{12})$
and $e_{12}(x_{12})$ while $D_2$ was embedded and extrapolated to data points and errors denoted as $d_{21}(x_{12})$ and $e_{21}(x_{12})$, respectively. Note that the domain of both of these extended data sets is the same $X_{12}$ domain. The index
``12'' indicates that $D_1$ was extended to $X_{12}$, while index ``21'' indicates that $D_2$ was extended to domain $X_{12}$.

Now, we are done with the preparations to compare the two data sets, using the following $\chi^2$ definition:
\begin{equation}
    \chi^2 \equiv \chi^2_A \, = \, \sum_{j=1}^{n_{12}} \frac{(d_{12}(j) - d_{21}(j))^2}{e_{12}^2(j) + e_{21}^2(j)}.
    \label{e:chi2-data}
\end{equation}
In this comparison, there are no free parameters, so the number of degrees of freedom is NDF $= n_{12} = n_1+f_2-i_1+1$,
the number of data points in the unified data sample.

Based on the above Eq.~(\ref{e:chi2-data}) we get the value of $\chi^2$ and NDF, which can be used to evaluate the $p$-value, or the confidence level (CL), of the hypothesis that the two
data sets represent the same $H(x)$ scaling function. If CL satisfies the criteria that CL $ > 0.1\%$, the two data sets do not
differ significantly. In the opposite case, if CL $ < 0.1\%$ the hypothesis that the two different measurements correspond to the same a priori $H(x)$ scaling function, can be rejected.

The advantage of the above $\chi^2$ definition by Eq.~(\ref{e:chi2-data}) is that it is straightforward to implement it, however, it has a drawback that it does not specify how to deal with the correlated $t$ or $x = -t B$ dependent errors, and horizontal or $x$ errors. The $t$ measurements at $\sqrt{s}=7$ TeV are published with their horizontal errors according to Table 5 of Ref.~\cite{Antchev:2013gaa}. These errors should be combined with the published errors on the nuclear slope parameter $B$ to get a horizontal error on $x$ indicated as $\delta x$. Such a horizontal error has to be taken into account in the final calculations of the significance of the Odderon observation.

Regarding the correlations among the measured values, and the measured errors, the best method would be to use the full covariance matrix of the measured differential cross section data. However, this covariance matrix is typically unknown or unpublished, with an exception of the $\sqrt{s} = 13$ TeV elastic $pp$ measurement by TOTEM~\cite{Antchev:2018edk}. Given that
this TOTEM measurement of $d\sigma/dt$ at 13 TeV indicates already  the presence of small scaling violating terms in $H(x)$ according to Fig.~\ref{fig:scaling-LHC}, this 13 TeV dataset cannot be used directly in our Odderon analysis,
that is based on the $s$-independence of the scaling function of the differential elastic $pp$ cross section $H(x) \ne H(x,s)$
in a limited range that includes  $\sqrt{s} = $ 2.76 and 7 TeV, but does not extend up to 13 TeV.
However, we can utilize this TOTEM  measurement of $d\sigma/dt$ at  13 TeV, to test the method of diagonalization of the covariance matrix that we apply in our final analysis of the Odderon significance.

Our analysis of the covariance matrix relies on a method developed by the PHENIX Collaboration and described in detail in Appendix A of Ref.~\cite{Adare:2008cg}. This method is based on the following separation of the various types of experimental uncertainties:

Type A errors are point-to-point uncorrelated systematic uncertainties.

Type B errors are point-to-point varying but 
correlated systematic uncertainties, for which the point-to-point correlation is 100 \%, as the uncorrelated part is separated and added to type A errors in quadrature.

Type C systematic errors are point-independent, overall systematic uncertainties, that scale all the data points up and down by exactly the same, point-to-point independent factor.

Type D errors are point-to-point varying statistical errors. These type D errors are uncorrelated statistical errors, hence they can be added to the also uncorrelated, type A systematic errors in quadrature.

In this paper, where we apply this method to compare two different $H(x)$ scaling functions, we also consider a fifth kind of error, type E that corresponds to the theoretical uncertainty, which we identify with the error of the interpolation of 
one of the (projected) data sets to the $x$ values that are compared at some (measured) values of $x$ to a certain measured
data point at a measured $x$ value. This type E error is identified with the value calculated from the linear interpolation,
described above, as given for each A, B, C and D type of errors similarly by Eq.~(\ref{e:error-E}). Type D errors are added in quadrature to type A errors, and in what follows we index these errors with the index of the data point as well as with subscripts $a$, $b$ and $c$, respectively.

Using this notation, Eq.~(A16) of Ref.~\cite{Adare:2008cg} yields the following $\chi^2$ definition, suitable for 
the  projection of dataset $D_2$ to $D_1$, or $2 \rightarrow 1$:
\begin{eqnarray}
\tilde{\chi}^2 (2 \rightarrow 1) & = &\sum_{j=i_1}^{f_1}
    \frac{(d_{1}(j)- d_{21}(j) +\epsilon_{b,1} e_{b}(j) +\epsilon_{c,1} d_{1}(j) e_{c} )^2} {\tilde e_{a,1}^2(j)} \nonumber \\ 
     \null & \null & \qquad + \epsilon_{b,1}^2 +\epsilon_{c,1}^2 \, , 
\label{e:chi2-final-without-horizontal-errors}
\end{eqnarray}
where $\tilde e_{a,12}(j)$ is the type A uncertainty of the data point $j$ of the united data set $D_{12}$ scaled by a multiplicative factor such that the fractional uncertainty is unchanged under multiplication by a point-to-point varying factor:
\begin{equation}
\tilde{e}_{a,1}(j)= e_{a,1}(j) \left( \frac{d_{1}(j) +\epsilon_{b,1} e_{b}(j)  + \epsilon_{c,1} d_{1}(j) e_{c}}{d_{1}(j)}\right) \,. 
\label{eq:tildesigma}
\end{equation}
In these sums, there are NDF$_1 = f_1 - i_1 - 1$ number of data points in the overlapping acceptance from dataset $D_1$.
A similar sum describes the one-way projection $1 \rightarrow 2$, but there are  NDF$_2 = f_2$ points in the common acceptance.
For the two-way projections, not only the number of degrees of freedom add up, ${\rm NDF}_{12} = {\rm NDF}_1+{\rm NDF}_2$, but also
the $\chi^2$ values are added as $\chi^2 (1 \leftrightarrow 2) = \chi^2(1  \rightarrow 2) + \chi^2( 2  \rightarrow 1)$.

Let us note at this point, that $H(x)$ is a scaling function that is proportional to the differential cross section normalized by the integrated cross section. In this ratio, the overall, type C point-independent normalization errors multiply both the numerator and the denominator, hence these type C errors cancel out in $H(x)$. Given that these type C errors are typically rather large, for example, 14.4 \% for the D0 measurement of Ref.~\cite{Abazov:2012qb}, it is an important advantage in the
significance computation that we use a normalized scaling function $H(x)$. So in what follows, we set $\epsilon_{c,1} = 0$ and rewrite the equation for the $\chi^2$ definition accordingly. This effect increases the significance of a $H(x)$-scaling test.

The price we have to pay for this advantage is that we have to take into account the horizontal errors on $x$ in order to not overestimate the significance of our $\chi^2$ test.
In this step, we follow the propagation of the horizontal error to the $\chi^2$ as utilized by the so-called effective variance method of the CERN data analysis programme ROOT. This yields the following $\chi^2$ definition that we have utilized in our significance analysis for the case of symmetric errors in $x$:
\begin{eqnarray}
\tilde{\chi}^2 (2 \rightarrow 1) & = &\sum_{j=1}^{n_{12}}
    \frac{(d_{1}(j)- d_{21}(j) +\epsilon_{b,1} e_{b}(j))^2 }{\tilde e_{a,1}^2(j) +
    (\delta x_{1}(j) d^{\prime}_{1}(j))^2} +
      \epsilon_{b,1}^2 \,, 
\label{e:chi2-final}
\end{eqnarray}
where $\delta x_{12}(j)$ is the (symmetric) error of $x$ in the $j$-th data point of the data set $D_{1}$, and $d^{\prime}_{1}(j))^2$ is the numerically evaluated derivative of the extrapolated value of the projected data point obtained with the help of a linear interpolation using Eq.~(\ref{e:interpolation}).
Such definition is valid when the type B errors are known and are symmetric for the data set $D_1$ and the errors on $x$ are also symmetric. When the data set $D_1$ corresponds to the D0 measurement of elastic $p\bar p$ collisions,
Ref.~\cite{Abazov:2012qb}, we have to take into account that D0 did not publish the separated statistical and $|t|$-dependent
systematic errors, but decided to  publish their values added in quadrature. 
So we use these errors as type A errors and with this method, we underestimate the significance of the results as we neglect the correlations among the errors of the data points in the D0 dataset.
The TOTEM published the $|t|$-dependent statistical type D errors and the $|t|$-dependent systematic errors both for the 
2.76 TeV and 7 TeV measurements of the differential cross sections~\cite{Antchev:2011zz,Antchev:2013gaa,Antchev:2018rec}, with the note that the $|t|$-dependent systematic errors are almost fully correlated. In these works, TOTEM did not separate the point-to-point varying uncorrelated part of the $|t|$-dependent systematic errors. We thus estimate the type A errors by the statistical errors of these TOTEM measurements, we then slightly underestimate them, hence overestimate the $\chi^2$ and the difference between the compared data sets. Given that they are almost fully correlated, we estimate the type B errors by the point-to-point varying almost fully correlated systematic errors published by the TOTEM. We have tested this scheme by evaluating the $\chi^2$ from a full covariance matrix fit and from the PHENIX method of diagonalizing the covariance matrix at $\sqrt{s} = 13$ TeV, using the L\'evy expansion method of Ref.~\cite{Csorgo:2018uyp}. We find that the fit with the full covariance matrix results in the same minimum within one standard deviation of the fit parameters, hence the same significance as the fit with the PHENIX method of Appendix A of Ref.~\cite{Adare:2008cg}.

We have thus validated the PHENIX method of Ref.~\cite{Adare:2008cg} for the application of
the analysis of differential cross section at $\sqrt{s} = 13 $ TeV, together with the effective variance method of the ROOT package. 
This validation is important
as the full covariance matrix of the $\sqrt{s} = 2.76 $ TeV and $7$ TeV measurements by TOTEM is not published, but the PHENIX method appended with the ROOT method of effective variances can be used to effectively diagonalize the covariance matrix and to get similar results within the errors of the analysis.
In Section~\ref{s:Odderon-significance}, we employ the preliminary $\chi^2$ definition of Eq.~(\ref{e:chi2-final}) to estimate the significance of the Odderon signal in comparison of the $H(x)$ scaling functions for elastic $pp$ and $p\bar p$ collisions. Our final $\chi^2$ definition and the corresponding final results are described in \ref{app:A}.

\section{Extrapolation of the differential cross-sections}
\label{s:extrapolations}

In this section, we discuss how to extrapolate the data points to energies where measurements are missing. We emphasize that this method is not
our best method to evaluate the significance of the Odderon signal, but we include this section for the sake of completeness, as other groups follow this method. The obvious reason for this is that a large, 14.4 \% overall correlated, type C error of the D0 measurement does not cancel from the
differential cross-sections, and their significances, while it simply cancels from the $H(x)$ scaling functions, that are normalized to the integral of the differential cross-section. A quantitative estimate of the importance of this effect is shown in \ref{app:A}, and we detail the results from
the comparison of the $H(x)$ scaling functions starting from the next section. We recommend this section to those readers, who are
motivated to understand how to extrapolate the differential cross-sections to a new, not measured energy in a domain of $(s,t)$ where the $H(x)$ scaling
is known to be valid from already performed measurements.

We have found, for example, that in the ISR energy range of $\sqrt{s} = 23.5$ -- $62.5 $ GeV the $H(x)$ scaling function is 
approximately independent of $\sqrt{s}$ within errors, and with a possible exception at a small region around the diffractive minimum. 
We show how to extrapolate data points to unmeasured energies, under the condition that in a given energy range, $H(x)$ is independent of the collision energy, $H(x) \neq H(x,s) $. In general, such a feature has to be established or cross-checked experimentally. This case is important, given that we have shown before, for example in Fig.~\ref{fig:scaling-LHC-7-vs-2.76-Hx}, that $H(x)$ for $pp$ collisions stays energy-indepen\-dent within errors between the LHC energies of $2.76 $ TeV $\le \sqrt{s}\le $ $7$ TeV. Furthermore, we have already shown that for $p\overline{p}$ collisions, $H(x) = H(x,s)$ in the energy range of 0.546 $\le \sqrt{s} \le 1.96$ TeV, as indicated in Fig.~\ref{fig:scaling-antiprotons}.

Let us denote two different center-of-mass energies between which $H(x)={\rm const}(\sqrt{s})$ within the experimental errors as $\sqrt{s_1}$ and $\sqrt{s_2}$. Analogically, we denote various observables as $B_i \equiv B(s_i)$, $\sigma_i \equiv \sigma_{el,i}\equiv \sigma_{el}(s_i)$, $x_i \equiv B_i t$.

The energy independence of the $H(x)$ scaling function formally can be written as
\begin{equation}
    H_1(x_1) = H_2(x_2) = H(x) \qquad \mbox{\rm if}\quad x_1 = x_2 \, .
\end{equation}
This simple statement has tremendous experimental implications. The equality $x_1 = x_2$ means that the scaling function is the same, if at center-of-mass energy $\sqrt{s_1}$ it is measured at $t_1$
and at energy $\sqrt{s_2}$ it is measured at $t_2$, so that
\begin{equation}
    t_1 B_1 = t_2 B_2 \qquad \mbox{\rm if}\quad x_1 = x_2 \, .
\end{equation}
The equality $H_1(x_1) = H_2(x_2) = H(x)$ is expressed as
\begin{equation}
    \frac{1}{B_1 \sigma_1}\left. \frac{d\sigma}{dt} \right\rvert_{t_1 = x/ B_1}
     = 
     \frac{1}{B_2 \sigma_2}\left. \frac{d\sigma}{dt} \right\rvert_{t_2 = x/ B_2} \,.
\end{equation}
Putting these equations together, this implies that the experimental data can be scaled to other energies in an energy range where $H(x)$ is found to be independent of $\sqrt{s}$ as follows:
\begin{equation}
    \left. \frac{d\sigma}{dt} \right\rvert_{t_1}
     = 
     \frac{B_1 \sigma_1}{B_2 \sigma_2}\left. \frac{d\sigma}{dt} 
     \right\rvert_{t_2 = t_1 B_1 / B_2} \, . \label{e:dsdt-rescaling}
\end{equation}
With the help of this equation, the data points on differential cross sections can be scaled to various different colliding energies, if in a certain energy region the $H(x)$ scaling holds within the experimental errors. In other words, the differential cross section can be rescaled from $\sqrt{s_1}$ to $\sqrt{s_2}$ by rescaling the $|t|$-variable using the ratio of $B_1/B_2=B(s_1)/B(s_2)$, and by multiplying the cross section with the 
ratio $\frac{B_1 \sigma_1}{B_2 \sigma_2}$.

\section{Results}
\label{s:results}

In this section, we present our results and close the energy gap, as much as possible without a direct measurement, between the TOTEM data on elastic $pp$ collisions at $\sqrt{s} = 2.76$ and $7.0$ TeV and D0 data on elastic $p\bar p$ collisions at $\sqrt{s} = 1.96 $ TeV. 
This section is based on the application of Eq.~(\ref{e:dsdt-rescaling}) in this energy range.
After the rescaling procedure, the resulting data set at the new energy is compared with the measured data quantitatively with the help of Eq.~(\ref{e:chi2-data}).

We have used the rescaling equation, Eq.~(\ref{e:dsdt-rescaling}) first to test and to cross-check, if the rescaling of the $\sqrt{s} = 23.5 $ GeV ISR data to other ISR energies works, or not.
The left panel of Fig.~\ref{fig:rescaling-of-dsigma-dt-at-ISR-and-LHC} indicates that such a rescaling of the differential cross sections from the lowest ISR energy of $\sqrt{s} = 23.5$ to the highest ISR energy of $62.5$ GeV actually works well. The level of agreement of the rescaled 23.5 GeV $pp$ data with the measured 62.5 GeV $pp$ data has been evaluated with the help of Eq.~(\ref{e:chi2-data}). We found an agreement with a $\chi^2/{\rm NDF} = 111/100$, corresponding to a CL = 21.3 \% and a difference is at the level  of 1.25$\sigma$ only. This result demonstrates that our rescaling method can also be used to get the differential cross sections at other energies, provided that the nuclear slope and the elastic cross sections are known at the new energy as well as at the energy from where we start the rescaling procedure.

Subsequently, one can also rescale the TOTEM data at $\sqrt{s} = 2.76$ or $7$ TeV to $1.96$ TeV, given that $H(x)$ is (within errors) energy independent in the range of $2.76 - 7 $ TeV, corresponding to nearly a factor of 2.5 change in $\sqrt{s}$, while the change in $\sqrt{s}$ from $1.96$ to $2.76$ TeV is only a factor of 1.4. The right panel of Fig.~\ref{fig:rescaling-of-dsigma-dt-at-ISR-and-LHC} indicates that 
rescaling of the differential elastic $pp$ cross section from $\sqrt{s} = 2.76$ to $1.96$ TeV also gives valuable results. We have evaluated the confidence level 
of the comparison of the rescaled 2.76 TeV $pp$ data with the 1.96 TeV $p\bar p$ data with the help of Eq.~(\ref{e:chi2-data}). As was already mentioned above, we have found a surprising agreement with a $\chi^2/{\rm NDF} = 18.1/11$, corresponding to a CL = 7.93 \%, and a difference at the level of 1.75 $\sigma$ only.

Another important result is illustrated in Fig.~\ref{fig:rescaling-from-7-to-1.96TeV}. This comparison indicates a difference between the rescaled $\sqrt{s} = $ 7 TeV elastic $pp$ differential cross-section~\cite{Antchev:2011zz,Antchev:2013gaa} to the $\sqrt{s} = $ 1.96 TeV energy and to the corresponding $p\bar p$ data measured at $\sqrt{s}= 1.96 $ TeV~\cite{Abazov:2012qb}. To obtain a first estimate, this difference is quantified with the help of Eq.~(\ref{e:chi2-data}) yielding a CL of $5.13\cdot 10^{-7}$ \%, which corresponds to a difference at the 5.84 $\sigma$ level. As this method adds the statistical and the point-to-point varying systematic errors in quadrature, it underestimates the actual significance of the difference between the two data sets. Although this estimate already provides a significant, greater than 5$\sigma$ effect for the Odderon observation, corresponding to a significant, 5.84$\sigma$ difference between the $pp$ dataset and the 1.96 TeV $p\overline p$ dataset, however, the evaluation of this significance does not yet take into account the rather large overall normalization error of 14.4 \% that has been published by the D0 collaboration. 

This Fig.~\ref{fig:rescaling-from-7-to-1.96TeV} indicates that not only the diffractive interference, the dip and the bump may carry an Odderon signal, but 
also the so called swing region, where the $pp$ differential cross-section bends below the straight exponential diffractive cone of the $p\bar p$ result. 
See also Fig.~\ref{fig:rescaling-dsigmadt-from-7-to-1.96TeV-with-type-C} of \ref{app:A} for more details on how the type C errors reduce the significance of the Odderon 
signal to a 3.64$\sigma$, if the comparison is done directly at the level of the differential cross-sections and if these type C, overall correlated errors 
are added in quadrature to the point-to-point correlated, type A errors. This is only a lower bound of the significance as type C errors are not point-to-point 
fluctuating, but shift the whole dataset up or down in a correlated way, see the end of  \ref{app:A}  for more details on this lower bound. The point is that
it is advantageous to use the $H(x)$ scaling function instead of the differential cross-sections, as the rather large type C errors cancel from 
$H(x)$ while they may lead to an important reduction of the significance of the signal when they are considered on the differential cross-sections.

It can be seen in Fig.~\ref{fig:rescaling-from-7-to-1.96TeV} that in the swing region, before the dip, the rescaled $pp$ differential cross section differ significantly from that of $p\bar p$ collisions. 
Looking by eye, the swing and the diffractive interference (dip and bump) regions both seem to provide an important contribution.
We have dedicated ~\ref{app:E} to evaluate the significanes of various regions, to determine precisely how much do they contribute to the significance of this Odderon signal. 

The estimates of statistical significances given in the present 
Section are based on a $\chi^2$ test that includes the $|t|$-dependent statistical errors 
and the $|t|$-dependent systematic errors added in quadrature. Thus the values of $\chi^2/$NDF and significances given in this Section
can  only be considered as estimates. Indeed, although the $|t|$-dependent systematic errors on these $\sqrt{s} = 7$ TeV data are known to be almost 
fully correlated,  the covariance matrix is not publicly available at the time of closing this manuscript from the TOTEM measurement at $\sqrt{s} = 7$ TeV. 
It is clear that the $\chi^2$ is expected to increase if the covariance matrix is taken into account, and this effect would increase the 
disagreement between the measured $p\bar p$ and the extrapolated $pp$ differential cross sections at $\sqrt{s} = 1.96$ TeV. 

So this indicates that we have to consider the proposed rescaling method as conservatively as possible, that allows us to take into account the statistical and $|t|$-dependent correlated systematic errors, as well as the $|t|$-independent correlated systematic errors. Such an analysis is presented in the next section, where we quantify the differences between the scaling functions $H(x)$ of elastic $pp$ and $p\bar p$ collisions using the fact that $H(x)$ is free of $|t|$-independent normalisation errors, and our final results are summarized in \ref{app:A}.

\section{A significant Odderon signal from the $pp$ and $p\bar p$ scaling functions}
\label{s:Odderon-significance}

In this section, we estimate  a preliminary, 6.55$\sigma$ significance for the  Odderon signal, while \ref{app:A} determines and summarizes our final Odderon signal of an at least 6.26$\sigma $ effect. Both results are obtained by comparing the $H(x)$ scaling functions of $pp$ and $p\bar p$ collisions. 

We have found a significant Odderon signal by comparing the $H(x)$ scaling functions of the differential cross section of elastic $pp$ collisions with $\sqrt{s} = 7$ TeV to that of $p\bar p$ collisions with $\sqrt{s} = 1.96$ TeV,
as indicated in Fig.~\ref{fig:rescaling-from-7-to-1.96TeV-and-back}. The comparison is made in both possible ways, by comparing the $pp$ data to the $p\bar p$ data, and vice versa. The difference between these two datasets corresponds to at least 
a $\chi^2/{\rm NDF} = 84.6/17$, giving rise to a CL of $5.8 \times 10^{-9}$ \% and to a preliminary, 6.55$\sigma$ significance, obtained
with the help of Eq.~(\ref{e:chi2-final}). The overall, $|t|$-independent normalization error of 14.4 \% on the D0 data set cancels  from this $H(x)$, and does not propagate to our conclusions.

These results are obtained for the $\sigma_{\rm el} = 17.6 \pm 1.1$ mb value of the elastic $p\bar p$ cross section at $\sqrt{s} = 1.96 $ TeV, and for the linear-exponential interpolation in $(x, H(x))$.
Using this method of interpolation, the nearest points were connected with a linear-exponential line, that corresponds to a straight line on a linear-logarithmic plot in $(x, H(x))$. We have used the published values of the differential cross sections $\frac{d\sigma}{dt}$, that of the nuclear slope parameter $B$ and the measured value of the elastic cross section $\sigma_{\rm el}$ for 7 TeV $pp$ elastic collisions. For the elastic cross section of $p\bar p$ collisions at $\sqrt{s}$ $ = 1.96$ TeV, we have numerically integrated the differential cross section with an exponential approximation at very low-$|t|$ that provided us with $\sigma_{\rm el} = 20.2 \pm 1.4$ mb.

We have systematically checked the effect of variations in our interpolation method by switching from 
the (linear-exponential) in $(x, H(x))$ interpolation to a linear-linear one and by changing the value 
of the elastic $p\bar p$ collisions from the numerically integrated differential cross-section value of  $\sigma_{\rm el} = 20.2 \pm 1.4$ mb, which is an unusually large value, but equals within the quoted 14.4\% systematic error  to the 
$\sigma_{\rm el} = 17.6 \pm 1.1$ mb value, that corresponds to the trend published by the Particle Data Group, see
the Fig. 51.6, bottom  panel, yellow line of Ref.~\cite{Tanabashi:2018oca}.
The input values of the nuclear slope parameter $B$ and  the elastic cross-section $\sigma_{\rm el}$
are summarized in Table~\ref{table:B-sigma-summary}, the corresponding results are shown in 
Tables~\ref{table:7-to-1.96-TeV-one-way-comparison},
\ref{table:2.76-to-1.96-TeV-one-way-comparison},
~\ref{table:7-to-1.96-TeV-two-way-comparison} and
~\ref{table:2.76-to-1.96-TeV-two-way-comparison}.

\begin{table}[htb]
\begin{center}
\scalebox{0.95}{
\begin{tabular}{llll}
Energy & $\sigma_{el}$ & $B$  & Reference \\
 (GeV) &    (mb)  &  (GeV$^{-2}$)     &           \\
\hline
 1960   &  17.6 $\pm$ 1.1   &     \null                & Fig. 51.6 of Ref.~\cite{Tanabashi:2018oca} \\
   ($p\bar{p}$)                 &  20.2 $\pm$ 1.4   &                     & from low $-t$  fit to data~\cite{Abazov:2012qb} \\
                    &                   & 16.86 $\pm$ 0.2     & ~\cite{Abazov:2012qb} \\ \hline
 2760           &  21.8 $\pm$ 1.4   &                     & \cite{Nemes:2017gut} \\
  ($pp$)                  &                   & 17.1  $\pm$ 0.3     & \cite{Antchev:2018rec} \\ \hline
 7000           &  25.43$\pm$ 1.02  &                     & \cite{Antchev:2013iaa} \\
    ($pp$)                &                   & 19.89 $\pm$ 0.272   & \cite{Antchev:2013gaa} \\
\end{tabular} }
\end{center}
\caption { Summary table of the elastic cross-sections $\sigma_{\rm el}$, the nuclear slope parameters $B$, and their sources or references.  }
\label{table:B-sigma-summary}
\end{table}

As part of our systematic studies, we have also  changed the direction of the projection. The results are summarized in Table~\ref{table:7-to-1.96-TeV-one-way-comparison}.
They indicate that the improved version of  Fig.~\ref{fig:rescaling-from-7-to-1.96TeV}, shown as the top left panel of Fig.~\ref{fig:rescaling-from-7-to-1.96TeV-and-back} and evaluated with the help of our improved $\chi^2$ definition of
Eq.~(\ref{e:chi2-final}) corresponds to a conservative 
case of Odderon observation based on the $\sqrt{s} = 7$ TeV TOTEM and the $\sqrt{s} = 1.96$ TeV D0 data sets.
This panel indicates that the Odderon signal is observed in this comparison with  a preliminary, at least a 6.55$\sigma$ significance, 
indicating the power of our method of Odderon observation. 
In addition to this, our final result includes a symmetry requirement and a robustness test described in ~\ref{app:A}. These effects decreased the significance of our Odderon observation, from a preliminary, $\ge $ 6.55 $\sigma$ effect 
to  a final and {\it statistically significant}, $\ge$ 6.26 $\sigma$ effect.

We have checked the robustness of this result for several possible variations of the $\chi^2$ definition. The consideration that was most successful in decreasing this significance was related to the fact that unlike the original PHENIX method of Ref.~\cite{Adare:2008cg}, that was worked out for a theory to data comparison, in this manuscript we compare data to data. So we have adapted the PHENIX method of Ref.~\cite{Adare:2008cg},
from a situation where there was a theoretical function without errors compared to data with errors to a situation where we compare two datasets
and both of these datasets have the same type of errors. This slightly decreased the significance of the Odderon signal, 
from the value of a preliminary, at least 6.55 $\sigma$ to the final value of 6.26 $\sigma$, as detailed in ~\ref{app:A} of this manuscript. Given that both significances of the preliminary 6.55 $\sigma$, 
detailed in this section, and 6.26 $\sigma$, detailed in  ~\ref{app:A} are clearly and safely above the 
5 $\sigma$ discovery threshold, this robustness test did not change our conclusions.

\begin{figure*}[hbt]
\begin{center}
\begin{minipage}{1.0\textwidth}
 \null   \hspace{1truecm}
\includegraphics[width=0.8\textwidth]{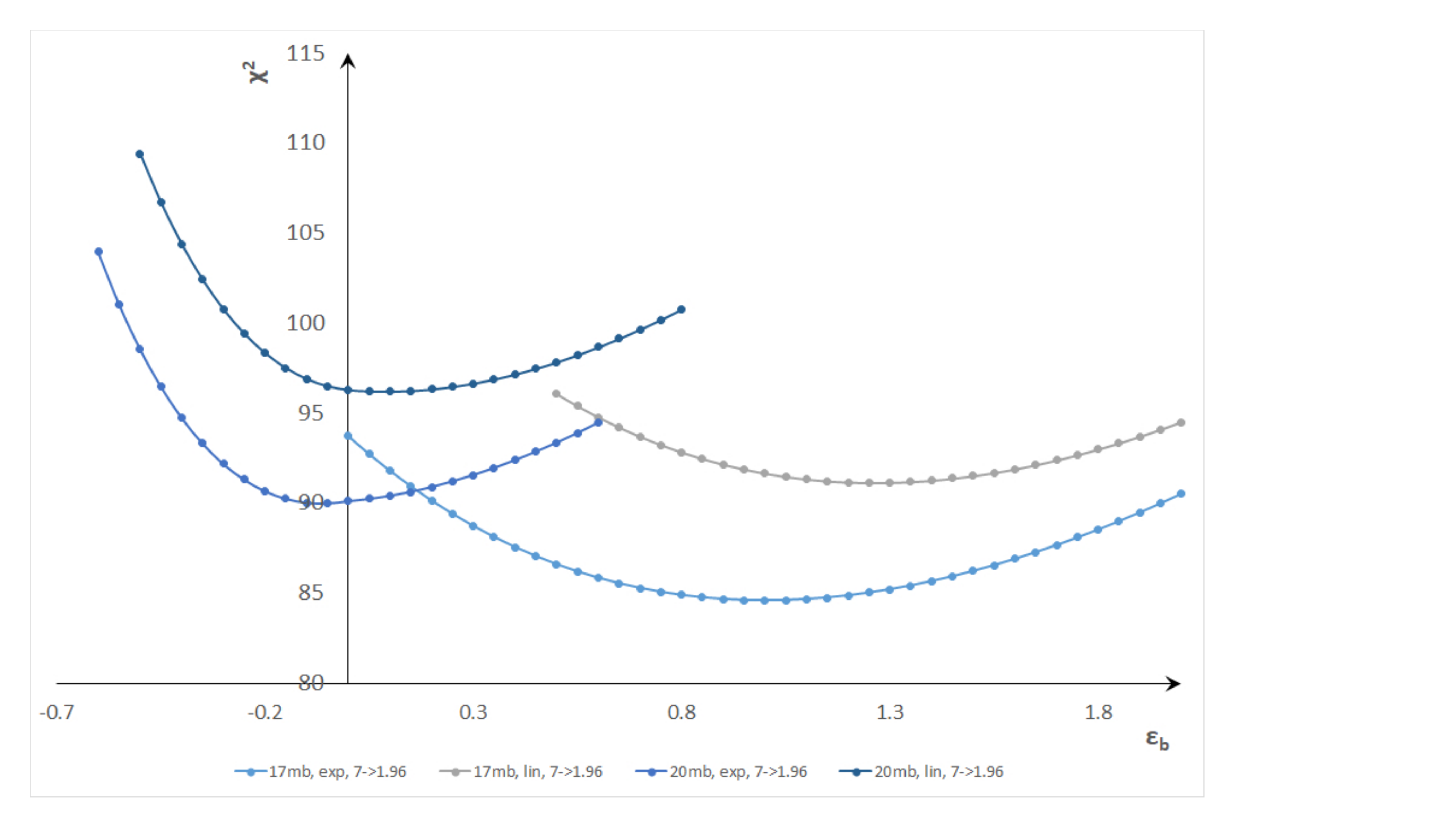}\\
 \null   \hspace{1truecm}
\includegraphics[width=0.8\textwidth]{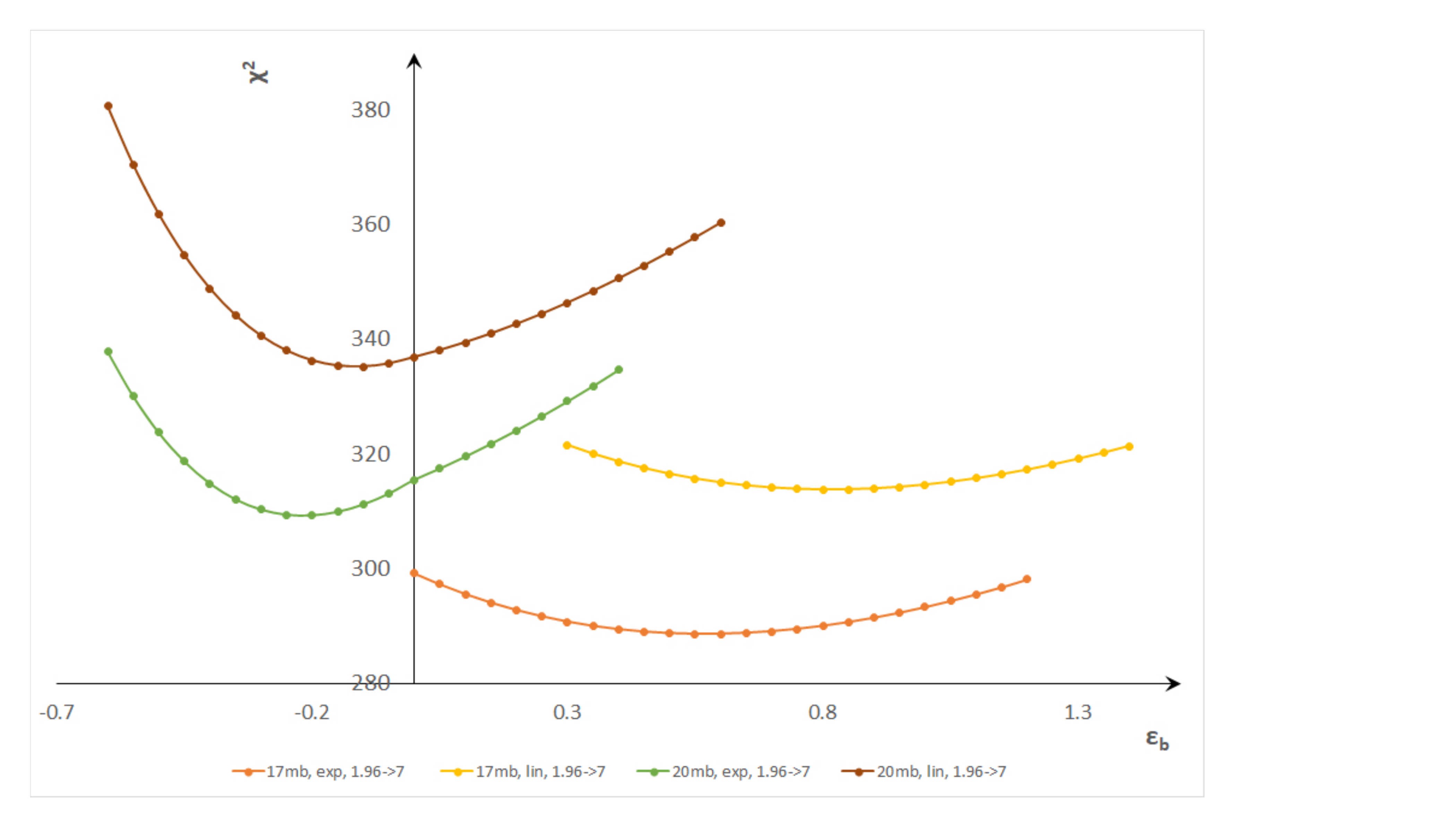}
\end{minipage}    
\end{center}
\caption{
Dependence of $\chi^2$ on the coefficient of the correlated but point-to-point varying systematic errors, $\epsilon_b$, for the comparison of the $H(x)$ scaling functions of elastic $p\bar p$ collisions at $\sqrt{s} = 1.96$ TeV with that of $pp$ collisions at $\sqrt{s} = 7.0$ TeV. 
Each of the four cases are shown together corresponding to the direction of the projection. 
{\it Upper panel} indicates the results of the 1.96 $\rightarrow$ 7.0 TeV projection. 
{\it Lower panel} indicates the results of the  7.0  $\rightarrow$ 1.96 TeV projection. Both cases indicate four $\chi^2(\epsilon_b)$ curves corresponding to the choice of linear-linear or linear-exponential interpolations in $(x,H(x))$, as well as to the choice of the elastic $p\bar p$ cross section at $\sqrt{s} = 1.96$ TeV (20.2 $\pm$ 1.4 mb vs 17.6 $\pm$ 1.1 mb). A parabolic structure is seen in each case with a clear minimum, and the fit quality corresponding to these minima in $\epsilon_b$ is summarized in Table~\ref{table:7-to-1.96-TeV-one-way-comparison}.
}
\label{fig:chi2-vs-epsilon-b-for-1960-vs-7000-GeV.png}
\end{figure*}

\begin{figure*}[ht]
\begin{center}
\begin{minipage}{0.95\textwidth}
 \centerline{\includegraphics[width=0.49\textwidth]{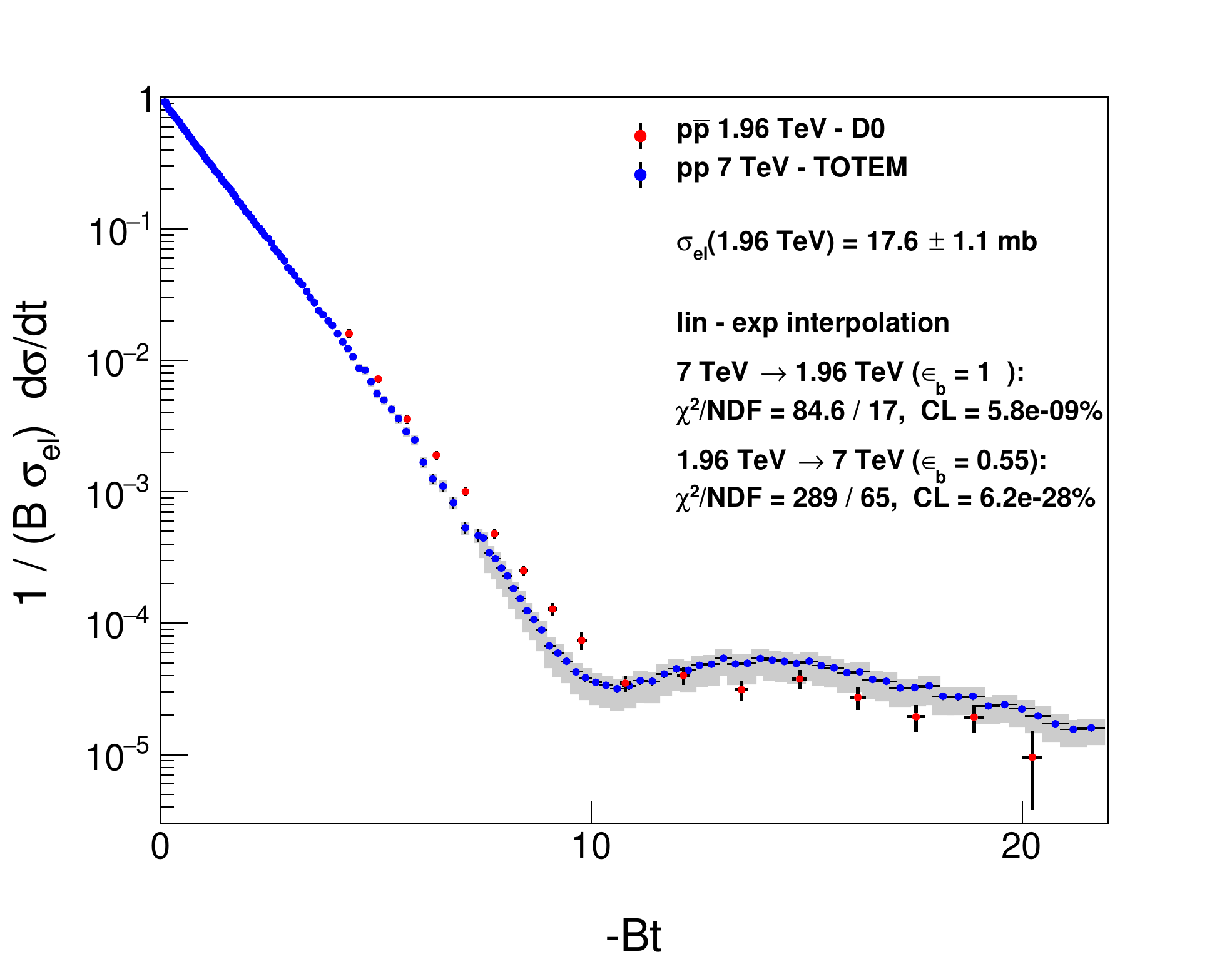}
 \includegraphics[width=0.49\textwidth]{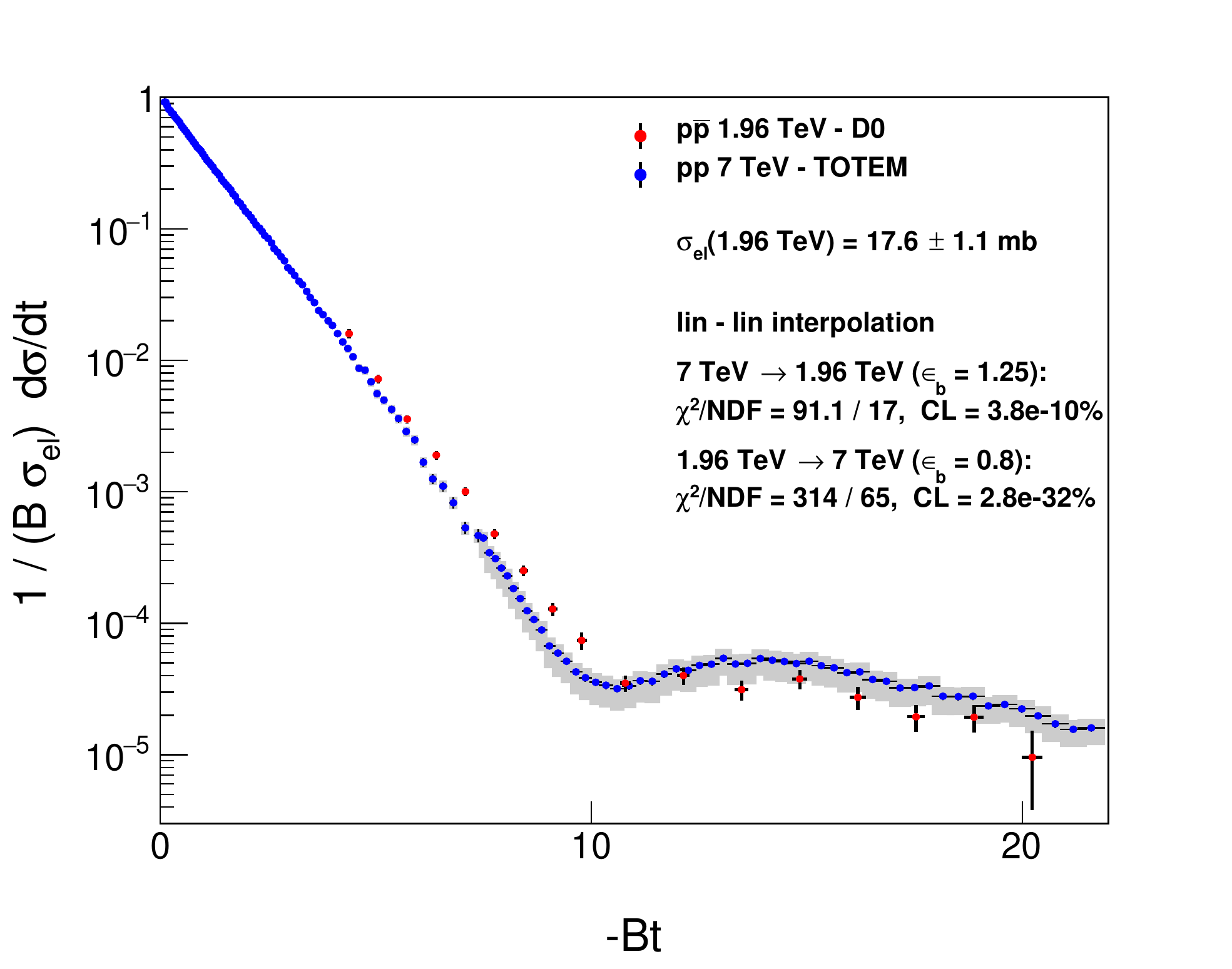}}
 \centerline{\includegraphics[width=0.49\textwidth]{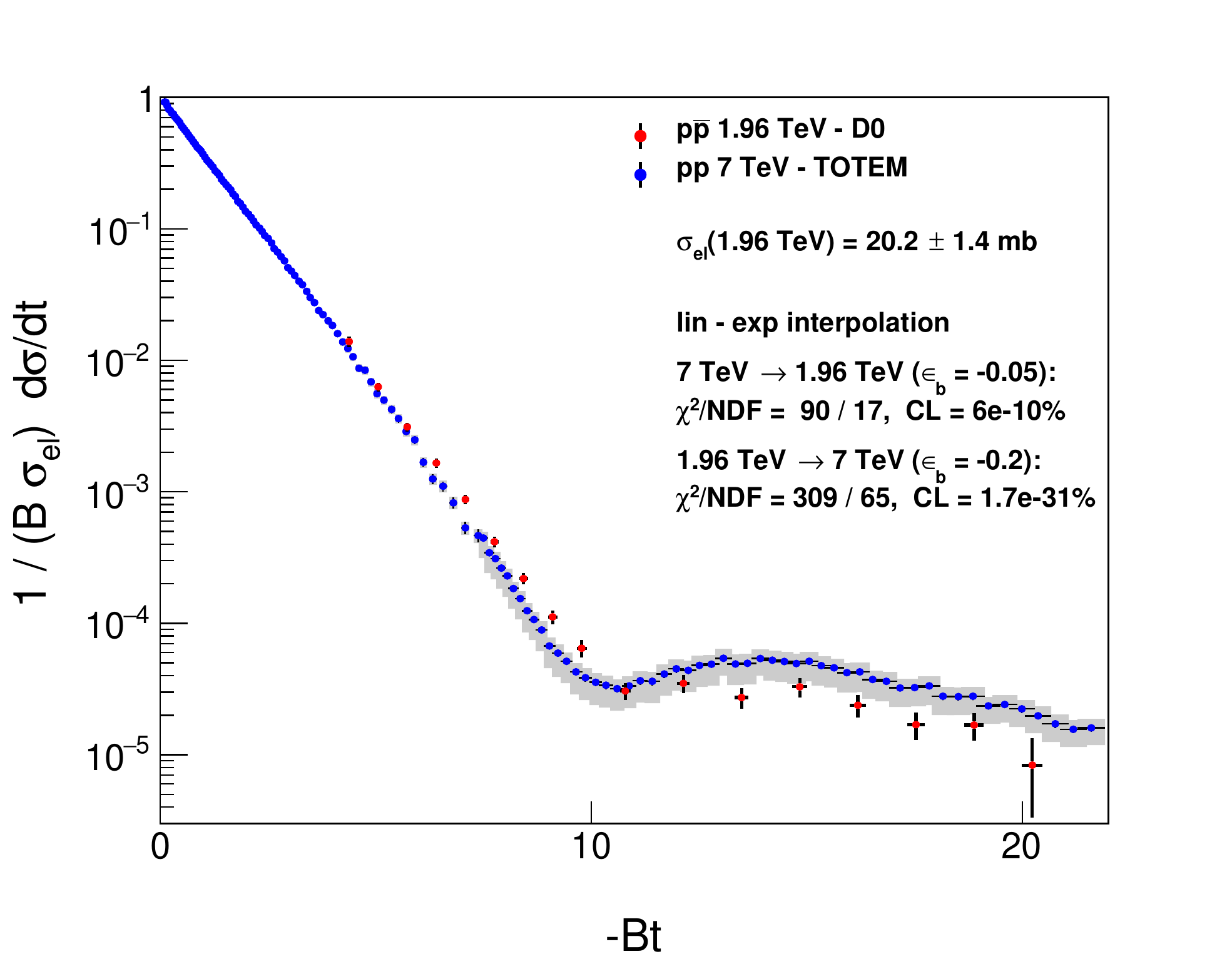}
 \includegraphics[width=0.49\textwidth]{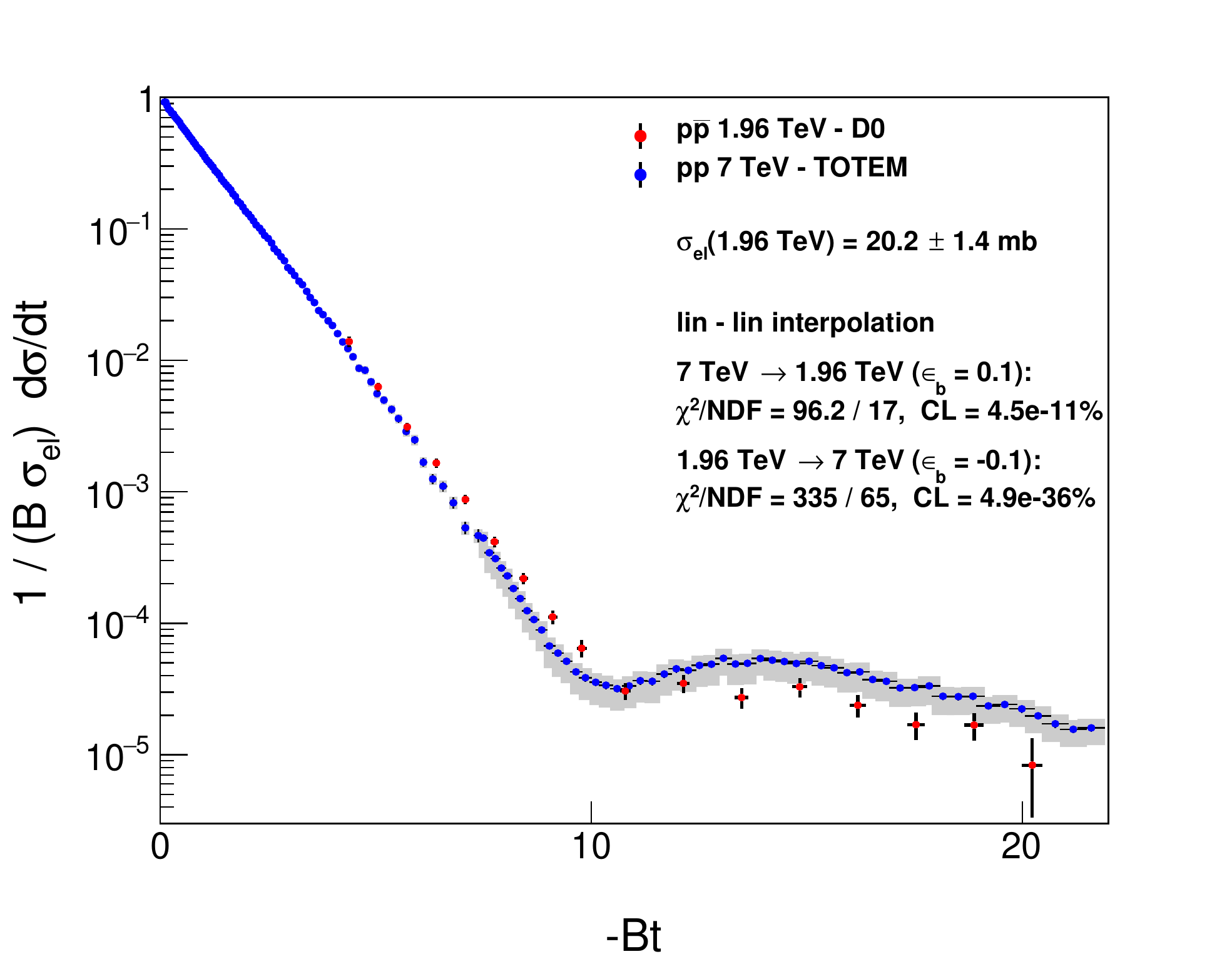}}
\end{minipage}    
\end{center}
\caption{Odderon signal in the comparison of the $H(x)$ scaling functions of $pp$ collisions at $\sqrt{s} = 7$ TeV, measured by the TOTEM experiment at the LHC~\cite{Antchev:2011zz,Antchev:2013gaa}, and $p\bar p$ elastic collisions at $\sqrt{s} = 1.96$ TeV measured by the D0 experiment at Tevatron~\cite{Abazov:2012qb}. The results of this preliminary Odderon observation are indicated on the plots, where the CL is evaluated without the rounding of the $\chi^2$ values to the printed level of precision. The rounded values of $\chi^2 $ and the corresponding CL values are summarized in Table~\ref{table:7-to-1.96-TeV-one-way-comparison}. The final Odderon significance  results are given in \ref{app:A}.
{\it Top-left panel:} This comparison uses 17.6 $\pm$ 1.1 mb for the elastic $p\bar p$ cross section at $\sqrt{s} = 1.96 $ TeV, and a linear-exponential interpolation technique in $(x, H(x))$. This corresponds to the smallest difference between the two data sets.
{\it Top-right panel:} 
Same as the top-left panel but for linear-linear interpolations in the
horizontal and vertical directions. For these interpolations, the nearest data points are connected with lines that correspond to a straight line on a linear-linear plot. 
{\it Bottom-left panel:}
Same as the top-left panel but now using 20.2 $\pm$ 1.4 mb for the elastic $p\bar p$ cross section at $\sqrt{s} = 1.96 $ TeV.
{\it Bottom-right panel:}
Same as the bottom-left panel but using a linear-linear interpolation method.
}
\label{fig:rescaling-from-7-to-1.96TeV-and-back}
\end{figure*}

\begin{figure*}[hbt]
\begin{center}
\begin{minipage}{1.0\textwidth}
  \null \hspace{1truecm}\includegraphics[width=1.0\textwidth]{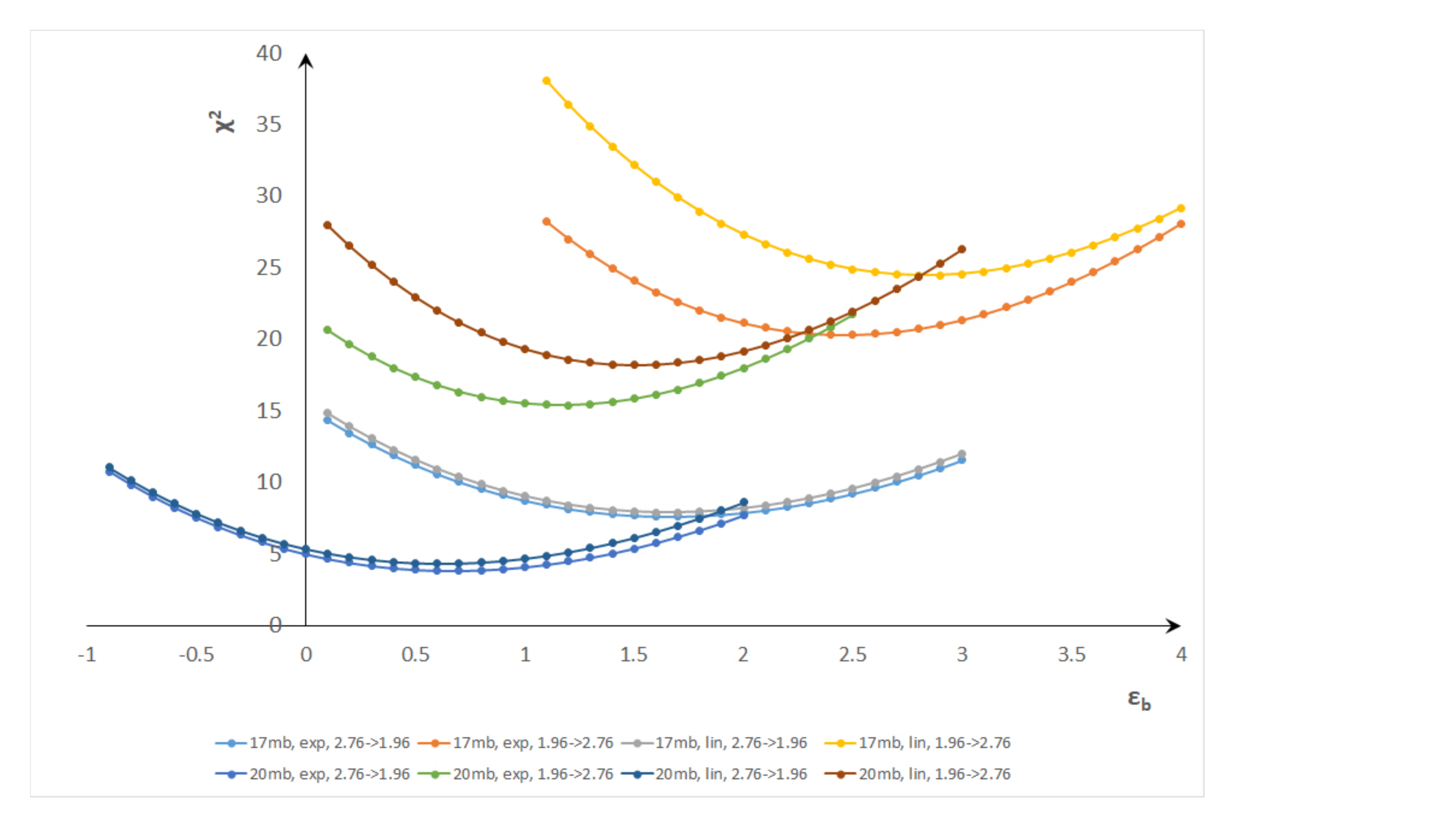}
\end{minipage}    
\end{center}
\caption{
Dependence of $\chi^2$ on the coefficient of the correlated but point-to-point varying systematic errors, $\epsilon_b$, for the comparison of the $H(x)$ scaling functions of elastic $p\bar p$ collisions at $\sqrt{s} = 1.96 $ TeV, measured by the D0 experiment at Tevatron~\cite{Abazov:2012qb}, with that of elastic $pp$ collisions at $\sqrt{s} = 2.76 $ TeV, measured by the TOTEM experiment at the LHC~\cite{Antchev:2018rec}. All the eight cases are shown together corresponding to the choice of linear-linear or linear-exponential interpolations in $H(x)$, to a different choice of the elastic cross section of $p\bar p$ collisions
at $\sqrt{s} = 1.96$ TeV (20.2 $\pm$ 1.4 mb vs 17.6 $\pm$ 1.1 mb), and to the direction of the projection (1.96 $\rightarrow$ 2.76 TeV, or 2.76 TeV $\rightarrow$ 1.96 TeV). A clear parabolic structure is seen in each case and the fit quality of the results that belong to these minima in $\epsilon_b$ is summarized in Table~\ref{table:2.76-to-1.96-TeV-one-way-comparison}.
}
\label{fig:chi2-vs-epsilon-b-for-1960-vs-2760-GeV.png}
\end{figure*}

The detailed figures, that show the $\chi^2(\epsilon_b)$ functions for each of these cases are summarized in the left and right panels of Fig.~\ref{fig:chi2-vs-epsilon-b-for-1960-vs-7000-GeV.png} for the comparison of the 7 TeV TOTEM data set with the 1.96 TeV D0 data set. Each plot indicates a clear, nearly quadratic minimum. The values of $\chi^2$ at the minima are summarized in Table~\ref{table:7-to-1.96-TeV-one-way-comparison}, together with other characteristics of significance, like the confidence level and the significance in terms of standard variations. Similarly, the  $\chi^2(\epsilon_b)$ functions for the comparison of the 2.76 TeV TOTEM data set with the 1.96 TeV D0 data set are summarized in Fig.~\ref{fig:chi2-vs-epsilon-b-for-1960-vs-2760-GeV.png}. The values of $\chi^2$ at the minima are given in Table~\ref{table:2.76-to-1.96-TeV-one-way-comparison}, together with other relevant characteristics.

\begin{table*}[ht]
\begin{center}
\begin{minipage}{1.0\textwidth}
\centerline{
     \includegraphics[width=0.95\textwidth]{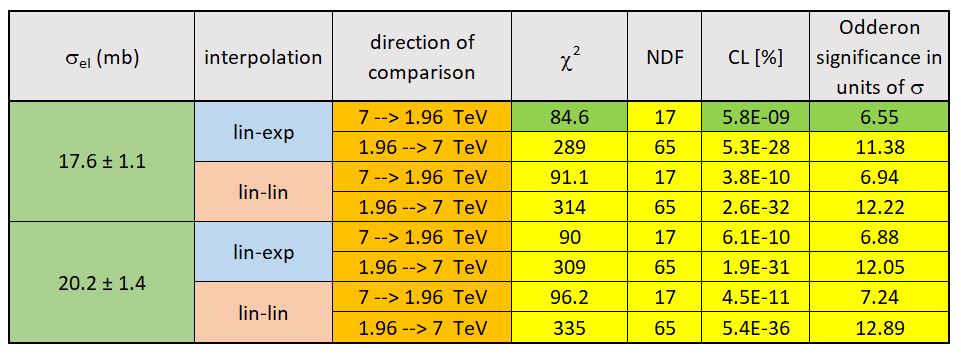}
     }
\end{minipage}    
\end{center}
\caption{Summary table of the significant Odderon signal in the one-way comparison of the $H(x)$ scaling functions of $pp$ collisions at $\sqrt{s} = 7$ TeV measured by the TOTEM experiment at the LHC, and $p\bar p$ elastic collisions at $\sqrt{s} = 1.96$ TeV measured by the D0 experiment at Tevatron. For the projection $1.96$ $\rightarrow$ $7.0$ TeV, very small confidence levels are obtained with $CL < 10^{-27}$ \%, and due to different rounding errors of the two different softwares that we utilized (Root vs Excel), tiny and negligible deviations are also seen between this Table and the more precise values indicated in Fig.~\ref{fig:rescaling-from-7-to-1.96TeV-and-back}.
This table indicates that the Odderon signal is observed in this comparison with at least a 6.55$\sigma$ significance.
In ~\ref{app:A} this is decreased to a significance of at least 6.26 $\sigma$. These significances are
robustly above the 5 $\sigma$ discovery treshold, corresponding to a statistically significant Odderon discovery.
}
\label{table:7-to-1.96-TeV-one-way-comparison}
\end{table*}

As summarized in Fig.~\ref{fig:rescaling-from-7-to-1.96TeV-and-back}, a significant Odderon signal is found in the comparison of the $H(x)$ scaling functions of the differential elastic $pp$ (at $\sqrt{s} = 7.0$ TeV) vs $p\bar p$ ($\sqrt{s} = 1.96$ TeV) cross sections. The horizontal error bars are indicated by a properly scaled horizontal line or ``$-$'' at the data point. The statistical (type A, point-to-point fluctuating) errors are indicated by the size of the vertical error bars ($|$), while shaded boxes indicate the size of the (asymmetric) type B (point-to-point varying, correlated) systematic errors. The overall normalization errors ($|t|$-independent, type C errors) cancel from the  $H(x)$ scaling functions since they multiply both the numerator and the denominator of $H(x)$ in the same way. The correlation coefficient of the $|t|$-dependent systematic errors, $\epsilon_b$, is optimized to minimize the $\chi^2$ based on Eq.~(\ref{e:chi2-final}), and the values indicated in Fig.~\ref{fig:rescaling-from-7-to-1.96TeV-and-back} correspond to the minimum of the $\chi^2(\epsilon_b)$. The location of these minima and the best values of 
$\epsilon_b$ depend on the domain in $x$ or $x$-range from where the contributions to the $\chi^2(\epsilon_b)$ are added up. The stability of our final results with respect to the variation of the $x$-range, together with the correlations between the best value of the  $\epsilon_b$ and the $x$-range are detailed in \ref{app:E}.
These $\chi^2$ values, as well as the numbers of degrees of freedom (NDFs) and the corresponding confidence levels (CLs) are indicated on both panels of Fig.~\ref{fig:rescaling-from-7-to-1.96TeV-and-back}, for both projections. The $\chi^2(\epsilon_b)$ functions are summarized in Fig.~\ref{fig:chi2-vs-epsilon-b-for-1960-vs-7000-GeV.png}. The 7 TeV $\rightarrow $ 1.96 TeV projection has a preliminary statistical significance of 6.55$\sigma$ of an Odderon signal, corresponding to a $\chi^2/{\rm NDF} = 84.6 / 17$ and CL = $5.78 \times 10^{-9}$ \%.   
~\ref{app:A} presents the robustness test of this result, and summarizes the result of our tests of various possible modifications of our $\chi^2 $ definition. It turns out that the symmetry requirement
discussed in  ~\ref{app:A} slightly reduces this significance from a 6.55 $\sigma$ level to a 6.26 $\sigma$ level, safely above the
5.0 $\sigma$  discovery threshold, corresponding to a $\chi^2/{\rm NDF}$ $=$ $80.1/17$
and CL = $3.7 \times 10^{-8}$ \%. Thus the probability of Odderon observation in this analysis is at least $P = 1-CL = 0.99999999963$.

Fig.~\ref{fig:rescaling-from-7-to-1.96TeV-and-back} illustrates some of the results of our systematic studies in four different panels described as follows. The top-left panel of this figure uses a linear-exponential interpolation in the $(x, H(x))$ plane and uses the value of 17.6 $\pm$ 1.1 mb for the elastic $p\bar p$ cross section at $\sqrt{s} = 1.96$ TeV. This case gives the lowest (6.55$\sigma$) significance for the Odderon observation from among the possible cases that we have considered in Fig.~\ref{fig:rescaling-from-7-to-1.96TeV-and-back}. The top-right panel is similar but for a linear-linear interpolation in the $(x, H(x))$. The bottom-left panel is similar to the top-left panel, but now using 20.2 $\pm$ 1.4 mb for the elastic $p\bar p$ cross section at $\sqrt{s} = 1.96 $ TeV and also using a linear-exponential interpolation in $(x, H(x))$. The bottom-right panel is similar 
to the bottom-left panel, but using a linear-linear interpolation method.

\begin{figure*}[hbt]
\begin{center}
\begin{minipage}{0.95\textwidth}
 \centerline{\includegraphics[width=0.49\textwidth]{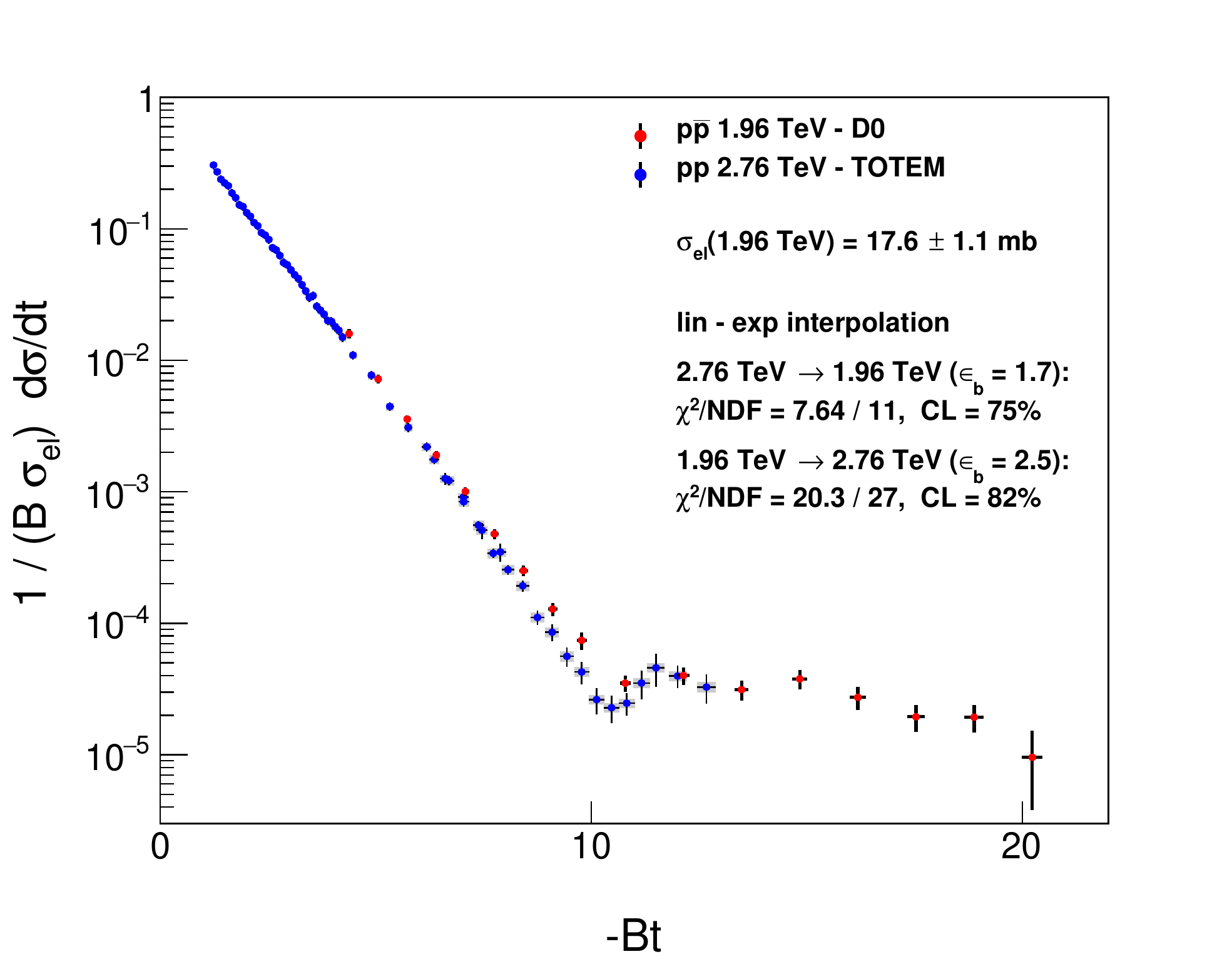}
 \includegraphics[width=0.49\textwidth]{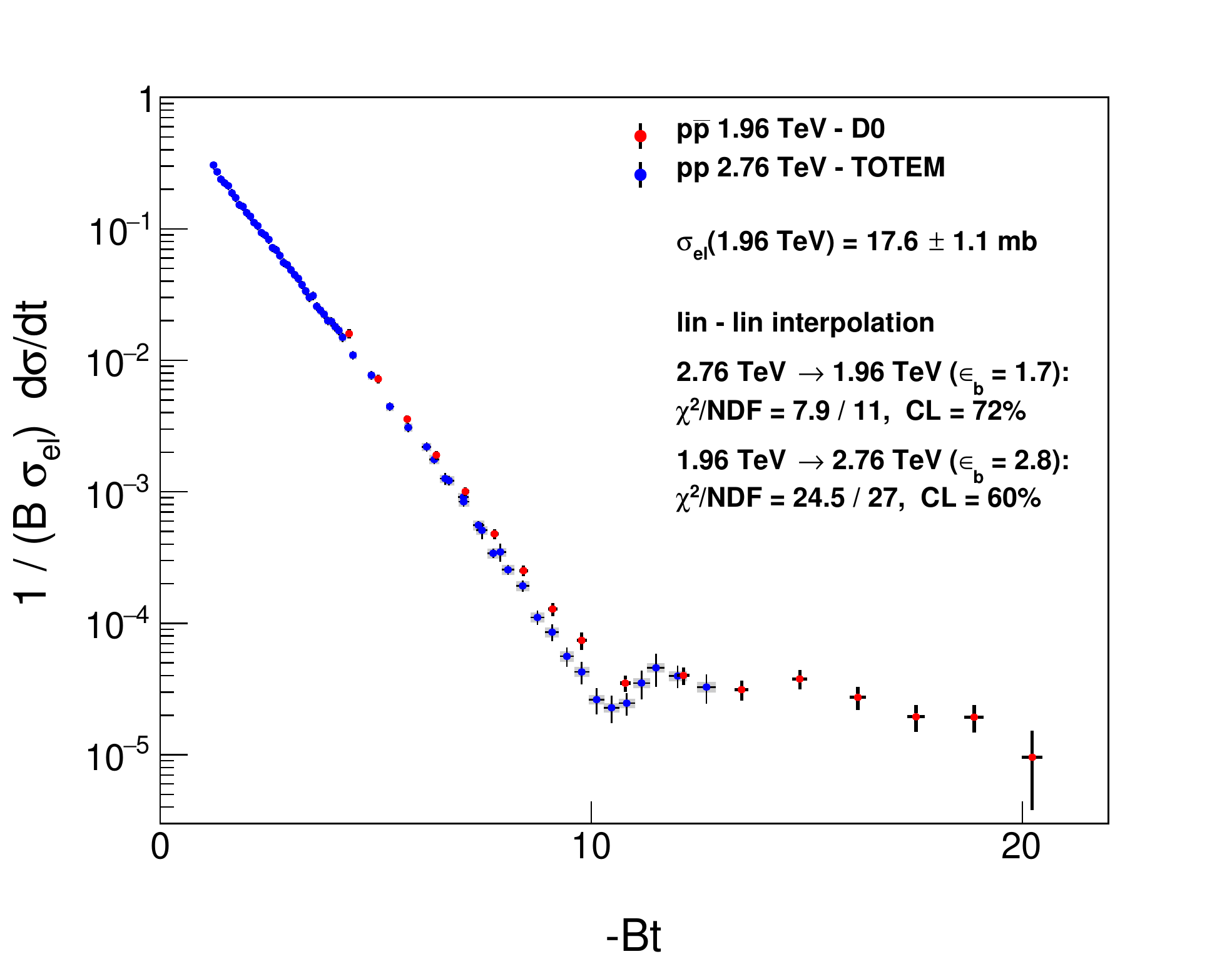}}
 \centerline{\includegraphics[width=0.49\textwidth]{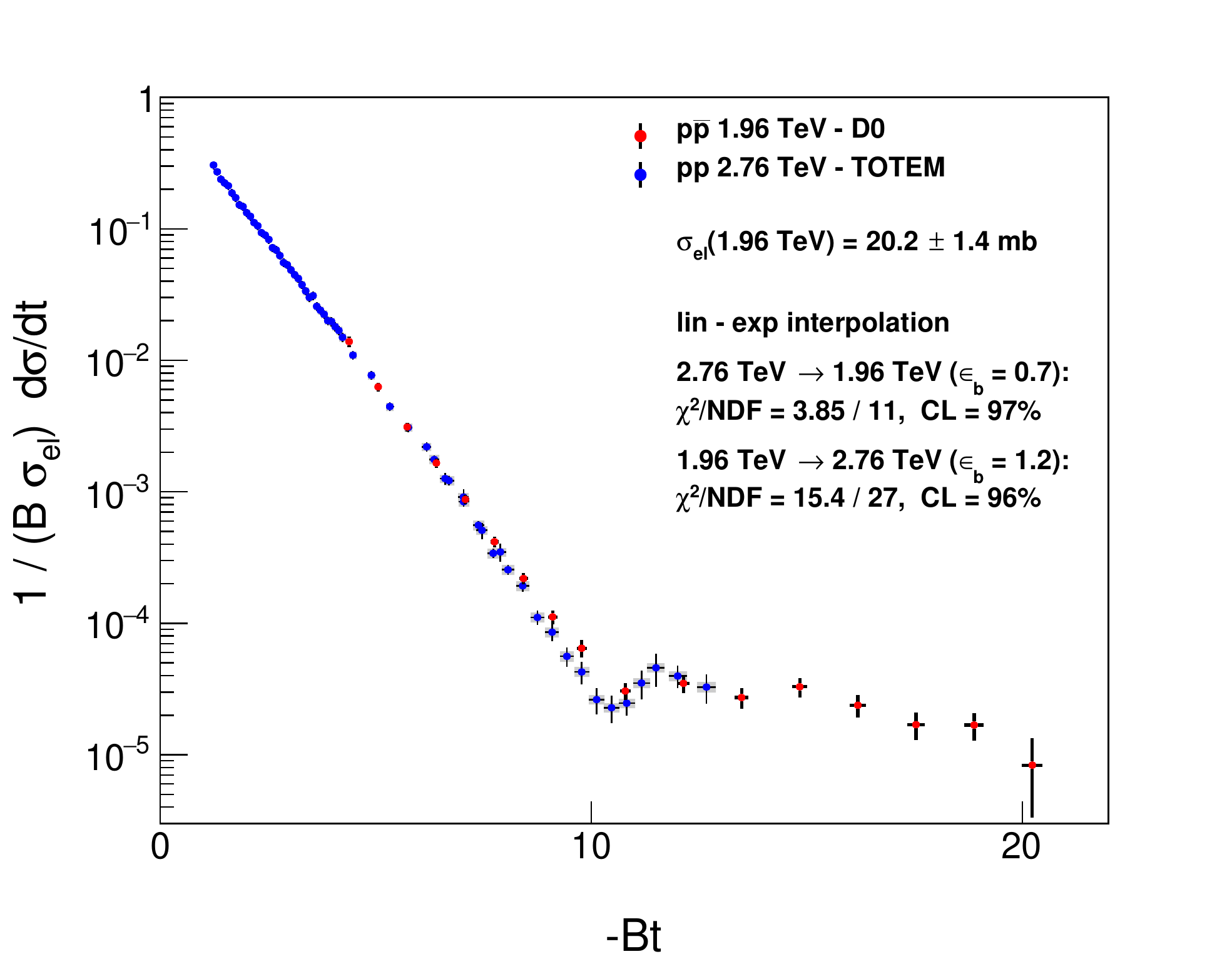}
 \includegraphics[width=0.49\textwidth]{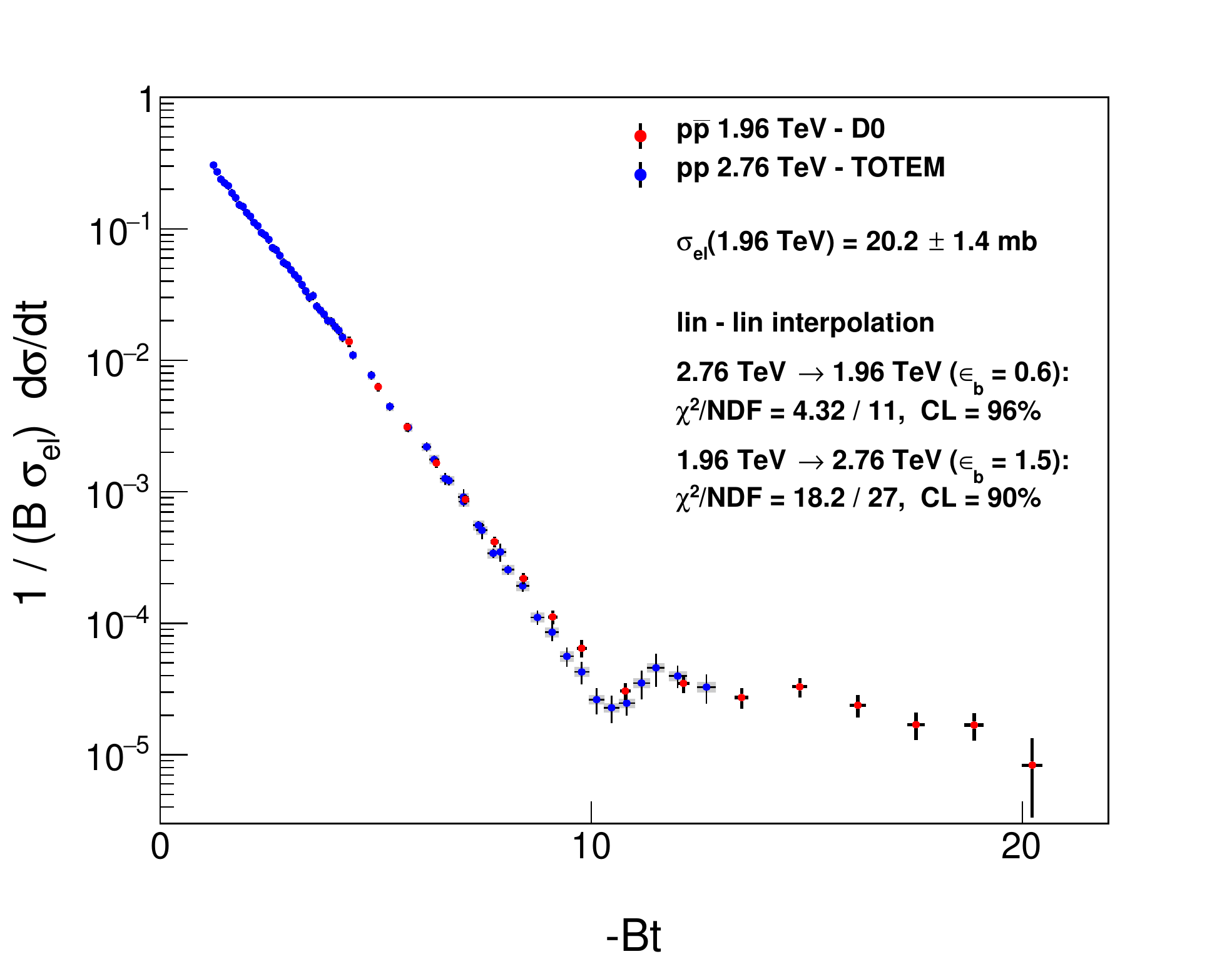}}
\end{minipage}    
\end{center}
\caption{
Lack of a significant Odderon signal in the comparison of the $H(x)$ scaling functions of the differential cross section of elastic $pp$ collisions with
$\sqrt{s} = 2.76$ TeV, measured by the TOTEM~\cite{Antchev:2018rec}, to that of $p\bar p$ collisions with $\sqrt{s} = 1.96$ TeV, measured by D0~\cite{Abazov:2012qb}. The correlation coefficient of the $|t|$-dependent systematic errors, $\epsilon_b$, is optimized to minimize the 
$\chi^2$ based on Eq.~(\ref{e:chi2-final}), and the value indicated on the plot corresponds to the minimum of $\chi^2(\epsilon_b)$. The results of our Odderon search are summarized in Table~\ref{table:2.76-to-1.96-TeV-one-way-comparison}. See also Table~\ref{table:2.76-to-1.96-TeV-two-way-comparison} for a summary of the results of the two-way comparisions of these $H(x)$ scaling functions.
{\it Top-left panel:} Using $\sigma_{\rm el} = 17.6 \pm 1.7$ mb and a linear-exponential interpolation method.
{\it Top-right panel:} Same as the top-left panel but for a linear-linear interpolation in $(x, H(x))$.
{\it Bottom-left panel:} Same as the top-left panel but for $\sigma_{\rm el} = 20.2 \pm 1.4$ mb.
{\it Bottom-right panel:} Same as the bottom-left panel but for a linear-linear interpolation in $(x, H(x))$.
}
\label{fig:rescaling-from-2.76-to-1.96TeV-and-back-17mb-lin-exp}
\end{figure*}

The results of the scaling studies for a comparison of elastic $pp$ collisions at $\sqrt{s} = 2.76 $ TeV, measured by the TOTEM experiment at the LHC~\cite{Antchev:2018rec} to that of $p\bar p$ collisions at $\sqrt{s} = 1.96 $ TeV, measured by D0 at the Tevatron~\cite{Abazov:2012qb} are summarized in Figs.~\ref{fig:chi2-vs-epsilon-b-for-1960-vs-2760-GeV.png} and~\ref{fig:rescaling-from-2.76-to-1.96TeV-and-back-17mb-lin-exp}. The top-left panel of Fig.~\ref{fig:rescaling-from-2.76-to-1.96TeV-and-back-17mb-lin-exp} uses $\sigma_{\rm el} = 17.6 \pm 1.1$ mb and a linear-exponential interpolation method in $(x, H(x))$. The top-right panel is the same as the top-left panel, but for a linear-linear interpolation in $(x, H(x))$. The bottom-left panel is nearly the same as the top-right panel,  but for $\sigma_{\rm el} = 20.2 \pm 1.4$ mb. The bottom-right panel is the same as the bottom-left panel, but for a linear-linear interpolation in $(x, H(x))$. Neither of these comparisions shows a significant difference between the $H(x)$ scaling function of elastic $pp$ collisions at $\sqrt{s} = 2.76 $ TeV as compared to that of $p\bar p$ collisions at 
$\sqrt{s} = 1.96 $ TeV. It seems that the main reason for such a lack of significance is the acceptance limitation of the TOTEM dataset at $\sqrt{s} = 2.76$ TeV, which extends up to $x = - t B \approx 13$, in contrast to the acceptance of the 7 TeV TOTEM measurement that extends up to $x = - tB \approx  20$. We have cross-checked this by limiting the 7 TeV data set also to the same acceptance region of $4.4 < -Bt < 12.7$ as that of the 2.76 TeV data set. This artificial acceptance limitation has resulted in a profound loss of significance, down a to $\chi^2/{\rm NDF} = 25.7 / 11$, that corresponds to a CL = 0.71\% and to a deviation at the 2.69 $\sigma$ level only. This result indicates that if we limit the acceptance of the 7 TeV TOTEM measurement to the acceptance of the 2.76 TeV TOTEM measurement, the significance of the Odderon observation decreases well below the 5$\sigma$ discovery treshold. This result can be understood if we consider, that the diffractive maximum (``bump") is located, if the $H(x)$ scaling is valid, at $x \approx 13$, which is very close but slightly above the
value of the $x_{max} = 12.7$ upper limit of the acceptance in $x$ of the TOTEM data published in Ref.~\cite{Antchev:2018rec}. Fig. 8 of  Ref.~\cite{Antchev:2018rec} indicates that indeed the precise location of the diffractive maximum can not be determined from these TOTEM data, it may be just close to the upper limit of the TOTEM acceptance at $\sqrt{s} = 2.76$ TeV.

We have thus dedicated ~\ref{app:E} to the scrutiny of the $x$-range dependence of the Odderon signal. In particular, we have investigated how important is the contribution from the large values of $x$. We developed and tested  our most conservative $\chi^2$ definition in ~\ref{app:A}.
We have investigated the domain of validity of the $H(x)$ scaling with the help of a model and detailed in
\ref{app:D}, that at $\sqrt{s} = 2.76$ TeV, the $H(x)$ scaling is expected to hold up to $x = 15.1$, well above the TOTEM acceptance of $x < 12.7 $. 
In \ref{app:E}, we find  that it is sufficient to include  a small shift to the investigated $x$ range: already for only 10 datapoints from the  D0 acceptance the significance of the Odderon signal is greater than $5$ $\sigma$ in the $5.1 < x \leq 13.1$ domain at $\sqrt{s} = 1.96$ TeV. In ~\ref{app:E} we also show that the minimum size of subsequent D0 datapoints for a greater than 5$\sigma$ Odderon signal is actually 8 out of 17, corresponding to the $7.0 \le x \le 13.5$ range.

We have performed  several cross-checks: this topic is detailed in the next section.

\begin{table*}[hbt]
\begin{center}
\begin{minipage}{0.97\textwidth}
 \centerline{
    \includegraphics[width=0.95\textwidth]{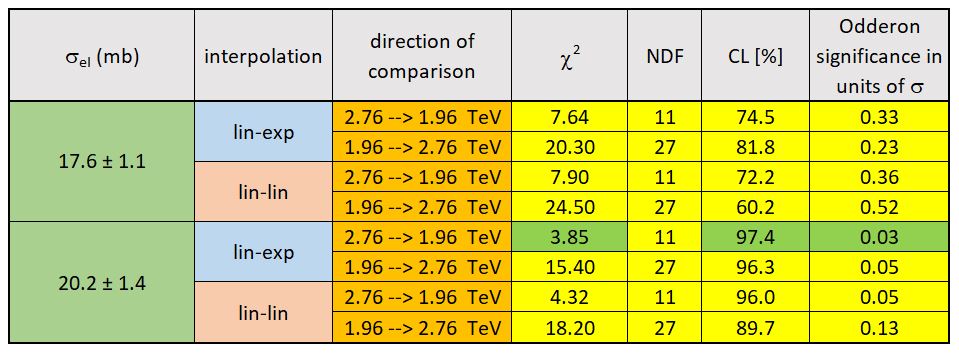}
 }
\end{minipage}    
\end{center}
\caption{
Summary table of the search for an Odderon signal in the one-way comparison of the $H(x)$ scaling functions of $pp$ collisions at $\sqrt{s} = 2.76$ TeV measured by the TOTEM experiment at the LHC, and $p\bar p$ elastic collisions at $\sqrt{s} = 1.96$ TeV measured by the D0 experiment at Tevatron.
}
\label{table:2.76-to-1.96-TeV-one-way-comparison}
\end{table*}

\begin{table*}[ht]
\begin{center}
\begin{minipage}{0.97\textwidth}
\centerline{
\includegraphics[width=0.95\textwidth]{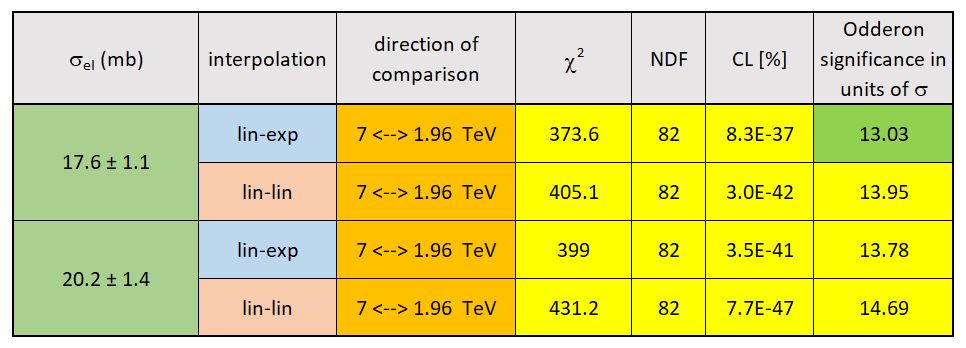}
}
\end{minipage}    
\end{center}
\caption{Summary table of the search for an Odderon signal in the two-way comparison, for the significance of an Odderon signal in the comparison of the $H(x)$ scaling functions of $pp$ collisions at $\sqrt{s} = 7$ TeV, measured by the TOTEM experiment at the LHC, and $p\bar p$ elastic collisions at $\sqrt{s} = 1.96$ TeV, measured by the D0 experiment at Tevatron.
This table indicates that the Odderon signal is observed with at least a 13$\sigma$ significance, when both projections are combined from the previous Table~\ref{table:7-to-1.96-TeV-one-way-comparison}, by adding the $\chi^2$ and the NDF values of both directions of the comparisons. These results are remarkably stable with respect to the choice of the unknown integrated elastic cross section at $\sqrt{s} = 1.96 $ TeV, and also with respect to the choice of the linear-exponential or linear-linear interpolations. This effectively indicates that the combined significance of the Odderon discovery is at least a 13$\sigma$ effect.
}
\label{table:7-to-1.96-TeV-two-way-comparison}
\end{table*}

\begin{table*}[hbt]
\begin{center}
\begin{minipage}{0.97\textwidth}
\includegraphics[width=0.95\textwidth]{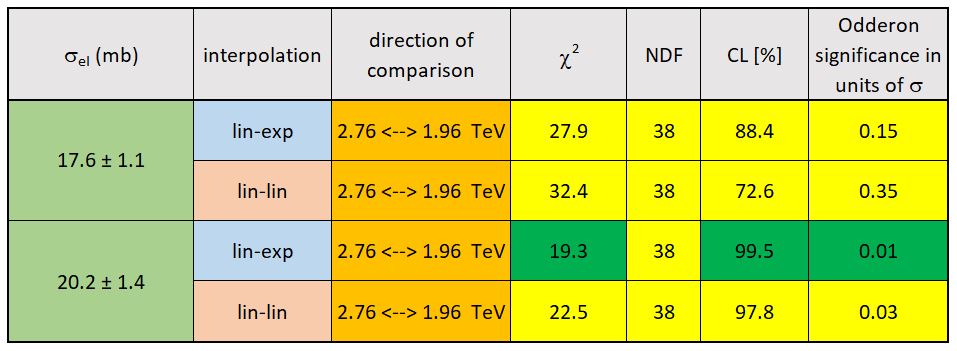}
\end{minipage}    
\end{center}
\caption{
Summary table of the search for an Odderon signal in the two-way comparison of the $H(x)$ scaling functions of $pp$ collisions at $\sqrt{s} = 2.76$ TeV, measured by the TOTEM experiment at the LHC, and $p\bar p$ elastic collisions at $\sqrt{s} = 1.96$ TeV, measured by the D0 experiment at Tevatron. The lowest value of significance in this comparison is found to be 0.01$\sigma$, which means that the $H(x)$ scaling functions of 1.96 TeV $p \bar p$ and 2.76 TeV $pp$ elastic collisions are nearly the same within errors. The level of maximal difference is much less than a 3$\sigma$ effect which does not reach the statistical significance of a discovery effect in this comparison.
}
\label{table:2.76-to-1.96-TeV-two-way-comparison}
\end{table*}

\section{A summary of cross-checks}
\label{s:cross-checks}

In this section, we summarize some of the most important cross-checks that we performed using our methods and results.

We have cross-checked what happens if one rescales the differential cross section of elastic $pp$ scattering form the lowest ISR energy of $\sqrt{s} = 23.5$ GeV to the top ISR energy of $\sqrt{s} = 62.5$ GeV. As can be expected based on the approximate equality of all the $H(x)$ scaling functions at the ISR energies, as indicated on the left panel of Fig.~\ref{fig:rescaling-of-dsigma-dt-at-ISR-and-LHC}, the rescaled 23.5 GeV $pp$ data coincide  with the measured 62.5 GeV $pp$ data. The resulting $\chi^2/{\rm NDF} = 111/100$ corresponds to a CL = 21.3 \%, or a lack of significant difference -- a 1.3$\sigma$ effect. Within errors, our quantitative analysis thus indicates that the two data sets at the ISR energies of 23.5 and 62.5 GeV correspond to the same $H(x)$ scaling function, but with possible small deviations in a small $x$-region around the dip position. 
This indicates that the method that we applied
to extrapolate the 2.76 and 7 TeV data sets to lower energies satisfied the cross-checks at the ISR energies, i.e. our method works  well. As one of the critical cross-checks of these calculations, two different co-authors coded the same formulae with two different codes using two different programming languages, and these codes were cross-checked against one another until both provided the same values of significances.

We have validated the PHENIX method of Ref.~\cite{Adare:2008cg} implemented in the form of the $\chi^2$ definition of Eq.~(\ref{e:chi2-final}) for the diagonalization of the covariance matrix on fits to the $\sqrt{s} = 13 $ TeV TOTEM data
of Ref.~\cite{Antchev:2018edk}. This  PHENIX method resulted, within one standard deviation, the same minimum, hence the same significances,  as the use of the full covariance matrix at $\sqrt{s} = 13 $ TeV elastic $pp$ collisions. At the lower LHC energies of $\sqrt{s} = 2.76$ and $7.0$ TeV, due to the lack of publicly available information on the covariance matrix, only the PHENIX method of Ref.~\cite{Adare:2008cg} was available for our final significance analysis.

We have also explored the main reason of the observation of a significant Odderon signal in the comparision of the $H(x)$ scaling functions of elastic $pp$ collisions at $\sqrt{s} =  7 $ TeV with that of the elastic $p\bar p$ collisions at $\sqrt{s} = 1.96 $ TeV. The question was rather intriguing as we have found no significant difference between the $H(x)$ scaling functions of elastic $pp$ collisions at $\sqrt{s}$ = 2.76 TeV and 7 TeV. At the same time, we also see that the comparison of the 2.76 TeV $pp$ dataset to the 1.96 TeV $p\bar{p}$ dataset does not indicate a significant Odderon effect. We have found that the Odderon signal vanishes from the comparison of the 7 TeV $pp$ and the 1.96 TeV $p\bar{p}$ datasets too, if we limit the acceptance of the 7 TeV dataset to the acceptance in $x = -tB$ as that of the 2.76 TeV $pp$ dataset: the significance of the Odderon observation decreased from an at least 6.26 $\sigma$ discovery effect, detailed in ~\ref{app:A}, to a 2.69$\sigma$ level agreement. We may note that a similar observation was made already in Ref.~\cite{Ster:2015esa} that pointed out a strong $|t|$ dependence of the Odderon contribution.

Table~\ref{table:7-to-1.96-TeV-two-way-comparison} summarises the search for an Odderon signal in the two-way comparison, for the significance of an Odderon signal in the comparison of the $H(x)$ scaling functions of $pp$ collisions at $\sqrt{s} = 7$ TeV and $p\bar p$ collisions at $\sqrt{s} = 1.96$ TeV. Applying this method the Odderon signal is observed with at least a 13$\sigma$ significance, when both projections are combined from Table~\ref{table:7-to-1.96-TeV-one-way-comparison}, by adding the $\chi^2$ and the NDF values of both directions of the comparisons.

\section{Discussion}
\label{s:discussion}

We have explored the scaling properties of the elastic differential cross sections at various energies, from the ISR up to the highest LHC energy. We have recalled that the earlier proposals for the $F(y)$ and $G(z)$ scaling functions were useful to explore if elastic scattering of protons in the LHC energy range is already close to the black-disc limit or not. After investigating several possible new dimensionless scaling variables and scaling function candidates, we have realized that in order to look for scaling violations in the low $|t|$ kinematic range, corresponding to the diffractive cone it is advisable to scale all the diffractive cones to the same dimensionless scaling function, $H(x) \approx \exp(-x)$. This function can be obtained as the differential cross section normalized to its value at the  $x = -tB = 0$ optical point, which also for nearly exponential distributions equals to the elastic cross section $\sigma_{\rm el}$ multiplied by the slope parameter $B$. Both are readily measurable in elastic $pp$ and $p\bar p$ collisions, while other scaling variables that we have investigated may depend on $t_{\rm dip}$ values -- the location of the diffractive minimum. The latter however is not readily accessible neither in elastic $p\bar p$ collisions (where there is no significant dip) nor in the acceptance limited elastic $pp$ differential cross section (where the diffractive minimum or maximum may be located outside the acceptance of the experiment for that particular data set).

The scaling function $H(x)$ of elastic proton-(anti)proton scattering transforms out the energy dependence of the elastic slope $B(s)$ and the elastic cross section $\sigma_{\rm el}(s)$, and due to the relation
$[ 1 + \rho_0^2(s) ] \sigma^2_{\rm tot}(s) = 16  \pi \sigma_{\rm el} (s)$
they also scale out a combination of 
the total cross sections and the real-to-imaginary ratio. As was discussed above, for analytic scattering amplitudes and for differential cross-sections starting with a diffractive cone
at low values of $x = - tB$ the scaling function will have a universal, $H(x) \approx \exp(-x)$ shape.
The price for the removal of these trivial $s$-dependencies from the scaling function is paid by an $s$-depended domain of 
validity, $x_{\rm max}(s)$ which is found to be typically above the position of the diffractive interference region.
Without direct experimental observations, or without theoretical, model-dependent calculations, it is not possible
to determine model-independently this $x_{\rm max}(s)$ function, the $s$-dependent upper limit of the domain of validity of this $H(x) $ scaling.

Figs.~\ref{fig:rescaling-from-7-to-1.96TeV} and ~\ref{fig:rescaling-from-7-to-1.96TeV-and-back} clearly indicate a crossing-odd component of the elastic scattering amplitude. At the $\sim$ 2 TeV energy scale, where the Reggeon contributions to the scattering amplitude are suppressed by their power-law decays, this is apparently a clear Odderon effect, a characteristic difference in the shape of the scaling function of elastic scattering between $pp$ and $p\bar p$ collisions at the logarithmically similar energies of 7 and 1.96 TeV, respectively.

The effects due to the energy-induced difference between TOTEM and D0 data sets can be estimated by the lack of change of the $H(x)$ scaling function for $pp$ scattering between 2.76 TeV and 7 TeV, within the {\it statistical} errors of these TOTEM data sets.
However, the $H(x)$ scaling function of elastic $pp$ scattering at $\sqrt{s} = 7.0$ TeV 
is significantly different from the corresponding result of elastic $p\bar p$ scattering at $\sqrt{s} = 1.96$ TeV. These qualitative and quantitative differences, first, show up well below the diffractive minimum of the $pp$ elastic scattering, namely, the $H(x)$ function for $pp$ collisions indicates a strong ``swing'' or faster than exponential decrease effect, before developing a characteristic interference pattern consisting of a diffractive minimum and subsequent maximum. In contrast, the D0 data on $p\bar p$
elastic scattering features a structureless exponential decrease that in turn changes to a plateaux or a shoulder-like structure at higher values of the scaling variable $x$. No clear indication of a diffractive maximum is seen in the $p\bar p$ elastic scattering data~\cite{Abazov:2012qb}, while the TOTEM data sets at each LHC energies of
2.76, 7 and 13 TeV clearly indicate a diffractive minimum followed by an increasing part of the differential cross section before the edge of the TOTEM acceptance is reached, respectively~\cite{Antchev:2018rec,Antchev:2011zz,Antchev:2018edk}.

These qualitative and quantitative differences between the $H(x)$ scaling functions of elastic $pp$ and $p\bar p$ scatterings provide a clear-cut and statistically significant evidence 
for a crossing-odd component in the scattering amplitude in the TeV energy range. This corresponds to the observation of the Odderon exchange in the $t$-channel of the elastic scattering. The Odderon in this context is a crossing-odd component of the amplitude of elastic $pp$ and $p\bar p$ scattering, that remains significant even in the large $s$ limit. In Regge phenomenology, the Odderon is a trajectory that at $J=1$ contains a $J^{\rm PC} = 1^{--}$ vector glueball as well as other glueball states with higher angular momentum. Hence, one of the implications of our result is that not only one but several glueball states should exist in Nature \cite{Szanyi:2019kkn}.

Due to the presence of the faster-than exponentially decreasing (swing) region in elastic $pp$ scatterings, high-statistic $pp$ elastic scattering data at $\sqrt{s} = 1.96 $ TeV may be taken as an additional measurement clearly closing the energy gap. However, the aperture limitation of the LHC accelerator is already resulting in a loss of significance of the comparison of the $H(x)$ scaling function at 2.76 TeV with that of the D0 data at 1.96 TeV. Due to this reason, we propose an additional measurement of the dip and bump region of elastic $pp$ collisions in the domain where the $H(x)$ scaling was shown to work, in between 2.76 TeV and 7 TeV, if that can be harmonized with the LHC running schedule and scenarios.

The current TOTEM acceptance (the upper end of the last bin of the published TOTEM data) ends at $-t B \approx 13$ at $\sqrt{s} = 2.76$ TeV. This value almost coincides with the bump position of the $H(x)$ scaling function. It seems that
including at least one D0 point to the comparision of the $H(x)$ scaling functions of $pp$ and $p\bar p$ data above the $x =13$ bump position is
sufficient for reaching an at least 5 $\sigma$ significance for the Odderon observation, as detailed in \ref{app:E}.

New elastic $pp$ scattering data around $\sqrt{s} \approx$ 4 -- 5 TeV could be particularly useful to determine more precisely any possible residual dependence of these Odderon effects as a function of $\sqrt{s}$.

The current significance of the Odderon observation may be further increased from the 6.26$\sigma$ effect, but only  by a tedious experimental re-analysis of some of the already published data, for example, by separating the point-to-point uncorrelated statistical and systematic errors (type A errors) from the point-to-point correlated systematic errors in elastic $p\bar p$ collisions by D0, or, by the publication of the covariance matrix of the elastic cross section measurement of $pp$ collisions at 2.76 and 7 TeV colliding energies by TOTEM.  So taking more TOTEM data in special runs at new energies between $\sqrt{s} = 2.76$ and $7.0$ TeV seems to be a more enlightening and inspiring scenario, 
if it can be harmonized with LHC schedule and other ongoing experimental efforts. 

\subsection{Discussion of some of our model-dependent results}

As noted above, the upper limit of the domain of validity of the $H(x)$ scaling, the $x_{max}(s)$ as a function of $s$ cannot be determined model independently, it has to be taken either from extrapolations between different measured points, or from theoretical, model-dependent and validated calculations. Such calculations are presented in ~\ref{app:A},~\ref{app:C} and~\ref{app:D}. One of the most interesting characteristic features of elastic $pp$ scattering at TeV energies is the presence of a single diffractive minimum and maximum in the experimental data on the differential cross-section of elastic $pp$ scattering  at TeV energies. In terms of the theory of multiple diffraction
a single diffractive minimum is obtained if the scattering structures have a two-component internal structure~\cite{Czyz:1969jg}.
The model that we have utilized for the evaluation of $x_{max}(s)$ is based on Refs.~\cite{Bialas:2006kw,Bialas:2006qf,Bialas:2007eg},
where the proton is assumed to have a quark-diquark structure, $p = (q,d)$ and in one variant of this picture, the diquark is further
resolved as a correlated $d = (q,q) $ structure, corresponding to the $p = (q, (q,q))$ case. As detailed in Ref.~\cite{Nemes:2015iia},
this scenario indeed gives too many diffractive minima in the experimental acceptance, so it can be excluded.
Thus our model-dependent results actually also reveal the effective sizes of constituent (dressed) quarks and diquarks inside the protons.

Concerning the quark and diquark sizes, let us note that our values are in qualitative agreement with those obtained first by Bialas and Bzdak in Refs.~\cite{Bialas:2006kw,Bialas:2006qf,Bialas:2007eg} at the ISR energies.
Already in those papers, the binding energy of the diquark was found to be negligibly small,
corresponding to the mass ratio of quarks to diquarks as 1:2. In the model, this mass ratio is  reflected in a fixed
value for the $\lambda = \frac{1}{2}$ parameter, that determines the location of their center of mass to the center of the proton.
The correlated motion of the quark and the diquark gives an important contribution
to the description of the differential cross-section of elastic $pp$ scattering, as the $p = (q,q,q)$ model of three uncorrelated
quarks inside the proton is in a disagreement with the experimental data.

However, the size and existence of the diquarks is a well-known controversy in the literature, related to the interpretation of diquarks.
Many scientists theorize that diquarks should not be considered as particles. Even though they may contain two correlated quarks, they are not colour neutral, and therefore cannot exist as isolated bound states. So instead they tend to float freely inside protons as composite entities; while free-floating they have a size of the order of 1 fm. This also happens to be the same size as the proton itself.
Other theorists analyzing elastic $pp$ scattering in the energy range of $\sqrt{s} = 23.5 - 62.5$ GeV
suggest~\cite{Dosch:2002ai}, that the size of the diquark is much smaller as compared to the size of the protons.

From our Levy studies, published recently in~\cite{Csorgo:2019egs}, it follows, that inside the protons the substructure increases in size, when going from the ISR energies of $\sqrt{s} = 23.5 - 62.5$ GeV to the LHC energy of 7.0 TeV, see Fig. 3 and Tables 1, 2 of that paper. 
So part of the difference of the diquark size in our current manuscript and the sizes obtained in Ref. ~\cite{Dosch:2002ai} 
might be the difference of the investigated energy range, $\sqrt{s} = 23.5 - 62.5$ GeV versus our results on the TeV energy scale.
Another part of these quantitative differences  might be due to the more precise, quantitative, statistically significant level of data description as  presented in our paper.  The comparison of the diquark sizes is a quantitative question, 
and it is difficult to make a quantitative comparison with models that were used to describe certain qualitative features of the experimental data, without aiming at a data description on statistically acceptable, significant level.

\section{Summary and conclusions}
\label{s:summary}

We have introduced a new, straightforwardly measurable scaling function $H(x)$ of elastic 
proton-(anti)proton scattering. This scaling function transforms out the trival energy-depen\-dent factors, in particular, the effects due to the $s$-de\-pend\-encies stemming from the elastic slope $B(s)$, from the real-to-imaginary ratio $\rho_0(s)$, as well as from the total and elastic cross sections, $\sigma_{\rm tot}(s)$ and $\sigma_{\rm el}(s)$, respectively. In our numerical re-analysis of already published TOTEM data, the $H(x)$ scaling is observed from a comparison of the $pp$ elastic scattering data at $\sqrt{s} = 2.76$ and  $7$ TeV,
without theoretical assumptions. TOTEM preliminary data at $\sqrt{s} = 8$ TeV are also in the scaling limit, however, published TOTEM
data at $\sqrt{s} = 13$ TeV indicate significant  violations of this new scaling. The theoretical background of this scaling law
is simple and straightforward in the diffraction cone, where $H(x) \approx \exp(-x)$, as detailed in Subsection~\ref{ss:Hxcone}. 
However, the range of the validity of this scaling extends well beyond the diffraction cone already at ISR energies, as shown in Fig.~\ref{fig:scaling-ISR-x}. A straightforward theoretical derivation for non-exponential $H(x)$ scaling functions was presented in
Subsection~\ref{ss:Hx-dip-bump}.

When comparing the $H(x)$ scaling function of the differential cross section of elastic $pp$ collisions at $\sqrt{s} = 2.76$ and $7.0$ TeV colliding energies, we find no qualitative differences. At ISR energies, in a limited energy region of $23.5 \leq \sqrt{s} \leq 62.5$ GeV,
the $H(x)$ scaling curves are approximately $s$-independent, with a possible small scaling violation in the region of the diffractive minimum.

Such a lack of energy evolution of the $H(x)$ scaling function 
of the $pp$ collisions, even outside the diffractive cone, is in a qualitative contrast with the evolution of the $H(x)$ scaling functions 
of $p\bar p$ collisions at energies of $\sqrt{s} = 0.546 - 1.96$ TeV, where a qualitative and significant energy 
evolution is seen in the $x = -t B > 8 $ kinematic range for all the investigated energies. 
This way, we have found a qualitative difference between elastic $pp$ and $p\bar p$ collisions in terms of their 
$H(x,s)$ scaling functions: these functions are not $s$-independent outside the diffractive cone for $p\bar p$ collisions,
while they are approximately $s$-independent in elastic $pp$ collisions even outside the diffractive cone.

Such a lack of energy evolution of the $H(x)$ scaling function  of the $pp$ collisions, 
the $H(x)$ scaling as a property of the data in the few TeV energy range provides a strong constraint on model-building.
Several simple models, like the simple eikonal amplitude of one-Pomeron-exchange lead to the violation of such a $H(x)$ scaling. It follows that in the   few TeV energy range, where the $H(x)$ scaling is found to be valid, one-Pomeron exchange cannot be the only contribution to the scattering amplitude. 

The main part of our manuscript deals with the quantification of this qualitative  Odderon signal, to determine if it is statistically significant, or not.

Figs.~\ref{fig:rescaling-from-7-to-1.96TeV} and ~\ref{fig:rescaling-from-7-to-1.96TeV-and-back} clearly illustrate a qualitative and a quantitative difference between the scaling properties of the elastic $pp$ and $p\bar p$ collisions, corresponding to a crossing-odd component of the elastic scattering amplitude at the TeV energy scale. As in this kinematic region the Reggeon contributions to the scattering amplitude are suppressed by their power-law decays, a significant characteristic difference between the $H(x)$ scaling functions of elastic $pp$ and $p\bar p$ collisions at the logarithmically similar energies of 7, 2.76 and 1.96 TeV is a clear-cut Odderon effect, because the trivial energy dependences of $\sigma_{\rm el}(s)$ and $B(s)$ as well as that of $(1+\rho^2_0(s))\sigma^2_{\rm tot}(s)$ are scaled out from $H(x)$ by definition.

A comparison in Fig.~\ref{fig:rescaling-from-7-to-1.96TeV-and-back} indicates a significant difference between the rescaled 7 TeV $pp$ data set down to 1.96 TeV with the corresponding $p\bar p$ data measured at $\sqrt{s} = 1.96 $ TeV.
Thus the re-analyzed D0 and TOTEM data, taken together with the verified energy independence of the $H(x)$ scaling function in the $\sqrt{s} = 2.76 - 7.0$ TeV energy range amount to the closing of the energy gap between 2.76 and 1.96 TeV in model-independent way, as much as reasonably possible without a direct measurement, provided that the $H(x)$ scaling is valid for $pp$ scattering in the kinematic range,
where D0 measured the differential cross-section of elastic $p\bar p$ scattering at $\sqrt{s} = 1.96 $ TeV.

We have dedicated~\ref{app:B} to relate the crossing-odd and crossing-even contributions to the 
elastic $pp$  and $p\bar p$ scattering to a model independent and  unitary framework, formulated in the impact parameter space.
We have specialized the results of~\ref{app:B} using the model of Ref.~\cite{Csorgo:2020wmw} in~\ref{app:C}
and demonstrated how the $H(x)$ scaling and $H(x,s)$, the collision energy dependent  violations of this scaling
can be evaluated with the help of this model~\cite{Csorgo:2020wmw}. Note that these calculations are based on
R. J. Glauber's multiple diffraction theory of elastic scattering, assuming that elastic scattering reveals a quark-diquark structure 
inside the scattered protons. 
In~\ref{app:D},  we have determined, model dependently, the domain of validity of the $H(x)$ scaling in $x = -tB$ 
at $\sqrt{s} = 1.96$ TeV. According to Fig.~\ref{fig:App-D-20.2-reBB_model_hx_s-cross-check-at-1.96-TeV}, the upper limit for the
domain of validity of the $H(x)$ scaling at $\sqrt{s} = 1.96$ TeV may include the whole D0 acceptance, with $x_{max}(s) \geq 20.2$ .
This plot is directly obtained by fitting the $H(x)$ scaling limit of the ReBB model to D0 data. In this fit, three out of the four model parameters are constrained by the $H(x)$ scaling, so only one physical parameter had to be fitted to achieve a beautiful agreement 
in a statistically acceptable manner. Another estimate for the upper limit for the domain of validity of the $H(x)$ scaling was
obtained by comparing the predicted 1 standard deviation error bands of the $H(x)$ scaling limit of the ReBB model with 
the same error band, obtained from the full model calculations that included scaling violations too. This result gave a more conservative
upper limit, $x_{max}(s) = 15.1$ at $\sqrt{s} = 1.96$ TeV.
Due to these theoretical uncertainties of the upper limit in $x$ of the domain of validity of the $H(x)$ scaling at
$\sqrt{s} = 1.96$ TeV, the highest colliding energy, where $p\bar p$ elastic scattering data are available,  we have investigated the stability of the Odderon signal within the theoretically determined domain of
validity of the $H(x)$ scaling in \ref{app:E}.


Our final significance analysis is presented in ~\ref{app:A}, resulting in an at least 6.26$\sigma$, discovery level Odderon effect, if  the $H(x)$ scaling is valid in the full kinematic range of the D0 measurement, $0 < x = -tB \leq 20.2$, corresponding to a $\chi^2/{\rm NDF}$ $=$ $80.1/17$ and CL = $3.7 \times 10^{-8}$ \%. 
The probability of this  Odderon signal   is at least $P = 1-{\rm CL} = 0.99999999963$. 
According to our model dependent calculations, as presented in
Fig.~\ref{fig:App-D-20.2-reBB_model_hx_s-cross-check-at-1.96-TeV}, the assumption  that the domain of validity
of the $H(x)$ scaling at $\sqrt{s} = 1.96$ TeV includes the whole D0 acceptance with $x_{max}(s) \geq 20.2$, is consistent with the D0
data at the 2.69 $\sigma$ level.

We have also performed an $x$-range stability analysis, also
model independently, in \ref{app:E}. We established,  that the significance of the Odderon is greater than $5$ $\sigma$,
if the $H(x)$ scaling is valid in the $7 < x = - tB \leq 13.5$ kinematic domain at $\sqrt{s} = 1.96$ TeV. 
However, we could not determine model independently, what is
the domain of validity of the $H(x)$ scaling at this energy, as there are no measured $pp$ data at $\sqrt{ s} = 1.96$ TeV.
So we have included a model dependent estimate of this $x$-range.
Using the model of Refs.~\cite{Bialas:2006kw,Nemes:2015iia,Csorgo:2020wmw}, that was shown to describe all the experimental data in elastic
$pp$ and $p\bar p$ scattering in the $0.546 \leq \sqrt{s} \leq 8$ TeV energy interval and in the $4.4 < x = -tB $ domain, we
found that the validity in $x$ of the $H(x)$ scaling at  $\sqrt{s} = 1.96$ TeV may extend up to $x \leq 15.1$, as detailed in~\ref{app:D}. This interval or domain of validity  of the $H(x) $ scaling includes the smaller $7 < x \leq 13.5$ domain, 
where the signal is larger than 5 $\sigma$. Thus the model independent and at least 5 $\sigma$ discovery level Odderon signal
is remarkably stable for the variations of the domain of validity of the $H(x)$ scaling at $\sqrt{s} = 1.96$ TeV: 9 out of the 17 D0 datapoints can be discarded, 5 at large $x$ and 4 points at low $x$, and the signal still remains significant enough for a discovery.
When we have fitted the $H(x)$ scaling limit of the same model to $p\bar p$ experimental data, as presented in
Fig.~\ref{fig:App-D-20.2-reBB_model_hx_s-cross-check-at-1.96-TeV}, we found that  $x_{max}(s) \geq 20.2$ is also consistent with the
description of the D0 data at the 2.69 $\sigma$ level. In general, the $H(x)$ scaling is not valid for $p\bar p$, but in the Real Extended Bialas-Bzdak model, the $H(x)$ scaling of elastic $pp$ collisions induces a one parameter fit to elastic $p\bar p$ collisions, as
three out of four physical model parameters are the same in this model for $pp$ and $p\bar p$ collisions.

Our $x$-range stability analysis, detailed in ~\ref{app:E}, indicates that part of the statistically significant
contribution to this Odderon signal is coming from the kinematic range of $x < 10 $: excluding this region  decreases
the significance of the crossing-odd signal below the discovery level. 
It is thus important to measure elastic scattering cross-sections at LHC at large $-t$, well beyond the diffractive cone. Elastic $pp$ scattering data in a vicinity of $\sqrt{s}\approx $ 2 TeV as well as in between 2.76 and 7 TeV would  be most useful for further detailing the Odderon properties.  Similarly, part of the signal is coming from large $x$ region, as excluding the kinematic range $x > 12.1$  also results in a loss of significance.

In order to determine where the important contributions to this signal are originating from, we have divided the $0< x \leq 20.2$ kinematic range of acceptance to four regions,
the diffractive cone, the swing, the diffractive interference and the tail regions, corresponding to
$0 < x \leq 5.1$, $5.1 < x \leq 8.4$, $8.4 < x \leq 13.5$ and $13.5 < x \leq 20.2$, respectively, with 2 D0 points associated with the diffractive cone, and 5-5 D0 points in each of the remaining three regions. 
We have shown that the type B, point dependent but overall correlated errors and their correlation coefficient plays an important role in this  analysis and that the best value of the correlation coefficient is $x$-range
dependent. This means that locally shifting the datapoints up or down in a specific interval the agreement between the $pp$ and $p\bar p$ measurements can be improved in that particular interval. We have performed the interval dependent optimizations and found that the locally optimized contributions from the swing and from the tail are not as important as the contributions from the  diffractive interference region, that includes the diffractive minimum and the maximum. The second most important contribution comes from the
swing region, where the $H(x)$ scaling function
for elastic $pp$ collisions bends below the exponential. The combination of the swing and the diffractive interference region already provided a more than 5 $\sigma$ effect, as detailed in ~\ref{app:E}.
If we consider that the interpolations needed to compare
the $H(x)$ scaling functions in a model-independent manner at different energies behave as theoretical curves (i.e. have only one kind of error) then we obtain the significance of at least 6.55$\sigma$ as discussed in the body of this manuscript. This significance further decreases to 6.26$\sigma$ if we consider that these interpolation lines do not have theoretical type of errors but both type A and type B, point-to-point fluctuating and point-to-point correlated systematic experimental errors as well. The only way to further decrease the significance of the Odderon signal is to limit the kinematic range of the comparison to narrower and narrower ranges in $x =-tB$.

Our final significance for an Odderon signal of $6.26\sigma$ is presented from the model-independent analysis in ~\ref{app:A},
which relies on the validity of the $H(x)$ scaling in the $0 < x \leq 20.2$ kinematic range at $\sqrt{s} = 1.96 $ TeV.
The self-consistency of this assumption is shown in Fig.~\ref{fig:App-D-20.2-reBB_model_hx_s-cross-check-at-1.96-TeV}.

Let us emphasize, for the sake of completeness, that we find an interesting hierarchy of significances.

Based on model dependent considerations, the $x$-range of the $p\bar p$ data of the D0 experiment
might be narrowed and correspondingly, the significances may decrease, as more and more
datapoints are removed. Model dependently, we estimate, that the $H(x)$ scaling may be valid at the D0 energy of $\sqrt{s} = 1.96 $ TeV
up to $x_{max} = 15.1$. In the $0 < x \leq 14.8$, theoretically limited $x$-range the Odderon signal remains greater than $5.3$ $\sigma$,
according to \ref{app:E}. We have investigated how far this domain can be narrowed down under the condition that the Odderon signal
remains greater than a 5 $\sigma$ effect? We found that in the very much narrowed from below and from above domain, 
corresponding to the $7.0 < x \leq 13.5$ interval that includes
only 8 out of the 17 D0 points, the Odderon signal that we analyzed has a significance that is greater than a 5 $\sigma$, discovery level effect. This range is well below the theoretically estimated $x_{max} = 15.1$ limit.

In~\ref{app:B}, we discuss the model-independent properties of the Pomeron and Odderon exchange at the TeV energy scale, under the condition that this energy is large enough that the Pomeron and Odderon exchange can be identified with the crossing-even and the crossing-odd contributions to the elastic scattering, respectively. We demonstrated here that $S$-matrix unitarity strongly constrains the possible form of the impact parameter dependence of the Pomeron and Odderon amplitudes.

In \ref{app:C}, we demonstrated how the $H(x)$ scaling emerges within a specific model, defined in Ref.~\cite{Nemes:2015iia}. This model is one of the possible models in the class discussed in~\ref{app:B}. We have demonstrated that four conditions must be satisfied simultaneously. One of the conditions for the validity of the $H(x)$ scaling is found to be the $s$-independence of the experimentally measured $\rho(s)$ parameter. The decrease of $\rho(s)$ at the currently maximal LHC energy of $\sqrt{s} = 13 $ TeV as measured by the TOTEM Collaboration in Ref.~\cite{Antchev:2017yns} thus provides a natural explanation for the violation of the $H(x)$ scaling at these energies.

In \ref{app:D} we estimate the domain of validity of the $H(x)$ scaling also in a model-dependent manner, using the same model of Ref.~\cite{Nemes:2015iia}. Surprisingly, we found that after carefully taking into account the possible quadratic in $\ln(s)$ energy dependencies of the scale parameters of the model of Ref.~\cite{Nemes:2015iia} and after taking into account the correlations between these model parameters,
the kinematic range for the domain of validity of this new $H(x)$ scaling may extend to a very broad range of $200$ GeV $\leq \sqrt{s}\leq$ 8 TeV,
however, with a range that gradually narrows in $-t = x/B(s)$ with decreasing energies.

 Another key point to recognize is that if we allow for a model-dependent analysis, the significance goes further up to
7.08$\sigma$ as detailed in Ref.~\cite{Csorgo:2020wmw} and summarized in \ref{app:E}. There is a trade-off effect in the background of this.
Model dependent calculations lead to a reduction of significance at $1.96$ TeV, as the extrapolation of $pp$ differential cross-section
becomes more uncertain, as compared to the extrapolation with the help of the $H(x)$ scaling. However, this loss in significance is overcompensated by the gain in the possibility to extrapolate the $p\bar p$ differential cross-sections up in energy. If we utilize only
the $H(x)$ scaling, it allows only to compare $pp$ data with $p\bar p$ data at decreasing energies, but $p\bar p$ data do not obey a $H(x)$ scaling law, so it cannot be used to compare them with $pp$ data at 2.76 TeV. But this extrapolation becomes possible with the help of a model calculation and it results in a huge increase in the Odderon significance, as detailed in \ref{app:E} and in ref.~\cite{Csorgo:2020wmw}.

TOTEM  data on an approximately energy independent ratio  of the differential cross-section at the diffractive maximum and minimum
~\cite{Nemes:2019nvj} indicate, that the expected upper limit for the domain of validity of the $H(x)$ scaling is at least 13.5
at $\sqrt{s} = 1.96$ TeV, as detailed at the end of \ref{app:A}. This experimental insight also suggests, that
the domain of validity of $H(x) $ scaling at $\sqrt{s} = 1.96$ TeV  includes the  $7.0 < x \le 13.5$ domain. 
These  observations combined with our model independent
$x$-range stability studies in \ref{app:E} indicate, that in this $7.0 < x \le 13.5$  domain, the significance
of the Odderon observation is at least 5.0 $\sigma$.

This $7.0 < x \leq 13.5$ interval physically begins with the ``swing", where the differential cross-section of $pp$ elastic scattering starts to bend below the exponential shape and ends just after the diffractive maximum or ``bump", located at $x_{bump} \approx 13.0$ .
For the theoretically expected domain of validity, the limited $0 < x \leq 15.1$ range, we find that our method provides an Odderon significance of at least 5.3 $\sigma$, as detailed in \ref{app:E}.

Recently, the STAR collaboration measured the differential cross-section of elastic $pp$ collisions at $\sqrt{s} = 200$ GeV~\cite{Adam:2020ozo}. This measurement resulted in a straight exponential differential cross-section in the range of $ 0.045 \leq -t \leq 0.135$ GeV$^2$. This range is the range where $H(x) = \exp(-x)$ and the conditions of the validity of the $H(x)$ scaling are indeed satisfied by this dataset, that is however limited to a rather low $-t$ range. It is thus a very interesting and most important experimental cross-check for the validity of the $H(x)$ scaling to push forward the experimental data analysis of elastic $pp $ collisions at the top RHIC energy of $\sqrt{s} = 510$ GeV including if possible a larger $-t$ range extending to the non-exponential domain of $\frac{d\sigma}{dt}$ as well.
 
In conclusion, we find from a model-independent re-analysis of the scaling properties of the differential cross sections of already published D0 and TOTEM data sets a statistically significant, more than a 6.26$\sigma$ Odderon effect, based on the assumption of the 
validity of the $H(x)$ scaling at $\sqrt{s} = 1.96$ TeV in the $0 < x \leq 20.2$ kinematic range.
Based on theoretical considerations we estimated that the domain of validity of the $H(x)$ scaling at this
particular energy might be actually smaller, $0 < x \leq 15.1$. So we have also determined
what is the minimum size of the domain of validity of the $H(x)$ scaling that corresponds to the 5$\sigma$
 level Odderon significance. As detailed in ~\ref{app:E}, any interval that fully includes the $7.0 < x \leq 13.5$ range
results in a greater than 5 $\sigma$, discovery level Odderon signal. The experimentally observed~\cite{Antchev:2018rec,Csorgo:2019fbf} energy independence of the 
diffractive maximum - to - minimum ratio $R(s)$ also supports that the domain of the $H(x)$ scaling at $\sqrt{s} = 1.96$ TeV
extends above the diffractive maximum, which is at $x_{bump} = 13$ in this scaling limit.

We thus find  a statistically significant, greater than 5$\sigma$
signal of $t$-channel Odderon exchange, that is robust for variation of the lower or upper limit of the domain of validity
of the $H(x) $ scaling at $\sqrt{s} = 1.96 $ TeV. We have highlighted a hierachy of significances, including experimental and theoretical, model dependent results too. If theoretical modelling is also taken into account, the combined significance of Odderon observation increases to 
at least 7.08$\sigma$, as shown in \ref{app:E}.

Whatever we tried the significance of the Odderon observation remained safely above the 5$\sigma$ discovery threshold, with the most conservative significance estimate detailed in~\ref{app:A}. An $x$-range stability analysis, summarized in \ref{app:E} indicates, that the only way to decrease this signal is to decrease the $-t = x/B(s)$ range of the comparison i.e. deleting data from the signal region.

We have validated the surprisingly large domain of $H(x)$ scaling with already published data both in elastic proton-proton and in proton-antiproton collisions and are eagerly waiting for upcoming results from the STAR collaboration to test our new scaling
in elastic proton-proton collisions at the top RHIC energy of $\sqrt{s} = 510$ GeV.

\appendix

\section{Cross-checking the $\chi^2$ definition: symmetric treatment}
\label{app:A}

This Appendix summarizes our final,  conservative and robust estimate of the significance of the Odderon observation 
in the compared D0~\cite{Abazov:2012qb} and
TOTEM~\cite{Antchev:2018rec,Antchev:2011zz,Antchev:2013gaa} 
datasets, at
$\sqrt{s} = 1.96$ TeV for $p\bar p$ and at $\sqrt{s} = 2.76$ and $7$ TeV for $pp$ elastic scattering.
Here we  compare the considered data sets in a symmetric 
manner, and also mention some of the several robustness and quality tests that we have performed.   

As a cross-check and a robustness test, we have validated the method with the help of a Levy-fit of Ref.~\cite{Csorgo:2018uyp}, confirming that both methods (the fit with the full covariance matrix and the method described below) gave within one standard deviation the same minima with MINOS errors, error matrix accurate, fit in converged status and a statistically acceptable confidence level of CL $ \ge $ 0.1 \% . As a robustness test, the same analysis was repeated by two different co-authors of this manuscript using two different programming codes written in two different programming languages, providing the same results. In order to test the robustness of the results, we have 
tried different possible definitions of $\chi^2$ and the  values reported in this Appendix correspond to the lowest possible significances, that we obtained when we used all the measured data in the signal region. Further reduction seems to be possible only by removing data from the Odderon signal region.

Let us also stress that our investigations were model-independent, as they were based on the direct comparison of the
$H(x)$ scaling functions of various experimentally measured data sets. However, we could not determine the domain of validity of the
$H(x)$ scaling, as a function of $s$. We have shown in the beginning of Section~\ref{s:Scalings}, that at low values of $x$, in the diffractive cone, for an analytic scattering amplitude $H(x) \approx \exp(-x)$, and in these cases, the lower limit of validity of the $H(x)$ scaling
corresponds to $x_{min}(s) \equiv 0$. However, $x_{max}$, the upper limit of the domain of validity of the $H(x)$ scaling at a given value of $s$ is an $s$-dependent function, $x_{max} \equiv x_{max}(s)$, that can be determined only in a model dependent manner. We have evaluated
$x_{max}(s)$ at $\sqrt{s} = 1.96$ and $0.546$ TeV, where $p\bar p$ data are available, based on the evaluation of the $pp$ differential cross-sections from the model of Ref.~\cite{Csorgo:2020wmw} and the $H(x)$ scaling limit of the same model.

\begin{figure*}[hbt]
\begin{center}
\begin{minipage}{1.0\textwidth}
\centerline{
\includegraphics[width=0.6\textwidth]{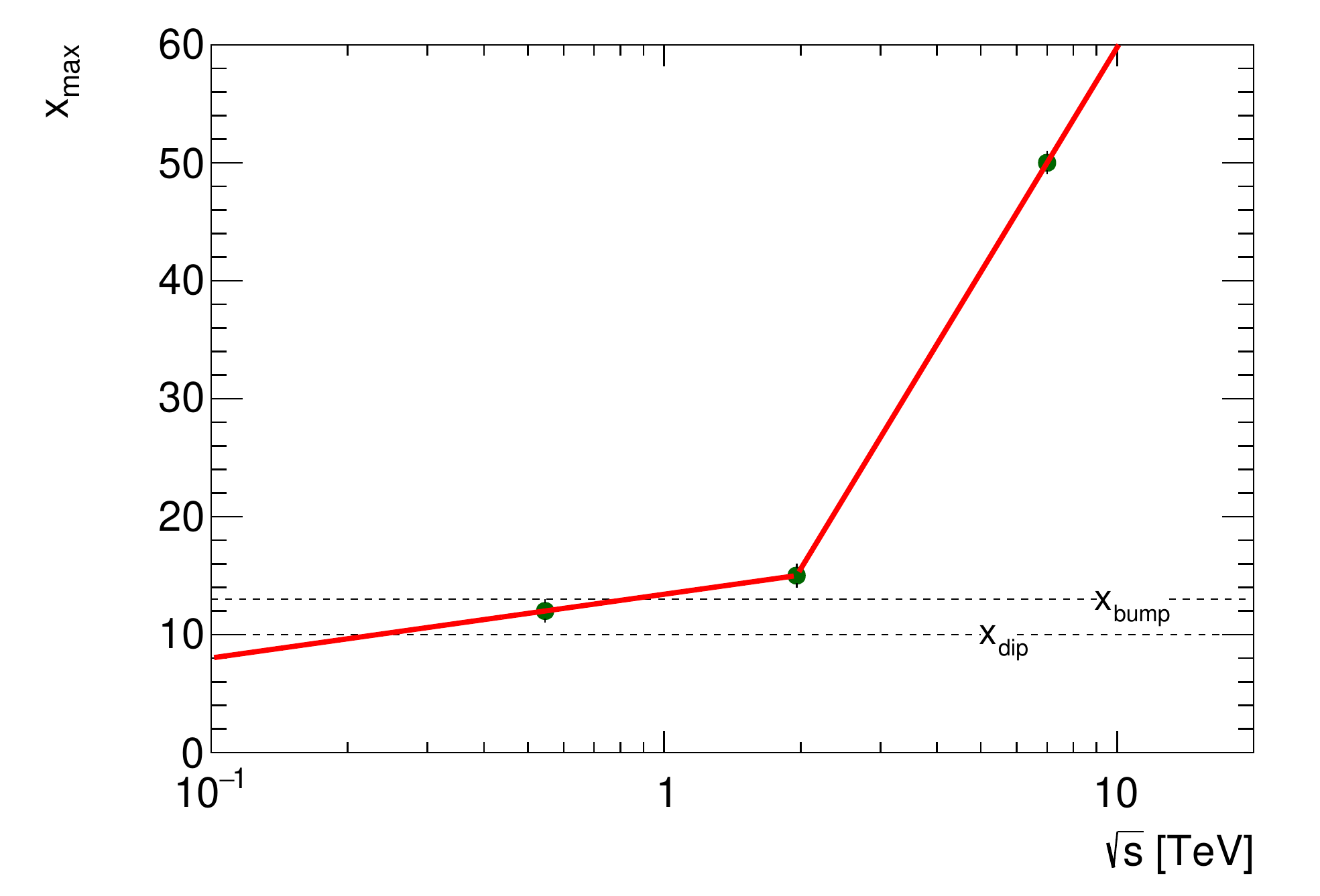}
}
\end{minipage}    
\end{center}
\caption{
The calculated upper limit of the $s$-dependent domain of validity of the $H(x)$ scaling, $x_{max}(s)$, is indicated by the dots and their corresponding error bars. Their $s$-dependent  extrapolation, linear in $\ln(s)$ is also shown, while the points are determined assuming that $H(x,s) = H(x, s_0)$ with $\sqrt{s_0} = 7 $ TeV, and comparing the resulting differential cross-sections from the $H(x)$ scaling limit of the ReBB model of Ref.~\cite{Csorgo:2020wmw} with the validated $s$-dependent calculations that include scaling violations, too. If the $H(x)$ scaling is valid, the dip position $x_{dip} \approx 10$ and the position
of the diffractive maximum or bump, $x_{\rm bump} \approx 13$ are independent of $s$, as indicated on the same plot.
}
\label{fig:App-A-xmax-vs-ln-s}
\end{figure*}

\begin{figure*}[hbt]
\begin{center}
\begin{minipage}{1.0\textwidth}
\centerline{
\includegraphics[width=0.5\textwidth]{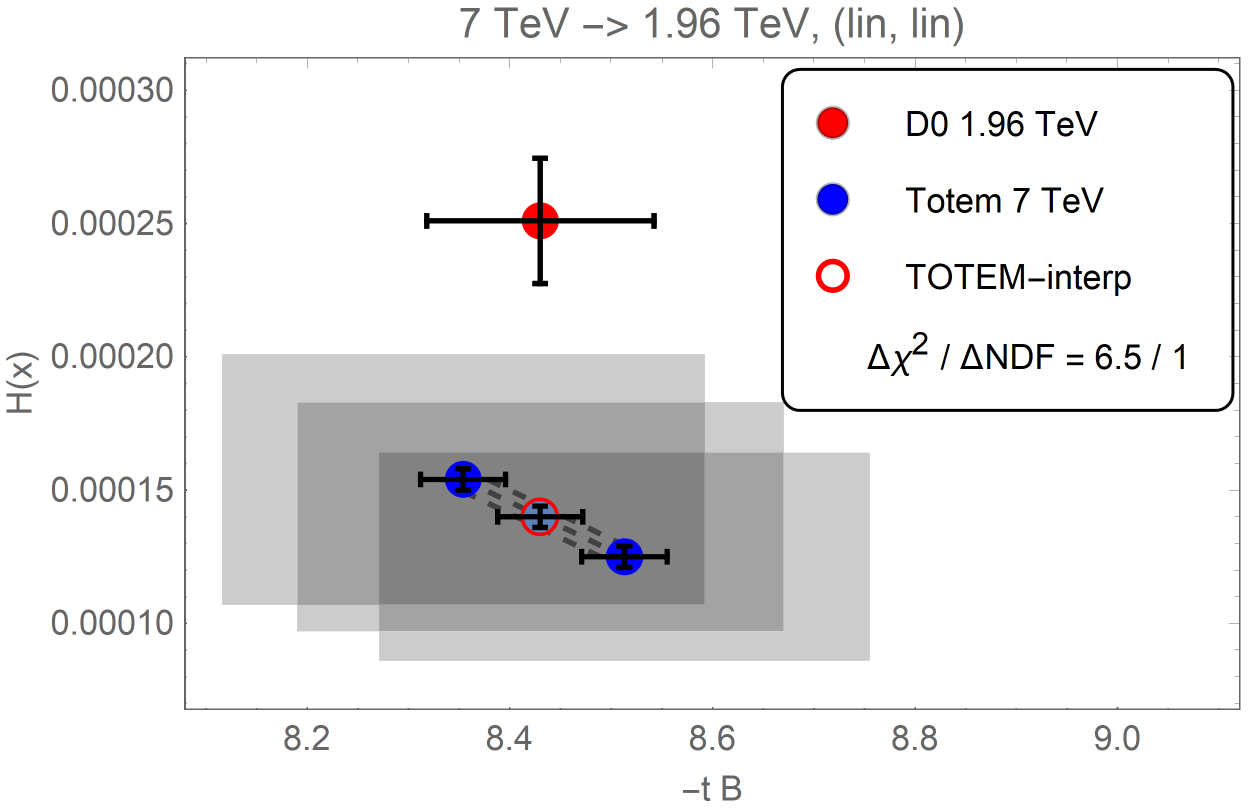}
\includegraphics[width=0.5\textwidth]{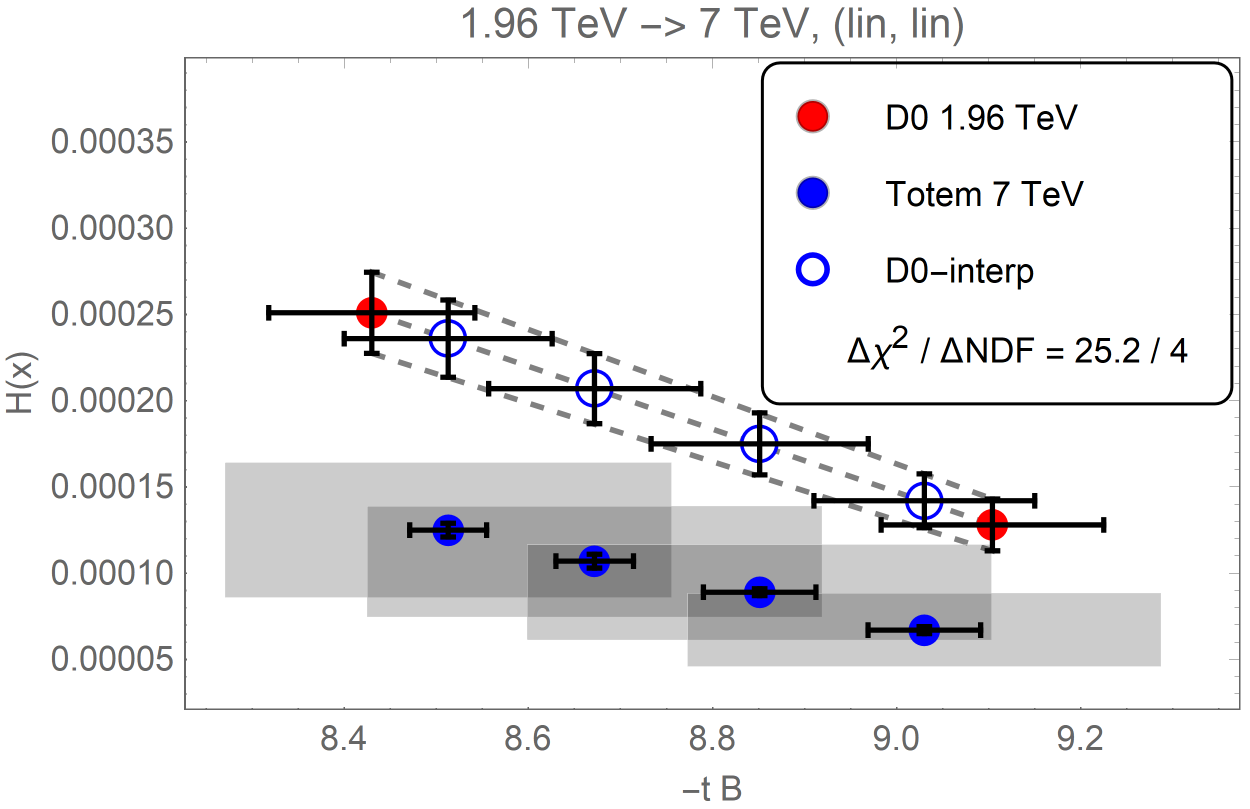}
}
\end{minipage}    
\end{center}
\caption{
Left panel indicates the interpolations that are needed for the $7$ $\rightarrow$ $1.96$ TeV
projection of the TOTEM $H(x)$ scaling function of elastic $pp$
collisions from $\sqrt{s} = 7.0$ TeV to that of elastic $p\bar p$ collisions of  D0 at $\sqrt{s} = 1.96$ TeV, taking only one particular point
of D0 and shown on a linear, linear horizontal and vertical scale. This D0 point is in between two nearby TOTEM points, that are interpolated to the same value of $x$, including both type A, point-to-point fluctuating vertical and horizontal errors (indicated by vertical
and horizontal lines, respectively) and type B, point-to-point 100 \% correlated vertical and horizontal errors,
indicated by the horizontal and  vertical size of the shaded boxes around the central values of the data.
The plot illustrates the lines of interpolations for these central values and the vertical type A errors only,
but the vertical type B errors and the horizontal type A and B errors are interpolated similarly, using straight line
segments, for a linear horizontal and  a linear vertical scale. As part of the systematic checks, the linear interpolation
is repeated also for a linear horizontal and a logarithmic vertical scale, as the low $x$ part of the $H(x)$ distribution
is nearly exponential, that looks like a straight line on such a linear-logarithmic scale. 
The right panel indicates the projection in the reversed direction: the interpolations during the $1.96$ $\rightarrow$ $7$ TeV projection. In this case, the D0 data do not have type B uncertainties, hence the shaded boxes are not shown, neither around the D0 points,  nor around their interpolated values. However, between two D0 datapoints, more than one TOTEM datapoints are measured, so the linear interpolation is utilized to evaluate the value of the D0 measurement at more than one values of $x$. Due to the larger linear segments in $x$, this method yields larger deviations hence more significant differences between the datasets. To estimate the final significances, we have utilized the smallest possible differences,
namely linear extrapolations in the $7$ $\rightarrow$ $1.96$ TeV on a linear horizontal and logaritmic vertical scales.
The partial contribution to 
the change of $\chi^2$ and $NDF$ is illustrated on both panels.
}
\label{fig:illustrations}
\end{figure*}

As the measurements were performed at different values of the 
horizontal axis $x = - Bt$, some interpolations were however inevitable. The technical aspects of these interpolations were
detailed in section~\ref{s:quantification}. Let us illustrate these interpolations by the two panels of Fig.~\ref{fig:illustrations}. We hope, that 
these pictures illustrate the model independent nature of our analysis more clearly than the technical description of these interpolations.

Our final quantification of the Odderon significance is based on a method developed by the PHENIX 
collaboration in Ref.~\cite{Adare:2008cg} using a specific $\chi^2$ definition that effectively 
diagonalizes the covariance matrix. We utilized the measured differential cross-section of elastic $pp$ scattering
and its published covariance matrix at $\sqrt{s} = 13$ TeV, as
measured by TOTEM in Ref.~\cite{Antchev:2018edk}, for a validation of this method. 
We have adopted the PHENIX method of the diagonalization of the covariance matrix using type A, B and C errors~\cite{Adare:2008cg}.

In its original form, the experimental 
data that have statistical and systematic errors are compared to a theoretical calculation that is assumed to be 
a function of the fit parameters. In our final analysis, presented in this Appendix, we adapted the PHENIX 
method for comparison of  a dataset that contains data with errors directly to another  dataset, that also contains central data values with errors. So our method is defined without referring to any theoretical model or parameter dependent function. Due to this reason, the most conservative
definition described in this Appendix is defined to be {\it symmetric for the exchange of the two datasets}.

We classify the experimental errors of a given data set into three different types: (i) type A, 
point-to-point fluctuating (uncorrelated) systematic and statistical errors, (ii) type B errors 
that are point-to-point dependent, but 100\% correlated systematic errors, and (iii) type C errors, 
that are point-to-point independent, but fully correlated systematic errors~\cite{Adare:2008cg}
to evaluate the significance of correlated data, when the covariance matrix is not publicly available. 

Suitably generalizing the method of Ref.~\cite{Adare:2008cg}, that was developed originally for a comparison of a dataset with a theoretical, parametric fit curve,
for a comparison of two data sets in this case, where the datasets have type A, B and cancelling type C errors, we obtain the significance of a projection of the data set $D_{2}$ to data 
set $D_{1}$ determined by the following $\chi^2$ definition:
\begin{eqnarray}
\nonumber
\tilde{\chi}^2_{21} & = &\sum_{j=1}^{n_{21}}
    \frac{
    (d_{1}(j) +\epsilon_{b,1} e_{B,1}(j)  
    - d_{21}(j) - \epsilon_{b,2 1} e_{B,21}(j) )^2 }
    {\tilde e_{A,1}^2(j) + \tilde e_{A,21}^2(j)} \\ &+&
      \epsilon_{b,1}^2 + 
            \epsilon_{b,21}^2 \, .
\label{e:chi2-final-Appendix}
\end{eqnarray}
In this equation, $\tilde e_{A,1}(j)$ is the type A uncertainty of the data point $j$ of the data set $D_{1}$ in the united acceptance, while $\tilde e_{A,21}(j)$ is the same for the $D_{21}$ data set obtained from the $D_2$ dataset by interpolation to point $j$ of dataset $D_1$. Both uncertainties are scaled by a multiplicative factor such that the fractional uncertainty is unchanged under multiplication by a point-to-point varying factor:
\begin{eqnarray}
\tilde{e}_{A,1}(j) & = & e_{A,1}(j) \left( \frac{d_{1}(j) + \epsilon_{b,1}
e_{B,1}(j)}{d_{1}(j)}\right),  \, \\
\tilde{e}_{A,21}(j) & = & e_{A,21}(j) 
\left( \frac{d_{21}(j) +\epsilon_{b,21} 
e_{B,21}(j)}{d_{21}(j)}\right) .
\label{eq:tildesigma-Appendix}
\end{eqnarray}
In these equations, $\epsilon_{b,1}$ and  $\epsilon_{b,21}$ stand for the overall correlation coefficient of the $j$-dependent, point-to-point correlated type B $D_{2}$ to the measured values in data set $D_{1}$, $e_{B,21}(j)$ of the projected data set $D_{21}$. Note that $\epsilon_{b,1}$ and $\epsilon_{b,21}$ are independent of the point $j$, while the B-type errors have a point-to-point changing values $e_{B,1}(j)$ and $e_{B,21}(j)$ in both $D_1$ and in the projected dataset in $D_{21}$. For our comparison of $H(x)$ scaling functions, where the absolute normalization and type C errors cancel, we have $\epsilon_{c,1} = \epsilon_{c,2} = 0$, so have not indicated these terms for the sake of simplicity.
For the sake of clarity and to demonstrate the importance of scaling out these type C errors,
we have also included Fig.~\ref{fig:rescaling-dsigmadt-from-7-to-1.96TeV-with-type-C}. That plot indicates
that if the overall correlated, type C errors are added (incorrectly) in quadrature with the point-to-point
fluctuating type A errors, the significance of the Odderon signal is decreased from 
6.26 $\sigma$ to 3.64 $\sigma$.

\begin{figure*}[hbt]
\begin{center}
\begin{minipage}{1.0\textwidth}
\centerline{
\includegraphics[width=0.75\textwidth]{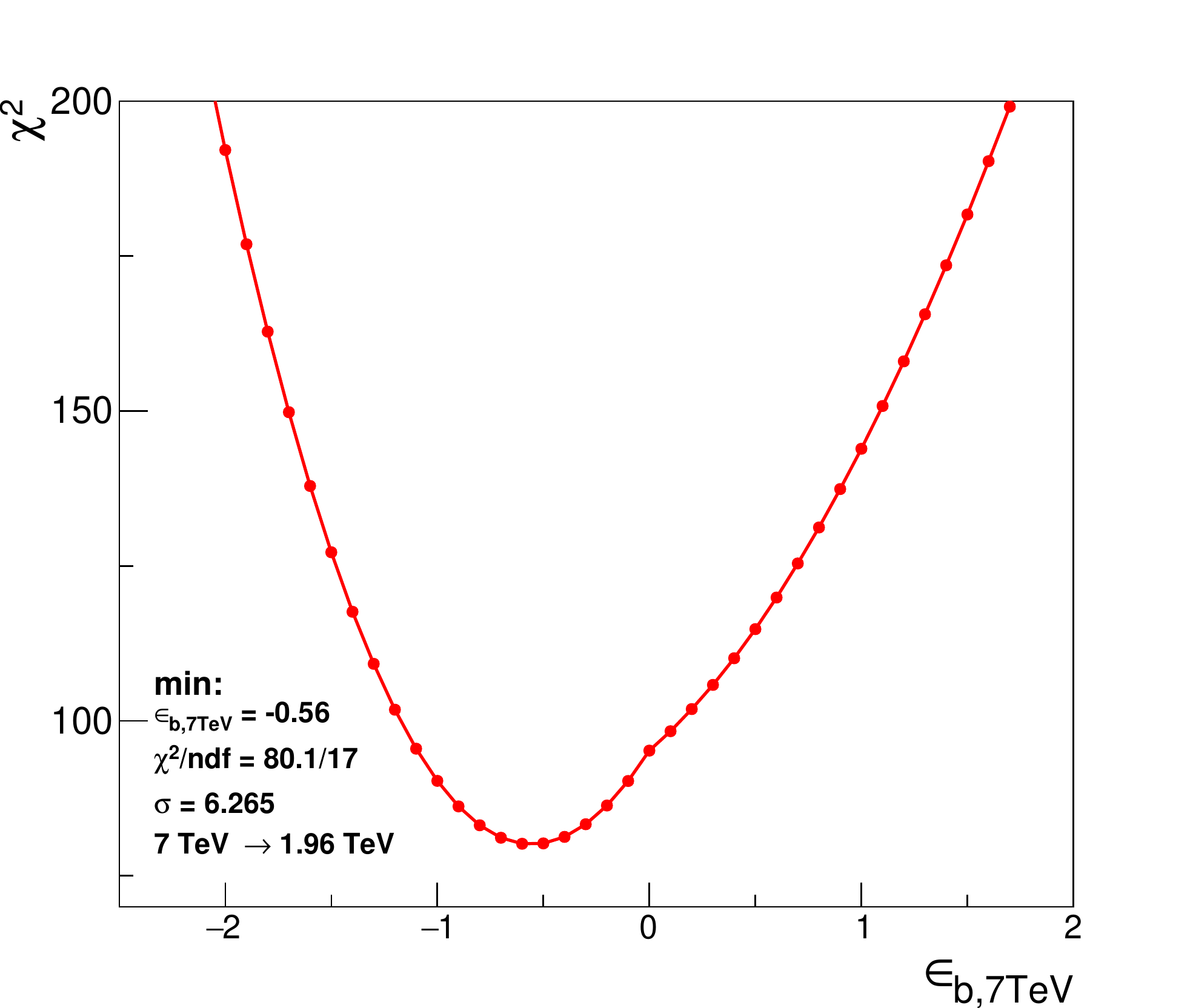}}
\centerline{
\includegraphics[width=0.75\textwidth]{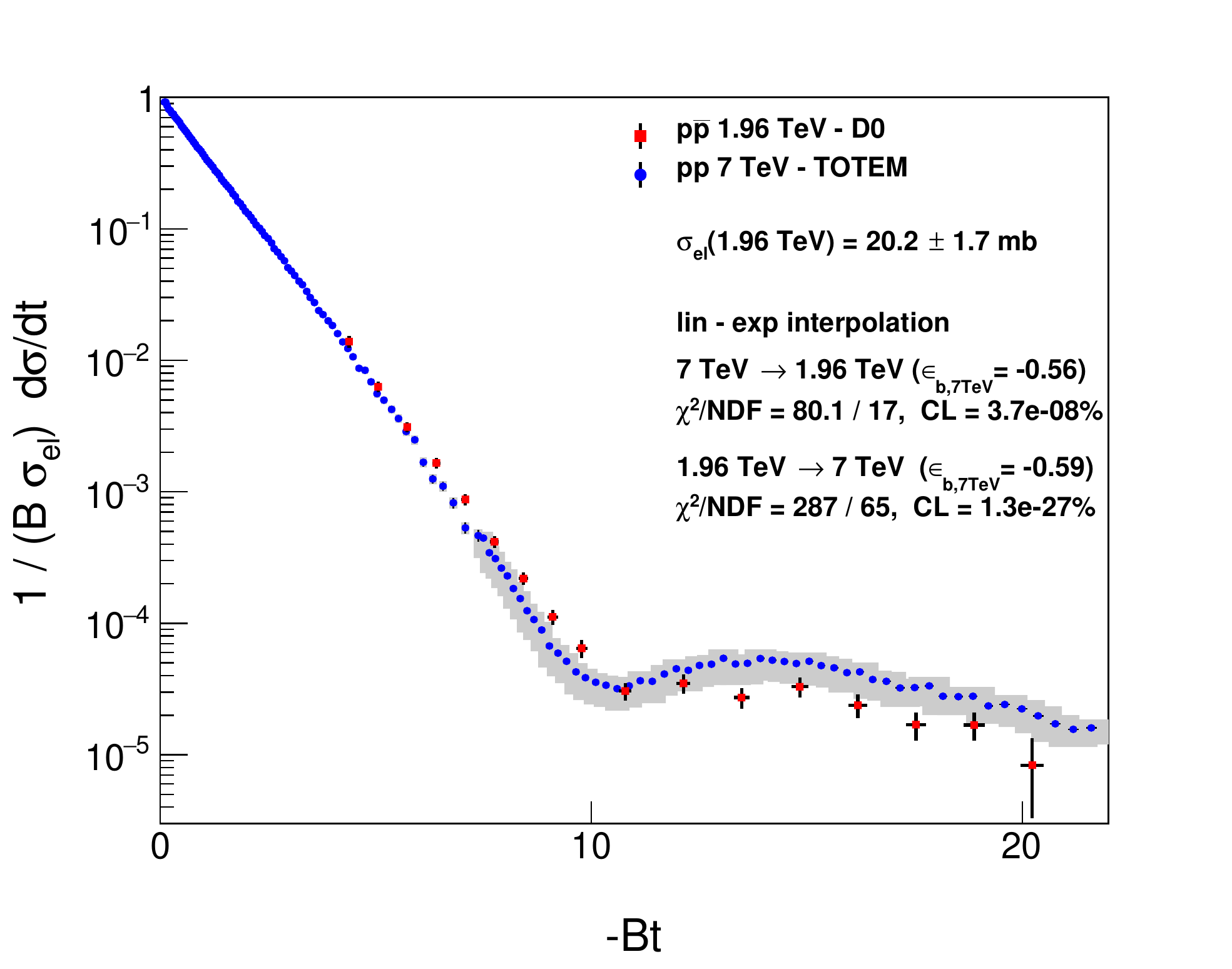}}
\end{minipage}    
\end{center}
\caption{
Top panel indicates that as a function of  $\epsilon_{b, \, \mbox{\rm\tiny 7\, TeV}}$, 
the $\chi^2 \equiv \tilde{\chi}^2_{21}$ distribution has a unique nearly quadratic minimum. The minimum value is 
$\chi^2/{\rm NDF}$ $=$ $80.1/17$, 
corresponding to a statistically significant difference between the $pp$ and $p\bar p$ $H(x)$ scaling functions
at the level of $6.26\sigma$. The bottom panel shows the comparison of the $H(x)$ data using the values of $\epsilon_{b,\, \mbox{\rm\tiny 7\,\, TeV}}$ 
corresponding to such a minimum, both for the case of the $7$ $\rightarrow$ $1.96$ TeV and for the case of $1.96$ $\rightarrow$ $7$ TeV projections.
}
\label{fig:rescaling-from-7-to-1.96TeV-and-back-17mb-lin-exp-new}
\end{figure*}

\begin{figure*}[hbt]
\begin{center}
\begin{minipage}{1.0\textwidth}
\centerline{
\includegraphics[width=0.5\textwidth]{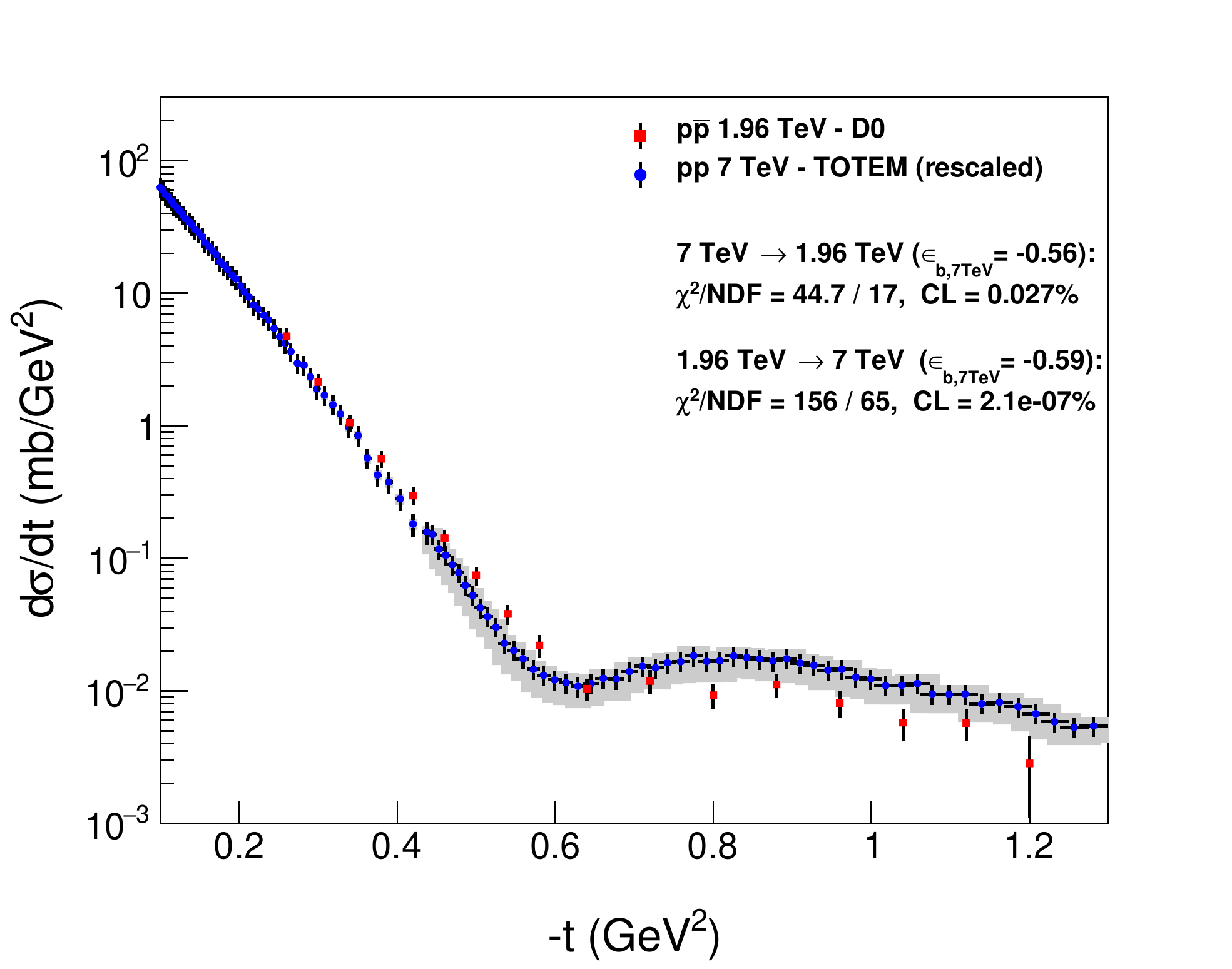}}
\end{minipage}    
\end{center}
\caption{
The artificially reduced Odderon significance, when type $C$ errors are included to the uncertainty of the vertical scale
of the differential cross-sections and are added, as a test, quadratically to the type A, point-to-point fluctuating errors,
just to estimate the magnitude and the importance of their effect. 
Due to the large, 14.4 \% type C errors of the D0 measurement,
the vertical scale becomes rather uncertain and 
the significance of the Odderon signal is reduced substantially. This demonstrates the power of the study of the $H(x)$ scaling function:
type C errors cancel from $H(x)$ and the Odderon signal has a significance of at least 6.26 $\sigma$, as compared to the extrapolation of the
differential cross-sections, where type C errors do not cancel. If we were allowed to add, incorrectly, the type C and type A errors quadratically,
the significance of the Odderon signal would be reduced to 3.64 $\sigma$, corresponding to the $\chi^2/NDF = 44.7/17$ shown above.
}
\label{fig:rescaling-dsigmadt-from-7-to-1.96TeV-with-type-C}
\end{figure*}
\begin{table*}[hbt]
\begin{center}
\begin{tabular}{l l l}
$\sqrt{s}$ (GeV)  &\,\,\,\, $\sigma_{\rm el}$ (mb)  &\,\,\, $B$ (GeV$^{-2}$) \\
\hline
 1960 ($p\bar{p}$)  &\,\,\,\, 20.2  $\pm$ $1.7^{A} $ $\pm$ $14.4\%^{C}$                [*]  & \,\,\, 16.86 $\pm$ $0.1^{A}$  $\pm$ $0.2^{A}$  ~\cite{Abazov:2012qb} \\
 2760 ($pp$)          &\,\,\,\, 21.8  $\pm$ $1.4^{A} $  $\pm$ $6.0\%^{C}$   ~\cite{Nemes:2017gut,Antchev:2018rec}
                     & \,\,\, 17.1  $\pm$ $0.3^{A}$ ~\cite{Antchev:2018rec}     \\
 7000 ($pp$)          &\,\,\,\, 25.43 $\pm$ $0.03^{A}$ $\pm$ $0.1^{B}$ $\pm$ $0.31^{C}$ $\pm$ $1.02^{C}$ ~\cite{Antchev:2013gaa}  & \,\,\, 19.89 $\pm$ $0.03^{A}$ $\pm$ $0.27^{B}$ ~\cite{Antchev:2013gaa} \\
\hline
\end{tabular}
\end{center}
\caption {Summary table of the elastic cross-sections $\sigma_{\rm el}$, the nuclear slope parameters $B$, with references.
We have indexed with superscript A the type A,  point-to-point fluctuating systematic and statistical errors, that can be added in quadrature,
while type B errors (point-to-point changing, fully correlated systematic errors) are indicated with superscript B and type C errors (overall correlated, but $-t$ independent
errors) are indicated with superscript C.
Note that the value and the type A error of the elastic cross-section $\sigma_{\rm el}$ 
at $\sqrt{s} = 1.96$ TeV
[*] is obtained from a low $-t$ exponential fit to data of Ref.~\cite{Abazov:2012qb}, 
while the type C error is directly taken from the publication~\cite{Abazov:2012qb}.
 }
\label{table:B-sigma-new}
\end{table*}
\begin{table*}[hbt]
\begin{center}
 \centering
\begin{tabular}{ccccccc}
 $\sigma_{\rm el}$ \mbox{\rm (mb)}& \mbox{\rm interpolation} &  \mbox{\rm direction of projection} & $\chi^2$ & \mbox{\rm NDF} & \mbox{\rm CL} (\% ) & \mbox{\rm Significance} [$\sigma$]  \\ \hline
20.2 $\pm $ 1.7  & \mbox{\rm lin-exp}  & 7 $\rightarrow$ 1.96 $\mbox{\rm TeV}$  & 80.1 & 17 & $3.7 \times 10^{-8}$& 6.26 \\ \hline
\end{tabular}
\end{center}
\caption{Summary table of the significant Odderon signal in the one-way comparison of the $H(x)$ scaling functions of proton-proton collisions at 
$\sqrt{s} = 7$ TeV, as measured by the TOTEM experiment at CERN LHC~\cite{Antchev:2011zz,Antchev:2013gaa},
and proton - antiproton elastic collisions at $\sqrt{s} = 1.96$ TeV as measured by the D0 experiment at Tevatron~\cite{Abazov:2012qb}.
This table indicates that the Odderon signal is observed in this comparison with  at least a 6.26 $\sigma$ significance, 
corresponding to an Odderon discovery.
}
\label{table:7-to-1.96-TeV-one-way-comparison-new} 
\end{table*}

We have utilized the scaled variance method of ROOT to include the horizontal errors, adding in quadrature to the type A errors also the type A error coming form the type A uncertainty of $x$, denoted as $\delta_A x$. Similarly, we have added in quadrature to
the type B error of the type B uncertainty of $x$, denoted by $\delta_B x$. Using a notation where $M$ may stand for any of $A$ or $B$, the errors are given as
\begin{eqnarray}
\label{e:chi2-final-scaled-variance-A}
e_{M,1}^2(j)   & = & \sigma_{M,1}^2(j) +  [d^{\prime}_1(j) \delta_{M,1} x(j) ]^2 \,, \\
e_{M,21}^2(j)   & = & \sigma_{M,21}^2(j) +  [d^{\prime}_{21}(j) 
\delta_{M,21} x(j) ]^2 \,,
\label{e:chi2-final-scaled-variance-B}
\end{eqnarray}
where $\sigma_{M}(j)$ indicates the type $M \in \left\{ A, B\right\}$ error of the 
value of the {\it vertical} error on data point $j$, and it is added in quadrature to
$d^{\prime}(j) \delta_M x(j)$, the corresponding vertical error that is associated
with the same uncertainty of type $M$ originating from the measurement error on the
horizontal axis $x$ in Eq.~(\ref{e:chi2-final-scaled-variance-A}). The notations 
$ d^{\prime}_1(j)$ and $ d^{\prime}_{21}(j)$ stand for the numerical derivative of the data points  
at point $j$ in the datasets $D_1$ and $D_{21}$, respectively. 

The errors on the projected data set ($D_{21}$) are also obtained by a linear-exponential interpolation between 
the projections of data set 2 ($D_2$) to data set 1 ($D_1$). Their type A and type B errors, indicated by 
$e_{A,21}(j)$ and $ e_{B,21}(j)$ are also added in 
quadrature with the other A or B type of errors. These errors 
on the  interpolated and on the measured values of $(x,H(x))$ through equations~(\ref{e:chi2-final-scaled-variance-A}) and (\ref{e:chi2-final-scaled-variance-B}) provided our most stringent significance estimate for the Odderon effects.
We have cross-checked that several variations on the $\chi^2$ definition, for example the frequently adopted
negligence of the horizontal errors and their contribution to the vertical errors through the scaled variance method,
or perturbing the central values or the errors of the elastic cross-sections within the allowed limits, may only increase
the significance reported in this Appendix.

We have evaluated Eq.~(\ref{e:chi2-final}) as a function of $\epsilon_{b,1}$ and $\epsilon_{b,21}$. However,
for the critical test of the projection of the $\sqrt{s} = 7.0$ TeV TOTEM data on $H(x)$ to that of D0 at $\sqrt{s} = 1.96$ TeV,
we found that D0 did not publish any error on $-t$ and cross-checked that the D0 value on $B$ contains only type A, uncorrelated
statistical and systematic errors only. We also noticed that there are no published type-B errors
on the published differential cross-section data of D0~\cite{Abazov:2012qb}. Hence for the D0 dataset, 
all the type B errors are zero and as a consequence, we have fixed the correlation coefficient of type B errors to zero
for the $\sqrt{s} = 1.96$ TeV D0 dataset. 

Let us now denote by subscript $21$ the projection of the $H(x) $ scaling function at  $\sqrt{s} = 7$ TeV measured by TOTEM for $pp$
reaction to the D0 dataset on $p\bar p$ elastic scattering at $\sqrt{s} = 1.96$ TeV. We found a minimum
for $\epsilon_{b,21}\equiv\epsilon_{b, \, \mbox{\rm\tiny 7\, TeV} }$ 
within the $-1 \le \epsilon_{b,21} \le 1$ domain, with the best value of $\epsilon_{b, \, \mbox{\rm\tiny 7\, TeV} }$ and the lowest value of $\chi^2 \equiv \tilde{\chi}^2_{21}$ of Eq.~(\ref{e:chi2-final-Appendix})
indicated in Fig.~\ref{fig:rescaling-from-7-to-1.96TeV-and-back-17mb-lin-exp-new}.
Table~\ref{table:B-sigma-new} summarizes the input values and the appropriate references to 
the utilized elastic cross-section $\sigma_{\rm el}$ and the nuclear slope $B(s)$.
The final and most stringent result of this cross check, corresponding to the lowest values of significance
for the Odderon observation is summarized in Table~\ref{table:7-to-1.96-TeV-one-way-comparison-new}.
We found that the significance of the Odderon observation in the 7 TeV $\rightarrow $ 1.96 TeV 
projection is at least $6.26\sigma$, corresponding to a $\chi^2/{\rm NDF} = 80.1 / 17$ and CL 
of not larger than $3.7 \times 10^{-8}$ \%. We notice that all variations of the procedure may only 
increase this significance. We conclude, that  the probability of the Odderon observation 
in this $t$-channel mode is statistically significant, at least $P = 1-{\rm CL} = 0.99999999963$,
if the $H(x)$ scaling is valid at $\sqrt{s} = 1.96$ TeV in the range of $0 < x \leq 20.2$ . As Fig.~\ref{fig:App-A-xmax-vs-ln-s} indicates
that model dependently this range might be smaller, only $0 < x \leq 15.1 = x_{max}(s_1)$ for $\sqrt{s_1} = 1.96$ TeV, we have also performed several $x$-range stability studies in \ref{app:E}.

From the experimental side, a very strong argument to support for the domain of validity of the $H(x)$ scaling can be obtained from the observation that the bump/dip ratio is $s$-independent, if the $H(x)$ scaling holds up to the bump position, $x_{bump}$.
If the $H(x)$ scaling is valid for $x_{max}(s) \geq x_{bump} = 13$, we find that
\begin{equation}
\frac{\frac{d\sigma (s)}{dt}|_{bump} }{\frac{d\sigma(s)}{dt}|_{dip}}  =\frac{ H(x_{bump})}{H(x_{dip})} = \mbox{\rm const}(s) .
\end{equation}
Recent TOTEM data indicate that this ratio is, within the energy range available for TOTEM, and within experimental errors,
is indeed independent of $s$~\cite{Nemes:2019nvj} and a diffractive minimum-maximum with an approximately $s$-dependent ratio
of $R = 1.7 \pm 0.2$, $1.7 \pm 0.1$ at $\sqrt{s} = 2.76$ and $7.0$ TeV, respectively.  Apparently, an $s$-independent $R(s)$ is
within errors a permanent feature of elastic $pp$ scattering at these energies~\cite{Nemes:2019nvj}. 
This experimental result supports, that the $H(x)$ scaling holds up to at least $x = x_{bump} = 13$ at energies close to the 
$2.76 \leq \sqrt{s} \leq 7$ TeV range. Indeed, at 13 TeV, $R = 1.77 \pm 0.01$~\cite{Nemes:2019nvj}, so we expect a similar value
of $R(s)$ at  $\sqrt{s} = 1.96$ TeV, too, as this scale is logarithmically close to 2.76 TeV.
Due to the observation of a diffractive cone in $pp$ collisions both at 2.76 and 7.0 TeV~\cite{Nemes:2019nvj},
we expect a similar diffractive cone in $pp$ elastic scattering at $\sqrt{s} = 1.96$ TeV, too, which correspods to $H(x) \simeq \exp(-x)$.
Thus the available experimental data also suggests that the expected domain of validity of $H(x) $ scaling at
$\sqrt{s} = 1.96$ TeV includes the  $7.0 < x \le 13.5$ domain. 
These experimental observations combined with our model independent
$x$-range stability studies in \ref{app:E} indicate, that in this domain the significance
of the Odderon observation is at least 5.0 $\sigma$.
This $7.0 < x \leq 13.5$ interval physically begins with the ``swing", where the differential cross-section of $pp$ elastic scattering starts to bend below the exponential shape and ends just after the diffractive maximum or ``bump", located at $x_{bump} \approx 13.0$ .
For one of the theoretically expected domains of validity, the limited $0 < x \leq 15.1$ range, we find that our method provides an Odderon significance of at least 5.3 $\sigma$, as detailed in \ref{app:E}. Another theoretical argument is shown in Fig.~\ref{fig:App-D-20.2-reBB_model_hx_s-cross-check-at-1.96-TeV}, where the $H(x)$ scaling limit of the ReBB model of Ref.~\cite{Csorgo:2020wmw} is shown to describe the experimental data in the whole D0 acceptance, corresponding to the domain of validity
of $0 < x \leq 20.2$ range and a 6.25 $\sigma$ Odderon effect. This is one of the indications of the robustness of our results.

For the sake of completeness, we also evaluate the asymmetry parameter from the $H(x)$ scaling functions to demonstrate the level of agreement between the experimental data and our full model calculations and also to have a better insight on the magnitude of the scaling violations. 
In order to measure the size of the Odderon effect, let us define the following  asymmetry measure or asymmetry parameter:

\begin{eqnarray}
    A(x|p\bar p,s_1| pp,s_2) & = &
            \frac{H(x| p\bar p,s_1) - H(x| pp,s_2)}{H(x| p\bar p,s_1) + H(x| pp,s_2)}, \\
    A(x|pp,s_1| pp,s_2) & = &  \frac{H(x| pp,s_1) - H(x| pp,s_2)}{H(x| pp,s_1) + H(x| pp,s_2)}.
\end{eqnarray}

\begin{figure*}[hbt]
\begin{center}
\begin{minipage}{1.0\textwidth}
\centerline{
\includegraphics[width=0.5\textwidth]{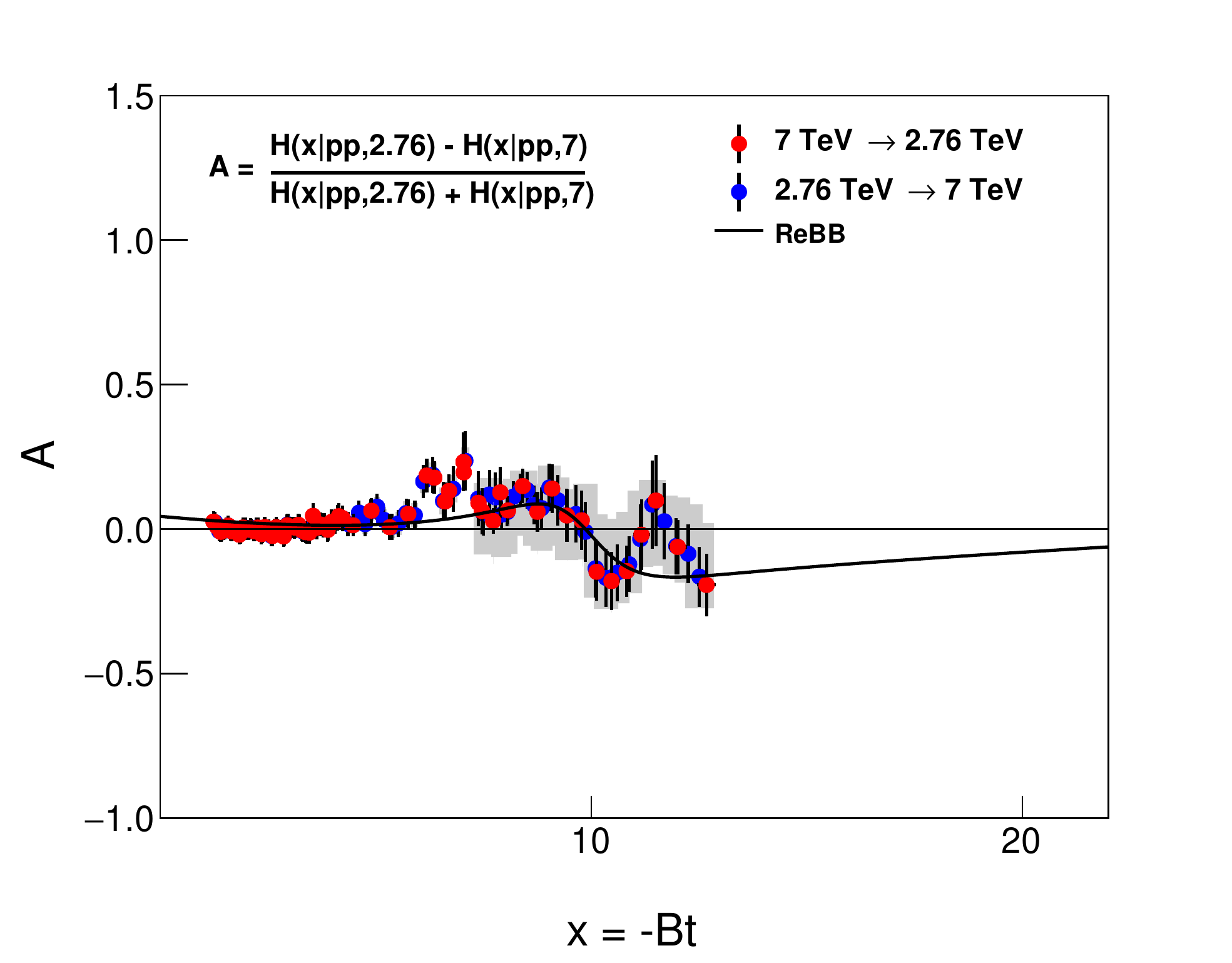}} 
\end{minipage}    
\end{center}
\caption{
Within experimental errors, the asymmetry parameter of elastic $pp$ scattering vanishes for $\sqrt{s} = 2.76$ and $7$ TeV, indicating,
that $H(x) $ scaling is valid within the experimental errors. The solid line is the result of a model calculation based on the full version of the Real Extended Bialas-Bzdak model of Ref.~\cite{Csorgo:2020wmw}, it indicates that the scaling violations are well under theoretical control in this calculation.
}
\label{fig:App-A-asymmetry-pp}
\end{figure*}

If the crossing-odd component of elastic scattering amplitude vanishes at high energies, it implies that 
$H(x| p\bar p,s_1) $ $=$ $ H(x| pp,s_1)$. In this case, we find that $A(x|p\bar p,s_1| pp,s_1)  $ $ \equiv $$ 0$ for $\sqrt{s_1} \geq 1$ TeV.
Additionally, if  $H(x)$ scaling holds for elastic $pp$ scattering at  high energies, then $H(x| pp,s_1) $ $ = $ $ H(x| pp,s_2)$,
hence $A(x|pp,s_1| pp,s_2)$ $ = $ $ 0$. Indeed, as Fig.~\ref{fig:App-A-asymmetry-pp} indicates, this asymmetry parameter
vanishes within experimental errors. There are small systematic deviations that are well described by theoretical model calculations
based on the Real Extended Bialas-Bzdak model of Ref.~\cite{Csorgo:2020wmw}. The agreement between the small asymmetries and the theoretical calculations indicates that the scaling violations are under precise theoretical control.

\begin{figure*}[hbt]
\begin{center}
\begin{minipage}{1.0\textwidth}
\centerline{
\includegraphics[width=0.5\textwidth]{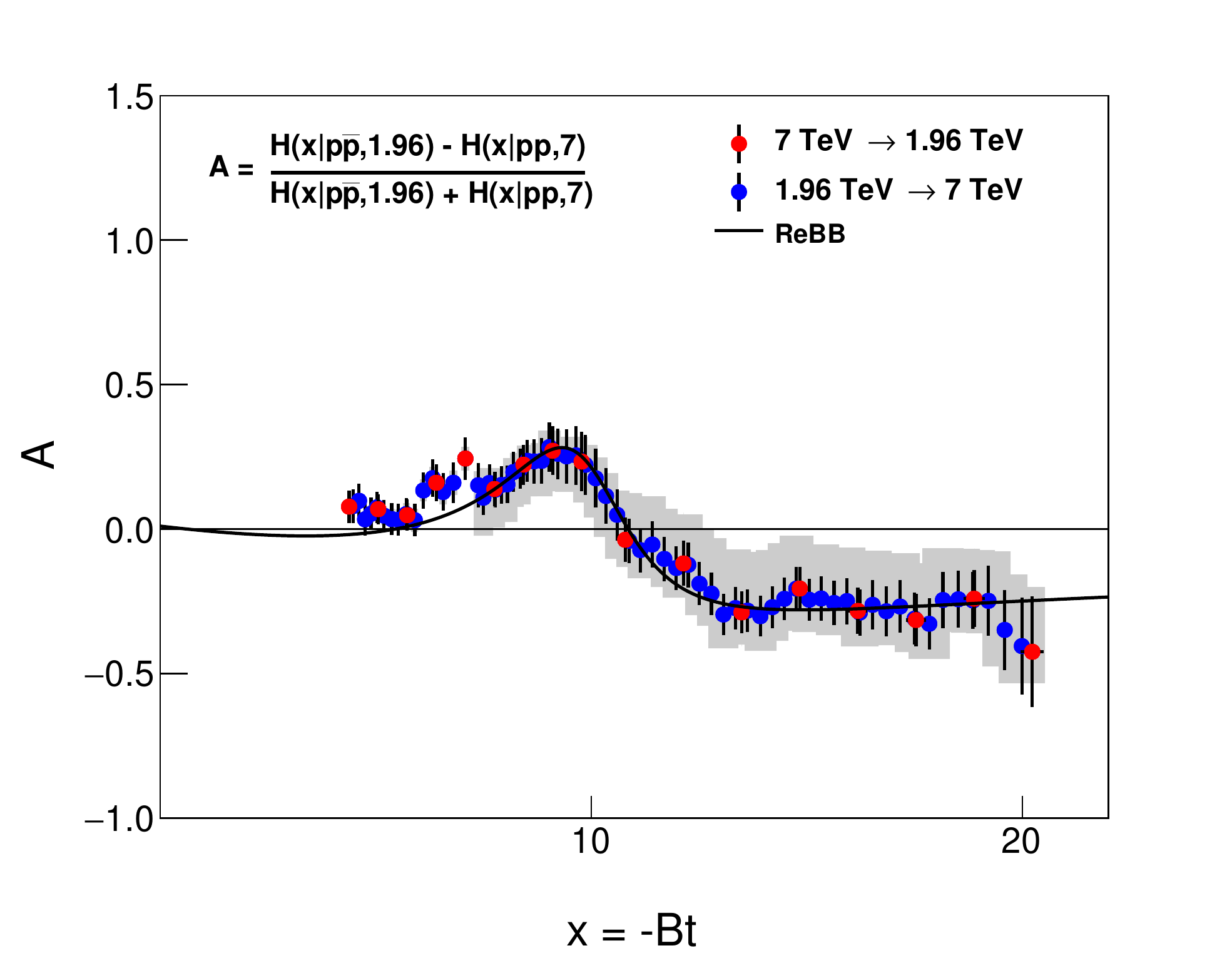}} 
\end{minipage}    
\end{center}
\caption{
Within experimental errors, the asymmetry parameter of elastic $p\bar p$ versus $pp$ scattering does not vanishe for $\sqrt{s_1} = 1.96$ and $7$ TeV, indicating, that the Odderon contribution is significant, larger than the experimental errors. The solid line is the result of a model calculation based on the full version of the Real Extended Bialas-Bzdak model of Ref.~\cite{Csorgo:2020wmw}, it indicates that the $H(x)$ scaling violations are well under theoretical control in this calculation, not only for the $pp$ case, demonstrated in Fig.~\ref{fig:App-A-asymmetry-pp}, but also for the $p\bar p$ case, demonstrated on this plot.
}
\label{fig:App-A-nonvanishing-pp-vs-pbarp}
\end{figure*}

On the other hand, if the $H(x) $ scaling holds for elastic $pp$ scattering and the  crossing-odd component of elastic scattering amplitude is not vanishing at high energies, then $H(x| p\bar p,s_1) \neq H(x| pp,s_1) $, 
hence $A(x|p\bar p,s_1| pp,s_2) \neq 0$ for $\sqrt{s_1}, \sqrt{s_2} \geq 1$ TeV.
Indeed, as shown in Fig.~\ref{fig:App-A-nonvanishing-pp-vs-pbarp}, this asymmetry parameter is significantly different from zero.
Similarly to Fig.~\ref{fig:App-A-asymmetry-pp}, the  solid line in Fig.~\ref{fig:App-A-nonvanishing-pp-vs-pbarp} is the result of a model calculation based on the full version of the Real Extended Bialas-Bzdak model of Ref.~\cite{Csorgo:2020wmw}. The solid line describes the experimental data well, within errors, which indicates that the scaling violations are well under theoretical control in this calculation.
\vfill

\section{Pomeron and Odderon at the TeV energy scale: model independent properties}
\label{app:B}

In this Appendix we discuss some model independent properties of the Pomeron and Odderon that utilize their correspondence with the crossing-even and crossing-odd components of the elastic scattering amplitude at the TeV energies. In the TeV energy range, we identify the crossing-even and crossing-odd components with the Pomeron and the Odderon amplitude, given that the Reggeon contributions in this energy range are generally expected to be negligibly small, as confirmed also by explicit calculations for example in Ref.~\cite{Broniowski:2018xbg}. These results are obtained by a straightforward generalization, from a model dependent to a model independent class, of the Pomeron and Odderon properties obtained in Ref.~\cite{Csorgo:2020wmw} for the Real Extended Bialas-Bzdak model of Ref.~\cite{Nemes:2015iia}.

The proton-proton ($pp$) as well as the proton-antiproton ($p\bar p$) elastic scattering amplitudes can always be written as 
the difference as well as the sum of the the $C$-even and $C$-odd amplitudes, as detailed in 
Eqs.~(\ref{e:tel-pp},\ref{e:tel-pbarp}).
These amplitudes depend on Mandelstam variables $s = (p_1 + p_2)^2$ and $t = (p_1 - p_3)^2$,
but we suppress these dependencies in our notation throughout this~\ref{app:B}.

With the help of the $pp$ and the $p\bar p$ scattering amplitudes, the crossing even and the crossing odd components of the elastic scattering amplitude can be expressed as Eqs.~(\ref{e:Tel-P},\ref{e:Tel-O}).
The $pp$ and the $p\bar p$ scattering amplitudes can be evaluated based for example on R. J. Glauber's theory of multiple diffractive scattering~\cite{Glauber:1955qq,Glauber:1970jm,Glauber:2006gd}, and various models of the structures inside the protons. However, in this Appendix we focus on the model-independent properties so we do not specify a model yet. Model dependent calculations are subject of 
\ref{app:C} and \ref{app:D}.

The differential elastic cross section is defined by Eq.~(\ref{e:dsigmadt-Tel}). 
The elastic, the total and the inelastic cross sections are given by Eqs.~(\ref{e:sigmael},\ref{e:sigmatot}) and Eq.~(\ref{eq:inelastic_cross_section}),
respectively.
The real to imaginary ratio is given by Eq.~(\ref{e:rhos}),
and the nuclear slope parameter is given by the model independent relation of Eq.~(\ref{e:Bs}).

These measurable quantities are utilized to characterize experimentally the $(s,t)$-dependent elastic scattering amplitudes, $T_{\rm el}(s,t)$ discussed above. These quantities can be evaluated for a specific process like elastic $pp$ or elastic $p\bar p$ collisions. If we can evaluate the elastic scattering amplitude for both $pp$ and $p\bar p$ scattering in the TeV energy range, that  straightforwardly yields also the $(s,t)$-dependent elastic scattering amplitude also for the Pomeron and the Odderon exchange.

Let us focus on the model independent properties of the crossing-even Pomeron ($\mathbb P$) and for the crossing-odd Odderon ($\mathbb O$) exchange.

The scattering amplitude $T_{\rm el}(s,t)$, for spin independent processes, is related to a Fourier-Bessel transform of the impact parameter dependent elastic scattering amplitudes $t_{\rm el}(s,b)$:
\begin{equation}
T(s,t) = 2\pi \int\limits_0^{\infty}{J_{0}\left(\Delta\cdot b\right)t_{\rm el}(s,b)b\, {\rm d}b}\,.
\label{eq:elastic_amplitude}
\end{equation}
Here, $b=|{\vec b}|$ is the modulus of the impact parameter vector $\vec b$, $\Delta$ stands for the modulus of the transverse momentum vector and $J_{0}$ is the is the zeroth order Bessel-function of the first kind. In the considered high energy limit, $\sqrt{s} \gg m_p$ and in this case the modulus of the transverse momentum transfer is calculated as $\Delta(t) \simeq \sqrt{-t}$.

This impact parameter dependent amplitude is constrained by the unitarity of the scattering matrix $S$,
\begin{equation}
   S S^{\dagger} = I \,,
\end{equation}
where $I$ is the identity matrix. Its decomposition is $S = I + iT$, where the matrix $T$ is the
transition matrix. In terms of $T$, unitarity leads to the relation
\begin{equation}
   T - T^{\dagger} =  i T T^{\dagger} \,,
\end{equation}
which can be rewritten in terms of the impact parameter or $b$ dependent amplitude $t_{\rm el}(s,b)$ as
\begin{equation}
2 \, {\mathcal Im} \, t_{\rm el}(s, b) = |t_{\rm el}(s, b)|^2 + \tilde\sigma_{\rm inel}(s,b) \,,
\end{equation}
where $\tilde\sigma_{\rm inel}(s,b)$ is a non-negative contribution. If the process is completely
elastic, this quantity is zero, and the elastic amplitude lies on the Argand circle, while in case
there are also inelastic collisions present, the elastic amplitude lies within the Argand circe~\cite{Block:2006hy}.
It can be equivalently expressed from the above unitarity relation as
\begin{equation}
\tilde\sigma_{\rm inel}(s,b)  = 1 - ({\mathcal Re} \, t_{\rm el}(s, b))^2 - 
({\mathcal Im} \, t_{\rm el}(s, b) - 1 )^2 \,.
\end{equation}
It follows that 
\begin{eqnarray}
0\leq\tilde\sigma_{\rm inel}(s,b) & \leq &  1 \, \qquad  0\leq | 1 + i t_{el}(s,b)|^2  \, \leq  \,  1 
\end{eqnarray}
as a consequence of unitarity. Thus probabilistic interpretation can be given only to the inelastic scattering
and to the sum of the elastic scattering amplitude and the amplitude for propagation without interaction.
This is why elastic scattering is a genuine quantum or wave-like process, and this is also the reason
why elastic scattering, in contrast to inelastic scattering, has no classical interpretation.
Thus $\tilde\sigma_{\rm inel}(s,b) $ is interpreted as the impact parameter and $s$-dependent probability 
of inelastic scattering.

The elastic scattering amplitude has the  unitary form
of Eq.~(\ref{e:tel-eikonal}) as the function of the center-of-mass energy squared $s$ and impact parameter $b$.
This form is expressed with the help of the  opacity (or, eikonal) function, denoted by $\Omega(s,b)$, which in the general case is a complex valued function.

The imaginary part of $\Omega(s,b)$ determines the real part of the scattering amplitude, 
while the real part of $\Omega(s,b)$ determines the dominant, imaginary part of the scattering amplitude.
Let us introduce $\tilde\sigma_{\rm inel}(s,b)$ with the help of the real part of the complex opacity function 
as follows:
\begin{eqnarray}
\sqrt{1 - \tilde\sigma_{\rm inel}(s,b)} & = & \exp\left(- {\mathcal Re} \, \Omega(s,b)\right) \,.
\end{eqnarray}
This leads to Eq.~(\ref{e:telsb-Omega}).

The shadow profile function is defined with the help of the opacity function, which yields 
\begin{equation}
    P(s,b) = 1 - |\exp(-\Omega)|^2 = \tilde \sigma_{\rm inel}(s,b) \,.
\end{equation}
This relation indicates that $\tilde \sigma_{\rm inel}(s,b)$ is interpreted as the probability of inelastic collisions at a given value of the
squared center of mass energy $s$ and impact parameter $b$. The inelastic profile function can in general be evaluated with the help of Glauber's multiple diffraction theory~\cite{Glauber:1970jm}, using various model assumptions, for example the assumptions of Ref.~\cite{Nemes:2015iia}. 

The impact parameter dependent elastic scattering amplitudes for elastic $pp$ and $p\bar p$ scatterings are given in terms of the complex opacity or eikonal functions $\Omega(s,b)$. The defining relations are
\begin{eqnarray}
    t_{\rm el}^{pp}(s,b)  & = & i \, (1 - \exp\left(-\Omega^{pp}(s,b) \right) \,, \\  
    t_{\rm el}^{p\bar p}(s,b)  & = & i \, (1 - \exp\left(-\Omega^{p\bar p}(s,b) \right) \,.  
\end{eqnarray}

The explicit expressions  for the $C$-even and the $C$-odd components of the impact parameter dependent elastic scattering amplitude
are detailed below. These relations are explicit consequences of unitarity and do not depend on model details. However, it is important to note that at the TeV energy range in $\sqrt{s}$, the $C$-even exchange corresponds to the Pomeron exchange while the $C$-odd amplitude corresponds to the Odderon exchange, while the corrections due to the exchange of Reggeons or hadronic resonances is smaller than the experimental errors as detailed recently in Ref.~\cite{Broniowski:2018xbg}. Thus, in the TeV energy range, the $S$-matrix unitarity and the dominance of the Pomeron and Odderon amplitudes constrains their form as follows:
\begin{eqnarray*}
    t_{\rm el}^{\mathbb P}(s,b)  & = & i \, \left(1 - \frac{1}{2} \left(\exp\left(-\Omega^{pp}(s,b) \right) + \exp\left(-\Omega^{p\bar p}(s,b) \right)\right)\right) \,, \\  
    t_{\rm el}^{\mathbb O}(s,b)  & = & i \, \frac{1}{2}
                            \left(\exp\left(-\Omega^{pp}(s,b) \right) - \exp\left(-\Omega^{p\bar p}(s,b) \right)\right) \,.
\end{eqnarray*}
It is obvious to note that the Pomeron amplitude given above is crossing-even, while the Odderon amplitude is crossing-odd: 
$C t_{\rm el}^{\mathbb P}(s,b) = t_{\rm el}^{\mathbb P}(s,b)$ and 
$C t_{\rm el}^{\mathbb O}(s,b) = - t_{\rm el}^{\mathbb O}(s,b)$ .

Under two assumptions, these relations can be further simplified for a broad class of models as follows.
\begin{itemize}
    \item {\it i)} If the imaginary part of the complex opacity function in elastic  $pp$ and  $p\bar p$ collisions has the same $b$-dependent factor, denoted here in general by $\Sigma(s,b)$, and
    \item {\it ii)} if these imaginary parts of the complex opacity function also include
    an $s$-dependent but $b$ independent prefactor that is a different function in elastic $pp$ 
    and in $p\bar p$ collisions,
\end{itemize}
then these assumptions correspond mathematically to the following relations:
\begin{eqnarray}
    {\mathcal Im} \Omega^{pp}(s,b) & = & - \alpha^{pp}(s) \Sigma(s,b) \,, \\
    {\mathcal Im} \Omega^{p\bar p}(s,b) & = & - \alpha^{p\bar p}(s) \Sigma(s,b) \,.
\end{eqnarray}
yielding the following simple expressions for the impact parameter dependent elastic $pp$ 
and $p\bar p$ scattering amplitudes
\begin{eqnarray}
    t_{\rm el}^{pp}(s,b) = i\left(1-e^{i\, \alpha^{pp}(s) \, \Sigma(s,b)} \, 
    \sqrt{1-\tilde\sigma_{\rm in}(s,b)}\right) \,, \\
    t_{\rm el}^{p\bar p}(s,b) = i\left(1-e^{i\, \alpha^{p\bar p}(s) \, \Sigma(s,b)} \, 
    \sqrt{1-\tilde\sigma_{\rm in}(s,b)}\right) \,.
\end{eqnarray}

These relations can be equivalently rewritten for the Po\-meron amplitude, using
$\tilde\sigma_{\rm in}\equiv \tilde\sigma_{\rm inel}(s,b)$  and $\Sigma \equiv \Sigma(s,b)$ as shorthand notations while also
suppressing the $s$-dependence of $\alpha^{pp}(s)$ and $\alpha^{p\bar p}(s)$:
\begin{eqnarray*}
        {\mathcal Im }\, t_{\rm el}^{\mathbb P}(s,b)  & = & 
        1 - \sqrt{1-\tilde\sigma_{\rm in} } \times \nonumber\\
         & \times &
        \cos\left( 
        \frac{\alpha^{pp} + \alpha^{p\bar p}}{2} \Sigma \right)
            \cos\left( 
            \frac{\alpha^{p\bar p} - \alpha^{pp}}{2} \Sigma  \right) \,, \\
        {\mathcal Re}\,  t_{\rm el}^{\mathbb P}(s,b)  & = & 
        \sqrt{1-\tilde\sigma_{\rm in} } \, \times \nonumber \\
        & \times &
        \sin\left( \frac{\alpha^{pp} + \alpha^{p\bar p}}{2}\Sigma  \right) 
        \cos\left( \frac{\alpha^{p\bar p} - \alpha^{pp}}{2}\Sigma  \right) \,.
\end{eqnarray*}
This form of the Pomeron amplitude is explicitly $C$-even, and satisfies unitarity. 
Thus, if the difference between the opacity parameters 
$\alpha$ for $pp$ and $p\bar p$ elastic collisions is small, the Pomeron is 
predominantly imaginary, with a small real part that is proportional to 
$\sin\left( \frac{\alpha^{pp} + \alpha^{p\bar p}}{2}\tilde\Sigma \right)$.
Similarly, for the Odderon we have, under the conditions {\it i)} and {\it ii)}, 
real and imaginary parts of the following amplitude
\begin{eqnarray}
        {\mathcal Re }\, t_{\rm el}^{\mathbb O}(s,b)  & = & \sqrt{1-\tilde\sigma_{\rm in} } 
        \times \label{e:Re-Odderon} \\
        &\times &
        \sin\left( \frac{\alpha^{p\bar p} - \alpha^{pp}}{2}\Sigma  \right)
            \cos\left( \frac{\alpha^{p\bar p} + \alpha^{pp}}{2}\Sigma  \right) \,, \nonumber\\
        {\mathcal Im}\, t_{\rm el}^{\mathbb O}(s,b)  & = & \sqrt{1-\tilde\sigma_{\rm in} } 
        \times \label{e:Im-Odderon} \\ 
        & \times & 
        \sin\left( \frac{\alpha^{p\bar p} - \alpha^{pp}}{2}\Sigma  \right) 
        \sin\left( \frac{\alpha^{pp} + \alpha^{p\bar p}}{2}\Sigma  \right) \,. \nonumber
\end{eqnarray}
This form of the Odderon amplitude is explicitly $C$-odd and satisfies unitarity, 
if the above two assumptions {\it i)} and {\it ii)} are satisfied, without any further reference 
to the details of the model. In this class of models, if the difference between the opacity 
parameters $\alpha$ for $pp$ and $p\bar p$ elastic collisions becomes vanishingly small, 
both the real and the imaginary parts of the Odderon amplitude vanish, as they are both 
proportional to
$\sin\left( \frac{\alpha^{p\bar p} - \alpha^{pp}}{2}\Sigma  \right)$.
If this term is non-vanishing, but $(\alpha^{p\bar p} + \alpha^{pp}) \Sigma$ remains small,
the above Odderon amplitude remains predominantly real, with a small, linear in 
$(\alpha^{p\bar p}+\alpha^{pp}) \Sigma$ at the leading order, imaginary part. 


\section{Emergence of the $H(x)$ scaling  from the Real Extended Bialas-Bzdak model}
\label{app:C}

With the help of the ReBB model of Ref.~\cite{Nemes:2015iia}, we have recently described 
the $pp$ and $p\bar p$ differential cross-sections in a limited kinematic range of 
$0.546 \le \sqrt{s} \le 8$ TeV and $ 0.372 \le -t \le 1.2$ GeV$^2$,
in a statistically acceptable manner with CL $\ge $ 0.1 \%, 
as detailed in Ref.~\cite{Csorgo:2020wmw}. 

It is important to realize that within the ReBB model, the $pp$ elastic scattering dependence 
on $s$ comes only through four energy-dependent quantities, as specified recently
in Ref.~\cite{Csorgo:2020wmw}. Let us recapitulate the general formulation, 
for the sake of clarity, denoting the $s$-dependent quantities
as $R_q^{pp}(s)$, $R_d^{pp}(s)$, $R_{qd}^{pp}(s)$ and $\alpha^{pp}(s)$:
\begin{equation}
    T_{\rm el}^{pp}(s,t) = F(R_q^{pp}(s), R_d^{pp}(s), R_{qd}^{pp}(s),\alpha^{pp}(s); t) \,. \label{e:Telpp-parameters}
\end{equation}

Similarly, the scattering amplitude of the elastic $p\bar p$ scattering 
is found in terms of four energy-dependent quantities, that we denote here for the sake 
of clarity as $R_q^{p\bar p}(s)$, $R_d^{p\bar p}(s)$, $R_{qd}^{p\bar p}(s)$ 
and $\alpha^{p\bar p}(s)$:
\begin{equation}
    T_{\rm el}^{p\bar p}(s,t) = F(R_q^{p\bar p}(s), R_d^{p\bar p}(s), R_{qd}^{p\bar p}(s),
    \alpha^{p\bar p}(s); t) \,.  \label{e:Telpbarp-parameters}
\end{equation}
Here, $F$ stands for a symbolic short-hand notation for a function, that indicates how 
the left hand side of the $pp$ and $p\bar p$ scattering amplitude depends on $s$ through 
their $s$-dependent quantities. Among those, $R_q$, $R_d$, and $R_{qd}$ 
correspond to the Gaussian sizes of the constituent quarks, diquarks and their separation
in the scattering (anti)protons. These scales are physically expected to be the same 
in $pp$ and in $p\bar p$ elastic collisions. 

Indeed, the trends of $R_q(s)$, $R_d(s)$ and $R_{qd}(s)$ follow, within errors, the same excitation 
functions in both $pp$ and $p\bar p$ collisions, as indicated on panels a, b and c of Fig.~6 of Ref.~\cite{Csorgo:2020wmw}.
Due to this reason, let us denoted these -- in principle, different -- scale parameters 
with the same symbols in the body of the manuscript, as they are found to be independent 
of the type of the collision, i.e.
\begin{eqnarray}
    R_q(s) & \equiv & R_q^{pp}(s) \, = \, R_q^{p\bar p}(s) \,, \label{e:Rqs} \\
    R_d(s) & \equiv & R_d^{pp}(s) \, = \, R_d^{p\bar p}(s) \,, \label{e:Rds}\\
    R_{qd}(s) & \equiv & R_{qd}^{pp}(s) \, = \, R_{qd}^{p\bar p}(s) \,. \label{e:Rqds}
\end{eqnarray}
On the other hand, the opacity or dip parameter $\alpha(s)$ is different for elastic $pp$ 
and $p\bar p$ reactions: if they were the same too, then the scattering amplitude 
for $pp$ and $p\bar p$ reactions were the same, and correspondingly the differential 
cross-sections were the same in these reactions, while the experimental results indicate 
that they are qualitatively different \cite{Csorgo:2020wmw}. Hence,
\begin{eqnarray}
     \alpha^{pp}(s) & \neq & \alpha^{p\bar p}(s) \,.
\end{eqnarray}

The ReBB model \cite{Nemes:2015iia} provides a statistically acceptable description 
of the elastic scattering amplitude, both for $pp$ and $p\bar p$ elastic scattering, 
in the kinematic range that extends to at least $ 0.372 \leq -t \leq 1.2$ GeV$^2$ and 
$0.546 \leq \sqrt{s} \leq 8$ TeV. For the sake of clarity, let us also note that the 
$s$-dependence of the Pomeron and Odderon components of the scattering amplitude thus 
happens through the $s$-dependence of five parameters only. 
Based on Ref.~\cite{Csorgo:2020wmw}, we write
\begin{eqnarray}
     T_{\rm el}^{\mathbb P}(s,t)  & = &
          G(R_q(s), R_d(s), R_{qd}(s),\alpha^{pp}(s), \alpha^{p\bar p}(s);t) \,,  \nonumber \\ 
  &  & \null 
    \label{e:pomeron-amplitude-in-terms-of-5}  \\ 
     T_{\rm el}^{\mathbb O}(s,t)  & = &
          H(R_q(s), R_d(s), R_{qd}(s),\alpha^{pp}(s), \alpha^{p\bar p}(s);t) \,.  \nonumber \\ 
   &  & \null 
     \label{e:odderon-amplitude-in-terms-of-H}
\end{eqnarray}
Here, $G$ and $H$ are just symbolic short-hand notations that summarize how the left hand sides 
of the above equations depend on $s$ through their $s$-dependent parameters.

As detailed in Ref.~\cite{Csorgo:2020wmw}, within the ReBB model there is a deep connection between 
the $t=0$ and the dip region. This supports the findings that the recently observed decrease 
in $\rho_0(s)$ around $\sqrt{s}=$13 TeV, the dip-bump structure in $pp$ scattering and its absence 
in $p\bar p$ scattering are both the consequences of the Odderon contribution.
In the ReBB model, this Odderon contribution is encoded in the difference between 
$\alpha^{pp}(s)$ and $\alpha^{p\bar p}(s)$. This conclusion is supported also by the detailed 
calculations of the ratio of the modulus-squared Odderon to Pomeron scattering amplitudes.
Thus, if $\rho_0^{pp}(s)\neq\rho_0^{p\bar p}(s)$, within the ReBB model it follows that $\alpha^{pp}(s)\neq\alpha^{p\bar p}(s)$ or, equivalently, $t_{\rm el}^{\mathbb O}(s,b)\neq 0$ in the TeV region.

Within the framework of the ReBB model, we have proved in Ref.~\cite{Csorgo:2020wmw} 
an interesting Odderon theorem. The weaker, original form of this theorem was formulated 
above in Sec.~\ref{s:Odderon-search} as follows:\\
{\bf Theorem 1}
If the $pp$ differential cross sections differ from that of $p\bar p$ scattering 
    at the same value of $s$ in a TeV energy domain, then the Odderon contribution to 
    the scattering amplitude cannot be equal to zero, i.e.
     \begin{equation}
        \frac{d\sigma^{pp}}{dt} \neq  \frac{d\sigma^{p\bar p}}{dt} \,\,\, 
        \mbox{ \rm for}\,\, \sqrt{s}\ge 1 \,\, \mbox{\rm TeV}
        \implies 
                T_{\rm el}^O(s,t) \neq 0  \, .
    \end{equation}
This theorem is model-independently true as it depends only on the general structure of 
the theory of elastic scattering. Within the ReBB model, this theorem has been sharpened in Ref.~\cite{Csorgo:2020wmw} as follows:\\
{\bf Theorem 2:}
In the framework of the unitary ReBB model, the elastic $pp$ differential cross sections differ 
from that of elastic $p\bar p$ scattering at the same value of $s$ in a TeV energy domain, 
if and only if the Odderon contribution to the scattering amplitude is not equal to zero. 
This happens if and only if $\alpha^{pp}(s) \neq \alpha^{p\bar p}(s)$ and as a consequence, 
if and only if $\rho_0^{pp} \neq \rho_0^{p\bar p}$:
     \begin{eqnarray}
        \frac{d\sigma^{pp}}{dt}  & \neq  & \frac{d\sigma^{p\bar p}}{dt} 
        \iff 
        T_{\rm el}^O(s,t) \neq 0  \, \nonumber \\  &\iff&  
        \rho_0^{pp}(s) \neq \rho_0^{p\bar p}(s) \nonumber \\
        & \iff &  \alpha^{pp}(s) \neq \alpha^{p\bar p}(s) \nonumber \\
        & & \,\,\, \mbox{ \rm for}\,\, \sqrt{s}\ge 1 \,\, \mbox{\rm TeV} \,. \nonumber
    \end{eqnarray}

In this work, we extend these theorems to the emergence of $H(x)$ scaling within the ReBB model, 
as follows:\\
{\bf Theorem 3:}
In the framework of the unitary ReBB model, the elastic $pp$ differential cross sections
obey a $H(x)$ scaling in a certain kinematic region, if and only if in that region the opacity 
parameter is approximately energy independent, $\alpha^{pp}(s) \approx {\rm const}$ and the geometrical 
scale parameters evolve with the same $s$-dependent, but radius parameter independent factor $b(s)$.
Thus, the conditions of validity of $H(x)$ scaling in elastic $pp$ collisions, within the framework 
of the ReBB model are the simultaneous validity of the following four equations: 
    \begin{eqnarray}
             R_q(s) & = & b(s) R_q(s_0) \,, \label{eq:Hx-Rq}\\
             R_d(s) & = & b(s) R_d(s_0) \,, \label{eq:Hx-Rd}\\ 
             R_{qd}(s) & = & b(s) R_{qd}(s_0) \,,\label{eq:Hx-Rqd}\\
             \alpha^{pp}(s) & = & \alpha^{pp}(s_0) \,, \label{eq:Hx-alpha}
    \end{eqnarray}
 where $b(s)$ is the same function of $s$ for each of $R_q$, $R_d$ and $R_{qd}$.
 
The key point of the proof of Theorem 3 is that $B(s)$ is related, for an analytic amplitude, 
to the variance of the scattering amplitude in the impact parameter space. 
If this variance depends on the evolution of the scale parameters $R_q(s)$, $R_d(s)$ and $R_{qd}(s)$ only, as $\alpha(s)$ is within the experimental errors a constant, and if these scale parameters all evolve with the same $s$-dependent factor, then the nuclear slope parameter must also scale as 
\begin{equation}
B(s) = b^2(s) B(s_0) \,. 
\label{eq:B-scaling}
\end{equation}
At the same time, the elastic and the differential cross-sections  must also scale as 
\begin{eqnarray}
\sigma_{\rm el}(s) &= & b^2(s) \sigma_{\rm el}(s_0) \,,  \label{eq:sigma-scaling}\\ 
\frac{d\sigma}{dt}(s,t) &=& b^2(s) \frac{d\sigma}{dt}(s_0, t_0 = \frac{t}{b(s)^2}) \,.
\end{eqnarray}
Hence, in such an $s$ and $x = - t B$, range the $H(x)$ scaling, 
defined by Eq.~(\ref{e:Hx-general}) must hold: $H(x,s) = H(x, s_0)$ and vice versa.

We have cross-checked the scaling properties of the ReBB model at both ISR and LHC energies.
At the ISR energies of 23.5 $\leq \sqrt{s} \leq 62.5$ GeV, $\rho_0 \equiv \rho_0(s)$ is not a constant as reviewed recently in~\cite{Antchev:2017yns}, so our Theorem 3
suggests, that the $H(x,s) \equiv H(x,s_0) $ scaling cannot be interpreted in terms of the ReBB model, and in particular
we expect that $R(s)$, the bump-to-dip ratio decreases with increasing  values of $s$ as ref.~\cite{Antchev:2017yns} suggests that  $\rho_0(s)$ is an increasing function of $s$ at ISR energies. Thus, the approximate $H(x)$ scaling, indicated in Fig.~\ref{fig:scaling-ISR-x} actually is expected to be violated in the dip regions  and also perhaps also in the tail region at ISR energies.
A more detailed investigation
of the scaling violations at ISR energies goes well beyond the scope of this manuscript.

At LHC energies, let us  summarize the main results of the ReBB model studies, as given in  Ref.~\cite{Csorgo:2020wmw} in greater details. If we do not 
utilize the validity of the $H(x)$ scaling at the LHC energies, we obtain Figs.~16, 17 and 18 of Ref.~\cite{Csorgo:2020wmw}. In these figures a yellow band indicates the uncertainty of the model prediction, without the assumption of the validity of the $H(x)$ scaling. Fig.~16 of Ref.~\cite{Csorgo:2020wmw} indicates that the extrapolation without the $H(x)$ scaling is rather uncertain in the region of the diffractive shoulder as compared to $\sqrt{s} = 1.96$ TeV $p\bar p$ elastic scattering data and correspondingly, no significant difference is observed in this model comparison at this energy. However, the model allows for the investigation of $H(x)$ scaling violations and the extrapolation of $p\bar p$ data to the LHC energies. As the $p\bar p$ data do not obey a $H(x)$ scaling, their extrapolation to the LHC energies without such a model is not possible: the $H(x)$ scaling works only for $pp$ but not for $p\bar p $ collisions. Comparing the ReBB model extrapolations of $p\bar p$ differential cross-sections with TOTEM data on $pp$ differential cross-sections at $\sqrt{s} = 2.76$ TeV, we obtained in Ref.~\cite{Csorgo:2020wmw} an Odderon effect with a significance of 7.12$\sigma$, as indicated on 
Fig.~17 of Ref.~\cite{Csorgo:2020wmw}. Combining this value with the model dependent results at $\sqrt{s} = 1.96$ TeV, the combined significance is hardly reduced, changes only to 7.08 $\sigma$. On the other hand, if we extrapolate $p\bar p$ data also up to $\sqrt{s} = 7$ TeV, the significance increases further, to values greater than 10$\sigma$. In practical terms, 
extrapolating $p\bar p$ data theoretically up to 7 TeV, we obtain a certainty for the Odderon contribution.
The quoted 6.26 $\sigma$ model independent significance is thus a safe, model independent  lower limit for the observation of a
crossing-odd component of elastic $pp$ and $p\bar p$ scattering in the TeV energy range.
  
\section{Model-dependent estimation of the range of validity and violations of $H(x)$ scaling}
\label{app:D}

In this Appendix, we summarize our model-dependent results on the estimated range of validity of $H(x)$ scaling in elastic $pp$ collisions. We find that this scaling might be extended in $\sqrt{s}$ from 7 and 8 TeV at LHC down to $\sqrt{s} = 200$ GeV at RHIC, as detailed below.

\begin{figure}[htb]
	\centering
    \includegraphics[width=0.8\linewidth]{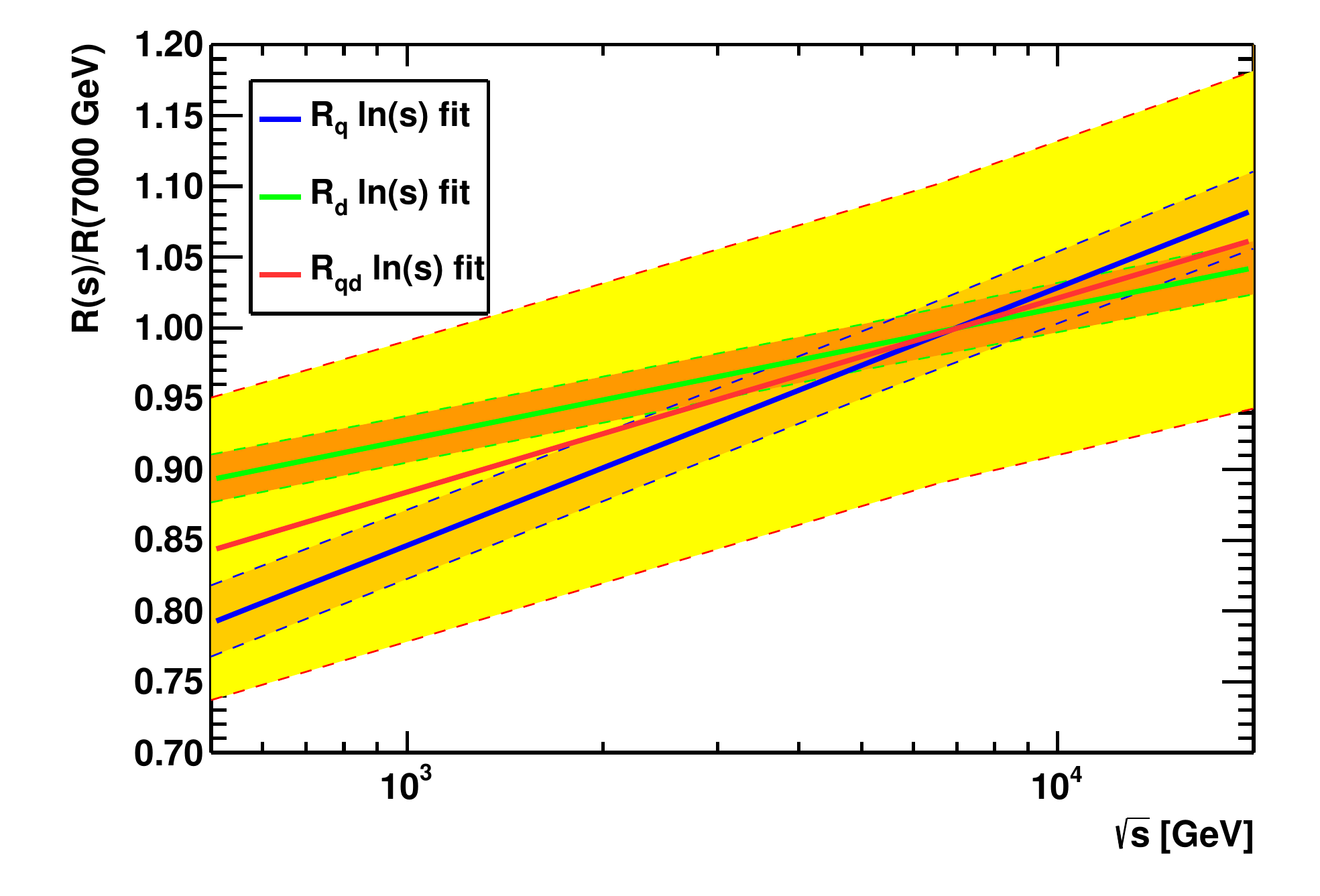}
	\caption{The scaled energy dependence of the ReBB model parameters by the parameter values at 7 TeV. The original logarithmic excitation functions of the parameters are presented in Ref.~\cite{Csorgo:2020wmw}. The lower limit of the $H(x)$ scaling from the linear logarithmic dependence is obtained at $\sqrt{s}=2436$ GeV.}
	\label{fig:reBB_model_hx_a0}
\end{figure}

The $R_q(s)$, $R_d(s)$ and $R_{qd}(s)$ parameters of the ReBB model were determined in Ref.~\cite{Csorgo:2020wmw} using both an affine linear and a quadratic dependence on $\ln(s)$. This allowed us to test if the scale parameters of the ReBB model obey, within one standard deviation, the same energy dependence or not. For the reference point, we have chosen $\sqrt{s_0} = 7$ TeV. Fig.~\ref{fig:reBB_model_hx_a0} indicates that such an affine linear scaled energy dependence of the ReBB model parameters $R_q(s)$, $R_d(s)$ and $R_{qd}(s)$ by the values of the same parameters at 7 TeV suggests that the one $\sigma$ systematic error-bands overlap down to $\sqrt{s}=2436$ GeV. However, this result does not yet take into account the possible quadratic dependence of these parameters
on $\ln(s)$ and it also neglects the correlations between the model parameters. 
However, the validation of this linear in $\ln(s)$ energy dependence of the $b(s)$ factor of the $H(x)$ scaling 
 by an explicit calculation failed,
indicating that a quadratic extrapolation in $\ln(s)$ is apparently necessary.

\begin{figure}[htb]
	\centering
    \includegraphics[width=0.8\linewidth]{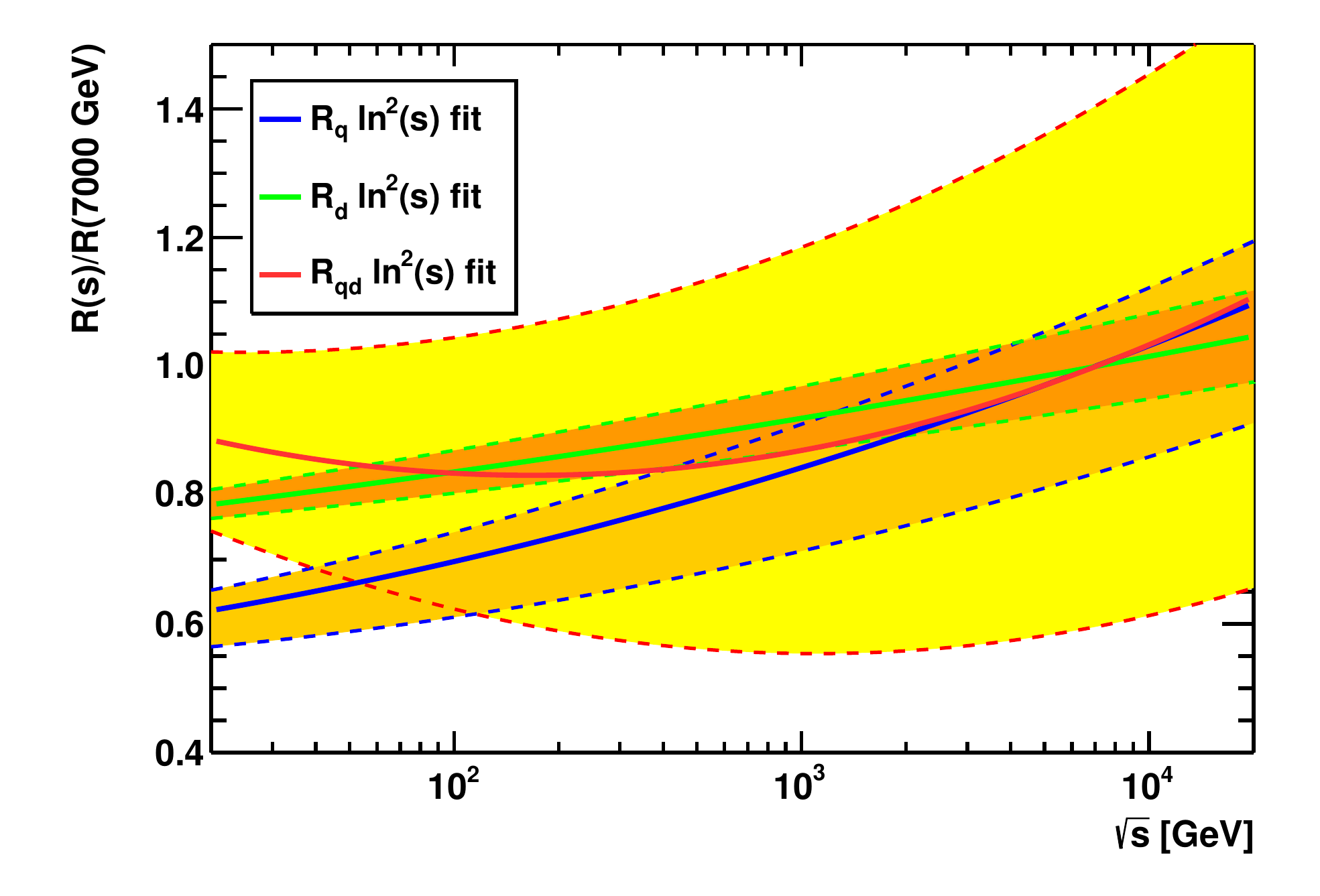}
	\caption{The scaled energy dependence of the ReBB model parameters by the parameter values at 7 TeV. The original squared logarithmic excitation functions of the parameters are presented in Ref.~\cite{Csorgo:2020wmw}. The lower limit of the $H(x)$ scaling is pushed further down to about $\sqrt{s}=500$ GeV, if the correlations between the parameters $R_q(s)$, $R_d(s)$ and $R_{qd}(s)$ are neglected. }
	\label{fig:reBB_model_hx_a}
\end{figure}

Taking into account the possible quadratic in $\ln(s)$ dependence of the excitation function of the ReBB model parameters $R_q(s)$, $R_d(s)$ and $R_{qd}(s)$ pushes this limit further down to $\sqrt{s}=500$ GeV, 
as demonstrated in Fig.~\ref{fig:reBB_model_hx_a}. This plot utilizes the parameters of quadratic 
dependence in $\ln(s)$ as indicated in Fig.~23 of Ref.~\cite{Csorgo:2020wmw}, and collected in Table 3 
of that manuscript, but without taking into account the correlations between these model parameters.
When we tried to validate such
a quadratic in $\ln(s)$ but uncorrelated dependence of the $b(s)$ parameter of the $H(x) $ scaling,
the validation plots did not result in acceptable confidence levels with CL $\geq 0.1$ \%
for $\sqrt{s} = 2.76$, $1.96$ and $0.546$ TeV. Hence we had to take into account the correlations
between $R_q$, $R_d$ and $R_{qd}$ together with their quadratic in $\ln(s)$ behaviour as detailed below.
	
Our final estimate for the range of the validity of the $H(x)$ scaling, in particular, the possible lowest value for the validity of this scaling is based on the quadratic in $\ln(s)$ dependence of the model parameters $R_q(s)$, $R_d(s)$ and $R_{qd}(s)$, taking also into account their correlations. This we have studied so that we determined these parameters at 5 different energies, at $\sqrt{s} = 23$ GeV, as well as at	0.546, 1.96, 2.76 and 7 TeV, and fitted the resulting 5 points with a 3-parameter quadratic formula of $R_i(s) = p_0 + p_1 \ln(s/s_0) + p_2 \ln^2(s/s_0)$. This line is our best estimate for the quadratic energy dependence for these parameters. However, the parameters are correlated so we have repeated these fits by shifting up (or down) by one standard deviation each of these model parameters at each energy and fixed their values, while re-fitting all the other parameters of the ReBB model at the same energy to find their best estimate. This way we perturbed in two different directions four model parameters at five different energies and re-fitted each set with the quadratic in $\ln(s/s_0)$ evolution, resulting in 2x4x5 = 40 curves around the central line. The area between these curves is our best estimate for the systematic error band of the energy evolution, that takes into account not only the errors of the ReBB model parameters but also the correlations between the ReBB model parameters at each energy. 

\begin{figure}[htb]
	\centering
    \includegraphics[width=0.8\linewidth]{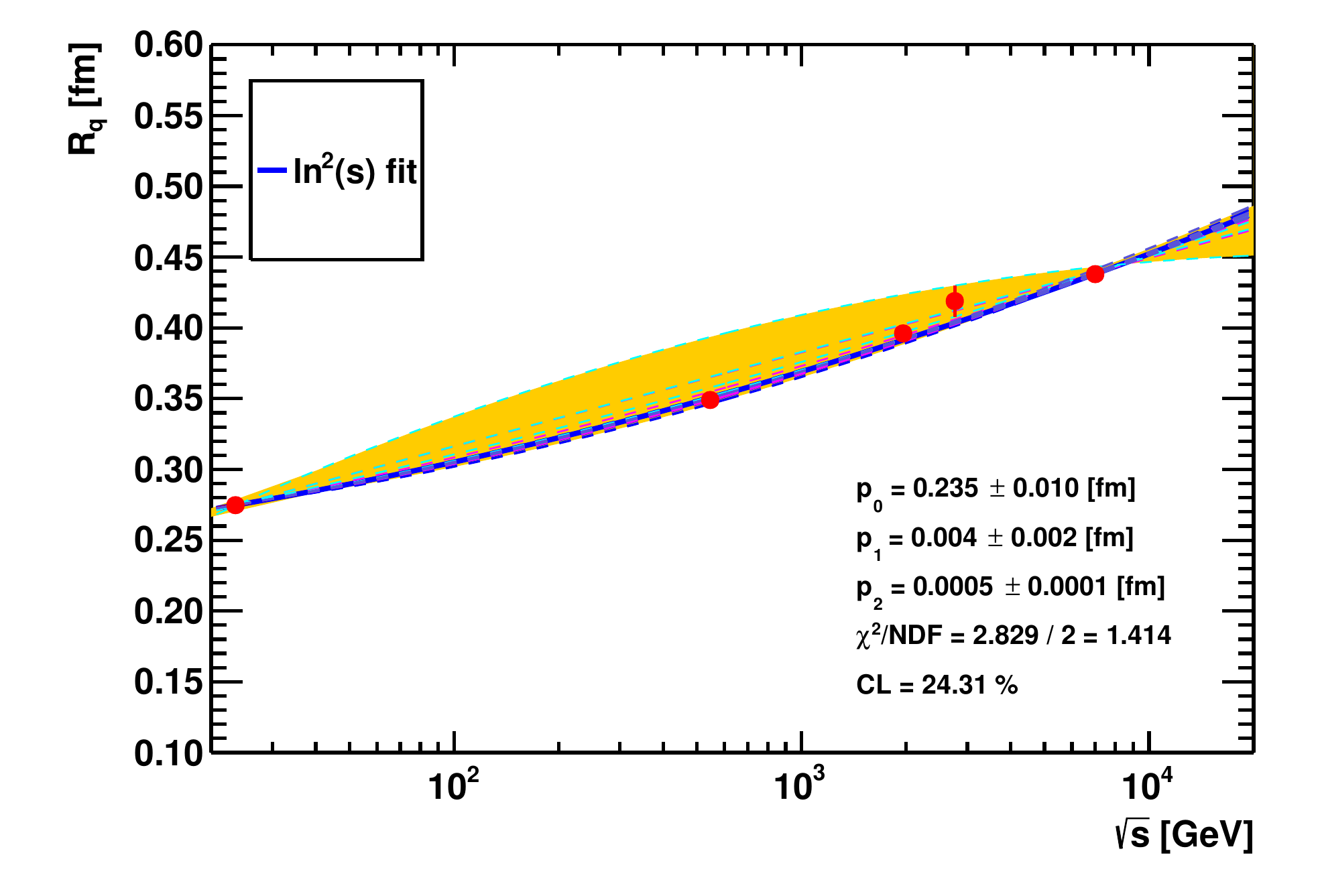}
	\caption{Systematic error band on the $R_q(s)$ parameter as obtained by systematically shifting and fixing the best $R_q$ values up and down by one standard deviation at each energy where they are determined and then re-fitting all the other parameters, to account for the correlations between the parameters of the ReBB model. The central line indicates the best estimate for the quadratic in $\ln(s)$ polynomial for $R_q(s)$. }
	\label{fig:reBB_model_hx_Rq-2nd-order}
\end{figure}

\begin{figure}[htb]
	\centering
    \includegraphics[width=0.8\linewidth]{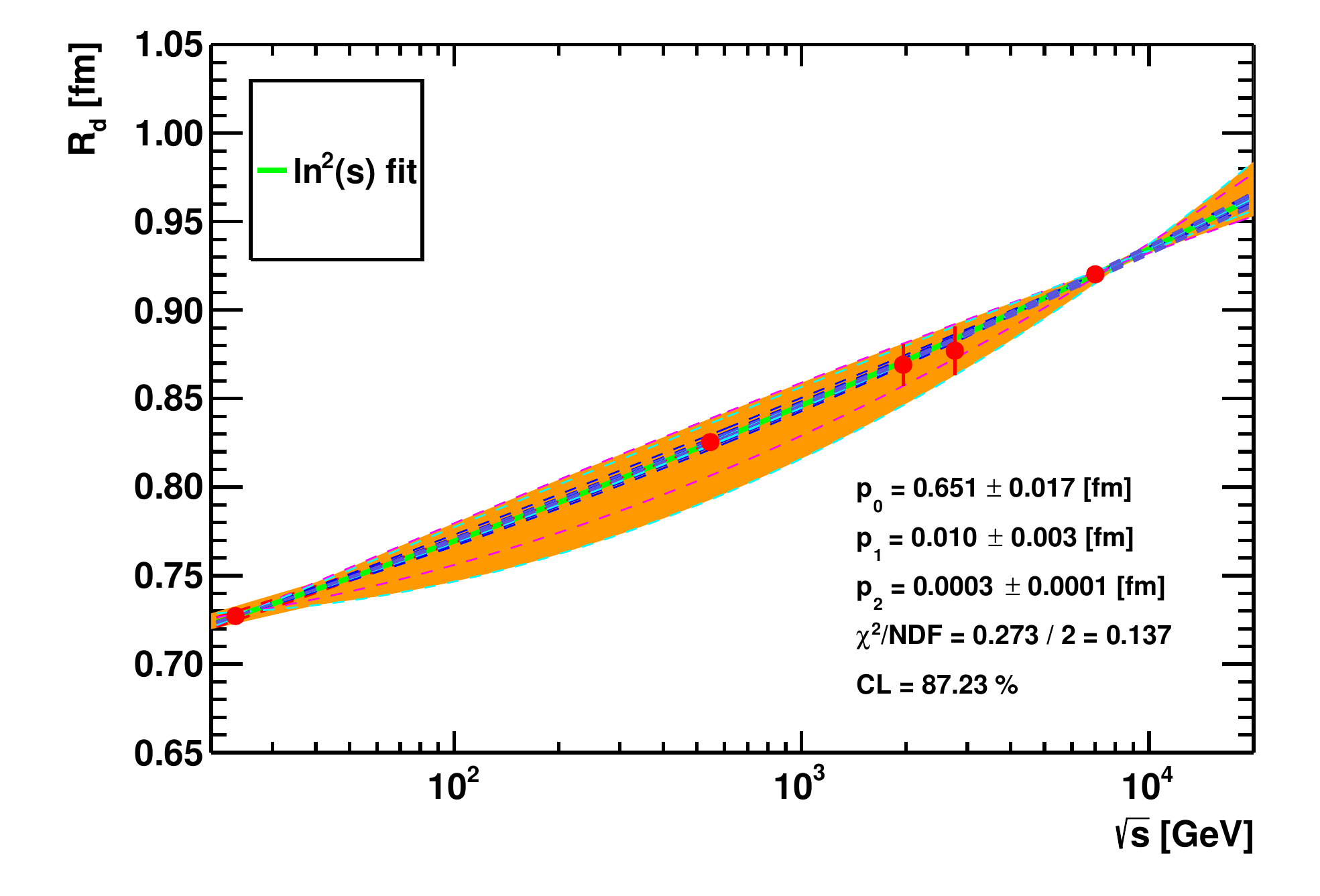}
	\caption{Systematic error band on the $R_d(s)$ parameter as obtained by systematically shifting and fixing the best $R_d$ values up and down by one standard deviation at each energy where they are determined ($\sqrt{s} = 7.0$, $2.76$, $1.96$, $0.546$ TeV and $23.5$ GeV) and then re-fitting all the other parameters to account for the correlations between the parameters of the ReBB model. The central line indicates the best estimate for the quadratic in $\ln(s)$ polynomial for $R_d(s)$.}
	\label{fig:reBB_model_hx_Rd-2nd-order}
\end{figure}

\begin{figure}[htb]
	\centering
    \includegraphics[width=0.8\linewidth]{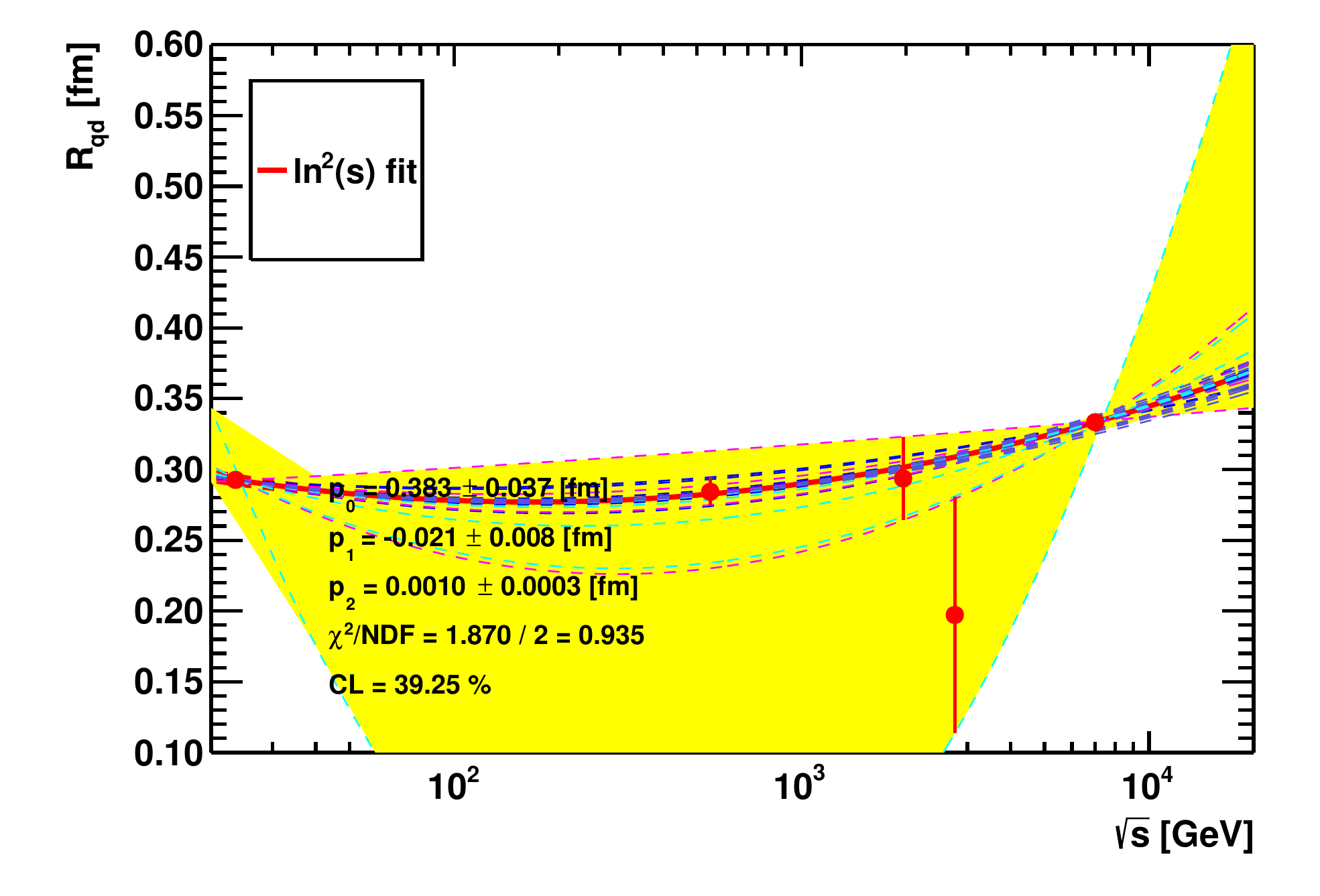}
	\caption{Systematic error band on the $R_{qd}(s)$ parameter as obtained by systematically shifting and fixing the best $R_{qd}$ values up and down by one standard deviation at each energy where they are determined ($\sqrt{s} = 7.0$, $2.76$, $1.96$, $0.546$ TeV and $23.5$ GeV) and then re-fitting all the other parameters to account for the correlations between the parameters of the ReBB model. The central line indicates the best estimate for the quadratic in $\ln(s)$ polynomial for $R_{qd}(s)$.}
	\label{fig:reBB_model_hx_Rqd-2nd-order}
\end{figure}

The one $\sigma$ systematic error band on the $R_q(s)$ parameter, that takes into account both the quadratic in $\ln(s)$ evolution and the correlations between the model parameters is presented as an orange band in 
Fig.~\ref{fig:reBB_model_hx_Rq-2nd-order}. Similarly, the one $\sigma$ systematic error band on the $R_d(s)$ parameter, that takes into account both the quadratic in $\ln(s)$ evolution and the correlations between the model parameters is presented as a darker orange band in Fig.~\ref{fig:reBB_model_hx_Rd-2nd-order}.
The same error band is shown for the $R_{qd}(s)$ parameter in yellow band in Fig.~\ref{fig:reBB_model_hx_Rqd-2nd-order}. These error bands are actually overlapping and the region of their overlap determines the domain of validity of the $H(x)$ scaling.

\begin{figure}[htb]
	\centering
    \includegraphics[width=0.8\linewidth]{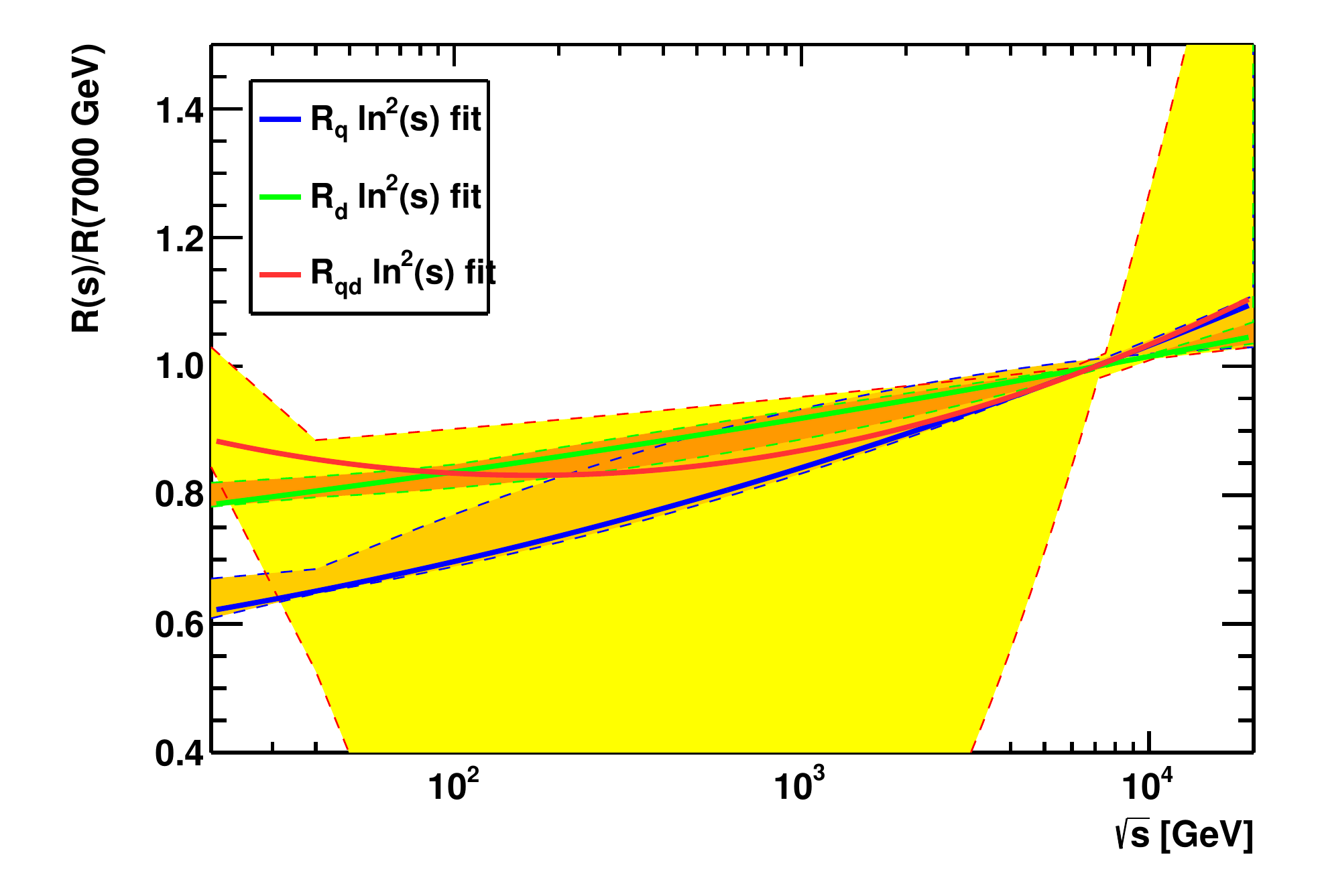}
	\caption{Overlaying the systematic error bands on the  $R_q(s)$, $R_d(s)$ and $R_{qd}(s)$ parameter as obtained by the previous three figures indicates that from $\sqrt{s} = 7$ TeV down to $200$ GeV these bands overlap within one standard deviation. This indicates that taking into account the quadratic  dependence on $\ln(s)$ and the correlations between the model parameters allows for an estimation of the lower limit of $\sqrt{s} = 200 $ GeV for the domain of the validity of the $H(x)$ scaling, which turns out to be experimentally testable in elastic $pp$ collisions at the RHIC acccelerator.}
	\label{fig:reBB_model_hx_s-limit-2nd-order}
\end{figure}

The overlaid one $\sigma$ systematic error-bands of the scale parameters of the ReBB model are shown in 
Fig.~\ref{fig:reBB_model_hx_s-limit-2nd-order}. This figure indicates that from $\sqrt{s} = 7$ TeV down to $\sqrt{s} = 200$ GeV these one standard deviation error-bands overlap within one standard deviation. This implies, that one of the necessary conditions for the validity of the $H(x)$ scaling in elastic $pp$ collisions is satisfied in the kinematic range of $0.2 \le \sqrt{s} \le 7.0$ TeV. The domain of validity of the ReBB model was limited in $-t$ as well, to $0.375 \le -t \le 1.2$ GeV$^2$ as detailed in Ref.~\cite{Csorgo:2020wmw}. Thus, this model-dependent study cannot be applied at very low or very high values of $-t$ and the additional condition for the domain of validity of the $H(x)$ scaling, the constancy of the parameter $\alpha^{pp}(s)$ can be cross-checked if experimental data on elastic $pp$ collisions are becoming available in the lower end of this energy range.

Let us first cross-check if indeed the $H(x)$ scaling can be valid in elastic $pp$ or $p\bar p $ collisions in such a broad energy range, or not. We have demonstrated that this $H(x)$ scaling is violated if we go with energy up to $\sqrt{s} = 13$ TeV. This is easily understood within the framework of the ReBB model. Condition {\it ii)} indicates that one of the necessary condition for the $H(x)$ scaling is the constancy, the approximately energy independence of the parameter $\alpha^{pp}(s)$. In Ref.~\cite{Csorgo:2020wmw} we have shown that within the ReBB model this corresponds to the energy independence of the real to imaginary ratio at $t=0$, the parameter $\rho_0(s)$. The TOTEM Collaboration recently demonstrated that at the top LHC energy of $\sqrt{s} = 13$ TeV, the $\rho_0$ parameter starts to decrease significantly~\cite{Antchev:2017yns}. This decrease increases the dip at these energies corresponding to Theorem 2 of Ref.~\cite{Csorgo:2020wmw} and thus the decrease of $\rho_0(s)$ leads also to a violation of the $H(x)$ scaling at the top LHC energies.

At the lower energies, the $H(x)$ scaling of elastic $pp$ collisions imposes a condition also on the differential cross-sections of elastic $p\bar p$ collisions. Although the value of $\alpha^{p\bar p}(s)$ is not constrained, the scale parameters in $pp$ and in $p\bar p$ elastic collisions were found to follow the same trends. Hence, if $H(x)$ scaling is valid down to $\sqrt{s} = 200$ GeV, then the scale parameters of elastic $p\bar p$
collisions at $\sqrt{s} = 0.546$ and $1.96$ TeV have also to follow the common energy dependencies as specified by Eqs.~(\ref{eq:Hx-Rq},\ref{eq:Hx-Rd},\ref{eq:Hx-Rqd}). So the validity of the $H(x)$ scaling in elastic $pp$ collisions constrains the possible shape of elastic $p\bar p$ collisions as well within the framework of the ReBB model and these constraints can be tested both theoretically and experimentally.

Let us present the tests of the upper limit of the domain of validity in $x$ of the $H(x)$ scaling on experimental data first.

\begin{figure}[htb]
	\centering 
    \includegraphics[width=0.8\linewidth]{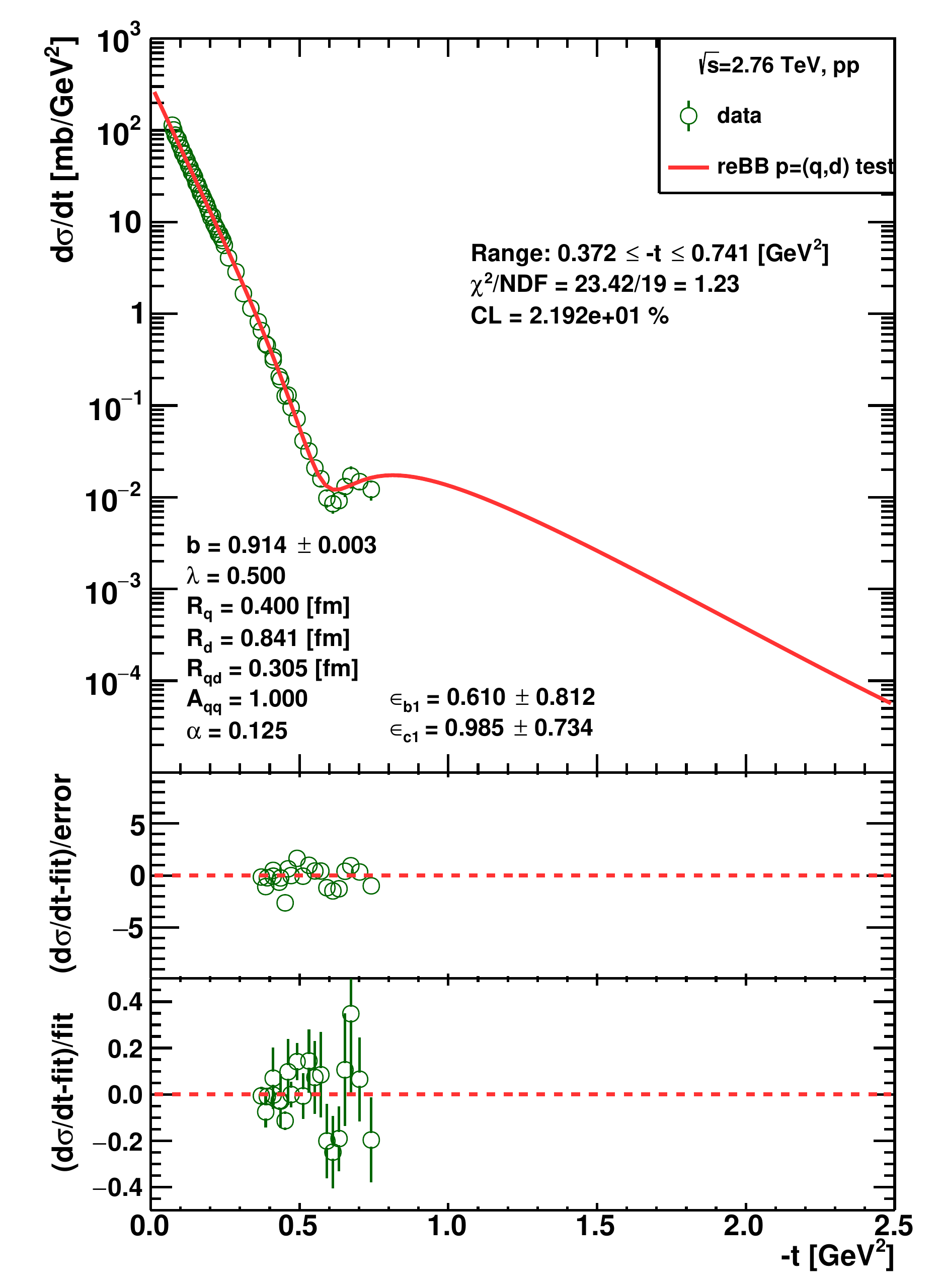}
	\caption{
	Validation plot of the $H(x)$ scaling at $\sqrt{s} = 2.76$ TeV within the framework of the ReBB model. As indicated on this plot such a constraint is satisfied with a CL = 21.9 \% $\gg$ 0.1 \% for the $pp$ data, in the $-t$ range of the measurement, $0.372 \le -t $
	due to the limited validity of the ReBB model at low values of $|t|$. At large values of $-t$, the dominant limitation is the experimental acceptance -- it is less than the limitation of the ReBB model.
	The good quality of the agreement between the TOTEM data and the $H(x)$ scaling limit of the ReBB model of Ref.~\cite{Nemes:2015iia} is also indicated on  the lower panels of this plot.}
	\label{fig:reBB_model_hx_s-cross-check-at-2.76-TeV}
\end{figure}

The test of the validity of the $H(x)$ scaling, using elastic $pp$ data at $\sqrt{s} = 2.76$ TeV LHC energy is shown in
Fig.~\ref{fig:reBB_model_hx_s-cross-check-at-2.76-TeV}. The agreement at 2.76 TeV is excellent and needs no comments or explanations.
The upper limit of the validity of the $H(x)$ scaling, $x_{max}$ at this $\sqrt{s} = 2.76$ TeV can not be determined from this plot, as apparently $x_{max}(2.76) \gg 12.7$ that is this upper limit is clearly larger, than the upper end of the  acceptance in $x$ of the TOTEM experiment at this energy.

\begin{figure}[htb]
	\centering
    \includegraphics[width=0.8\linewidth]{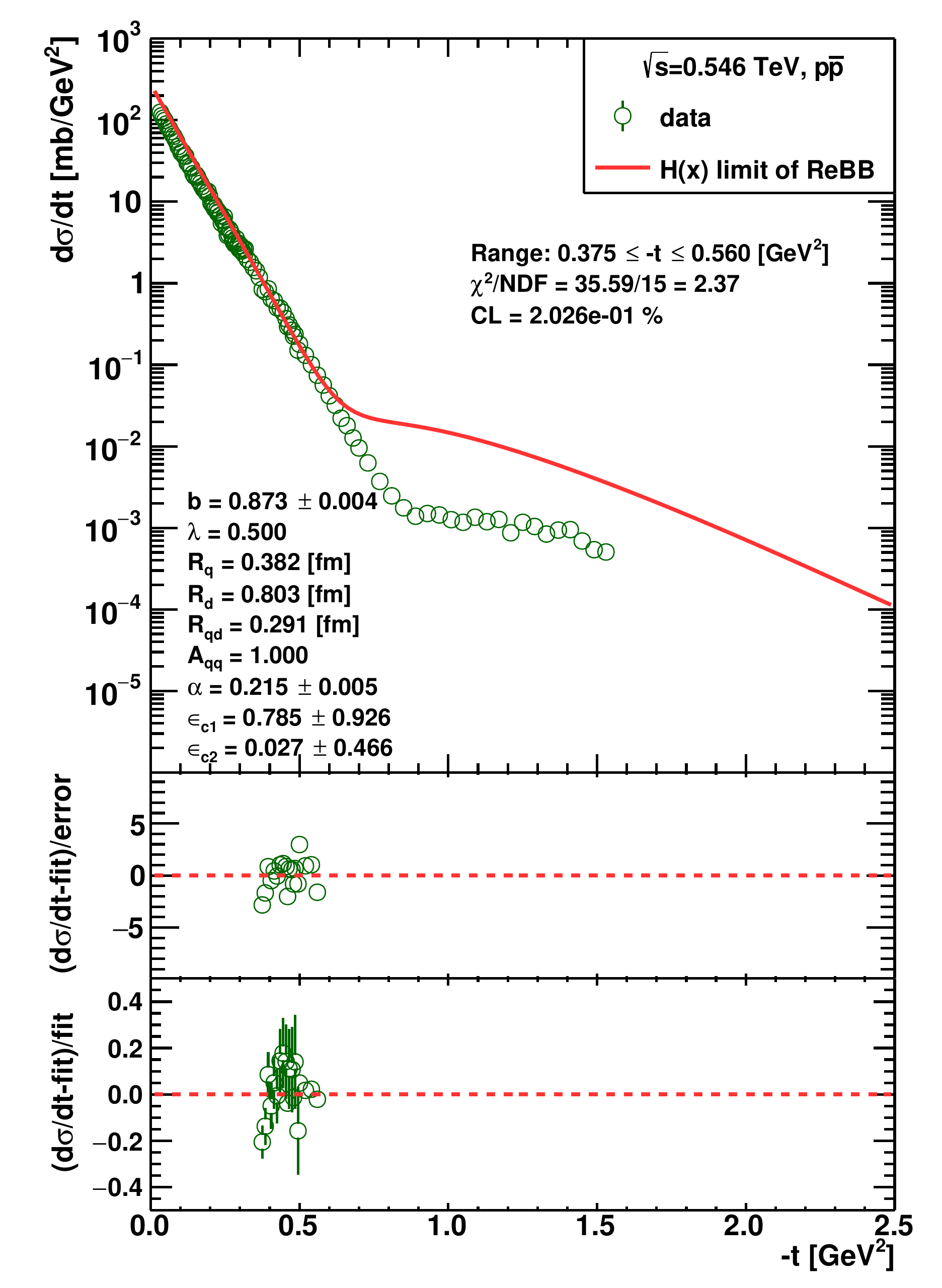}
	\caption{
	Validation plot of the $H(x)$ scaling at $\sqrt{s} = 0.546$ TeV within the framework of the ReBB model.
	As indicated on this plot this constraint is satisfied with a CL $\ge$ 0.1 \% for the measured  $p\bar p$ data, but only in a rather  limited $-t$ range of $0.375 \leq -t \leq 0.56$ GeV$^2$, which is less than the domain  of validity of the ReBB model.}
	\label{fig:reBB_model_hx_s-cross-check-at-0.546-TeV}
\end{figure}

At $\sqrt{s} = 0.546$ TeV, $x_{max}(s)$, the upper limit of 
the domain of the validity of the $H(x)$ scaling is investigated in Fig.~\ref{fig:reBB_model_hx_s-cross-check-at-0.546-TeV}.
For $p\bar p$ collisions, $\alpha^{p\bar p}$ remains the only free fit parameter,
except the overall normalization parameters.  In this case, the $p\bar p$ differential cross-sections are constrained,
because of the requirement $R_q(s)/R_q(s_0) =  R_d(s)/R_d(s_0)  = R_{qd}(s)/R_{qd}(s_0) = b(s)$ is a prescribed function of $s$
and the parameters at $\sqrt{s_0} = 7$ TeV are already determined at $\sqrt{s} = $ 7 TeV. 
As indicated in Fig.~\ref{fig:reBB_model_hx_s-cross-check-at-0.546-TeV}, these constraints are satisfied with a CL = 0.2 \% $>$ 0.1 \% for the measured  $p\bar p$ data, but only in a rather narrow kinematic region of  $0.375 \leq -t \leq 0.56$ GeV$^2$.

\begin{figure*}[htb]
	\centering 
    \includegraphics[width=0.8\linewidth]{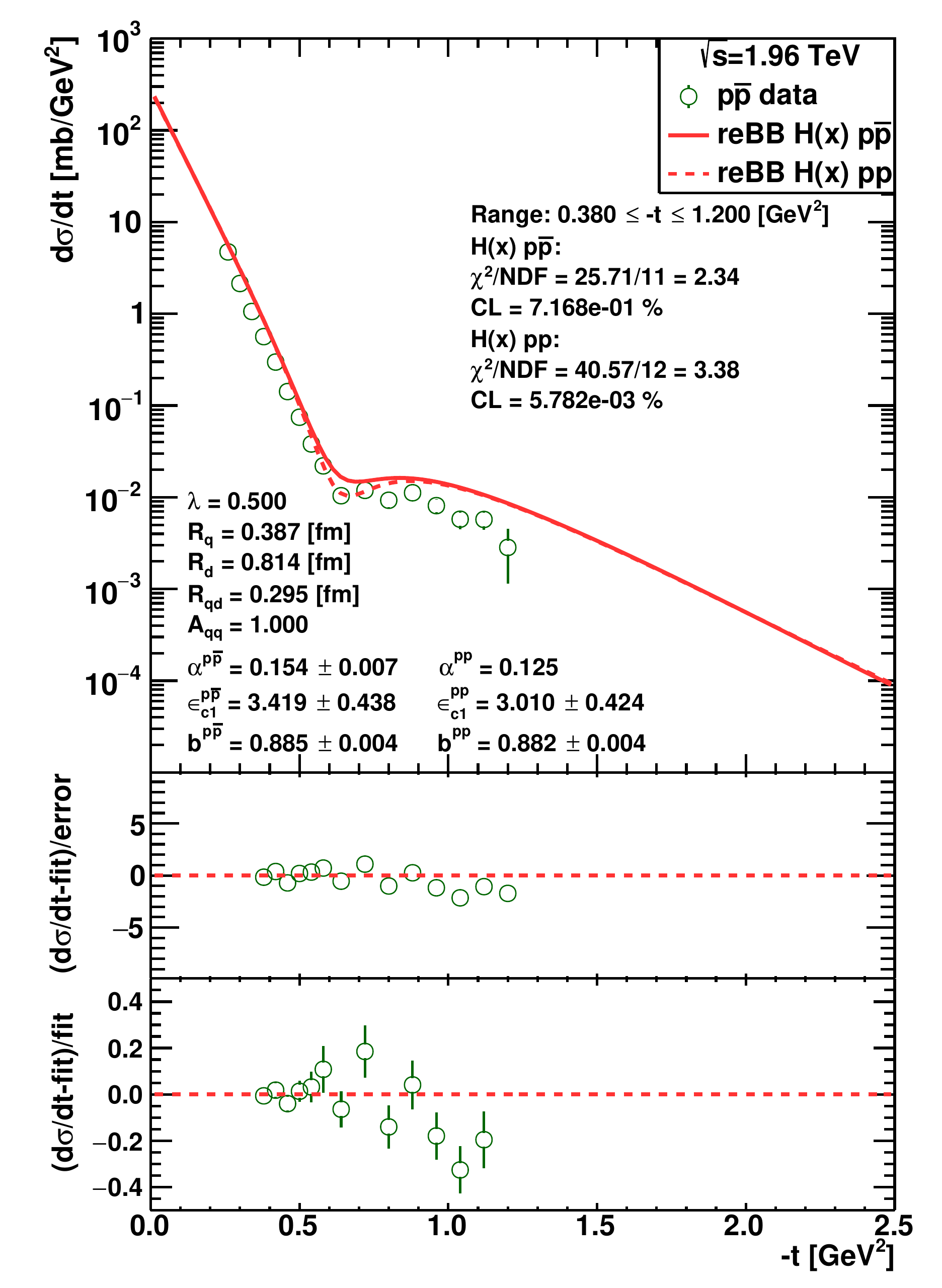}
	\caption{
	Validation plot of the $H(x)$ scaling at $\sqrt{s} = 1.96$ TeV within the framework of the ReBB model. As indicated on this plot this constraint is satisfied with a CL $\ge$ 0.1 \% for the measured  $p\bar p$ data, in the limited $-t$ range of validity of the ReBB model.
	Note that the rescaling (type C) coefficients for this comparison are rather large, outside their usual range of $(-1,1)$. This is why the
	data points are visibly below the best fitted curve, as in the upper panel of  this plot, the datapoints were not rescaled/shifted by the rescaling coefficients of $\epsilon_C$.
	However, the good quality agreement between the rescaled data and the solid red line is indicated on the two lower panels, by the low values of the pull plots on the (data-fit)/error and on the (data-fit)/fit distributions. The agreement is statistically acceptable, as reflected also by the value of CL = 0.7 \%.  
	The corresponding solid red line has only two free physical fit parameters, indicated by $\alpha(p\bar p)$ and $b(s)$, that are shown with their errors. A type C, overall normalization parameter is also allowed in this fit, that shifts all the datapoints up or down.  The other three physical parameters are constrained by the $H(x)$ scaling: $R_i(s) = b(s) R_i(s_0)$, with $i = (q,d,qd) $ and $\sqrt{s_0} =  7 $ TeV.
	The difference between the solid red line and the dashed red line corresponds to the Odderon signal at this energy as for the $pp$ case,
	this  parameter is also constrained, $\alpha^{pp}(s) = \alpha^{pp}(s_0) = 0.125$ is a constant of $s$.
		This plot thus indicates, that the $H(x)$ scaling at
	$\sqrt{s} = 1.96$ TeV may extend up to the full D0 acceptance with $x_{max}(s) = 20.2  $ at $\sqrt{s} = 1.96$ TeV.
	}
	\label{fig:App-D-20.2-reBB_model_hx_s-cross-check-at-1.96-TeV}
\end{figure*}

An important  test of the validity of the $H(x)$ scaling  $p\bar p$ collisions at the Tevatron energy of $\sqrt{s} = 1.96$ TeV is indicated in Fig.~\ref{fig:App-D-20.2-reBB_model_hx_s-cross-check-at-1.96-TeV}. In this plot, the model parameters $R_q(s)$, $R_d(s)$ and $R_{qd}(s)$ are constrained
by the $H(x)$ scaling. Solid line indicates that one parameter, $\alpha^{p\bar p}(s)$ can be fitted to describe the D0 differential cross-section in the whole acceptance of the D0 experiment in  $-t$. According to this plot, at the D0 energy of $\sqrt{s} = 1.96$, 
the domain of the $H(x)$ scaling extends to the   $-t \leq 1.2$ GeV$^2$ domain, which corresponds to $x_{max}(s) = 20.2 $ at $\sqrt{s} = 1.96$ TeV. 
In $pp$ collisions, the other condition of validity of the $H(x)$ scaling is that $\alpha^{pp}$ is independent of the energy of the collision in the $\sqrt{s} = 1.96 - 7.0$ TeV range, however, for $p\bar p$ collisions, $\alpha^{p\bar p}$ is a free fit parameter. For $p\bar p$ collisions,
the $H(x)$ scaling limit of the ReBB model of Ref.~\cite{Csorgo:2020wmw} describes the D0 data in a statistically acceptable manner, with a CL = 0.7 \% , corresponding to an agreement with a $\chi^2/NDF = 25.7/11$, and no significant deviation, an agreement  at the 2.69 $\sigma$ level.
	
As the diffractive minimum in $pp$ is deeper than in $p\bar p$ if the $H(x)$ scaling is valid, the plot also indicates that a signal of Odderon exchange is also present in the ReBB model if extrapolated with the $H(x)$ scaling for $pp$ collisions to $\sqrt{s} = 1.96 $ TeV. However, some significances are  lost, due to two reasons: {\it i)} the ReBB model-dependent extrapolations are limited to the $-t \geq 0.375$ GeV$^2$ region, while the model-independent comparisons can be utilized in the whole $-t$ region; and {\it ii)} the comparison is done on the level of the differential cross-sections so the overall correlated, type $C$ errors do not cancel. In this case, for the Odderon signal we find a $\chi^2/NDF = 40.57/12$, corresponding to a statistical significance of $4.02$ $\sigma$. 
	
\begin{figure*}[htb]
	\centering 
    \includegraphics[width=0.8\linewidth]{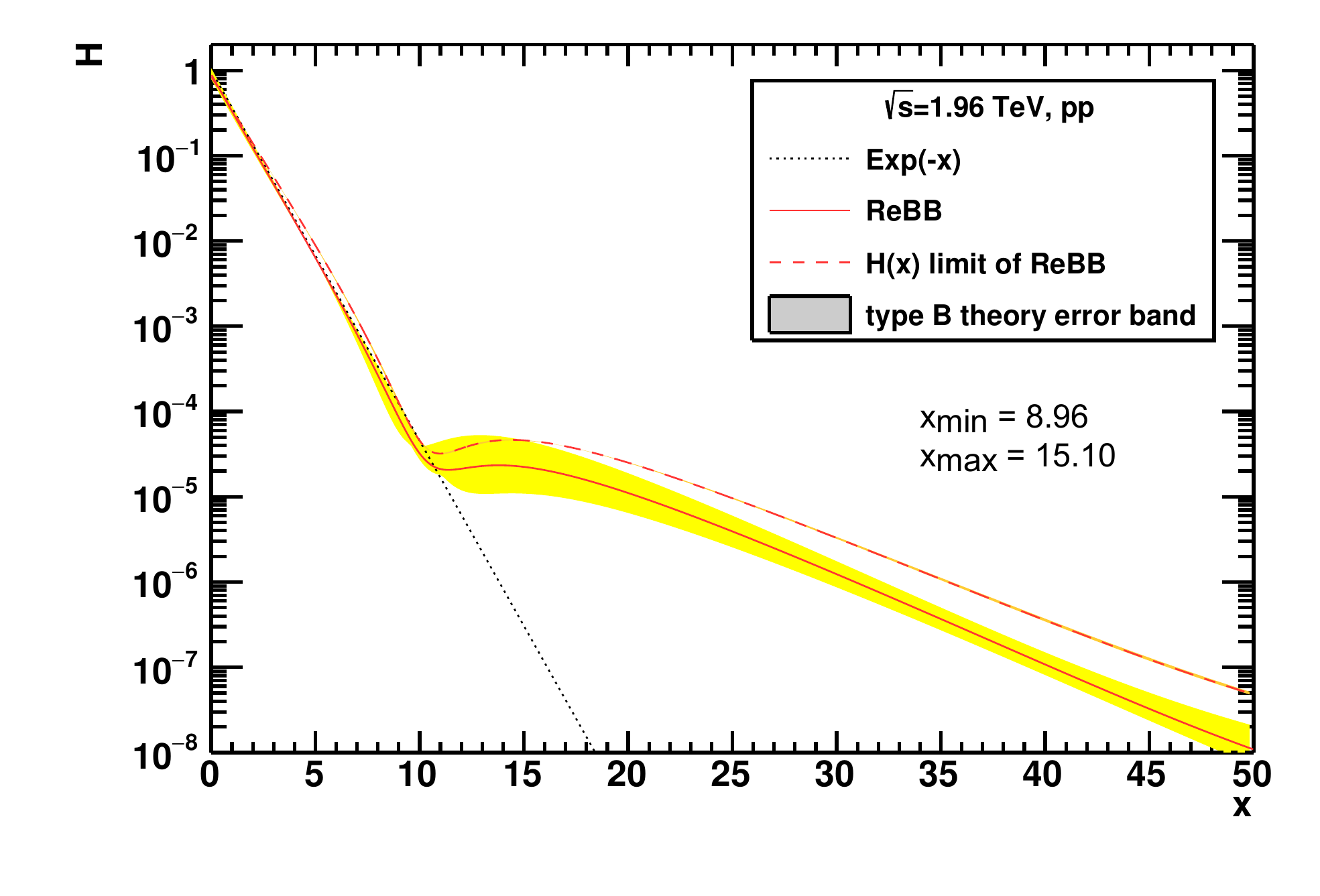}
	\caption{
	A theoretical determination  of the domain of validity of the $H(x)$ scaling at $\sqrt{s} = 1.96$ TeV within the framework of the ReBB model. As indicated on this plot, the differential cross-sections of elastic $pp$ collisions can be evaluated from the full model
	including the 1$\sigma$ theoretical uncertainties of these calculations.
	The corresponding yellow band can be compared with a similar calculation when the $H(x)$ scaling as a constraint  is also 
	implemented within the same model. This calculation does not include the possible compensation of the deviations by an overall vertical rescaling factor, so it gives a conservative estimate for the domain of validity of the $H(x)$ scaling as $x_{max} (s) = 15.1$ at $\sqrt{s}
	= 1.96$ TeV.
	}
	\label{fig:Appendix-D-direct-test-1.96TeV}
\end{figure*}

In addition to these experimental tests, that are obtained from fitting measured $p\bar p$ data with the help of one free parameter, $\alpha^{p\bar p}(s)$ and using the results for the $s$-dependence of the other 3 physical parameters of the ReBB model, $R_q(s)$, $R_d(s)$ and 
$R_{qd}(s)$, we can have a theoretical test as well. This corresponds to the evaluation of the differential cross-section of elastic $pp$ collisions from both the $H(x)$ scaling limit of the ReBB model of refs.~\cite{Nemes:2015iia,Csorgo:2020wmw} and from the fully fledged version of the same model, that includes also terms that result in scaling violations, and makes the full $H(x,s)$ functions
weakly $s$-dependent. Evaluating the error bands from the uncertainty of the model parameters, we can evaluate $x_{max}(s)$ at a given $s$ by determining the domain in $x$, where the two calculations agree within 1 standard deviations of the model parameters. Such a calculation is performed in Fig.~\ref{fig:Appendix-D-direct-test-1.96TeV}. This calculation does not allow for a compensation of the modification of the shape by a possible overall normalization factor and so it results in a more stringent upper limit for the domain of the validity of the $H(x)$ scaling,
$x_{max}(s) = 15.1$ at $\sqrt{s} = 1.96$ TeV.

Recently, the STAR collaboration measured the differential cross-section of elastic $pp$ collisions at the center-of-mass energy of $200 $ GeV~\cite{Adam:2020ozo}. This measurement has resulted in a straight exponential differential cross-section in the range of $ 0.045 \leq -t \leq 0.135$ GeV$^2$. This range is the range where $H(x) = \exp(-x)$ and the conditions of the validity of the $H(x)$ scaling are indeed satisfied by this dataset, that is, however, limited to a rather low $-t$ range. It is thus a very interesting and most important experimental cross-check for the validity of the $H(x)$ scaling to push forward the experimental data analysis of elastic $pp $ collisions at the top RHIC energy of $\sqrt{s} = 510$ GeV including if possible a larger $-t$ range extending to the non-exponential domain of $\frac{d\sigma}{dt}$ as well.

\clearpage
\section{Study of the stability of the Odderon signal for the variation of x-range}
\label{app:E}

In this Appendix, we summarize our $x = -tB$ range stability results. The central topic of this
Appendix is the clarification of the role of the correlation between the correlation coefficient $\epsilon_{b,21}$ and the domain of $x$, over which the $\chi^2$ or the statistical significance $\sigma $ is optimized. This correlation coefficient  shifts all the projected datapoints together, up or down, and its best value for all the datapoints is $\epsilon_{b,21} = -0.56$ if the $H(x)$ scaling function of the densest $pp$ dataset of $\sqrt{s} = 7$ TeV is projected to the $\sqrt{s} = 1.96$ TeV $p\bar p$ data. If we fixed this value, and started to limit the $x$-range of comparison by removing 1, 2, 3, 4, 5 and 6 D0 datapoints with the largest values of $x$, we obtain the  results summarized in Table~\ref{fig:App-E-1.png}.

\begin{table*}[htb]
	\centering
    \includegraphics[width=0.99\linewidth]{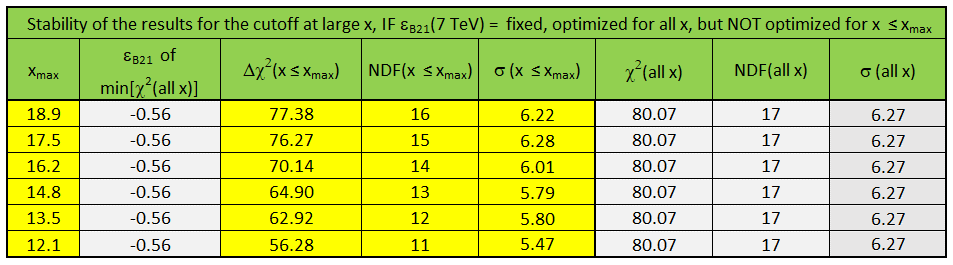}
	\caption{
Stability of the Odderon signal on the variation $x_{max}$, the upper limit of the domain of validity of the $H(x)$ scaling using $\sqrt{s} = 7$ TeV TOTEM pp data projected to $\sqrt{s} = 1.96$ TeV D0 $p\bar p $ data  in $x = -tB$ . 
If the correlation coefficient kept its value of $\epsilon_{b,21} = -0.56$,  corresponding to the “global” minimum of $\chi^2$, the Odderon signal would remain at least 5.4 $\sigma$,
even if the last 1,2, ... 5 and 6 D0 points were discarded (by hand) in the $x$-range stability analysis.
	}
	\label{fig:App-E-1.png}
\end{table*}

In this case of a constant, $x$-range independent $\epsilon_{b,21} = -0.56$ , the  $\chi^2$ for all the points and the 6.26 $\sigma$ overall significance are both constants, but the partial contributions to the  $\chi^2$ and NDF are $x$-range dependent. Note, that even for the $x \leq 10.8$ region (left from the dip) we obtain a significant contribution, with a statistical significance of 5.46 $\sigma$. This is seen visually in Fig.~\ref{fig:rescaling-from-7-to-1.96TeV} as well as on our final $H(x)$ comparison plots,
Fig.~\ref{fig:rescaling-from-7-to-1.96TeV-and-back-17mb-lin-exp-new} in ~\ref{app:A}.

However, if we  applied a more conservative approach, and re-optimized the correlation coefficent to obtain an $x$-range dependent $\epsilon_{b,21} $, that minimizes the  partial contributions to the $\chi^2$ in the investigated $x$-range, we obtain the  results summarized in Table~\ref{fig:App-E-2.png}.

\begin{table*}[htb]
	\centering
    \includegraphics[width=0.99\linewidth]{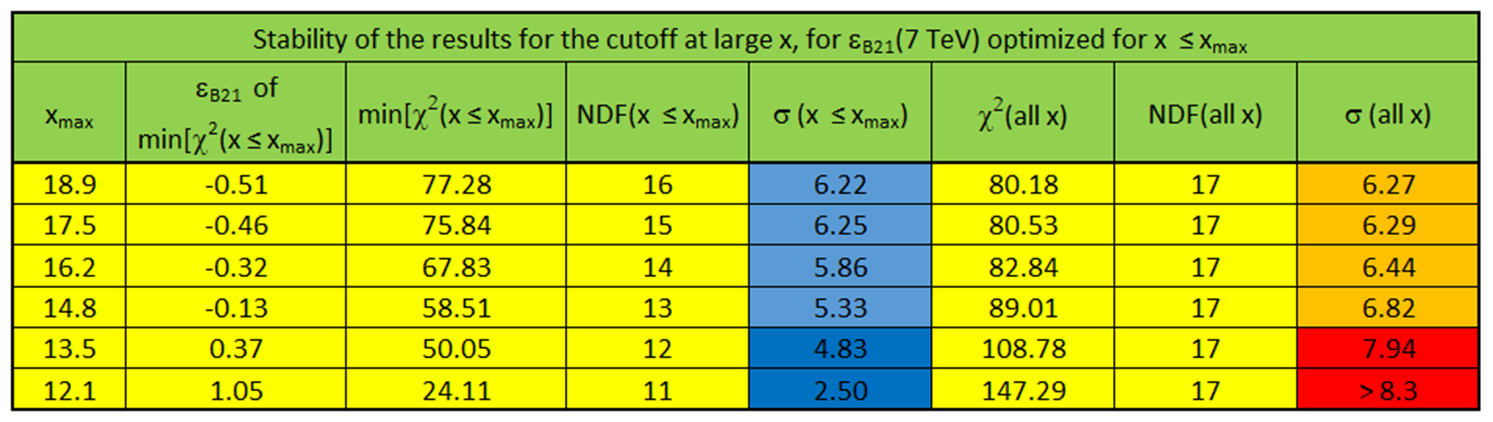}
	\caption{
Stability of the Odderon signal on the variation $x_{max}$, the upper limit of the domain of validity of the $H(x)$ scaling using $\sqrt{s} = 7$ TeV TOTEM pp data projected to $\sqrt{s} = 1.96$ TeV D0 $p\bar p $ data  in $x = -tB$, for a fit range dependent, minimized 
 correlation coefficient $\epsilon_{b,21}$,  corresponding to the minimum of $\chi^2$ in the considered $x$-range. The Odderon signal  remains at least 5.3 $\sigma$,
 if the last 1,2, 3 and 4 D0 points were discarded (by hand) in this $x$-range stability analysis. In this sense, the Odderon signal remains significant, if the
 $H(x)$ scaling is valid at last up to x = 14.8. The last column indicates that these locally optimized $\epsilon_{b,21}$ coefficients globally increase the Odderon significance.
	}
	\label{fig:App-E-2.png}
\end{table*}

As shown in Table~\ref{fig:App-E-2.png}, the correlation coefficient $\varepsilon_{B21}$ turns out to be strongly $x$-range dependent. The importance of the contribution from large values of $x$ can be formulated as follows: if we optimize the correlation coefficient $\varepsilon_{B21}$ 
for the range of investigation, the Odderon signal is stable for the removal of  the top 1, 2, 3 and 4 D0 datapoints with $x \geq  14.8$, as even with the $x$-range dependent minimization of the $\chi^2$, the statistical significance of the Odderon observation remains at least 5.33 $\sigma$ . 
However, if we remove 5 or more of the D0 datapoints at large values of $x = - tB$, then the remaining D0 $p\bar p$ data and the $H(x)$ scaling function of $pp$ at $\sqrt{s} = 7$ TeV can be renormalized to the top of each other. 
TOTEM data at $\sqrt{s} = 2.76 $ TeV extend only to the $x \leq 12.1$ region, where even a 2.5 $\sigma$ level, statistically not significant apparent agreement can be reached by totally distorting the value of the correlation coefficient 
$\varepsilon_{B21}$, even if using only the more dense and more precise TOTEM  7 TeV pp data. 
Orange colored fields indicate that if we apply a local optimalization to the correlation coefficient $\varepsilon_{B21}$, this increases the overall significance of the Odderon, if all the datapoints are used.
As indicated by the last column of Table  ~\ref{fig:App-E-2.png}, these values of the correlation coefficient $\varepsilon_{B21}$  are gradually becoming rather unreasonable, when all the datapoints are considered. As more and more D0 data points are removed at high $x$, the global statistical significance  increases drastically, even above the model-dependent limit of 7.08 $\sigma$,
as indicated by the red-colored fields in the rightmost column of the above table. 
This 7.08 $\sigma$ limit corresponds to the combined significance of 
$pp$ vs $p\bar p$ comparisons at both $\sqrt{s} =  1.96$ and $2.76$ TeV,  as summarized in Table~\ref{table:7-to-1.96-TeV-two-way-comparison} below,
and detailed in Ref~\cite{Csorgo:2020wmw}.

If we fully utilize the results of the model dependent analysis of Ref.~\cite{Csorgo:2020wmw}, we find  a larger, combined significance of 7.08 
$\sigma$  as shown in Table~\ref{fig:App-E-3.png}. This result is obtained by comparing simultaneously  the $pp$ and $p\bar p$ differential cross-sections at $\sqrt{s} = 1.96$ and $2.76$ TeV with the help of the ReBB model. The increased significance is due to the fact that with the help of the same model, the $p\bar p$ data can also be extrapolated to $2.76$ TeV and result in an overwhelming Odderon signal. Let us stress, that this model dependent significance is obtained  even without using the 7 TeV TOTEM dataset,  to evaluate the $\chi^2$ and the  Odderon significance and to bridge the energy gap  between 2.76 and 1.96 TeV.
If the $p\bar p$ differential cross-section is evaluated and extrapolated also up to 7 TeV with the help of the same ReBB model, the probability of Odderon observation becomes very much larger than a 7.08 $\sigma$ effect~\cite{Csorgo:2020wmw}, practically it becomes a certainty.

\begin{table*}[htb]
	\centering
    \includegraphics[width=0.5\linewidth]{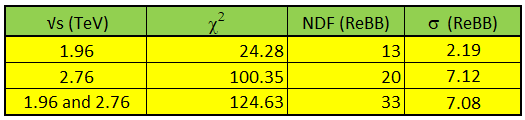}
	\caption{
	The trade-off effect of using  model dependent results: 
	If we utilize the ReBB model instead of $H(x)$ scaling, it decreases the significance of the $pp$ prediction vs $p\bar p$ data at $\sqrt{s} = 1.96$ TeV, from 6.26 down to 2.19 $\sigma$. However, as a trade-off, the same model allows for an extrapolation of the $p\bar p$ data to $\sqrt{s} = 2.76$ TeV,
	which was not possible with the help of the $H(x)$ scaling. This $p\bar p$ differential cross-section vs $pp$ data at $\sqrt{s} = 2.76$ TeV results in a 7.12 $\sigma$ effect. The combined significance of the ReBB model on both the  1.96 TeV D0 and 2.76 TeV TOTEM data is found to be 7.08 $\sigma$ . Thus the lower limit of significance from the ReBB model to data comparision, 7.08 is larger than the model independent estimate of 6.26 $\sigma$, that was obtained using the full $x$-range of D0 and the assumption of the validity of the $H(x)$ scaling in this range.
	}
	\label{fig:App-E-3.png}
\end{table*}

We have made another cross-check and divided the final result of our calculations to four different regions: We have  2 D0 datapoint in Region 0, the diffractive cone with $0 < x = -tB \leq  5.1$ . The remaining  15 D0 datapoints can be divided into 3 regions with 5-5 D0 data points as follows. Region I corresponds to the ``swing” region, just to the left of the dip, corresponding to $5.1 < x \leq  8.4$; Region II, including  the dip and the bump, to 
$8.4 < x \leq 13.5$; and Region III, the tail corresponds to $13.5 < x \leq 20.2$.
We evaluated their partial contributions to our final Odderon significance of 6.26 $\sigma$. For a cross-check we have also evaluated their combined significance and also the contribution of the first two D0 datapoint from the diffractive cone.

The results for a fixed $\varepsilon_{B21} = -0.56$
- optimized on all the 17 D0 datapoints - are shown in Table~\ref{fig:App-E-4.png}.
In this case, the greatest partial contribution to the Odderon significance comes from the swing region, 
$5.1 < x \leq  8.4$   the second most important contribution comes from the diffractive 
interference (dip and bump) region with $8.4 < x \leq 13.5$, and for this value of the correlation coefficient,
the tail with  $13.5 < x \leq 20.2$ has a relatively small contribution.

In contrast, similar results for a regionally optimized correlation coefficient, an $x$-range dependent
$\varepsilon_{B21} $ is shown on Table  ~\ref{fig:App-E-5.png}.
In this case, the greatest partial contribution to the Odderon significance comes from
the diffractive  interference (dip and bump) region with $8.4 < x \leq 13.5$, 
the second most important contribution comes from the swing region, 
$5.1 < x \leq  8.4$, while   the relatively least important contribution comes from 
the tail with  $13.5 < x \leq 20.2$. It is important to recognise, that the 10 D0 datapoints in the 
swing and diffractive interference region already 
provide a statistically significant, more than 5 $\sigma$ Odderon effect.
The interference and the tail together also indicate an Odderon effect, with a significance that is between
a 3 and a 5 $\sigma$ effect.
When all these three regions are combined together, they dominate the final Odderon significance, 
providing $6.23$ out of the $6.26$ $\sigma$ Odderon effect for all x.

\begin{figure*}[htb]
	\centering 
    \includegraphics[width=0.9\linewidth]{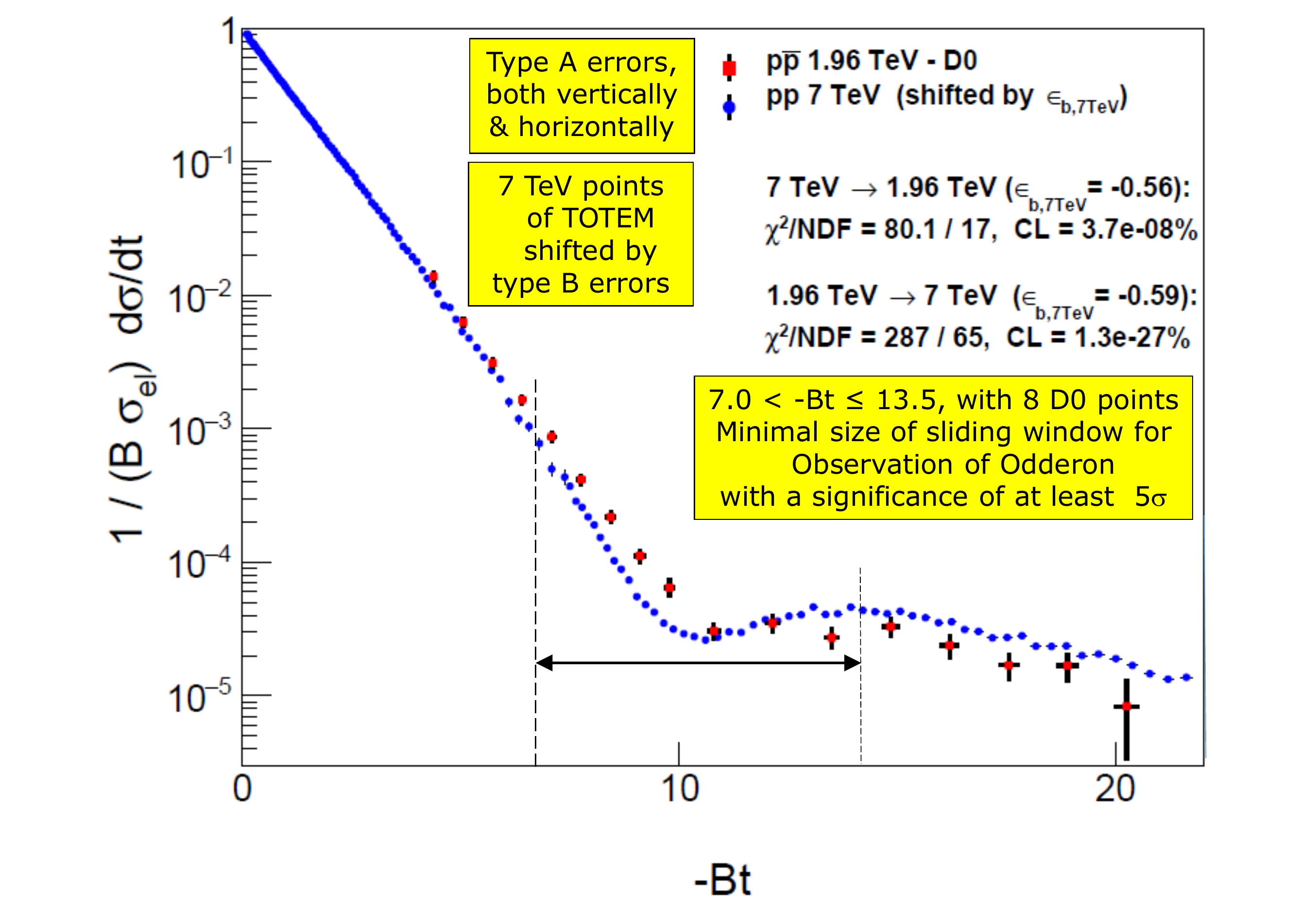}
	\caption{
	Model independent determination of the dominant source of the Odderon signal, using the minimal sized sliding window technique.
	We varied the number of subsequent D0 datapoints and also their locations in the D0 acceptance in all possible manner. 
	This way we determined the location of the smallest number of subsequent D0 datapoints, that provide  a statistically significant
	Odderon signal, that  remains above the 5$\sigma$ discovery threshold, 
	even after locally minimizing $\chi^2$
	and obtaining an interval dependent,  local value for $\epsilon_{B, 7 TeV}$ for that particular interval. The above plot indicates that the minimum number of subsequent D0 datapoints is 8, and that they come from the $7.0$ $< $ $x = -Bt $ $\leq$ $13.5$ kinematic domain. The result is consistent with Table ~	\ref{fig:App-E-6.png}.
	}
	\label{fig:Appendix-E-minimal-sized-sliding-window}
\end{figure*}

It is an intriquing problem if one can determine model independently the region, from where the dominant contribution
to the Odderon signal is coming. To reach that goal, we have developed a so called sliding window technique, and determined
the minumum size of this sliding window that still provides an Odderon signal on the discovery level of at least 5.0 $\sigma$.
Namely, D0 published 17 datapoints. We have taken the first $n$ of these datapoints, with $n$ varied from 2 to 17,
and then locally optimized the correlation coefficients $\epsilon_B$ for both projections of $7 $ $\rightarrow $ $1.96$ TeV and $1.96$ $\rightarrow$ $7$ TeV, and determined where the minimum sized sliding window is, wherein the observation of the Odderon is 
an at least 5 $\sigma$ effect. As one picture is worth ten thousand words according to a Chinese word of wisdom, we have summarized our results in
Fig.~\ref{fig:Appendix-E-minimal-sized-sliding-window}. For the sake of clarity, and only on this plot, we have shifted the TOTEM $7$ TeV datapoints by their type $B$ errors (properly multiplied by the correlation coefficient $\epsilon_{B,7 TeV}$ that minimalized $\chi^2$ for that given sliding acceptance window) and show only the type A vertical and horizontal errors.  Fig.~\ref{fig:Appendix-E-minimal-sized-sliding-window}
indicates that there are 8 D0 datapoints in the minimal sized sliding acceptance window, where an at least 5 $\sigma$, statistically 
significant Odderon signal is observed. Thus
9 out of the 17 datapoints can be removed (5 from the tail and 4 points from the diffractive cone region), 
without destroying the greater than 5 $\sigma$  level of the Odderon significance. 

\begin{table*}[htb]
	\centering
    \includegraphics[width=0.7\linewidth]{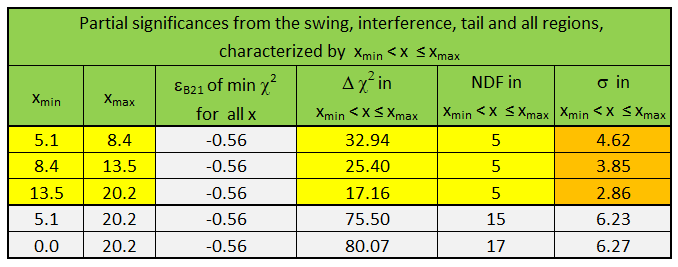}
	\caption{
Stability of the Odderon signal in various  regions of $x_{min}$ and $x_{max}$,
for a constant value of the correlation coefficient $\epsilon_{b,21}$, minimized on all the 17 available D0 data points.
In this case, the greatest partial contribution to the Odderon significance comes from the swing region, 
$5.1 < x \leq  8.4$   the second most important contribution comes from the diffractive 
interference (dip and bump) region with $8.4 < x \leq 13.5$, and for this value of the correlation coefficient,
the tail with  $13.5 < x \leq 20.2$ has a relatively small contribution.
	}
	\label{fig:App-E-4.png}
\end{table*}

\begin{table*}[htb]
	\centering
    \includegraphics[width=0.7\linewidth]{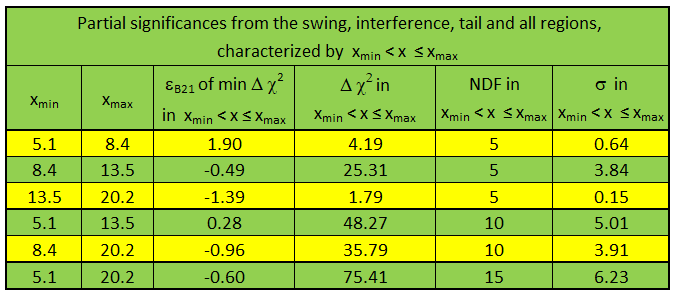}
	\caption{
	Stability of the Odderon signal in various  regions of $x_{min}$ and $x_{max}$,
in the case,  when the value of the correlation coefficient $\epsilon_{b,21}$ is locally minimized 
for the data in the $x_{min} < x \leq x_{max}$ range.
In this case, the greatest partial contribution to the Odderon significance comes from
the diffractive  interference (dip and bump) region with $8.4 < x \leq 13.5$, 
the second most important contribution comes from the swing region, 
$5.1 < x \leq  8.4$, while   the relatively least important contribution comes from 
the tail with  $13.5 < x \leq 20.2$. It is important to recognise, that the 10 D0 datapoints in the 
swing and diffractive interference region already 
provide a statistically significant, more than 5 $\sigma$ Odderon effect.
The interference and the tail, taken together,  also indicate an Odderon signal, as a 3.91 $\sigma$, indicative but statistically not yet sufficient effect.
When all these three regions are combined together, they dominate the final Odderon significance, 
providing $6.23$ out of the $6.26$ $\sigma$ Odderon effect for all x.
	}
	\label{fig:App-E-5.png}
\end{table*}

If we take into account the evolution of $H(x,s)$ as a function of $s$, this evolution becomes model dependent, but allows for the estimation of the domain of validity of the $H(x,s) = H(x,s_0)$ scaling law, as detailed in 
\ref{app:D}.

We obtained our model-dependent results within the framework a Glauber-type calculation using the ReBB model
of refs.~\cite{Csorgo:2020wmw,Nemes:2015iia}. This model is validated at 
$\sqrt{s} = 1.96 $ TeV in the  
$x_{min} = 4.4 < x$  region~\cite{Csorgo:2020wmw}. 
According to the calculations of \ref{app:D}, the $H(x)$ scaling at this 1.96 TeV energy is expected to be valid at least in the $9.0 < x = - tB \leq 15.1$ kinematic domain. 
As we cannot estimate reliably the validity of the lower limit of the $H(x)$ scaling at low values of x with this model and in addition, in the diffractive cone, $H(x) \approx \exp(-x)$ and the scaling is expected to hold, the 
$9.0 < x = - tB \leq 15.1$ kinematic domain  seems to give the worst possible, model dependent limit
for the domain of validity of the $H(x)$ scaling at 1.96 TeV. As shown on Table ~\ref{fig:App-E-6.png},
this interval corresponds to a significance of at least 3.82 $\sigma$.
In this very limited x range,  the corresponding best correlation coefficient, - 0.62, is rather close to the best correlation coefficient, -0.56, that minimizes significance for  the case of 
the complete, $0 < x = - tB \leq 20.2$ kinematic domain of all the D0 data, 
thus for this correlation coefficient, the significance for the $ 0 < x = - tB \leq 20.2$
kinematic domain  is nearly unchanged from 6.27 to 6.28 $\sigma$.
If we assume that the lower limit $x_{min}$ corresponds to the lower limit of the validation of the ReBB model
~\cite{Csorgo:2020wmw}, then we find that in the $4.4 < x \leq 15.1 $ kinematic domain the significance of the Odderon signal is at least 5.3 $\sigma$, as detailed on Table ~\ref{fig:App-E-6.png}. However, we know that in the low $x = -t B$ region,
there is a diffractive cone, where $H(x) \approx \exp(-x)$ for scattering amplitudes that are analytic at $t=0$, and for experimental data that are not indicating a non-exponential behaviour at low values of $|t|$. This is the case for the $\sqrt{s} = 1.96$ TeV  D0
$p\bar p$ and for the $\sqrt{s} = 2.76$ TeV TOTEM $pp$ data, so for closing the energy gap between 1.96 and 2.76 TeV, the lower limit of the applicability of the $H(x)$ scaling is actually $x_{min} = 0$.

If we fully utilize the results of this model dependent analysis 
we find that the model dependent, combined Odderon significance on $pp$ prediction versus $p\bar p$
data at $\sqrt{s} = 1.96$ TeV data and $p\bar p$ prediction versus $pp$
data at $\sqrt{s} = 2.76$ TeV is 7.08 $\sigma$, as shown on Table ~\ref{fig:App-E-3.png}.
This increased significance is due to the fact that with the help of the same ReBB model~\cite{Csorgo:2020wmw},
the $p\bar p$ data can also be extrapolated to the lowest TOTEM energy of $\sqrt{s} = 2.76$ TeV and they
result in a dominant Odderon signal, as detailed also in Table~\ref{fig:App-E-3.png}.

In this Appendix, we thus find a hierarchy of the Odderon significances. When we take theoretical modelling into account, we find that the significance of an Odderon observation on elastic $pp$ collisions at $\sqrt{s} = 2.76$
TeV versus elastic $p\bar p$ collisions at $\sqrt{s} = 1.96$ TeV, in the corresponding TOTEM and D0 acceptances,
is at least 7.08 $\sigma$. If we do not utilize fully the model dependent information, but use only the theoretical limits for the validity of the $H(x)$ scaling at $\sqrt{s} = 1.96$ TeV, we find that the Odderon signal is greater than 5 $\sigma$ if the $H(x)$ scaling is valid in the range of $5.1 < x \leq 13.1$ . We cannot reliably estimate the lower limit of the $H(x)$ scaling, but demonstrated on available data that in the diffraction cone, $H(x) \approx \exp(-x)$ is satisfied. We have validated the model of Ref.~\cite{Csorgo:2020wmw} in the range of $4.4 < x$ at 1.96 TeV,
and find that this model dependent domain of validity of the $H(x) $ scaling extends up to $x_{max} = 15.1$ .

The theoretically and model dependently  validated range of $H(x)$ scaling
at $\sqrt{s}  = 1.96$ TeV, $4.4 < x \leq 15.1$ includes the interval $7.0 < x \leq 13.5$,
where we find that the Odderon signal is greater than a 5 $\sigma$ effect. 

We conclude this $x$-range stability investigations as follows:

\begin{itemize}
    \item Model independently, we find that the significance of the Odderon is greater than $5$ $\sigma$
    in the $7 < x \leq 13.5$ domain at $\sqrt{s} = 1.96$ TeV.
    \item In our model independent analysis, we relied only on already published D0 and TOTEM datapoints, without relying on preliminary data. We have evaluated the $\chi^2$ of
    the Odderon signal using arithmetic operations only (addition, substraction, multiplication, division) but we did not impose any model dependent fits.
    \item In our model independent analysis, we utilized a newly introduced $H(x)$ scaling function, that is not sensitive to the dominant, and
    overall correlated normalization errors of the differential cross-sections. As a trade-off, the domain of validity of this $H(x)$ scaling became an energy dependent $(0, x_{max}(s)) $ interval. 
    \item This $H(x)$ scaling function scales out the trivial energy dependencies that appear due to the energy dependence of the 
    elastic cross-section $\sigma_{el}(s)$ and the nuclear slope parameter $B(s)$.
    \item Using a model, validated in the $4.4 < x$ domain, we find that the validity in $x$ of the $H(x)$ scaling at 
    $\sqrt{s} = 1.96$ TeV is extending up to $x \leq 15.1$. Using published D0 data, and the same model, we validated 
    in Fig.~\ref{fig:App-D-20.2-reBB_model_hx_s-cross-check-at-1.96-TeV} that the $H(x)$ scaling at $\sqrt{s} = 1.96$ is extending even up
    to the end of D0 acceptance, to $x \leq 20.2$ . 
    \item In the very limited interval $7.0 < x = -B t \leq 13.5$,
    we find that the Odderon signal is greater than a 5 $\sigma$ effect at $\sqrt{s} =  1.96$ TeV.
    \item Thus the model independent and at least 5 $\sigma$, discovery level Odderon signal
    is remarkably stable for the variations of the domain of validity of the $H(x)$ scaling at $\sqrt{s} = 1.96$ TeV.
\end{itemize}

\begin{table*}[htb]
	\centering
    \includegraphics[width=0.7\linewidth]{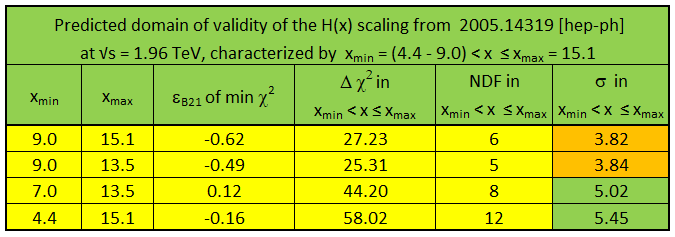}
	\caption{
In the $9.0 < x = - tB \leq 15.1 $ kinematic domain we obtain a lower, model dependent limit of significance of 
3.82 $\sigma$. Note, that the ReBB model is not validated in the $-t \le 0.372$ kinematic range~\cite{Csorgo:2020wmw},
corresponding to low values of $x$. At $\sqrt{s} = 1.96$ TeV, 
this  lower limit of validity of the ReBB model  is $x_{min} = 4.4$.  
In the corresponding  $4.4 < x  \leq 15.1$ domain, the lowest limit of Odderon significance is 
an 5.37 $\sigma$ effect. Within this domain, in the range of $7.0 < x  \leq 13.5$, the Odderon signal is still above
the 5 $\sigma$, discovery level, even when 9 out of the 17 D0 datapoints are removed from the analysis.
	}
	\label{fig:App-E-6.png}
\end{table*}

\section*{Acknowledgments}

We acknowledge inspiring  discussions with C. Avila, S. Giani, P. Grannis,  W. Guryn, G. Gustaf\-son, L. Jenkovszky, V. A. Khoze, E. Levin, L. L\"onnblad,  B. Nicolescu, K. \"Osterberg,  C. Royon, M. Strikman and M. \v{S}umbera. In particular, we thank our Referees for their valuable comments, suggestions and clarifications and for their suggestion to create an easier-to-read summary of these results, that we have provided in Ref.~\cite{Csorgo:2020rlb}.
R.P. is partially supported by the Swedish Research Council grants No. 621-2013-4287 and 2016-05996, by the European Research Council (ERC) under the European Union's Horizon 2020 research and innovation programme (agreement No. 668679), as well as by the NKFI grant K133046 (Hungary). 
T. Cs, T. N, I. Sz. and A. S. were partially supported by the NKFIH grants No. FK-123842, FK-123959 and K133046 as well as by the EFOP 3.6.1-16-2016-00001 grant (Hungary). 
Our collaboration was supported  by the framework of COST Action CA15213 
THOR: ``Theory of hot matter and relativistic heavy-ion collisions".

\clearpage
\bibliographystyle{utcaps}       

\bibliography{references}

\vfill
\end{document}